\documentclass[author-year, jmp, amsmath, amssymb, longbibliography, reprint]{revtex4-1}

\usepackage{graphicx}
\usepackage[caption=false]{subfig}
\usepackage{dcolumn}
\usepackage{bm}
\usepackage{xfrac}
\usepackage[toc,page]{appendix}
\usepackage{siunitx}
\usepackage{tikz}
\usepackage{pgfplots}
\usepackage{pgfplotstable}
\usepackage{adjustbox}
\usepackage{etoolbox}
\usepackage[breaklinks]{hyperref}
\usepackage{mdframed}
\usepackage{color}
\usepackage{array}
\usepackage{float}
\usepackage{bigints}
\usepackage{float}
\usepackage{subfloat}
\usepackage{tabularx}
\usepackage{booktabs}
\usepackage{upgreek}
\usepackage{xparse}
\usepackage{xstring}
\usepackage{ifthen}

\usepgfplotslibrary{external} 
\definecolor{linkcolor}{rgb}{0.0,0.3,0.5}
\definecolor{venetianred}{rgb}{0.78, 0.03, 0.08}
\hypersetup{colorlinks=true, linkcolor=linkcolor, citecolor=linkcolor, filecolor=linkcolor, urlcolor=linkcolor}

\AtBeginDocument{
\heavyrulewidth=.08em
\lightrulewidth=.05em
\cmidrulewidth=.03em
\belowrulesep=.65ex
\belowbottomsep=0pt
\aboverulesep=.4ex
\abovetopsep=0pt
\cmidrulesep=\doublerulesep
\cmidrulekern=.5em
\defaultaddspace=.5em
}

\definecolor{color1}{RGB}{202,0,32}
\definecolor{color2}{RGB}{244,165,130}
\definecolor{color3}{RGB}{146,197,222}
\definecolor{color4}{RGB}{5,113,176}

\newcommand\solidrulefordashed[1][1cm]{\rule[0.5ex]{#1}{0.35pt}}
\newcommand\solidrulefordotted[1][1cm]{\rule[0.5ex]{#1}{0.55pt}}

\newcommand\dottedrule{\mbox{\solidrulefordotted[0.15mm]\hspace{0.35mm}\solidrulefordotted[0.15mm]\hspace{0.35mm}\solidrulefordotted[0.15mm]\hspace{0.35mm}\solidrulefordotted[0.15mm]\hspace{0.35mm}\solidrulefordotted[0.15mm]\hspace{0.35mm}\solidrulefordotted[0.15mm]\hspace{0.35mm}\solidrulefordotted[0.15mm]\hspace{0.35mm}\solidrulefordotted[0.15mm]\hspace{0.35mm}\solidrulefordotted[0.15mm]}}
  
\newcommand\denselydashedrule{\mbox{\solidrulefordashed[1mm]\hspace{0.75mm}\solidrulefordashed[1mm]\hspace{0.75mm}\solidrulefordashed[1mm]}}

\newcommand\solidrule[1][1cm]{\rule[0.5ex]{#1}{0.35pt}}

\newcommand\looselydashedrule{\mbox{\solidrulefordashed[1mm]\hspace{1mm}\solidrulefordashed[1mm]}}

\newcolumntype{P}[1]{>{\centering\arraybackslash}p{#1}}

\makeatletter
    \renewcommand\@make@capt@title[2]{%
     \@ifx@empty\float@link{\@firstofone}{\expandafter\href\expandafter{\float@link}}%
      {#1}\@caption@fignum@sep#2\quad}%
\makeatother

\renewcommand{\thetable}{\arabic{table}}
   
\makeatletter 
	\renewcommand{\fnum@figure}{Fig.~\thefigure}
\makeatother

\makeatletter 
	\renewcommand{\fnum@table}{Table~\thetable}
\makeatother

\newcommand*\dd{\mathop{}\!\mathrm{d}}

\newcommand*\spi{\sqrt{\pi}}
\newcommand*\Sgn{\mathop{}\!\mathrm{sgn}}
\newcommand{\np}[1]{^{\!\!\!#1\,}}
\newcommand{\bigo}[1]{\mathcal{O}\!\left(#1\right)}
\newcommand{\bom}{\bm{\omega}}
\newcommand{\bphi}{\bm{\phi}}

\newcommand{\FrC}[1]{C\!\left(#1\right)}
\newcommand{\delpsi}{\delta\psi}
\newcommand{\delchi}{\delta\chi}

\newcommand{\vph}{\vphantom{\Bigg[\Bigg]}}
\newcommand{\vphbig}{\vphantom{\Bigg[\Bigg]\np{2}}}
\newcommand{\vphbigg}{\vphantom{\left[\Bigg[\Bigg]\np{2}\right]\np{2}}}
\newcommand{\vphbiggg}{\vphantom{\left[\left[\Bigg[\Bigg]\np{2}\right]\np{2}\right]\np{2}}}
\newcommand{\vphbigggg}{\vphantom{\left[\left[\left[\Bigg[\Bigg]\np{2}\right]\np{2}\right]\np{2}\right]\np{2}}}

\ExplSyntaxOn
\DeclareExpandableDocumentCommand{\RepQuad}{m}
{\int_compare:nT { #1 > 0 }
	{\quad \prg_replicate:nn { #1 - 1 } {\quad}}}
\ExplSyntaxOff

\newcommand{\sqrtbig}[1]{\sqrt{\vphantom{|}#1}}

\newcommand{\SlopeFrstOrdSM}{\pm \frac{1}{2}\,\lambda}
\newcommand{\SlopeScndOrdSM}{\pm\!\left[\frac{1}{2} \, \lambda - \frac{\sqrt{2}}{4} \, \lambda^{2}\right]}

\newcommand{\SlopeFrstOrdR}{\pm \frac{1}{2} \, \frac{\lambda_{r}}{\sqrt{r}}}
\newcommand{\SlopeScndOrdR}{\pm \sigma_{2}^{(r)} (\lambda_{r})}

\newcommand{\SlopeFrstOrdRS}{\pm \frac{1}{2}\!\left(\!\frac{\lambda_{r}}{\sqrt{r}} + \frac{\lambda_{s}}{\sqrt{s}}\!\right)\vphantom{\left[\frac{\lambda_{r}}{\sqrt{r}}\right]\np{2}}\!\!}
\newcommand{\SlopeFrstOrdL}{\pm \frac{1}{2}\!\left(\!\frac{\lambda_{l}}{\sqrt{l}} + \frac{\lambda_{\bar{l}}}{\sqrt{\bar{l}\phantom{\,}}}\!\right)\vphantom{\left[\frac{\lambda_{r}}{\sqrt{r}}\right]\np{2}}\!\!}
\newcommand{\SlopeFrstOrdRSSymb}{\pm \sigma_{1}^{(r,\,s)}}
\newcommand{\SlopeScndOrdRSSymb}{\pm \sigma_{2}^{(r,\,s)}}

\newcommand{\SlopeFrstOrdSMP}{\pm \frac{1}{2} \big(\!\cos(\delpsi) + \sin(\delpsi)\big)\lambda}
\newcommand{\SlopeFrstOrdSMPSymb}{\pm \varsigma_{1}}
\newcommand{\SlopeScndOrdSMPSymb}{\pm \varsigma_{2}}
\newcommand{\SlopeFrstOrdMMP}{\pm \bm{\varsigma}_{1}}
\newcommand{\SlopeScndOrdMMP}{\pm \bm{\varsigma}_{2}}
\newcommand{\SlopeThrdOrdMMP}{\pm \bm{\varsigma}_{3}}

\newcommand{\ConstMultiMode}[1]{\frac{1}{2}\sum_{#1} \frac{\lambda_{#1}}{\sqrtbig{#1}}}
\newcommand{\FirstOrderMultiMode}{\pm \bm{\sigma}_{1}}
\newcommand{\SecondOrderMultiMode}{\pm \bm{\sigma}_{2}}

\DeclareDocumentCommand{\LinearTerm}{O{} O{}}{\StrLen{#1}[\arglen]\ifthenelse{\equal{\arglen}{0}}{\left(\vphantom{\left[#2\right]\np{2}}\!\frac{x}{\spi} #2\!\right)}{\left(\vphantom{\left[\frac{x}{\spi}#2\right]\np{2}}\!\!\ifthenelse{\equal{\arglen}{1}}{\sqrt{#1}}{\sqrtbig{#1}} \!\left[\frac{x}{\spi}#2\right]\!\right)}}

\DeclareDocumentCommand{\FresnelCTerm}{O{} O{}}{\left(\!\ifthenelse{\equal{#1}{}}{\vphantom{\left[\left[\frac{x}{\spi}#2\right]\np{2}\right]}}{\vphantom{\left[\left[\frac{x}{\spi}#1#2\right]\np{2}\right]\np{2}}}C\!\LinearTerm[#1][#2]\!\mp \frac{1}{2}\!\right)}
\DeclareDocumentCommand{\FresnelSTerm}{O{} O{}}{\left(\!\ifthenelse{\equal{#1}{}}{\vphantom{\left[\left[\frac{x}{\spi}#2\right]\np{2}\right]}}{\vphantom{\left[\left[\frac{x}{\spi}#1#2\right]\np{2}\right]\np{2}}}S\!\LinearTerm[#1][#2]\!\mp \frac{1}{2}\!\right)}

\DeclareDocumentCommand{\SinTerm}{O{} O{} O{} O{}}{\sin\!\left[\frac{\pi}{2}#4\!\LinearTerm[#1][#2]\np{2}\!#3\right]}
\DeclareDocumentCommand{\CosTerm}{O{} O{} O{} O{}}{\cos\!\left[\frac{\pi}{2}#4\!\LinearTerm[#1][#2]\np{2}\!#3\right]}

\newcounter{matrix}
\renewcommand*{\thematrix}{F\arabic{matrix}}

\makeatletter
\def\@equationname{equation}
\newenvironment{m}[1]{%
    \def\mymathenvironmenttouse{#1}%
    \ifx\mymathenvironmenttouse\@equationname%
        \refstepcounter{matrix}%
    \else
        \patchcmd{\@arrayparboxrestore}{equation}{matrix}{}{}
        \patchcmd{\print@eqnum}{equation}{matrix}{}{}%
        \patchcmd{\incr@eqnum}{equation}{matrix}{}{}%
    \fi
    \csname\mymathenvironmenttouse\endcsname%
}{%
    \ifx\mymathenvironmenttouse\@equationname%
        \tag{\thematrix}%
    \fi
    \csname end\mymathenvironmenttouse\endcsname%
}
\makeatother

\newcounter{phase}
\renewcommand*{\thephase}{PH\arabic{phase}}

\makeatletter
\def\@equationname{equation}
\newenvironment{ph}[1]{%
    \def\mymathenvironmenttouse{#1}%
    \ifx\mymathenvironmenttouse\@equationname%
        \refstepcounter{phase}%
    \else
        \patchcmd{\@arrayparboxrestore}{equation}{phase}{}{}
        \patchcmd{\print@eqnum}{equation}{phase}{}{}%
        \patchcmd{\incr@eqnum}{equation}{phase}{}{}%
    \fi
    \csname\mymathenvironmenttouse\endcsname%
}{%
    \ifx\mymathenvironmenttouse\@equationname%
        \tag{\thephase}%
    \fi
    \csname end\mymathenvironmenttouse\endcsname%
}
\makeatother

\newcommand*{\Cdot}[1][1.25]{%
  \mathpalette{\CdotAux{#1}}\cdot%
}
\newdimen\CdotAxis
\newcommand*{\CdotAux}[3]{%
  {%
    \settoheight\CdotAxis{$#2\vcenter{}$}%
    \sbox0{%
      \raisebox\CdotAxis{%
        \scalebox{#1}{%
          \raisebox{-\CdotAxis}{%
            $\mathsurround=0pt #2#3$%
          }%
        }%
      }%
    }%
    \dp0=0pt %
    \sbox2{$#2\bullet$}%
    \ifdim\ht2<\ht0 %
      \ht0=\ht2 %
    \fi
    \sbox2{$\mathsurround=0pt #2#3$}%
    \hbox to \wd2{\hss\usebox{0}\hss}%
  }%
}

\newmdenv[leftline=false,rightline=false]{topbotfig}

\allowdisplaybreaks[4]

\begin{document}

\def\apjl{ApJ}
\def\prd{Phys.~Rev.~D}
\def\araa{Annual~Review~of~Astron.~and~Astrophys.}


\title[]{Transition of EMRIs through resonance: corrections to higher order in the on-resonance flux modification}

\author{Deyan P. Mihaylov}
\email{d.mihaylov@ast.cam.ac.uk}
\affiliation{Institute of Astronomy, University of Cambridge, Madingley Road, Cambridge, CB3 0HA}
\author{Jonathan R. Gair}
\email{j.gair@ed.ac.uk}
\affiliation{Institute of Astronomy, University of Cambridge, Madingley Road, Cambridge, CB3 0HA}
\affiliation{School of Mathematics, University of Edinburgh, Peter Guthrie Tait Road, Edinburgh, EH9 3FD}


\begin{abstract}
Extreme-mass-ratio inspirals (\textsc{emri}s) are candidate events for gravitational wave detection in the millihertz band (by detectors like \textsc{lisa}). These events involve a stellar-mass black hole, or a similar compact object, descending in the gravitational field of a supermassive black hole, eventually merging with it. Properties of the inspiralling trajectory away from resonance are well known and have been studied extensively, however little is known about the behaviour of these binary systems at resonance, when the radial and lateral frequencies of the orbit become commensurate. We describe the two existing models, the instantaneous frequency approach used by Gair, Bender, and Yunes, and the standard two timescales approach implemented by Flanagan and Hinderer. In both cases, the exact treatment depends on the modelling of the gravitational self-force, which is currently not available. We extend the results in Gair, Bender and Yunes to higher order in the on-resonance flux modification, and argue that the instantaneous frequency approach is also a valid treatment of the resonance problem. The non-linear differential equations which arise in treating resonances are interesting from a mathematical view point. We present our algorithm for perturbative solutions and the results to third order in the infinitesimal parameter, and discuss the scope of this approach.
\end{abstract}

\maketitle

\section{Introduction}
\label{sec:intro}
\noindent
Gravitational waves are one of the most fascinating predictions of general relativity. The theoretical underpinning for them was first presented by \cite{einstein1918}. In 1974, Taylor and Hulse discovered indirect evidence for their existence when they observed the first pulsar in a binary system, in which the object separation was gradually decreasing, and were the first to suggest that gravitational wave effects are at play \citep{hulsetaylor, taylorweis1982}. The pair were awarded the Nobel Prize in Physics in 1993 for their discovery -- until recently the only evidence for the existence of gravitational waves. Nearly a century after Einstein's paper, gravitational waves were directly observed for the first time in September 2015 by the two advanced \textsc{ligo} detectors, thus culminating a decades-long efforts to detect gravitational radiation \citep{ligo2016detect}. Currently, several experiments, including Advanced \textsc{ligo} and Advanced Virgo, which are searching for waves in the range \SIrange{E2}{E4}{\hertz} \citep{LIGO2015, virgo2015}, are being deployed to search for further signals. The first joint run is scheduled to take place in late 2016 or early 2017 \citep{abbott2016} and it is expected that a significant number of events will be observed in the near future \citep{rate2010, ligovirgo2016}. The European Space Agency has selected gravitational wave detection from space as the science theme to be addressed by the L3 large satellite mission, currently scheduled for launch in 2034. This space-based detector, called Laser Interferometer Space Antenna (\textsc{lisa}), will operate in the millihertz frequency range. The mission design is still being formulated, but it will be finalised by the end of this decade \citep{eLISA2012}. The need for multiple approaches to detection (currently there are more than a dozen projects, both ground- and space-based, see \cite{moore2015} for an overview) is driven by the broad spectrum of gravitational wave frequencies (the typical frequency of a gravitational wave system is given by \(f \sim (G M / L^{3})^{1/2}\), where \(M\) is the mass of the system, and \(L\) is its length scale), namely in the range \SIrange{E-16}{E4}{\hertz}, produced by a variety of potential gravitational radiation sources \citep{thorne1995}.

Systems which are expected to produce gravitational waves are those that are accelerating and exhibit asymmetry in their configuration: for instance a binary \citep{abbott2014}, a supernova \citep{gossan2015}, or a spinning star with a pronounced bulge. Among the range of binary celestial systems, we focus on the case of a compact object (of several solar masses, usually a stellar-mass black hole, or a neutron star) gradually spiralling towards a supermassive black hole (typically between \(10^{5}\) and \(10^{12}\) solar masses, although the ones of interest to us are between \(10^{5}\) and \(10^{7}\) \(M_{\odot}\), since these generate gravitational waves at frequencies in the range of sensitivity of \textsc{lisa} \citep{eLISA2012, babak2014}). Since the mass \(m\) of the compact object is in the range of \(10 - 10^{2} M_{\odot}\), the ratio of the masses is typically \(\epsilon = m / M \sim 10^{-6} \text{ to } 10^{-4}\). Termed extreme-mass-ratio inspirals (\textsc{emri}s), these binaries are expected to radiate gravitational waves with frequency in the range \SIrange{E-4}{E-2}{\hertz}. This fact places them in a position to be likely candidates for detection by the \textsc{lisa} mission. \textsc{emri}s reside near the centres of galaxies (where supermassive black holes are situated \citep{kormendyrichstone1995}) and clusters of stars and stellar remnants are available for capture by the black hole. Plenty of scientific effort has been invested into modelling the evolution of \textsc{emri}s \citep{barakcutler2004, gairglamp2006, drasco2006, chua2015}, computing their waveforms \citep{babakal2006}, predicting the rates and timescales of inspiral \citep{glampedakiskennefick2002, hopman2009}, and estimating the parameters of the orbit \citep{barakcutler2004, arun2009}. Applications of gravitational wave observations of \textsc{emri}s to inference on astrophysical populations \citep{gair2010}, cosmology \citep{macleodhogan2008} and fundamental physics~\citep{TestGRLivRev} have also been explored in detail.

The mass of the compact object in an \textsc{emri} is typically at least 4 orders of magnitude smaller than the mass of the supermassive black hole. This allows us to treat the presence of the compact body in the background space-time of the central object as a perturbation. According to the uniqueness theorem for black holes, they can be uniquely characterised by three parameters - mass, spin and charge. Since electromagnetic repulsion is much stronger than gravitational attraction (about 40 orders of magnitude), we can assume that astrophysical black holes have negligible net charge. Hence, we consider the Kerr metric, where black holes are described only by their mass and spin. The Kerr space-time is well-known, with a complete set of integrals of motion, and the geodesics can be characterised by three constants (\citet{carter1968} and Appendix \ref{sec:GeodesicConstants}). For a point mass, the trajectory would be that of a geodesic around the black hole, but while the point-mass notion is convenient mathematically, it does not apply in the context of the current problem. Therefore, we need to address the finite size and mass of the compact object. The deviation from geodesic motion can be quantified by allowing for the gravitational self-force on the compact object, \citep{nakamura1982, wardell2012}, also known as the gravitational radiation reaction force, since it is responsible for the gravitational radiation emission from the object. There is a radiative component of the self-force that leads to evolution of the orbital frequencies, as well as a conservative component that changes the instantaneous values of the orbital frequencies. The trajectory can be modelled as a sequence of geodesics, each one with lower energy than the previous one, and therefore (on average) closer to the central object. This method is called the osculating element approach \citep{poundpoisson2008, gairflanagan2011, sopuerta2012}. Individual orbits are well approximated as geodesics, however the defining parameters of these geodesics are continuously changing, and this is what shapes the inspiral orbit. This effect will be visible over a large number of orbits -- typically over a timescale of \(\bigo{\epsilon^{-1}}\).

As the test body progresses along its in-spiral, at certain points two of the three frequencies of motion will become rational multiples of each other, putting the system into resonance \citep{flanagan2012}. While in general the behaviour of the inspiral is well-studied and modelled \citep{barack2009}, during resonance the system exhibits unusual behaviour, accompanied by changes in the orbital parameters \citep{flanagan2012}. The evolution of \textsc{emri}s during resonances, and their effect on gravitational wave emission is currently an interesting topic in gravitational wave astronomy \citep{berry2016}.

There are two approaches in the literature that treat the behaviour of extreme-mass-ratio inspirals on resonance. On one hand is the instantaneous value treatment by \citep{gair2010}, with manifestations in both frequency and phase formalisms, and on the other is the approach by \citep{flanagan2012}, which separates the timescales associated with the in-spiral into small and large. We will describe both methods here, but focus on extending the former approach to higher order, as we anticipate the predictions of the latter to be consistent with the phase formalism in the former.

\subsection{The Gravitational Self-force}
\label{sec:force}
\noindent
The geodesic principle of General Relativity dictates that free point particles, of zero mass, traverse time-like geodesics of the space-time. Unfortunately, while treating bodies in General Relativity as point masses is mathematically convenient, in reality this approach is far from accurate, and introduces immense difficulties to the problem we are considering. As the size \(R\) of a body is scaled down to 0, its mass \(M\) should decrease accordingly, scaling to 0 at the same time as the size of the body \citep{gralla2008}, otherwise a black hole solution would form before the point-limit of the body is reached. We can account for the finite size and mass of the body if we consider the effect of these on the deviation from geodesic motion. At leading order, we ignore the presence of the space-time generated by the compact object in the background space-time, and in this idealised system, the test body follows a time-like geodesic, whose shape can be accurately computed (see Appendix~\ref{sec:KerrGeodesics} and Figure~\ref{fig:kerr_geodesic} for details). The next order accounts for the non-infinitesimal size and mass of the object, and this causes a deviation from the geodesic motion described above. We quantify this phenomenon by introducing the gravitational self-force, responsible for the inward spiral of the compact object \citep{nakamura1982, flanagan2012}. This force is at the same time responsible for dissipating the energy made available by the inspiral motion in the form of gravitational radiation.

A proper treatment of the gravitational self-force is not the objective of this paper. The reader may wish to consult \cite{poisson2011}, as well as the more concise review \cite{poisson2009}. Further recommended reading is \cite{nakamura1982} and \cite{wald2009}, or the more mathematical approach described in \cite{gralla2008}.

In the context of this article, it is worth knowing that since the self-force is the perturbation in the mass parameter to the geodesic, its leading order dependence on the mass of the test body is linear. Further, the self-force will depend on the shape of the orbit, which for the Kerr space-time is specified by three constants of the motion, for example the energy \(\mathcal{E}\), the 3\textsuperscript{rd} component of the angular momentum \(\mathcal{L}_{z}\), and Carter's constant, \(\mathcal{Q}\). Furthermore, the individual components of this self-force will depend on the position of the test body along the orbit, specified, for example, by the action-angle coordinates \(q_{\alpha}\), where the index \(\alpha \in \{t, r, \theta, \phi\}\) (this implies there are 4 separate action-angle coordinates, with 4 corresponding components of the self-force. In fact, only 3 of these force components are independent, as can be verified by differentiating the condition \(g_{\alpha\beta} u^{\alpha} u^{\beta} = 1\) with respect to the proper time \(\tau\) to obtain \(u^{\alpha} a_{\alpha} = 0\) \citep{gairflanagan2011}). With this in mind, we can assume that the linear order gravitational self-force can be written in the form \citep*{poisson2009, gair2012}
\begin{align}
a_{\alpha} = \frac{\dd^{2} x_{\alpha}}{\dd \tau^{2}} = \epsilon \, F_{\alpha} (\mathbf{q}, \mathbf{J}), \label{eq:self-force}
\end{align}
where \(\epsilon = m / M\) is the ratio of the compact object mass to the mass of the supermassive black hole, and the vector \(\mathbf{J}\) contains the three constants of motion defining the shape of the orbit. The values of \(\mathbf{q}\) and \(\mathbf{J}\) in general depend on the way the orbit evolved before the moment at which they are computed. Thus, it can be regarded that the instantaneous self-force is in fact given by an integral over the past history of the orbit.

\subsection{Orbital resonances}
\label{sec:orbit}
\noindent
We already reasoned why the gravitational self-force depends on the shape of the orbit, i.e. the values of the three constants of motion in the Kerr metric. These quantities can be mapped to the three fundamental frequencies of orbital motion \(\{\omega_{r}, \omega_{\theta}, \omega_{\phi}\}\) \citep{schmidt2002}. Therefore, instead of eq.~(\ref{eq:self-force}), we can write the expression for the self-force as \(\epsilon \, \mathbf{F} (\mathbf{q}, \bm{\omega})\). The fundamental frequencies are another valid way to label orbits of the Kerr space-time, and can be used in lieu of the geodesic constants of motion. It should be noted that the reverse substitution, from \(\bm{\omega}\) to \(\mathbf{J}\) is not always possible, on the account of the isofrequency pairs exhibited by the proper time frequencies \citep{warburton2013}. For the discussion in this paper, we can assume local invertibility, however it should be noted that this is not universal.

The three fundamental frequencies evolve independently according to their equations of motion. However, at certain points along the trajectory, two of the frequencies will become commensurate, and the system will temporarily exhibit a qualitatively different behaviour, a resonance, with a reduced number of degrees of freedom. Since the Kerr metric (Appendix~\ref{sec:KerrSpacetime} and Figure~\ref{fig:kerr}) does not depend explicitly on the azimuthal angle \(\phi\), the corresponding fundamental frequency \(\omega_{\phi}\) is not relevant for any resonant motion. While the azimuthal frequency is trivial, the radial and the polar frequencies will have independently varying values along the in-spiral. Resonance occurs when these two frequencies become commensurate with each other, or, in more formal terms, when their ratio is a rational number \citep{ruangsri2014}:
\begin{align}
\frac{\omega_{r}}{\omega_{\theta}} = \frac{n_{\theta}}{n_{r}}, \quad \{n_{r}, n_{\theta}\} \subset \mathbb{Z}. \label{eq:resonance_condition}
\end{align}
During resonance, the apparent effect of the self-force on the orbit is modified, in particular there is a fractional change of order \(\bigo{\epsilon^{0}}\) to the dynamics of the in-spiral~-- the evolution rate of parameters and the in-spiral rate \citep{flanagan2012}. A typical \textsc{emri} passes through several resonance points before the two objects merge. Obviously, this complicates the numerical analysis of the in-spiral, since one has to develop a formalism to account for the occurrence of resonances, and to quantify the effects these have on the orbital dynamics \citep{flanagan2012}.

In the current paper, we will first describe the two approaches to resonance modelling that have appeared in the literature, which were suggested by \citep{hinderer2008} and \citep*{gair2012}. We then describe what the differences in the model assumptions are, explain why there is a preferred case of physical relevance to resonances in \textsc{emri}s, and argue that full knowledge of the self-force is required to robustly identify the most relevant model. We will then extend the results of \citep*{gair2012} to obtain complete results describing the transition through resonance in that model, at leading order in the mass-ratio.

\section{Instantaneous frequency approach}
\label{sec:InstantaneousFrequencyApproach}
\noindent
This method is described in detail in \cite*{gair2012}, although here we will give an outline of the approach to aid further discussions. As explained above, individual orbits of the compact object around the black hole can be thought of as tri-periodic geodesics with fundamental frequencies \(\{\omega_{r}, \omega_{\theta}, \omega_{\phi}\}\). This encourages us to employ the ``osculating element'' formalism, where the in-spiral is modelled as a sequence of geodesics, with the constants of motion continuously evolving as the orbit approaches the black hole, and furthermore, the effect of the compact body on the trajectory can be treated as a perturbation \citep{poundpoisson2008}. Since these constants of the motion do not change on a given geodesic, at leading order, and only evolve (continuously) as the object spirals into the supermassive black hole, we can argue that, to leading order, this effect is driven by the self-force on the object. Hence, we can write the evolution equation for \(J_{\mu} = (\mathcal{E}, \mathcal{L}_{z}, \mathcal{Q})\) as follows:
\begin{align}
\frac{\dd J_{\mu}}{\dd \tau} = \epsilon\, \mathcal{F}_{\mu} (\mathbf{q}, \mathbf{J}, \mathbf{F}) + \bigo{\epsilon^{2}}, \label{eq:frequency_diff}
\end{align}
where \(\mathcal{F}_{\mu}\) is a function of the coordinates, the generalised coordinates, and the self-force, which describes the rate of change of orbital parameters as the object moves along the inspiral. We argued that the self-force is itself a function of the coordinates and the fundamental frequencies (see Section \ref{sec:orbit}), hence we can write \(\mathcal{F}_{\mu} = \mathcal{F}_{\mu} (\mathbf{q}, \bm{\omega})\). Similarly, the action-angle coordinates of the system evolve proportionally to the fundamental frequencies, with a correction owing to the self-force:
\begin{align}
\frac{\dd q_{\mu}}{\dd \tau} = \omega_{\mu} + \epsilon \, \mathcal{H}_{\mu} (\mathbf{q}, \bm{\omega}) + \bigo{\epsilon^{2}}.
\label{eq:coordinates_evolution}
\end{align}
We will not concern ourselves with the precise form of the functions \(\mathcal{F}_{\mu}\) and \(\mathcal{H}_{\mu}\), but rather will point out that they can be expanded in tri-harmonic Fourier series as discussed above, which is valid without any loss of generality. \(\omega_{\mu}\) is the instantaneous value of the frequency at this moment along the trajectory. This is calculated according to eq.~(\ref{eq:frequency_diff}), and the value is continuously updated to reflect the motion of the body along its inspiral and the effect of the self-force. Without the self-force, the frequencies will, of course, remain constant, since the trajectory would be that of a geodesic around the black hole, unaffected by gravitational wave emission.

Before carrying out the Fourier mode expansion, we recall that a geodesic in Kerr space-time is uniquely specified by three constants. Instead of the orbital parameters \(\mathcal{E}\), \(\mathcal{L}_{z}\), and \(\mathcal{Q}\), we could use the set of fundamental frequencies -- the radial frequency, \(\omega_{r}\), the polar frequency, \(\omega_{\theta}\), and the azimuthal frequency, \(\omega_{\phi}\), to specify the geodesics, as these are constant along a given geodesic. Since these can be mapped to the aforementioned integrals of the motion \citep{schmidt2002}, we can use these maps to argue that the fundamental frequencies evolve along the trajectory according to a differential equation of the same form as eq.~(\ref{eq:frequency_diff}):
\begin{align}
\frac{\dd \omega_{\mu}}{\dd \tau} = \epsilon \, \mathcal{G}_{\mu} (\mathbf{q}, \bm{\omega}) + \bigo{\epsilon^{2}}. \label{eq:asymptotic_frequency_expansion}
\end{align}

The two orbital frequencies relevant to the occurrence of resonances are the radial and the polar (lateral) frequencies. As noted previously, the third frequency \(\omega_{\phi}\), that of azimuthal motion, does not play a role in the occurrence of resonances due to the axisymmetric nature of the Kerr space-time. We use this property to write a bi-periodic expansion for the function of the self-force in Fourier modes of the two frequencies \(\{\omega_{r}, \omega_{\theta}\}\):
\begin{align}
\begin{split}
\frac{\dd \omega_{\mu}}{\dd \tau} = \epsilon \sum_{l,\,m} &\left(A_{lm} \cos\!\Big[\!\left(l \omega_{r} + m \omega_{\theta}\right) \tau\Big] \vphantom{\Big[\Big]\np{2}}\right.\\
&\left.\vphantom{\Big[\Big]\np{2}}\!\!\!+B_{lm} \sin\!\Big[\!\left(l \omega_{r} + m \omega_{\theta}\right) \tau\Big]\right)\!+ \bigo{\epsilon^{2}}.
\end{split} \label{eq:freq_general1}
\end{align}
We can regard this equation as an adiabatic evolution equation -- at leading order the terms on the right-hand side are constant in time since they are taken along the same geodesic, while the differential on the left-hand side expresses their evolution along the trajectory of the compact object towards the black hole. Of course, the quantities on the right-hand side are instantaneous, and their values are constantly updated through the evolution equations. The sum is over the non-negative integers \(l\) and \(m\). Since we have written the time-dependence explicitly in the Fourier series, the coefficients of the expansion depend smoothly on the constant parameters of the geodesic. Therefore, we can expand the coefficients in orders of the mass ratio, and neglect all terms sub-leading in \(\epsilon\).

Alternatively, instead of using the \(\{\omega_{\mu}\}\) as the basis for the Fourier expansion, we could rewrite the evolution equation in terms of the phase angle variables \(\{\phi_{\mu}\}\). Phase variables are related to the fundamental frequencies through \(\dot{\phi}_{\mu} = \omega_{\mu}\), and since on a geodesic the frequencies remain constant, \(\phi_{\mu} = \omega_{\mu} \tau\). Changing variables from frequencies to phase angles in eq.~(\ref{eq:freq_general1}), we arrive at a second-order differential equation:
\begin{align}
\begin{split}
\frac{\dd^{2} \phi_{\mu}}{\dd \tau^{2}} = \epsilon \sum_{l,\,m} &\,\Big(C_{lm} \cos\left(l \phi_{r} + m \phi_{\theta}\right) \Big. \\
&\Big.\!\!\!+ D_{lm} \sin \left(l \phi_{r} + m \phi_{\theta}\right)\!\Big)\!+ \bigo{\epsilon^{2}}
\label{eq:phase_general1}
\end{split}
\end{align}

Geodesic resonances occur when the two frequencies become commensurate, or, more precisely, when their ratio is a rational number (eq.~(\ref{eq:resonance_condition})). Therefore, near resonance the term \(\left(l_{0} \, \omega_{r} - m_{0} \, \omega_{\theta}\right)\) will approach 0 (for some \(l_{0}\) and \(m_{0}\) and their integer multiples), and the sum will contain both slowly-oscillating resonance terms, and rapidly oscillating off-resonance terms. Let us now restructure eq.~(\ref{eq:freq_general1}) to change variables to \(\bar{\omega} = l_{0} \, \omega_{r} - m_{0} \, \omega_{\theta}\), which is small near a resonance. Now the arguments of the oscillating modes on the right-hand side can be written as a linear combination of \(\bar{\omega}\) and \(\{\omega_{\mu}\}\) (we can use either of them, but choose \(\omega_{r}\) for definitiveness):
\begin{align}
\begin{split}
\frac{\dd \bar{\omega}}{\dd \tau} = \epsilon \sum_{l,\,m} &\left(A_{lm} \cos\!\Big[\!\left(l \bar{\omega} + m \omega_{r}\right) \tau\Big] \vphantom{\Big[\Big]\np{2}}\right.\\
&\!\!\!\left.\vphantom{\Big[\Big]\np{2}}+ B_{lm} \sin\!\Big[\!\left(l \bar{\omega} + m \omega_{r}\right) \tau\Big] \right)\!+ \bigo{\epsilon^{2}}.
\end{split} \label{eq:freq_general2}
\end{align}
Similarly, we can rewrite the phase equation in terms of \(\bar{\phi}\) and \(\phi_{r}\), where \(\bar{\phi} = l_{0} \, \phi_{r} + m_{0} \, \phi_{\theta}\) vanishes on resonance:
\begin{align}
\begin{split}
\frac{\dd^{2} \bar{\phi}}{\dd \tau^{2}} = \epsilon \sum_{l,\,m} &\,\Big(C_{lm} \cos \left(l \bar{\phi} + m \phi_{r}\right) \Big. \\
&\!\!\!\Big. + D_{lm} \sin \left(l \bar{\phi} + m \phi_{r}\right)\!\Big)\!+ \bigo{\epsilon^{2}}.
\end{split} \label{eq:phase_general2}
\end{align}
Let us change variables to absorb the factor of the mass ratio appearing on the right-hand sides, namely:
\begin{align*}
\hat{\omega} = \frac{\bar{\omega}}{\sqrt{\epsilon}}, \;\;\; \hat{x} = \sqrt{\epsilon} \, \tau \;\;\; \text{in eq.~(\ref{eq:freq_general2})}
\end{align*}
and
\begin{align*}
\hat{\phi} = \bar{\phi}, \;\;\; \hat{\phi}_{r} = \sqrt{\epsilon} \, \phi_{r}, \;\;\; \hat{x} = \sqrt{\epsilon} \, \tau \;\;\; \text{in eq.~(\ref{eq:phase_general2})}.
\end{align*}
Using this substitution, eqs.~(\ref{eq:freq_general2}) and (\ref{eq:phase_general2}) become
\begin{figure*}
\includegraphics[width=\textwidth]{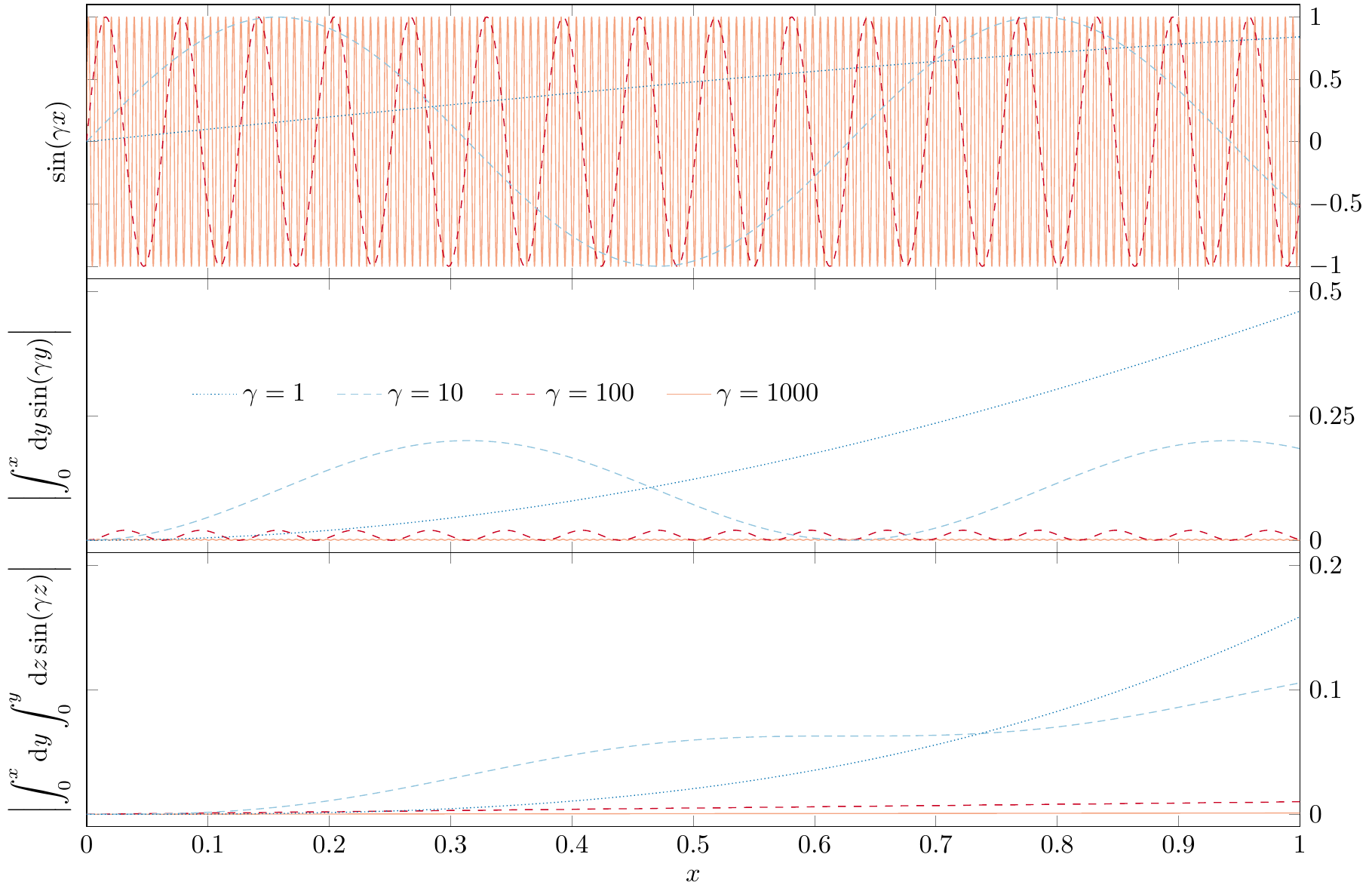}
\caption{\label{fig1}\emph{Upper panel:} the oscillating function \(\sin(\gamma x)\), for \(\gamma = 1, 10, 100\) and \(1000\). It is evident that as \(\gamma\) increases, the function oscillates more rapidly, crossing the horizontal axis more times per unit interval. \emph{Middle panel:} the integral of the functions in the upper panel, of the form \(\left|(1 / \gamma) (1 - \cos(\gamma x))\right|\), for the same values of \(\gamma\).  \emph{Lower panel:} the double integral of the functions in the upper panel. As we can see, both graphically and analytically, as \(\gamma\) increases, the maxima of the integrals become smaller, and their amplitude decreases as \(1 / \gamma\). This allows us to neglect terms with \(\gamma \gtrsim 10^{3}\) when compared to terms where \(\gamma \sim 1\).}
\end{figure*}
\begin{align}
\begin{split}
\frac{\dd \hat{\omega}}{\dd \hat{x}} = \sum_{l,\,m} &\left(A_{lm} \cos\!\Bigg(\!l \hat{x} \hat{\omega} + \frac{m \omega_{r} \hat{x}}{\sqrt{\epsilon}}\!\Bigg) \vphbig \right. \\
&\left. \vphbig \!\!+ B_{lm} \sin\!\Bigg(\!l \hat{x} \hat{\omega} + \frac{m \omega_{r} \hat{x}}{\sqrt{\epsilon}}\!\Bigg)\!\!\right)\!+ \bigo{\epsilon},
\end{split} \\[5pt]
\begin{split}
\frac{\dd^{2} \hat{\phi}}{\dd \hat{x}^{2}} = \sum_{l,\,m} &\left(C_{lm} \cos\!\left(\!l \hat{\phi} + \frac{m \hat{\phi}_{r}}{\sqrt{\epsilon}}\!\right) \vphbig \right. \\
&\left. \vphbig \!\!\! + D_{lm} \sin\!\Bigg(\!l \hat{\phi} + \frac{m \hat{\phi}_{r}}{\sqrt{\epsilon}}\!\Bigg)\!\!\right)\!+ \bigo{\epsilon}.
\end{split}
\end{align}
For any non-zero value of the index \(m\), corrections to the frequency or phase equations are of the form \(\cos(\gamma x + \delta)\), which oscillate rapidly owing to the \(\gamma \sim \epsilon^{-\sfrac{1}{2}} \sim10^{3}\) factor both on resonance and away from it (and therefore will average out to zero over the timescales under consideration --- this is illustrated in Figure~\ref{fig1}). Hence, the correction to the first order solution of the phase equation is of the form \((1/\gamma^{2}) \cos (\gamma x + \delta)\) and that of the frequency equation is \((1/ \gamma) \sin(\gamma x + \delta)\). These are terms of order \(\epsilon\) and \(\epsilon^{\sfrac{1}{2}}\), respectively. Hence, we can perform an expansion to establish that they are corrections of higher order in the mass ratio and hence can be ignored at the same level as the other higher order in mass ratio terms that we have omitted. We can therefore consider only terms with \(m = 0\), and relabel the remaining modes by \(n\), obtaining
\begin{align}
\frac{\dd \hat{\omega}}{\dd \hat{x}} &= \sum_{n}\left(\!A_{n} \cos \left(n \hat{x} \hat{\omega}\right)\! + \!B_{n} \sin \left(n \hat{x} \hat{\omega}\right)\!\!\vphantom{\Big(\Big)\np{2}}\right)\!+ \bigo{\epsilon} \\
\frac{\dd^{2} \hat{\phi}}{\dd \hat{x}^{2}} &= \sum_{n}\left(\!C_{n} \cos\!\left(n \hat{\phi}\right)\! + \!D_{n} \sin\!\left(n \hat{\phi}\right)\!\!\vphantom{\Big(\Big)\np{2}}\right)\!+ \bigo{\epsilon}.
\end{align}
We combine the two oscillating modes together, and this introduces a phase to the oscillating term. This could sometimes be set to 0 by suitable choice of initial coordinates, however it is not always possible, since the resonance phase depends on the prior evolution of the binary system. To preserve generality of the problem, we will include it in our equations. The presence of phase does not complicate the method of solution, only brings in additional terms, as we will demonstrate later in Section~\ref{sec:SolutionsResonanceEqs}.
\begin{align}
\frac{\dd \hat{\omega}}{\dd \hat{x}} &= \sum_{n}\,\mathcal{A}_{n} \cos \left(n \hat{x} \hat{\omega} - \delta \chi_{n}\right) + \bigo{\epsilon}\\
\frac{\dd^{2} \hat{\phi}}{\dd \hat{x}^{2}} &= \sum_{n}\,\mathcal{B}_{n} \cos\!\left(n \hat{\phi} - \delta \psi_{n}\right)\!+ \bigo{\epsilon},
\end{align}
where
\begin{align}
\begin{split}
&\quad\quad \mathcal{A}_{n} = \sqrtbig{A_{n}^{2} + B_{n}^{2}}\,, \;\;\; \mathcal{B}_{n} = \sqrtbig{C_{n}^{2} + D_{n}^{2}}\,,\\
&\delchi_{n} = \arctan\!\left(\!\frac{B_n}{A_n}\!\right) \;\; \text{and} \;\;\; \delpsi_n = \arctan\!\left(\!\frac{D_n}{C_n}\!\right).
\end{split} \label{eq:parameter_constant_relations}
\end{align}
Finally, we split the first term from the sum, \(n = 0\), which is constant (with respect to time or phase, respectively):
\begin{align}
\frac{\dd \hat{\omega}}{\dd \hat{x}}\! &=\! \mathcal{A}_{0} \cos (\delta \chi_{0}) +\!\sum_{n > 0}\! \mathcal{A}_{n} \cos\! \left(n \hat{x} \hat{\omega}\! -\! \delta \chi_{n}\right) + \bigo{\epsilon} \label{eq:result_8}\\
\frac{\dd^{2} \hat{\phi}}{\dd \hat{x}^{2}}\! &=\! \mathcal{B}_{0} \cos (\delta \psi_{0}) +\!\sum_{n > 0}\! \mathcal{B}_{n} \cos\! \left(n \hat{\phi}\! -\! \delta \psi_{n}\right)\!+ \bigo{\epsilon} \label{eq:result_9}
\end{align}
and make a suitable, final change of variables:
\begin{align*}
&\begin{split}
\omega &= \big(\mathcal{A}_{0}  \cos (\delta\chi_{0})\big)^{-\sfrac{1}{2}}\: \hat{\omega}\\
x_{\omega} &= \big(\mathcal{A}_{0}  \cos (\delta\chi_{0})\big)^{\sfrac{1}{2}}\: \hat{x}
\end{split} \quad\quad
&\text{in eq.~(\ref{eq:result_8})}
\end{align*}
and
\begin{align*}
&\begin{split}
x_{\phi} &= \big(\mathcal{B}_{0} \cos(\delta\psi_{0})\big)^{\sfrac{1}{2}}\: \hat{x} \\
\phi &= \hat{\phi}
\end{split} \quad\quad
&\text{ in eq.~(\ref{eq:result_9})}
\end{align*}
to bring the equations to their most simplified form:
\begin{subequations}
\label{eq:final_e}
\begin{align}
\omega^{\prime} &= 1 + \sum_{n} \kappa_{n} \cos \big(n x_{\omega}\,\omega - \delta \chi_{n}\big) \label{eq:frequency_final} \\
\phi^{\prime\prime} &= 1 + \sum_{n} \lambda_{n} \cos \big(n \phi - \delta \psi_{n}\big) \label{eq:phaseEquation},
\end{align}
\end{subequations}
where \(\kappa_{n}\!=(\mathcal{A}_{n}/\mathcal{A}_{0})\sec(\delta \chi_{0})\) and \(\lambda_{n}\!=(\mathcal{B}_{n}/\mathcal{B}_{0})\sec(\delta \psi_{0})\) are in principle known constants, dependent on the orbital parameters at resonance, prime denotes a derivative with respect to the rescaled time \(x\) and the quantities \(\{\delta \chi_{i}\}\) and \(\{\delta\phi_{j}\}\) are constants. The sums over \(n\) run from 1 to \(\infty\), however we will later consider finite number of modes on the right-hand side. In order to achieve convergence when solving these equations, we need to make the assumption that the series \(\{\kappa_n\}\) and \(\{\lambda_{n}\}\) follow a power law of the form \(n^{- \alpha}\), with \(\alpha \gtrsim 2\), this is discussed in detail in Section~\ref{sec:RegimeDomainOfVariables}. In this paper, we are attempting to find solutions to eqs.~(\ref{eq:frequency_final}) and (\ref{eq:phaseEquation}) in increasing order of complexity. It is worth noting that while one of them is a first-order equation, the other is of second order. They derive from equivalent representations of the geodesics, however the first and second order equations are no longer equivalent, because they make different assumptions about the expansion of the self-force.

Before proposing the respective solutions, we will focus on presenting an alternative approach to describing the behaviour of an \textsc{emri} near resonance.

\section{Two-timescale approach}
\label{sec:two}
\noindent
An alternative approach to describing the transition of \textsc{emri}s through resonances was first suggested by \citep{hinderer2008}, and is often called the ``two-timescale method'' since it differentiates between processes which occur on short (orbital) timescales, and those happening on long (orbit evolution) timescales. A good overview is offered by \cite{kevorkian1987}.

Here we will present this method as described by \cite{meent2014}, while it is also advisable that the reader consults \cite{berry2013d}, and the original article  by \cite{hinderer2008}. Again, we consider the motion of a compact object spiralling towards a supermassive black hole, and describe the evolution in terms of the perturbative parameter, the mass ratio \(\epsilon\). At leading order, we recover the geodesic equations of motion, as before. At next order, we need to introduce corrections due to the self-force acting on the compact object, in analogy with eq.~(\ref{eq:frequency_diff}) in the instantaneous frequency approach:
\begin{align}
\frac{\dd J_{\mu}}{\dd \tau} &= \epsilon \, \mathcal{I}_{\mu} (\mathbf{q}, \mathbf{F}) + \bigo{\epsilon^{2}} \label{eq:two_time_one}
\end{align}
While the equations in \cite{berry2013d} are presented in terms of the Mino time \(\lambda\) \citep{mino2003p}, we can map \(\lambda\) to the proper time \(\tau\) through a function involving the coordinates of the compact object. We can absorb this dependence into the functions \(\mathcal{I}_{\mu}\) on right-hand side of eq.~(\ref{eq:two_time_one}) and work in terms of the proper time. As before, we can exchange the integrals of the motion \(\mathbf{J}\) in favour of the fundamental frequencies \(\bm{\omega}\), since mappings between these exist \citep{schmidt2002}, albeit not invertable over the entire domain. Further we can implicitly write the self-force as a function of the coordinates and their corresponding frequencies (cf. eq.~(\ref{eq:asymptotic_frequency_expansion})) to obtain:
\begin{align}
\frac{\dd \omega_{\mu}}{\dd \tau} &= \epsilon \, \mathcal{J}_{\mu} (\mathbf{q}, \mathbf{J}) + \bigo{\epsilon^{2}} \label{eq:freq_evol_2}
\end{align}
Similarly, the action-angle variables evolve as per eq.~(\ref{eq:coordinates_evolution}):
\begin{align}
\frac{\dd q_{\mu}}{\dd \tau} = \omega_{\mu}\left(\mathbf{J}\right) + \epsilon \, \mathcal{K}_{\mu} (\mathbf{q}, \mathbf{J}) + \bigo{\epsilon^{2}}.
\label{eq:coord_evolve_2}
\end{align}
We construct the resonant frequency \(\bar{\nu} = l_{0}\, \omega_{r} - m_{0}\, \omega_{\theta}\), and the corresponding action-angle variable combination \(\bar{q} = l_{0}\, q_{r} - m_{0}\, q_{\theta}\). \(\bar{\nu}\) is the analogous variable to \(\bar{\omega}\) in Section~\ref{sec:InstantaneousFrequencyApproach}, but we use a different notation here to differentiate between the two methods. Again, we ignore the third fundamental frequency \(\omega_{\phi}\), since the motion is axisymmetric. Let \(\bar{J}\) denote the action conjugate to \(\bar{q}\), which can be mapped to \(\bar{\nu}\). Further, let us rescale the proper time to write \(x_{\nu} = \sqrt{\epsilon}\, \tau\). We now expand the frequency and coordinate variables in powers of \(\epsilon\):
\begin{subequations}
\label{eq:two_timescales_method_expansion}
\begin{align}
\bar{\nu} \left(x_{\nu}, \epsilon\right) &= \bar{\nu}_{0} + \epsilon^{\sfrac{1}{2}}\,\bar{\nu}_{1} (x_{\nu})+ \bigo{\epsilon^{2}} \label{eq:two_timescales_method_expansion_freq}\\
\bar{q} \left(x_{\nu}, \epsilon\right) &= \bar{q}_{0} (x_{\nu}) + \epsilon^{\sfrac{1}{2}}\,\bar{q}_{1} (x_{\nu}) + \bigo{\epsilon^{2}}.
\end{align}
\end{subequations}
Plugging these power laws into the (averaged) evolution equations (\ref{eq:freq_evol_2}) and (\ref{eq:coord_evolve_2}) yields the resonance equations:
\begin{subequations}
\label{eq:two_timescales_method_resonance_equations}
\begin{align}
\frac{\dd \bar{\nu}_{1}}{\dd x_{\nu}} &= \bar{\mathcal{J}}\!\left(\bar{q}, \bar{J}\right) + \bigo{\epsilon} \\
\frac{\dd \bar{q}_{0}}{\dd x_{\nu}} &= \bar{\nu}_{1} (x_{\nu}) + \bigo{\epsilon}.
\end{align}
\end{subequations}
This relays the leading terms in the expansions (\ref{eq:two_timescales_method_expansion}). Further members of the series \(\{\bar{\nu}_{n}\}\) and \(\{\bar{q}_{n}\}\) can be found iteratively by substituting the results from eqs.~(\ref{eq:two_timescales_method_resonance_equations}) back into the resonance evolution equations.
\begin{figure*}[t]
\includegraphics[scale=1]{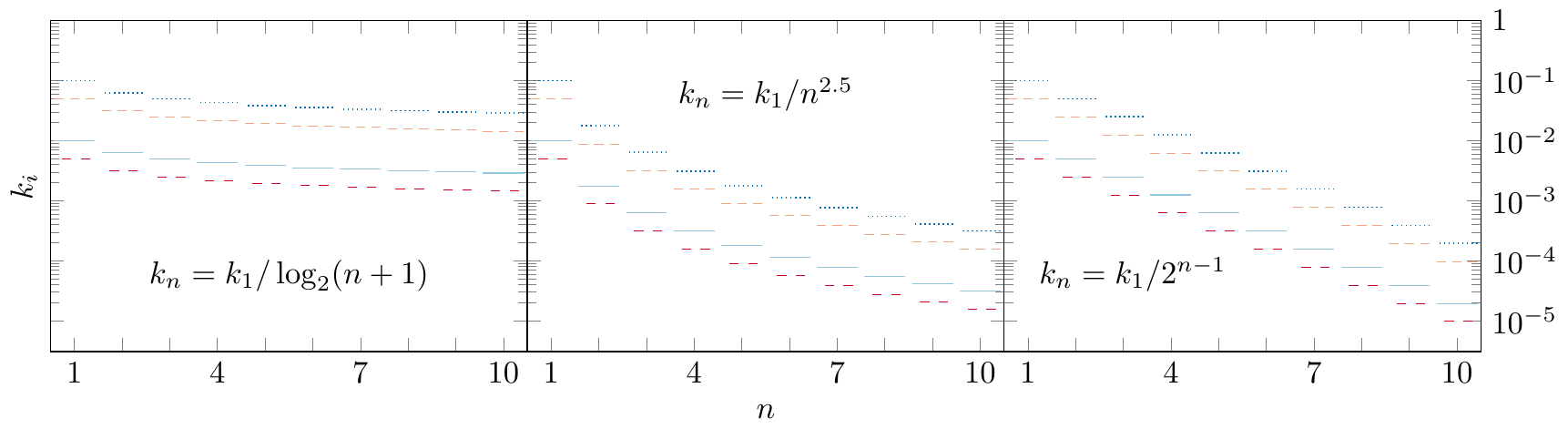}
\caption{\label{fig2}Visualisation of several types of series for the mode parameters \(k_{i} = \{\kappa_{i}, \lambda_{i}\}\). For each of them, we take 4 different values of the initial parameter: \(k_{1} = 0.1\) ({\color{color4} \protect\dottedrule}), \(k_{2} = 0.05\) ({\color{color2} \protect\denselydashedrule}), \(k_{3} = 0.01\) ({\color{color3} \solidrule[4.3mm]}),  and \(k_{4} = 0.005\) ({\color{color1} \protect\looselydashedrule}). \emph{Left panel:} Logarithmic progression with base 2, \(k_{n} = k_{1} / \log_{2} (n+1)\), which is too slow to satisfy our conditions for convergence. \emph{Centre panel:} Power law \(k_{n} = k_{1} / n^{\alpha}\), with \(\alpha = 2.5\), as suggested by \citep{drasco2006} (the lower bound is \(\alpha \gtrsim 2\)). \emph{Right panel:} Exponential series \(k_{n} = k_{1} / 2^{n-1}\). It is expected that in reality, this progression will start as a power law (centre panel) and will gradually evolve into an exponential series for large values of \(n\) (right panel). This data is represented numerically in Table~\ref{tab:ladder_values}.}
\label{fig:ladders_plot}
\end{figure*}
In~\cite{flanagan2012}, eq.~(\ref{eq:two_timescales_method_expansion_freq}) was further simplified by expanding \(\bar{\nu}_1\) as a series in \(\sqrt{\epsilon}\,x_\nu\). The leading term is then \(\nu_1 (x_\nu - x_0)\) for a constant \(\nu_1\). Proceeding with the two timescale expansion then yields a result accurate to first order in the resonant modification, \(\{\lambda_n\}\), which agrees with the result obtained from the asymptotic Fourier series approach in~\cite*{gair2012}. Corrections at higher order in \(\lambda_n\), but leading order in mass ratio, cannot be found in this way, since the assumption of a linearly evolving frequency is only valid close to the resonance point. The two-timescale approach can be extended to higher order by using a less restrictive ansatz for \(\bar{\nu}_1\), which is given implicitly by eqs~(23) and (25) in~\cite{meent2014}. Explicit solutions to higher order in the resonant modification at leading order in mass ratio have not been computed in the two-timescale approach. However, we anticipate that these will agree with the phase version of the asymptotic Fourier series approach. The purpose of this paper is to extend previous calculations to higher order in resonant modification and to include multiple modes in the resonance. To do this we will use the asymptotic Fourier series approach and this will be the focus for the remainder of this paper.

\section{Solutions of the resonance equations}
\label{sec:SolutionsResonanceEqs}
\noindent
We are now in a position to solve the instantaneous frequency approach equations that we derived earlier in Section~\ref{sec:InstantaneousFrequencyApproach}. Evolution of the action-angle coordinates according to the instantaneous frequency of the system yields the 2 non-linear differential eqs.~(\ref{eq:frequency_final}) and (\ref{eq:phaseEquation}). Due to the constant additive term on their right-hand sides, these equations cannot be solved exactly, and therefore we need to come up with an alternative approach. In this Section we propose solutions to these equations in an increasing order of complexity. Solutions for the single-mode versions of eqs.~(\ref{eq:final_e}) are provided in \cite*{gair2012}, we will start by re-deriving those to depict how the order-by-order solution algorithm works. Following this, we relax the assumptions which have been considered in \cite*{gair2012} in order to build up solutions to the generalised source equations.
\subsection{Regime and domain of the variables}
\label{sec:RegimeDomainOfVariables}
\noindent
Before attempting to solve the resonance equations, we need to specify their domains and any special limits we are taking. The time parameters \(x_{\{\omega,\,\phi,\,\nu\}}\) range from \(-\infty\) to \(+\infty\), however we are interested in the vicinity of \(x = 0\) as this is when resonance occurs, hence where we have specified the initial conditions. In this notation, \(-\infty\) corresponds to times long before resonance, and \(+\infty\) to times when the resonance has had an effect on the orbit. Therefore, in solving these equations, we will consider definite integrals of the form \(\int_{0}^{x} \dd \bar{x}\), where \(x\) can take large positive or negative values. Since the equations are not coupled, and each rescaled time variable appears in the differential equation for its corresponding quantity, we will drop the subscripts on \(x\), and will keep track of them by the subject of the equation we are solving.

More attention needs to be devoted to the fractional resonance modifications, \(\kappa_{i}\) and \(\lambda_{j}\), which are themselves functions of the orbital parameters. As discussed in~\cite*{gair2012}, if these modifications are of order unity, there can be a qualitative difference in the on-resonance behaviour, with the possibility of sustained resonances. These were explored in more detail in~\cite{meent2014}. A complete study of the size of on-resonance flux modifications has not yet been carried out. Some insight on the likely size of these corrections can be obtained by studying solutions to the Teukolsky equation, which provide a decomposition of the off-resonance energy flux into different Fourier modes. Such results can be found in~\cite{drasco2006}. Section I~C of that paper indicates that the size of the terms in this decomposition initially follows a power law of the form \(\sim n^{- \alpha}\) before eventually falling off exponentially (see Figure~\ref{fig:ladders_plot} and Table~\ref{tab:ladder_values} for an illustration different rates of fall off). From Fig. 3 of that paper it appears that \(\alpha \gtrsim 2\) for all \(n\), but as a lower limit in our calculations we will use the value of \(\alpha = 5/2\). \cite{drasco2006} also indicate that the size of the first term in the series is less than unity, and in fact of order \(\kappa_{1}\sim\lambda_{1}\sim 0.1\). This implies that the resonance is likely to be ``weak'', but the Teukolsky results do not directly give the size of the on-resonance flux change. The size of the resonance effect was assessed more carefully in~\cite{flanaganhughesruangsri2012}, who computed the flux change for a number of different resonances for a few systems and found that the resonant flux change was only \(\sim 1\%\) in all cases. This also suggests that \(\kappa_1\) and \(\lambda_1\) are small, typically \(\sim 0.01\). Based on this evidence it seems unlikely that strong resonances will occur in practice and so in this article we focus on the weak resonance case with \(\kappa_1 \ll 1\), which will allow us to perform an expansion in \(\kappa_1\). However, we should note that \cite{flanaganhughesruangsri2012} did not thoroughly explore the resonance parameter space and so it is still possible that there are regions of parameter space where strong sustained resonances may exist. If this is possible, it would be very interesting, as discussed in detail in~\cite{meent2014}.

Regarding the other relevant parameters, the mass-ratio \(\epsilon\) is typically in the range \(\sim{10^{-6}} - \sim{10^{-3}}\). The quantity \(\epsilon^{1/2}\) is therefore \(\bigo{10^{-3} - 10^{-1.5}}\). The phases on resonance, \(\delta \chi_{i}\) and \(\delta \psi_{j}\), are constants in the range \([0, 2\pi)\), which depend only on the coefficients of the Fourier series in eqs.~(\ref{eq:freq_general1}) and (\ref{eq:phase_general1}), and the index \(n\).

\begin{figure}[t]
\includegraphics[scale=1]{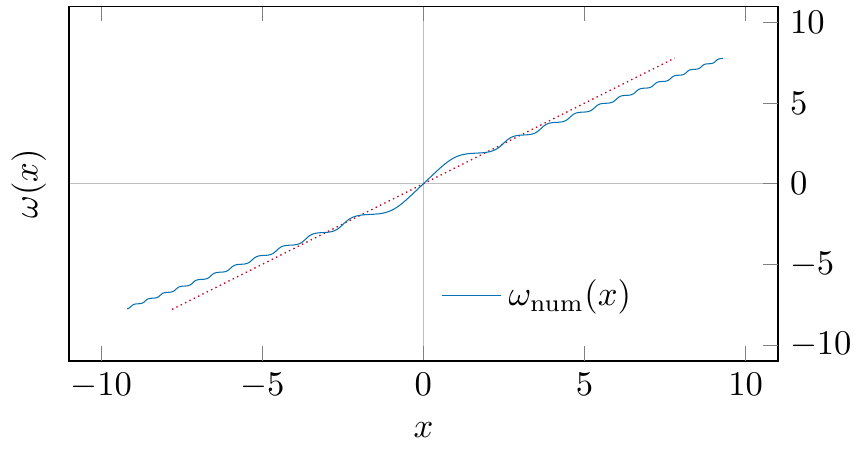}
\caption{Numerical solution of the equation \(\omega^{\prime} = 1 + \kappa \cos\left(x\omega\right)\), with \(\kappa = 0.9\), an exaggerated value is used to demonstrate the notable features of the solution, namely the slope and the decaying oscillations. The initial condition \(\omega (0)\) depends on the initial phase of the resonance, and has been set to 1 here. The line \(\omega(x) = 1 + x\) is plotted (in red dots) for reference.}
\label{fig:frequency_numerical}
\end{figure}

\subsection{Frequency resonance equation}
\label{sec:frequency_solution_ch}
\noindent
Consider the generalised frequency resonance equation:
\begin{align}
\omega^{\prime} &= 1 + \sum_{n} \kappa_{n} \cos \big(n x_{\omega}\,\omega - \delta\chi_{n}\big),
\tag{\ref{eq:frequency_final}}
\end{align}
where prime denotes differentiation with respect to the rescaled time parameter \(x_{\omega}\). Before attempting to solve this equation in the general case, we consider only the single oscillatory mode (\(n = 1\), besides the \(n = 0\) mode) on the right-hand side, and as agreed, drop the subscript on \(x\). The resulting problem is not much simpler to solve, however it will provide useful insight about the method:
\begin{align}
\omega^{\prime} (x) = 1 + \kappa \cos\left(x \omega\right), \label{eq:FrequencySMEq}
\end{align}
where \(\kappa \equiv \kappa_{0}\). Inspired by the numerical solution of eq.~(\ref{eq:FrequencySMEq}), shown in Fig.~(\ref{fig:frequency_numerical}), we split the solution in two parts: the first for the region where the oscillations are still growing, and the second where the oscillations gradually decay. After we establish appropriate solutions to eq.~(\ref{eq:FrequencySMEq}) in each of these regions, we will come up with a way to match them on the boundary region.

Let us initially look for a solution in the \(\kappa |x| \ll 1\) region. That is, find an expansion of the frequency \(\omega\) as a function of \(x\) in the region immediately around \(x = 0\), in powers of the fractional resonance modification \(\kappa\):
\begin{align}
\omega (x) = \sum_{n = 0}^{\infty} \kappa^{n} \varpi_{n} (x). \label{eq:frequency_ansatz_small_x}
\end{align}
\begin{figure}[b]
\vspace*{20pt}
\includegraphics[scale=1]{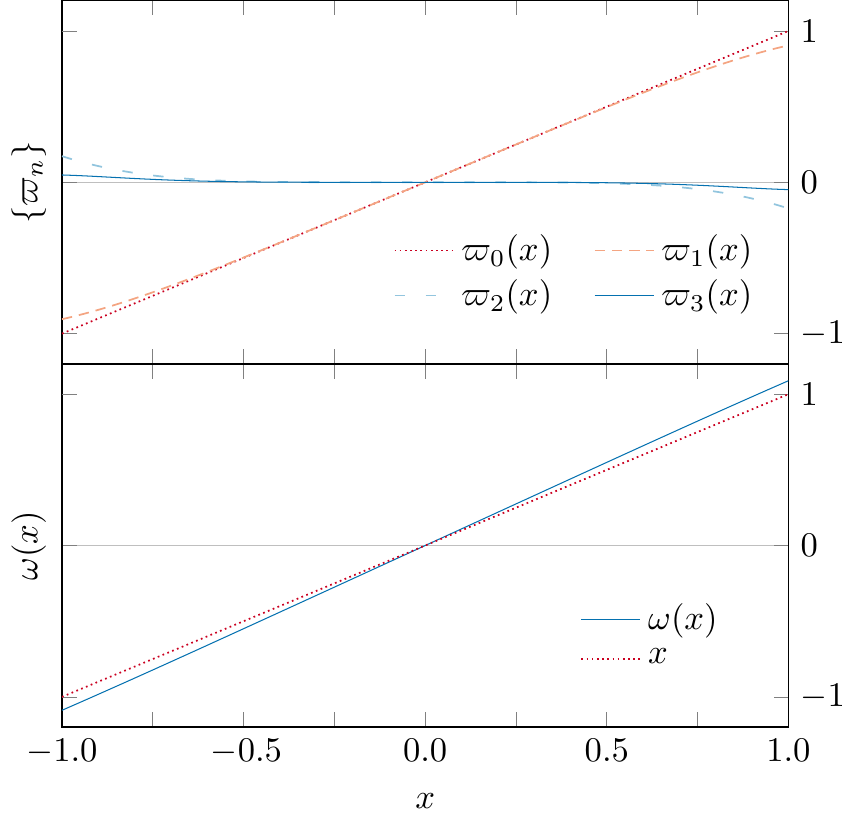}
\caption{Analytical solution to the frequency equation for \(|x| < 1\). \emph{Top panel:} The solutions at each order from 0 to 3 inclusive, plotted as a function of \(x\). \emph{Bottom panel:} The combined solution at third order, eq.~(\ref{eq:FrequencySolutionSmallX}).}
\label{fig:frequencySolutionSmallXPlot}
\vspace*{-10pt}
\end{figure}
We remind ourselves that the parameter \(\kappa \ll 1\), as per Section~\ref{sec:RegimeDomainOfVariables}. In order to find explicit forms of the functions \(\{\varpi_{n}\}\), we substitute the expansion~(\ref{eq:frequency_ansatz_small_x}) in eq.~(\ref{eq:FrequencySMEq}) and proceed to match terms on both sides. The results we now derive are plotted in Figure~\ref{fig:frequencySolutionSmallXPlot}. At lowest order, we disregard all terms of \(\bigo{\kappa}\) and higher:
\begin{align}
\varpi_{0}^{\prime} (x) = 1 \quad \Leftrightarrow \quad \varpi_{0} (x) = x,
\end{align}
where it is important to note that the integration has been carried out between 0 and \(x\). Further terms in the series~(\ref{eq:frequency_ansatz_small_x}) will be reduced by one or more factors of \(\kappa \ll 1\), hence we can treat them as a perturbation to \(\varpi_{0} (x)\):
\begin{align}
\sum_{n = 1}^{\infty} \kappa^{n} \varpi_{n} (x) = \kappa \cos\!\left[x^{2} + \sum_{n = 1}^{\infty} \kappa^{n} x\,\varpi_{n} (x)\right]. \label{eq:frequency_approximate_source_eq_small_x}
\end{align}
At the next higher order we find the equation for \(\varpi_{1} (x)\):
\begin{align}
\kappa \varpi_{1}^{\prime} (x) + \bigo{\kappa^{2}} = \kappa \cos\!\left(x^{2}\right) + \bigo{\kappa^{2}}.
\end{align}
integrating the right-hand side yields the solution:
\begin{align}
\varpi_{1} (x) = \int_{0}^{x} \dd y \cos\!\left(y^{2}\right) = \frac{\sqrt{2 \pi}}{2}\,C\!\left(\!\frac{\sqrt{2}\,x}{\spi}\!\right),
\end{align}
where \(C(\bullet)\) is the Fresnel cosine integral (see Appendix~\ref{sec:AppFresnelFuncs} and Figure~\ref{fig:fresnel} for details). At \(\bigo{\kappa^{2}}\), we find:
\begin{align}
\varpi_{2}^{\prime} (x) = -\,\frac{\spi}{\sqrt{2}} \, x\, \sin\!\left(x^{2}\right) C\!\left(\!\frac{\sqrt{2}\,x}{\spi}\!\right).
\end{align}
Solving this differential equation (which also involves a simple integration by parts) provides the solution:
\begin{align*}
\varpi_{2} (x) = \frac{\sqrt{2 \pi}}{4} \cos\!\big(x^{2}\big) C\!\left(\!\frac{\sqrt{2}\,x}{\spi}\!\right)\!- \frac{\spi}{8}\,C\!\left(\!\frac{\sqrt{4}\,x}{\spi}\!\right)\!-\frac{x}{4}.
\end{align*}

It is obvious that with increasing order the functional form of \(\varpi_{n} (x)\) become more and more involved. It will become apparent later on why we do not need solutions at \(\bigo{\kappa^{4}}\) and beyond. However, at third order, we find the following differential equation for the functional \(\varpi_{3} (x)\).
\begin{figure}[b]
\includegraphics[scale=1.0]{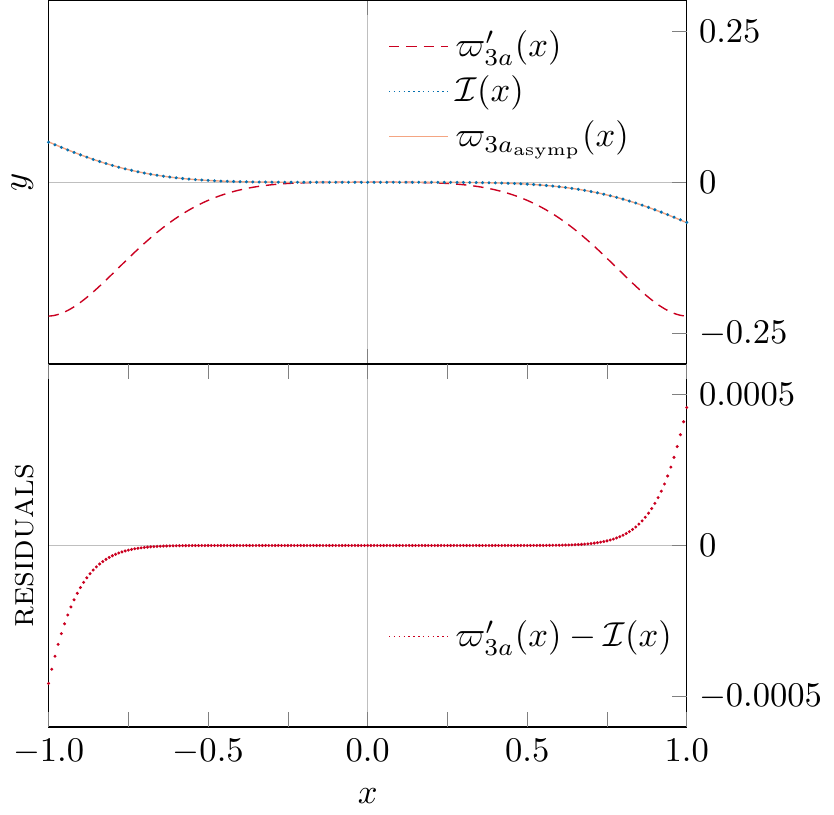}
\caption{Approximate integral of \(\varpi_{3a}^{\prime} (x)\) near the origin. \emph{Top panel:} The integrand, the numerical integral, and the approximation. \emph{Bottom panel:} Residuals between the approximation and the numerical solution, showing the validity of eq.~(\ref{eq:frequencySmallXThOrdAsymp}).}
\label{fig:frequencyApproxPlot}
\vspace*{-15pt}
\end{figure}
\begin{align*}
\varpi_{3}^{\prime} (x) = - \frac{\pi}{4}\,x^{2} \cos\!\left(x^{2}\right) C^{2}\!\!\left(\!\frac{\sqrt{2}\,x}{\spi}\!\right)\!- x \sin\!\left(x^{2}\right) \varpi_{2} (x).
\end{align*}
We divide \(\varpi_3\) into two parts, \(\varpi_3(x) = \varpi_{3a}(x)+\varpi_{3b}(x)\), defined via the two differential equations:
\begin{subequations}
\begin{align}
\varpi_{3a}^{\prime} (x) &= - \frac{\pi}{4}\,x^{2} \cos\!\left(x^{2}\right) C^{2}\!\!\left(\!\frac{\sqrt{2}\,x}{\spi}\!\right)\! \label{eq:frequencyThOrdAIntegral} \\
\varpi_{3b}^{\prime} (x) &= - x \sin\!\left(x^{2}\right) \varpi_{2} (x) \label{eq:om3b}.
\end{align}
\end{subequations}
Unfortunately, the term \(\varpi_{3a} (x)\) above is the only integral in our analysis which cannot be performed analytically. However, we can proceed by expanding the Fresnel term on the right hand side for small \(x\) and integrating the resulting terms. We then define \(\varpi_{3a_{\rm approx}} (x)\) to be the terms arising from this integration, given analytically by eq.~(\ref{eq:frequencySmallXThOrdAsymp}). This has the property \(\varpi_{3a}^{\prime}(x) - \varpi_{3a_{\rm approx}}^{\prime} (x)  \sim \bigo{x^{9}}\) and hence \(\varpi_{3a}(x)-\varpi_{3a_{\rm approx}}(x) \sim \bigo{x^{10}}\). Finally, we need to verify that this is a valid approximation; we can do this by tabulating the function
\begin{align}
\mathcal{I} (x) = \int_{0}^{x} \!\! \dd y \,  \varpi_{3a}^{\prime} (y)
\end{align}
for values of \(x\) between \(-1\) and 1, and comparing this with our (approximate) analytical result for the integral, eq.~(\ref{eq:frequencySmallXThOrdAsymp}). The integrand, both functions, and the residuals of the approximation are shown in Fig.~\ref{fig:frequencyApproxPlot}.

The term in eq.~(\ref{eq:om3b}) can be integrated analytically, and the result is given by expression (\ref{eq:frequency_small_x_y3_second_term}). Gathering the above results together, we can write down an approximate solution to the Frequency equation, which is given in eq.~(\ref{eq:FrequencySolutionSmallX}). We note that this is valid for \(\kappa |x| \ll 1\) up to \(\bigo{\kappa^{2}}\), while the result for \(\varpi_{3} (x)\) is only valid when \(|x| \ll 1\). Later on, as the need arises, we will find a solution for \(\varpi_{3} (x)\) which is valid in other regimes.

\begin{widetext}
\vspace*{-20pt}
\begin{align}
\begin{split}
&\varpi_{3a_{\rm approx}} (x) = \frac{33\sqrt{2\pi}}{64}\,C\!\left(\!\frac{\sqrt{2}\,x}{\spi}\!\right)\!- \frac{33}{32}\,x \cos\!\big(x^{2}\big) - \frac{11}{16}\,x^{3} \sin\!\big(x^{2}\big) + \frac{7}{40}\,x^{5}\cos\!\big(x^{2}\big) + \frac{1}{20}\,x^{7} \sin\!\big(x^2\big)
\end{split} \label{eq:frequencySmallXThOrdAsymp} \\
&\varpi_{3b} (x) = \frac{\spi}{16}\!\left(\!\vphbig\frac{\sqrt{2}}{2} \cos\!\left(2 x^{2}\right) C\!\left(\!\frac{\sqrt{2}\,x}{\spi}\!\right)\!- \cos\!\left(x^{2}\right) C\!\left(\!\frac{\sqrt{4}\,x}{\spi}\!\right)\!\!\right)\!+ \frac{\spi}{32}\!\left(\!\vphantom{\Bigg(\Bigg)\np{2}}\frac{\sqrt{6}}{6}\,C\!\left(\!\frac{\sqrt{6}\,x}{\spi}\!\right)\!+ \frac{5\sqrt{2}}{2}\,C\!\left(\!\frac{\sqrt{2}\,x}{\spi}\!\right)\!\!\right)\!-\frac{1}{8}\,x \cos\!\left(x^{2}\right)
\label{eq:frequency_small_x_y3_second_term}
\end{align}
\begin{m}{align}
\omega (x) = x + \kappa \, \frac{\sqrt{2 \pi}}{2}\,C\!\left(\!\frac{\sqrt{2}\,x}{\spi}\!\right)\!+ \kappa^{2}\!\left(\!\frac{\sqrt{2 \pi}}{4} \cos\!\big(x^{2}\big)\,C\!\left(\!\frac{\sqrt{2}\,x}{\spi}\!\right)\!- \frac{\spi}{8}\,C\!\left(\!\frac{\sqrt{4}\,x}{\spi}\!\right)\!-\frac{x}{4}\!\vphbig\right)\!+ \kappa^{3} \, \varpi_{3} (x) + \bigo{\kappa^{4}}
\label{eq:FrequencySolutionSmallX}
\end{m}
\end{widetext}

\begin{figure}[t]
\includegraphics[scale=1]{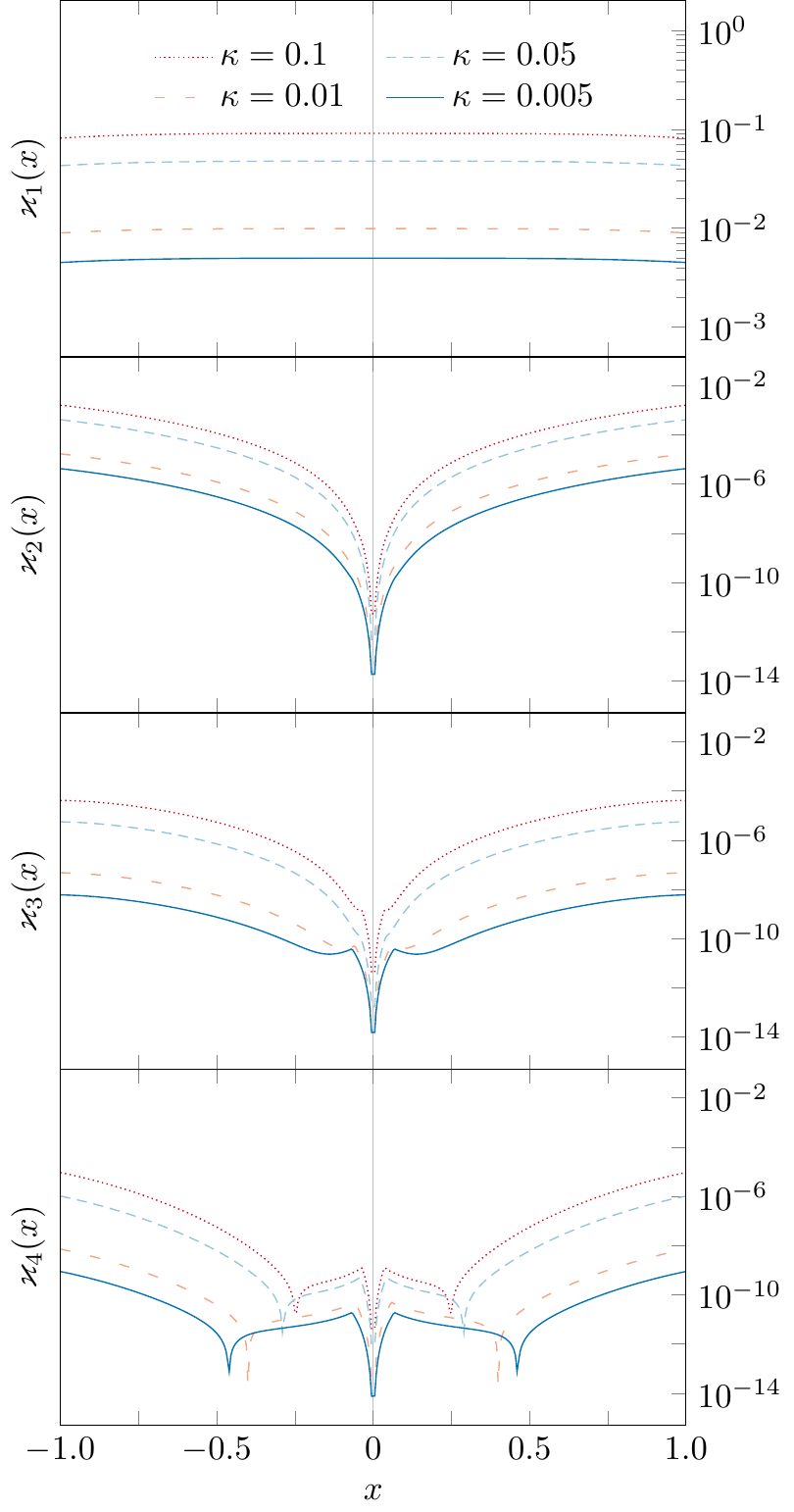}
\caption{Graphical demonstration of the validity of solution~(\ref{eq:FrequencySolutionSmallX}). It is evident that the residuals scale with the value of the resonance modification \(\kappa\), and for each order \(n\) of the solution, the remainder is of order \(\bigo{\kappa^{n+1}}\).}
\label{fig:frequencyValiditySmallX}
\vspace*{-15pt}
\end{figure}

Before we continue with our analysis, let us consider the asymptotic value of this solution in the limit of it's validity, i.e. for \(1 \ll |x| \ll \kappa^{-1}\). The non-oscillating part of solution~(\ref{eq:FrequencySolutionSmallX}) can be written as  
\begin{align}
\omega (x) \sim x  + f(\kappa).
\end{align}
We see that any constant term in the solution for \(\omega (x)\) is at least \(\bigo{\kappa}\), rather than \(\bigo{\kappa^{0}}\). To confirm the validity of solution~(\ref{eq:FrequencySolutionSmallX}) we compute the quantities
\begin{align}
\varkappa_{n+1} = \left|\vph\frac{\omega_{\text{num}} - \omega^{(n)} (x)}{\omega_{\text{num}}}\right|, \label{eq:frequencyValidityFormula}
\end{align}
where \(\omega^{(n)} (x)\) is eq.~(\ref{eq:FrequencySolutionSmallX}) restricted only to order \(n\) (e.g. \(\omega^{(1)} (x) = \varpi_{0} (x) + \kappa \varpi_{1} (x)\)). If our solution is correct, then we would find that the fractional error, computed by the above formula is \(\bigo{\kappa^{n+1}}\). This scaling is evident in the plot of these functions, shown in Figure~\ref{fig:frequencyValiditySmallX}.

Let us now seek a solution in the region of decaying oscillations, \(|x| \gg 1\). At lowest order, the solution is
\begin{m}{align}
\omega_{0} (x) = x. \label{eq:frequencySolution0}
\end{m}
At higher order, we attempt an ansatz, composed of a constant, a linear term, and a decaying oscillating term. We allow for three undetermined constants, \(a_{1}, b_{1}\) and \(c_{1}\), as suggested by \cite*{gair2012}:
\begin{align}
\omega_{1} (x) = a_{1} + \left(1 + b_{1}\right) x + \frac{c_{1}}{x} \, \sin\!\Big[a_{1} x + \left(1 + b_{1}\right) x^{2}\Big] \label{eq:ansatz_zero}
\end{align}
where the subscript ``1" on the slope correction term indicates that it pertains to the solution \(\omega_{1} (x)\), as the value of \(n\) will change for higher-order solutions. When we substitute this ansatz into the model eq.~(\ref{eq:FrequencySMEq}):
\begin{align}
\omega_{1}^{\prime} (x) \sim 1 + \kappa \cos\left(x \omega_{1}\right),
\end{align}
we ignore \(\bigo{x^{-1}}\) terms and beyond, and use our knowledge of the behaviour of each separate term to reason about the values of the undetermined constants in eq.~(\ref{eq:frequencyIntermediate1}) below. Using standard trigonometric identities, we can expand the second term on the right-hand side, as shown explicitly in eq.~(\ref{eq:frequencyIntermediate2}) below.
\begin{widetext}
\begin{align}
&1 + b_{1} + 2 \left(1 + b_{1}\right) c_{1} \cos\!\Big[a x + \left(1 + b_{1}\right) x^{2}\Big] + \bigo{x^{-1}} \sim 1 + \kappa \cos\!\left[ax + \left(1 + b_{1}\right) x^{2} + c_{1} \sin\!\Big[ x + (1 + b_{1}) x^{2}\Big]\vphantom{\Big(\Big)\np{2}}\!\right] \label{eq:frequencyIntermediate1} \\[5pt]
&\quad\;\,\begin{aligned}
b_{1} &+ 2 \left(1 + b_{1}\right)\, c_{1} \cos\!\Big[a x + \left(1 + b_{1}\right) x^{2}\Big] + \bigo{x^{-1}} \sim \\
&\sim \kappa \cos\!\Big[ax + \left(1 + b_{1}\right) x^{2}\Big] \cos\!\left[c_{1} \sin\!\Big[a x + \left(1 + b_{1}\right) x^{2}\Big]\vphantom{\Big(\Big)\np{2}}\!\right] - \kappa \sin\!\Big[ax + \left(1 + b_{1}\right) x^{2}\Big] \sin\!\left[c_{1} \sin\!\Big[a x + \left(1 + b_{1}\right) x^{2}\Big]\vphantom{\Big(\Big)\np{2}}\!\right]
\end{aligned} \label{eq:frequencyIntermediate2}
\end{align}
\end{widetext}
At this point we introduce a shorthand notation for the oscillating terms. Please refer to Appendix \ref{sec:ShorthandTrig} (and the plots in Figure~\ref{fig:sncn}) for details.
\begin{subequations}
\begin{align}
\,S_{n} (x) \equiv \sin\!\left[n\Big(a x + \left(1 + b\right) x^{2}\Big)\vphantom{\Big(\Big)\np{2}}\!\right] \label{eq:def_snx} \\
C_{n} (x) \equiv \cos\!\left[n\Big(a x + \left(1 + b\right) x^{2}\Big)\vphantom{\Big(\Big)\np{2}}\!\right] \label{eq:def_cnx}
\end{align}\label{eq:shorthand_trigonometric}
\end{subequations}
Using these, we cast our last equation (\ref{eq:frequencyIntermediate2})\ in the form:
\begin{align}
\label{frequency3}
\begin{split}
b_{1}& + 2\left(1 + b_{1}\right) c_{1} C_{1} (x) + \bigo{x^{-1}} \sim\\
&\sim \kappa \, C_{1} (x) \cos\!\left[c_{1} S_{1} (x)\vphantom{()\np{2}}\right] - \kappa \, S_{1} (x) \sin\!\left[c_{1} S_{1} (x)\vphantom{()\np{2}}\right]
\end{split}
\end{align}
Comparing terms on both sides of eq.~(\ref{frequency3}), we notice that in order for this identity to hold, it is necessary that \(c_{1} \sim \bigo{\kappa}\). We remind ourselves of the assumption that \(\kappa \ll 1\) which justifies the small angle approximation we used above. Since the range of the sine function is between \(-1\) and 1, and \(c_{1} \sim \bigo{\kappa} \ll 1\), we can safely assume that \(c_{1} \sin\!\left[a x + \left(1 + b\right) x^{2}\vphantom{()\np{2}}\right] \ll 1\) for all \(x\) and expand both trigonometric terms in eq.~(\ref{frequency3}):
\begin{align}
b_{1} + 2 \left(1 + b_{1}\right) c_{1} C_{1} (x) \sim \kappa \, C_{1} (x) - \kappa \, c_{1} S_{1}^{2} (x) \label{frequency4}
\end{align}
As we only retain the leading order terms in these expansions, we must disregard any \(\bigo{\kappa^{3}}\) or higher terms in our further calculations. Since we exclude them at this stage, such terms can no longer appear in the final solution for \(\omega_{1} (x)\). We match the coefficients of the two \(\bigo{\kappa}\) terms on either side of the relation~(\ref{frequency4}): \(2\left(1 + b\right) \, c_{1} \sim \kappa\). Therefore:

\begin{align}
c_{1} \sim \frac{\kappa}{2\left(1 + b_{1}\right)}.
\label{eq:c1_result_1}
\end{align}
By using the double-angle trigonometric identities on the remaining term on the right-hand side of (\ref{frequency4}), we find
\begin{align}
b_{1}\!\sim - \frac{\kappa c_{1}}{2} \big(1 - C_{2} (x)\big)\!\sim -\frac{\kappa c_{1}}{2}.
\label{eq:b_result_1}
\end{align}
Combining this with result (\ref{eq:c1_result_1}), we establish that \(b_{1}\!\sim \bigo{\kappa^{2}}\), i.e., 0 to linear order. This result follows solely from the linear independence of the \(\{C_{n}\}\) terms. While they are not orthogonal (this can be easily verified by analytical integration), they are linearly independent, and therefore the coefficients on both sides need to match for eq.~(\ref{eq:b_result_1}) to hold true. Similarly, we obtain the revised result \(c_{1} \sim \kappa/2 + \bigo{\kappa^{3}}\). The additive constant \(a \equiv \omega (0)\) can be determined by comparing the limit \(|x| \gg 1\) of solution~(\ref{eq:FrequencySolutionSmallX}) to the ansatz~(\ref{eq:ansatz_zero}) in the regime \(\kappa |x| \ll 1\):
\begin{align}
\begin{split}
x \pm \kappa \, \frac{\sqrt{2 \pi}}{4} & + \kappa \, \frac{1}{2} \, \frac{\sin\big(x^{2}\big)}{x} +\bigo{x^{-2}} \sim \\
& \sim a + x + \kappa \, \frac{1}{2} \, \frac{\sin\big(x^{2}\big)}{x} + \bigo{\kappa^{3}}
\end{split}
\end{align}
where on each side there are terms which we have ignored in the respective analysis. We find that \(a = \pm \big(\sqrt{2 \pi} / 4\big) \kappa\) to linear order, and we can write down the leading-order solution (\ref{eq:FrequencySolution1}) to the frequency equation (for a single oscillatory mode and no initial phase). Note that here, and for the rest of this article, whenever the symbols \(\sfrac{\pm}{\mp}\) are used, the upper sign is valid for \(x \ge 0\), and the lower for \(x \le 0\). The function in question is thus defined piecewise, but we will use this notation throughout for brevity.
\begin{m}{align}
\omega_{1} (x) = \pm \frac{\sqrt{2 \pi}}{4} \, \kappa + x + \kappa\,\frac{1}{2}\,\frac{1}{x}\, \sin\!\big(x^{2}\big)
\label{eq:FrequencySolution1}
\end{m}

Once we have established the suitability of this ansatz, we propose a further trial solution by adding an extra term to eq.~(\ref{eq:ansatz_zero}). It accounts for the terms in eq.~(\ref{eq:b_result_1}) we ignored when considering only first-order corrections:
\vspace*{-2pt}
\begin{align}
\frac{c_{2}}{x} \, \sin\!\left[2\Big(a x + \left(1 + b\right) x^{2}\Big)\vphantom{\Big(\Big)\np{2}}\!\right].
\end{align}
The full ansatz can now be written as:
\begin{align}
\omega_{2} \left(x\right) =  a + \left(1 + b_{2}\right)x + \frac{c_{1}}{x} \, S_{1}(x) + \frac{c_{2}}{x} \, S_{2}(x). \label{eq:freq_ansatz_order2}
\end{align}
We substitute this into the source equation~(\ref{eq:FrequencySMEq}) and carry out a similar procedure to establish the new values of the constants, as detailed below. On the left-hand side, we ignore terms of \(\bigo{x^{-1}}\) in the derivative:
\begin{align}
\omega_{2}^{\prime} \sim 1 + b_{2} + 2\left(1 + b_{2}\right) c_{1}\,C_{1} + 4\left(1 + b_{2}\right) c_{2}\, C_{2}\,, \label{eq:freq_order2_der}
\end{align}
while on the right-hand side, we need to unpack the argument of the cosine function. Previously we established that \(c_{1} \sim \bigo{\kappa}\), and since we expect that the new terms originating from \(\omega_{2} (x)\) will match the remaining term in eq.~(\ref{eq:b_result_1}), we expect \(c_{2} \sim \bigo{\kappa^{2}}\). This justifies the further use of the small-angle approximation for the latter two terms. We point out that we need to ignore non-constant terms of \(\bigo{\kappa^{3}}\) in this expansion, as they will not be matched by a suitable term in the ansatz.
\begin{align}
&1 + \kappa \cos\!\left[a x + \left(1 + b_{2}\right) x^{2} + c_{1} S_{1} + c_{2} S_{2}\vphantom{()\np{2}}\right] \sim \nonumber \\
&\quad\quad \sim 1 + \kappa\, C_{1} - \kappa \, c_{1} S_{1}^{2} - \kappa \, c_{2} \, S_{1} S_{2} = \label{eq:freq_order2_rhs} \\
&\quad\quad\quad\quad = 1 - \frac{\kappa c_{1}}{2} + \kappa\,C_{1}(x) - \frac{\kappa c_{1}}{2}\,C_{2}(x) + \bigo{\kappa^{3}} \nonumber
\end{align}
Matching the coefficients of terms in eqs.~(\ref{eq:freq_order2_der}) and (\ref{eq:freq_order2_rhs}), we obtain the coupled equations:
\vspace*{-2pt}
\begin{align}
b_{2}\!\sim - \frac{\kappa c_{1}}{2},~2\left(1 + b_{2}\right) c_{1} \sim \kappa,~4 \left(1 + b_{2}\right) c_{2} \sim \frac{\kappa c_{1}}{2}.
\end{align}
These imply that \(c_{1} \sim \kappa/2 + \bigo{\kappa^{3}}\) and \(b_{2}\!\sim -\kappa^{2}/4 + \bigo{\kappa^{4}}\). The last equation above determines the value of the newly introduced constant \(c_{2}\):
\vspace*{-2pt}
\begin{align}
c_{2} \sim \frac{\kappa c_1}{8\left(1 + b_{2}\right)} \sim \frac{\kappa^{2}}{16\left(1 + b_{2}\right)} \sim \frac{\kappa^{2}}{16} + \bigo{\kappa^{4}}.
\end{align}

To find the constant \(a\) at second order, we need to match this solution to the \(\bigo{\kappa^{2}}\) part of solution (\ref{eq:FrequencySolutionSmallX}) in the buffer zone where both solutions are valid, as per eq.~(\ref{eq:FrequencySecondOrderMatching}). Expanding (\ref{eq:FrequencySolutionSmallX}) for \(x \gg 1\) and the above ansatz for \(\kappa |x| \ll 1\) (both to order \(\bigo{x^{-1}}\) inclusive), we find that all functional term match on both sides. The constant term is the value of \(a\) at second order, \(\big(\!\mp \spi / 16\big) \kappa^{2}\).

Combining these results together, we establish the improved solution \(\omega_{2}(x)\), given by eq.~(\ref{eq:FrequencySolution2}).

\begin{widetext}
\vspace*{-5pt}
\begin{align}
\left. \frac{\sqrt{2 \pi}}{4}\,\cos\!\big(x^{2}\big)\,C\!\left(\!\frac{\sqrt{2}\,x}{\spi}\!\right)\!- \frac{\spi}{8}\,C\!\left(\!\frac{\sqrt{4}\,x}{\spi}\!\right)\!-\frac{1}{4}\,x\,\right|_{x \gg 1} \!\! = a - \frac{1}{4}\,x + \frac{\sqrt{2 \pi}}{8}\,\cos\!\big(x^{2}\big) + \frac{1}{16}\,\frac{\sin\!\big(2x^{2}\big)}{x^{2}} + \bigo{x^{-2}}
\label{eq:FrequencySecondOrderMatching}
\end{align}
\begin{m}{align}
&\omega_{2} (x) = \pm\,\frac{\sqrt{2 \pi}}{4}\,\kappa \mp \frac{\spi}{16}\,\kappa^{2} +\!\Bigg(\!1 - \frac{\kappa^{2}}{4}\!\Bigg) x + \kappa\,\frac{1}{2}\,\frac{1}{x} \, \sin\!\left[\!\vphbig\left(\!\pm\,\frac{\sqrt{2 \pi}}{4}\,\kappa\!\right)\!x + x^{2}\right]\!+ \kappa^{2}\,\frac{1}{16}\,\frac{1}{x} \, \sin\!\left(2 x^{2}\right) \label{eq:FrequencySolution2}
\end{m}
\end{widetext}

At the next order, the task of finding appropriate terms becomes more complicated, since we need to balance the terms in eq.~(\ref{eq:frequencySmallXThOrdAsymp}). At this order there is a logarithmic trend, as well as an oscillatory part, hence we propose the following ansatz which reflects the nature of the solution:
\begin{align}
\omega_{3} (x) =  a &+ \left(1 + b_{3}\right)x + c_{1} \frac{S_{1}(x)}{x} + c_{2} \frac{S_{2}(x)}{x} + c_{3} \frac{S_{3}(x)}{x} \nonumber\\
& +  d_{1} \frac{S_{1}(x)}{x^{2}} + d_{2} \frac{S_{2}(x)}{x^{2}} + l \ln\!|x|. \label{eq:FreqThirdOrderAnsatz}
\end{align}
The details of calculating the expressions for these constants are burdensome, hence we have left them in Appendix~\ref{sec:AppFrequencyThirdOrderSMNP}. The important message here is that we have still managed to find a set of terms which satisfy the governing equation~(\ref{eq:FrequencySMEq}), without conflicting the previous solutions (i.e. without producing new terms at \(\bigo{\kappa, x^{-1}}\)).
\begin{figure}[t]
\includegraphics[scale=1]{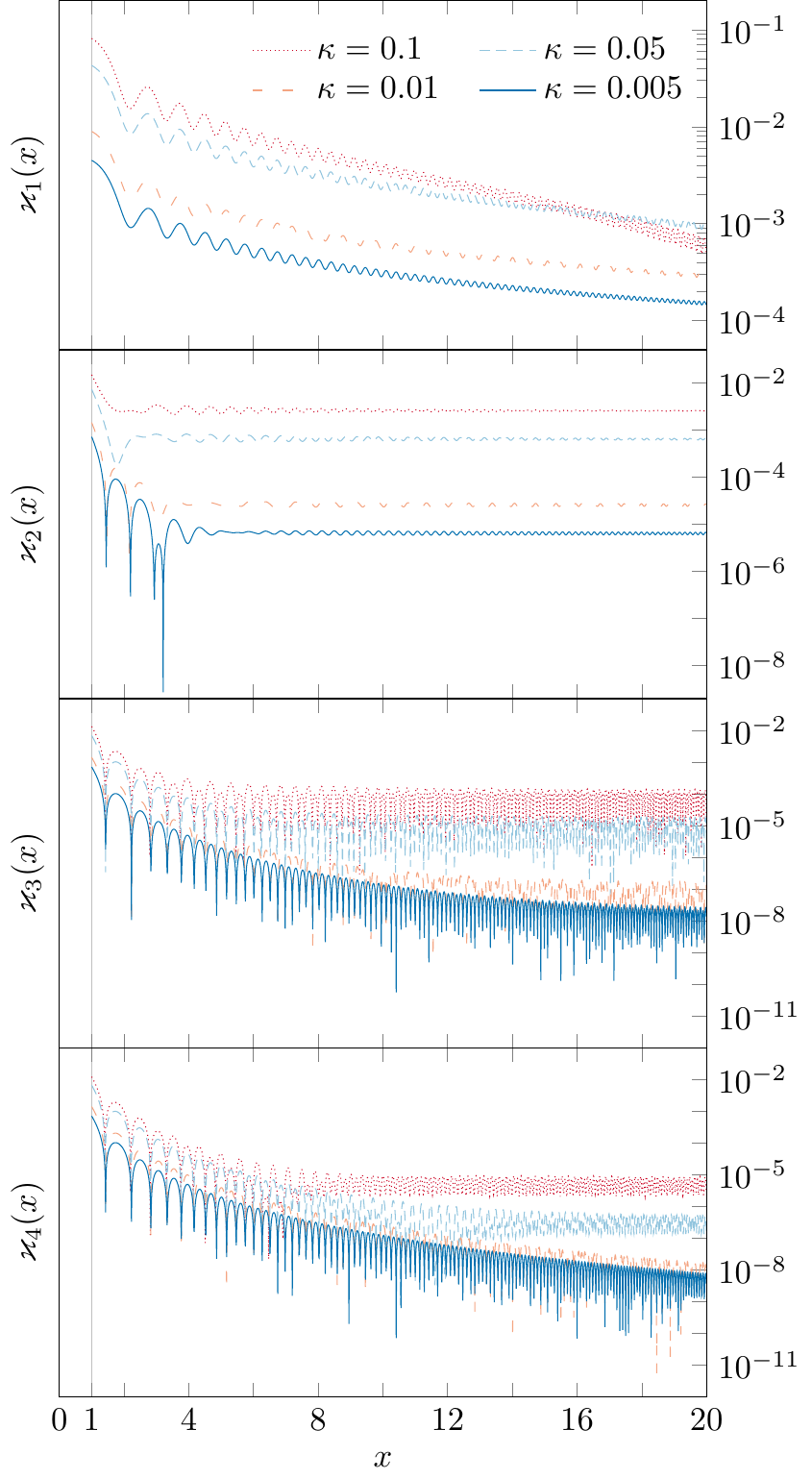}
\caption{Graphical demonstration of the validity of solutions~(\ref{eq:frequencySolution0}) -- (\ref{eq:FrequencySolution3}). The fractional errors scale with the value of the resonance modification \(\kappa\). At each order \(n\) of the solution, the remainder is of order \(\bigo{\kappa^{n+1}}\), as expected.}
\label{fig:frequencyValidity}
\vspace*{-10pt}
\end{figure}
To determine the new value of the constant \(a\), we again need to turn to our solution near the origin. In particular, here we need an expansion of \(\varpi_{3} (x)\) for large \(x\). However, we used a small-\(x\) approximation to obtain \(\varpi_{3a_{\rm asymp}} (x)\) earlier, which is not valid in the buffer zone where both solutions~(\ref{eq:FrequencySolutionSmallX}) and (\ref{eq:FreqThirdOrderAnsatz}) are valid. We need to go back to the original problem (\ref{eq:frequencyThOrdAIntegral}), and instead of expanding in the small-$x$ regime, we expand for \(|x| \gg 1\) (which is valid when \(\kappa |x| \ll 1\)). To do this, we expand the Fresnel term on the right hand side of eq.~(\ref{eq:frequencyThOrdAIntegral}) for large $|x|$, truncate the series at \(\bigo{x^{-5}}\) and then compute the antiderivative of this series, the function $\varpi_{3a_{\rm asymp}}$ given in eq. (\ref{eq:frequencyThOrdAAsymp}). This procedure ensures that \(\varpi_{3a} (x) - \varpi_{3a_{\rm asymp}} (x) \sim \alpha + \bigo{x^{-4}}\). The constant $\alpha$ is the asymptotic value of  \(\varpi_{3a} (x) - \varpi_{3a_{\rm asymp}} (x)\), which may be found by integrating  \(\varpi_{3a} (x) - \varpi_{3a_{\rm asymp}} (x)\) from $0$ to infinity
\begin{align}
\begin{split}
\alpha =& \int_{0}^{x_c} \!\!\!\! \dd y \,  \varpi_{3a}^{\prime} (y) - \varpi_{3a_{\rm asymp}}(x_c) \\
& \RepQuad{2} +\! \int_{x_c}^{\infty} \!\!\!\! \dd y\,\big(\varpi_{3a}^{\prime}(y) - \varpi_{3a_{\rm asymp}}^{\prime} (y)\big).
\end{split} \label{eq:FrequencyNumericalIntegralResult}
\end{align}
The quantity \(x_{c}\) is arbitrary, but by choosing a value \(x_c \sim 1/k_{0} \sim 10\) we may bound the value of the second integral by \(10^{-5}\). Numerical evaluation of the terms on the first line of eq.~(\ref{eq:FrequencyNumericalIntegralResult}) then give \(\alpha = \mp\,0.00991\). We deduce that, asymptotically, we may write \(\varpi_3(x) = \alpha +  \varpi_{3a_{\rm asymp}}(x) + \varpi_{3b}(x)\), with higher-order corrections which scale as \(x^{-4}\).

Combining these resulting terms with the ones coming from eq.~(\ref{eq:frequency_small_x_y3_second_term}), we arrive at the left-hand side of eq.~(\ref{eq:FrequencyThirdOrderMatching}). The right-hand side of eq.~(\ref{eq:FrequencyThirdOrderMatching}) is comprised of the terms in an expansion of the ansatz~(\ref{eq:FreqThirdOrderAnsatz}) for small \(\kappa |x|\). 

The terms match nicely as we would expect, apart from the coefficients of \(\sin\!\big(x^2\big)/x\) and \(\cos\!\big(x^2\big)/x\). The reason for this is that to determine these terms accurately we need to extend the asymptotic solution to the next order in \(1/x\). At that order we get a term of the form \(C_1(x)/x^{3}\) in the solution, and it is straightforward to deduce that, at leading order, the coefficient in front of this term is \(-\,\kappa/4\). This additional term contributes terms at order \(\kappa^3\) on the right hand side of eq.~(\ref{eq:FrequencyThirdOrderMatching}) and those terms are \(\pi \cos\!\big(x^2\big)/(64 x)\) and \(-\sin\!\big(x^2\big)/(16 x)\), which are exactly what is needed to match the terms from the \(\kappa |x| \ll 1\) solution on the left hand side of eq. (\ref{eq:FrequencyThirdOrderMatching}).

We can now read off the third-order piece of the coefficient \(a\) from eq. (\ref{eq:FrequencyThirdOrderMatching})
\begin{align}
a_{3} = \left(\!\alpha \pm \frac{\sqrt{2\pi}}{32} \pm \frac{\sqrt{6\pi}}{192} \pm \frac{\sqrt{2}\,\pi^{\sfrac{3}{2}}}{128}\!\right)\!\kappa^{3} \approx \pm\,0.15255\,\kappa^3
\end{align}
The third-order solution \(\omega_{3}(x)\) is given in eq.~(\ref{eq:FrequencySolution3}). This is the most accurate solution to the single-mode, zero-phase frequency equation that we will present. The lack of necessity for higher order iterations will be explained when discussing the multi-mode frequency equation.

Once we have obtained these solutions, it is worthwhile to investigate their properties visually. To confirm their validity, Figure~\ref{fig:frequencyValidity} demonstrates that these solutions are correct at their respective order by computing the quantities \(\varkappa_{n} (x)\) defined in eq.~(\ref{eq:frequencyValidityFormula}), and showing graphically that they have the expected magnitudes. Please note we have shown only the region \(x > 0\), with the region \(x < 0\) being a mirror image of this one.

Figure~\ref{fig:frequency_features_single_mode} shows how the intercept \(a\), the slope \((1 - b)\), and the coefficients of the oscillating modes \(\{c_{n}\}\) and \(\{d_{n}\}\) vary with the parameter \(\kappa\).
\begin{widetext}
\vspace*{-15pt}
\begin{align}
\varpi_{3a_{\rm asymp}} (x) =&\pm\frac{\sqrt{2\pi}}{32}\,\ln\!|x| - \frac{\sqrt{2\pi}}{64}\,C\!\left(\!\frac{\sqrt{2}\,x}{\spi}\!\right)\!+ \frac{\sqrt{2}\,\pi^{\sfrac{3}{2}}}{64}\,S\!\left(\!\frac{\sqrt{2}\,x}{\spi}\!\right)\!+ \frac{\sqrt{6 \pi}}{192}\,C\!\left(\!\frac{\sqrt{6}\,x}{\spi}\!\right)\!- \frac{\pi}{32}\,x \sin\!\big(x^{2}\big) \pm \frac{\sqrt{2\pi}}{64}\,\cos\!\big(2x^{2}\big)\nonumber\\
& \pm \frac{\sqrt{2\pi}}{128}\,\frac{\sin\!\big(2x^{2}\big)}{x^{2}} - \frac{1}{64}\,\frac{\cos\!\big(x^{2}\big)}{x^{3}} - \frac{1}{192}\,\frac{\cos\!\big(3 x^{2}\big)}{x^{3}} \mp \frac{\sqrt{2\pi}}{64}\,\frac{\cos\!\big(2x^{2}\big)}{x^{4}} - \frac{3}{256}\,\frac{\sin\!\big(x^{2}\big)}{x^{5}} - \frac{3}{256}\,\frac{\sin\!\big(3 x^{2}\big)}{x^{5}} \label{eq:frequencyThOrdAAsymp}
\end{align}
\vspace*{-15pt}
\begin{center}
\noindent\begin{tabular}{P{0.48\linewidth} m{0.05\linewidth} P{0.4\linewidth}}
\noindent\parbox[c]{\hsize}{\begin{align*}\alpha \pm \frac{\sqrt{2\pi}}{32} \pm \frac{\sqrt{6\pi}}{192} \pm \frac{\sqrt{2}\,\pi^{\sfrac{3}{2}}}{128} \pm \frac{\sqrt{2 \pi}}{32}\,\ln\!|x| - \frac{\pi}{32}\,x\,\sin\!\left(x^{2}\right) \\
- \frac{1}{8}\,x\,\cos\!\left(x^{2}\right) \mp \frac{\spi}{32}\,\cos\!\left(x^{2}\right) \pm \frac{\sqrt{2 \pi}}{32}\,\cos\!\left(2 x^{2}\right) \\
+ \frac{1}{32}\,\frac{\sin\!\left(x^{2}\right)}{x} + \frac{1}{96}\,\frac{\sin\!\left(3 x^{2}\right)}{x} - \frac{\pi}{64}\,\frac{\cos\!\left(x^{2}\right)}{x}
\end{align*}}
& \(\quad = \) &
\noindent\parbox[c]{\hsize}{\begin{align}&a_{3} \mp \frac{\pi}{32}\,x\,\sin\!\left(x^{2}\right) \mp \frac{\spi}{32}\,\cos\!\left(x^{2}\right) - \frac{1}{8}\,x\,\cos\!\left(x^{2}\right) \nonumber\\
& \pm \frac{\sqrt{2 \pi}}{32}\,\cos\!\left(2 x^{2}\right) + \frac{3}{32}\,\frac{\sin\!\left(x^{2}\right)}{x} + \frac{1}{96}\,\frac{\sin\!\left(3 x^{2}\right)}{x} \nonumber \\
& - \frac{\pi}{32}\,\frac{\cos\!\left(x^{2}\right)}{x} \pm \frac{\sqrt{2 \pi}}{32}\,\ln\!|x| \label{eq:FrequencyThirdOrderMatching} \end{align}}
\end{tabular}
\end{center}
\vspace*{-20pt}
\begin{m}{align}
&\omega_{3} (x) = \pm \left(\!\frac{\sqrt{2\pi}}{4}\,\kappa - \frac{\spi}{16}\,\kappa^{2} + 0.15255\,\kappa^3\!\right)\!+\!\left(\!\vph 1 - \frac{\kappa^{2}}{4}\!\right)\!x + \kappa\,\frac{1}{2}\,\frac{1}{x}\, \sin\!\left[\vphbig\!\left(\!\pm\frac{\sqrt{2\pi}}{4}\,\kappa \mp \frac{\spi}{16}\,\kappa^{2}\!\right)\!x +\!\Bigg(\!1 - \frac{\kappa^{2}}{4}\!\Bigg)x^{2}\right] \nonumber\\
& \RepQuad{4} + \kappa^{2}\frac{1}{16}\,\frac{1}{x} \, \sin\!\left[\vphbigg 2\!\left(\!\!\vphbig \left(\!\pm\,\frac{\sqrt{2 \pi}}{4}\,\kappa\!\right)\!x + x^{2}\!\right)\!\right]\!+ \kappa^{3}\,\frac{1}{x}\!\left(\!\vph\frac{3}{32}\,\sin\!\left(x^{2}\right) + \frac{1}{96}\,\sin\!\left(3 x^{2}\right)\!\right) \label{eq:FrequencySolution3}\\
& \RepQuad{8} \mp \kappa^{2}\frac{\sqrt{2 \pi}}{16}\,\frac{1}{x^{2}} \, \sin\!\left[\vphbig\!\!\vphbig \left(\!\pm\,\frac{\sqrt{2 \pi}}{4}\,\kappa\!\right)\!x + x^{2}\right]\!\pm \kappa^{3}\!\left(\!\vph\frac{\spi}{64}\,\frac{1}{x^{2}}\,\sin\!\left(x^{2}\right) - \frac{\sqrt{2 \pi}}{64}\,\frac{1}{x^{2}}\,\sin\!\left(2 x^{2}\right) + \frac{\sqrt{2 \pi}}{32}\,\ln\!|x|\!\right) \nonumber
\end{m}
\vspace*{-10pt}
\end{widetext}

\begin{table*}[t]
\includegraphics[scale=1]{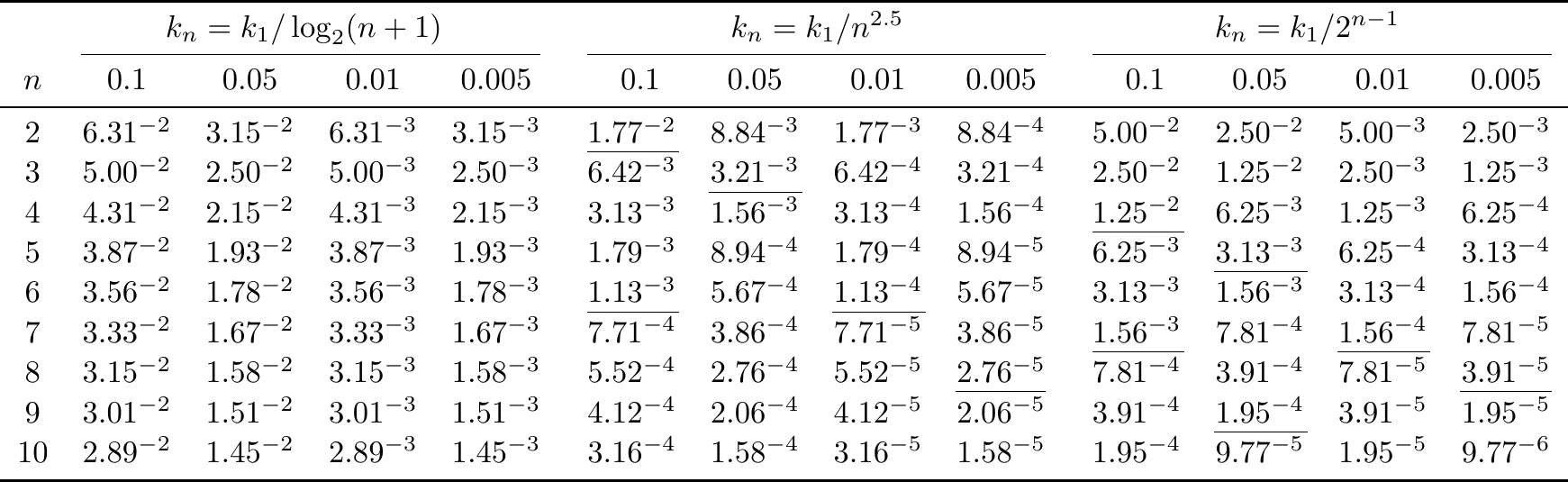}
\caption{The first 10 terms of the series \(\{k_{n}\}\) for the three different series of interest, for each of our 4 chosen numerical values: \(0.1, 0.05, 0.01\), and \(0.005\). These were plotted in Figure~\ref{fig:ladders_plot}. It should be noted that the numbers in this table are given using the scientific notation \(\overline{a.bc}^{\,d} \equiv \overline{a.bc}\times\!10^{d}\). The horizontal lines in the data represent numerical boundaries: numbers below the first line in a column are smaller than \(k_{i}^{2}\), and numbers below the second line are smaller than \(k_{i}^{3}\).}
\label{tab:ladder_values}
\end{table*}

The solution can be made more accurate, by adding terms at higher orders, of the form \(c_{n} S_{n} (x)/ x\) and \(d_{j} S_{n} (x) / x^{2}\) to the ansatz~(\ref{eq:freq_ansatz_order2}). Below we see the general case of having \(m\) oscillatory terms in the trial solution \(\omega_{m}(x)\), where \(m\) could be either a finite number or infinity:
\begin{align}
\omega_{m} (x) = a +\left(1 + b\right)x + \sum_{j = 1}^{m} \left(c_{j} \frac{S_{j} (x)}{x} +  d_{j} \frac{S_{j} (x)}{x^{2}} \right).
\end{align}
We shall discuss later (after we solve the multimode version of the phase equation) why there is no practical need for solutions beyond third order. If we seek to find lower-order corrections in \(\bigo{x^{-p}}\), we add terms of the form:
\begin{align}
\sim \frac{1}{x^{p}} \sum_{j = 1}^{m} \left(c_{j}^{(p)} S_{j} (x) +  d_{j}^{(p)} C_{j} (x)\right).
\end{align}

However, instead of following through with this procedure, we look at extensions of eq.~(\ref{eq:FrequencySMEq}) to find the general solution of eq.~(\ref{eq:frequency_final}). Initially, we allow for multiple oscillation modes, each with a different amplitude (parameter \(\kappa_{i}\)), and following this, we include a non-vanishing phase in our calculations.

In the following we will use bold quantities, e.g., $\bom(x)$, to indicate values of the coefficients in the multiple mode case. We note that these are not vectors, but the scalars that appear in the multiple mode solution. To solve for the evolution of the frequency resonances,
\begin{align}
\bom^{\prime} (x) = 1 + \sum_{n} \kappa_{n} \cos \left(n x_{\omega} \, \bom\right) \label{eq:freq_multi_modes}
\end{align}
we can again look at the problem from two scales: small and large \(|x|\). Near the origin, we can write down a series expansion similar to eq.~(\ref{eq:frequency_ansatz_small_x}):
\begin{align}
\bom (x) = \sum_{m = 0}^{\infty} \sum_{n = 0}^{\infty}\,\kappa_{n}^{m} \bm{\varpi}_{m, n} (x). \label{eq:frequencyMMSmallX}
\end{align}
Similarly to before, we can derive solutions for \(n = 0, 1\):
\begin{subequations}
\begin{align}
\bm{\varpi}_{0, n} (x) &= x \\
\bm{\varpi}_{1, n} (x) &= \frac{\spi}{\sqrt{2n}}\,C\!\left(\!\frac{\sqrt{2n}\,x}{\spi}\!\right)
\end{align}
\end{subequations}
Further solutions are not going to be pursued, as higher orders of \(\kappa_{n}\) (for small \(n\)) would be of the same size as \(\kappa_{n}\) for some larger \(n\). Hence, \(\bom (x)\) near the origin can be most completely written down as
\begin{m}{align}
&\bom (x) = x + \sum_{n} \kappa_{n} \frac{\spi}{\sqrt{2n}}\,C\!\left(\!\frac{\sqrt{2n}\,x}{\spi}\!\right)  \label{eq:frequencyMMSolutionSmallX}
\end{m}

For \(|x| \gg 1\), we propose an extended ansatz of the form:
\begin{align}
\bom_{1} (x) = \bm{a} +\left(1 + \bm{b}_{1}\right) x + \sum_{n} \bm{c}_{n} \, \frac{S_{n} (x)}{x} \label{eq:multi_mode_ansatz}
\end{align}
with derivative
\begin{align}
\bom_{1}^{\prime} = 1 + \bm{b}_{1} + \sum_{n} 2n\!\left(1 + \bm{b}\right) \bm{c}_{n} C_{n} (x) + \bigo{x^{-1}}.
\end{align}
We substitute this into the governing eq.~(\ref{eq:freq_multi_modes}) and proceed to match the relevant coefficients on both sides of this relation. It is important to explain here that the leading order dependence of each coefficient \(\bm{c}_{n}\) will be the same as if considering its corresponding mode on its own (see Appendix~\ref{sec:freq_extension_simple} for derivation), with higher-order corrections due to cross-terms emerging from the cosine term (see Appendix~\ref{sec:freq_extension_double}), and therefore \(\bm{c}_{n} \sim \bigo{\kappa_{n} / n} \sim \bigo{\kappa_{1} / n^{\alpha+1}}\). Thus, each coefficient \(\bm{c}_{n}\) is smaller than the previous one in the series, and we can expand the trigonometric terms on the right-hand side, to obtain the form (\ref{eq:frequency_multimode_expansion}). Evidently, to be able to match the constant term \(\bm{b}\) on the left-hand side, we need to find a constant term on the right-hand side too. Such a constant could only arise from the second term on the right-hand side, in the case \(n = m\), and yields eq.~(\ref{eq:frequency_multimode_b}). The remaining terms must combine to form an expression for \(\bm{c}_{n}\), and we match them using the linear independency of the functions \(\{C_{n} (x)\}\), yielding eq.~(\ref{eq:frequency_multimode_c_n_init}). On the right-hand side of this equation, we omitted a pre-factor of \((1+\bm{b})^{-1} \sim 1 + \bigo{\kappa_{1}^{2}}\), since we know that \(\bm{b} \sim \bigo{\kappa_{1}^{2}}\).
\begin{widetext}
\begin{minipage}{0.985\textwidth}
\begin{align}
\bm{b} + 2(1 + \bm{b}) \sum_{n} n \, \bm{c}_{n} C_{n} (x) \sim \sum_{n}  \kappa_{n} \, C_{n} (x) - \frac{1}{2} \sum_{n,\,m} n \, \kappa_{n} \, \bm{c}_{m} C_{n-m}(x) + \frac{1}{2} \sum_{n,\,m} n \, \kappa_{n} \, \bm{c}_{m} C_{n + m}(x) \label{eq:frequency_multimode_expansion}
\end{align}
\end{minipage}\\ [5pt]
\begin{minipage}{.25\textwidth}
\begin{align}
\bm{b} \sim -\,\frac{1}{2} \sum_{n} n \, \kappa_{n} \, \bm{c}_{n} \label{eq:frequency_multimode_b}
\end{align}
\end{minipage}
\begin{minipage}{.75\textwidth}
\begin{align}
\bm{c}_{n} \sim \frac{\kappa_{n}}{2n} - \frac{1}{4n} \sum_{m,\,p} m \, \kappa_{m} \, \bm{c}_{p} \, \delta_{|m-p|,\,n} + \frac{1}{4n} \sum_{m,\,p} m \, \kappa_{m} \, \bm{c}_{p} \, \delta_{(m+p),\,n} \label{eq:frequency_multimode_c_n_init}
\end{align}
\end{minipage}
\vspace*{5pt}
\end{widetext}
We now proceed to demonstrate that the second and third terms in the relation (\ref{eq:frequency_multimode_c_n_init}) are sub-leading compared to the first term, and therefore can be ignored in the final expression. To do this, we remind ourselves of our assumption that \(\kappa_{n} \sim \kappa_{1} / n^{\alpha}\), with \(\alpha \gtrsim 2\) and substitute this into our expression for \(\bm{c}_{n}\). From the first term we can see that \(\bm{c}_{n} \sim \bigo{\kappa_{n}/n}\). Using this, the formula for \(\bm{c}_{n}\) can be written down solely in terms of~\(\kappa_{1}\). After applying the action of the Kr\"{o}necker delta operators, the second and third terms of eq.~(\ref{eq:frequency_multimode_c_n_init}) become
\begin{align}
\Xi_{n} (\kappa_{1}) = -\,\frac{\kappa_{1}^{2}}{8n}\,\sum_{p = 1}^{\infty} \frac{1}{(n+p)^{\alpha - 1}\,p^{\alpha + 1}}\,\delta_{|m-p|,\,n} \label{eq:frequency_term_1}
\end{align}
and
\begin{align}
\Upsilon_{n} (\kappa_{1}) = \frac{\kappa_{1}^{2}}{8n}\,\sum_{m = 1}^{n-1} \frac{1}{m^{\alpha - 1} (n-m)^{\alpha + 1}} \, \delta_{(m+p),\,n}. \label{eq:frequency_term_2}
\end{align}
We need to inspect and bound each of these terms, in order to show they are sub-leading compared to the first term in eq.~(\ref{eq:frequency_multimode_c_n_init}), \(\kappa_{1} / n^{\alpha + 1}\). In eq.~(\ref{eq:frequency_term_1}), we split off the \(p=1\) term from the sum, and bound the remaining terms from above by an integral over the index \(p\):
\begin{align}
\sum_{p = 1}^{\infty} \frac{1}{(n+p)^{\alpha - 1}\,p^{\,\alpha + 1}} < \frac{1}{(n+1)^{\alpha - 1}} +\!\!\int_{1}^{\infty}\!\!\!\!\!\!\frac{\dd p}{(n+p)^{\alpha - 1}\,p^{\,\alpha + 1}}
\end{align}
We re-scale by writing \(p = n \tilde{p}\), and the integral becomes:
\begin{align}
\frac{1}{n^{2\alpha - 1}} \int_{\sfrac{1}{n}}^{\infty} \frac{\dd \tilde{p}}{(1+\tilde{p})^{\alpha - 1}\,\tilde{p}^{\,\alpha + 1}}.
\end{align}
We split the integration region in two intervals: from \((1/n)\) to \(1\), and from \(1\) to \(\infty\). The first of the resulting integrals can be bounded as follows
\begin{align}
\int_{\sfrac{1}{n}}^{\infty} \frac{\dd \tilde{p}}{(1+\tilde{p})^{\alpha - 1}\,\tilde{p}^{\,\alpha + 1}} < \int_{\sfrac{1}{n}}^{\infty} \frac{\dd \tilde{p}}{\tilde{p}^{\,\alpha + 1}} = \frac{n^{\alpha} - 1}{\alpha}, \label{eq:xi_bound}
\end{align}
while the second one can be solved and the solution expressed in terms of the hypergeometric function \(_{2}F_{1}\), independently of the integration variable \(\tilde{p}\). We find that the term \(\Xi_{n} (\kappa_{1})\) is bounded above by (\(\xi_{i}\) are constants):
\begin{align}
\Xi_{n} (\kappa_{1}) < -\,\frac{\kappa_{1}^{2}}{8n} \Bigg(\!\frac{1}{(n+1)^{\alpha - 1}} + \frac{\xi_{1}}{n^{2\alpha - 1}} + \frac{\xi_{2}}{n^{\alpha - 1}}\!\Bigg)\!\sim \frac{\kappa_{1}^{2}}{n^{\alpha}}.
\end{align}
Considering the term in eq.~(\ref{eq:frequency_term_2}), we can find a bound using a similar argument, albeit the reasoning is slightly longer, as the sum has definite bounds on both ends, and the integrand is not monotonic for the interval of integration. We strip off the first and last terms of the sum: \((n-1)^{-\alpha + 1}\) and \((n+1)^{-\alpha-1}\), respectively. The remainder can be bound from above by an integral,
\begin{widetext}
\vspace*{30pt}
\begin{minipage}{\textwidth}
\begin{minipage}[t]{0.45\textwidth}
\begin{figure}[H]
\includegraphics[scale=1]{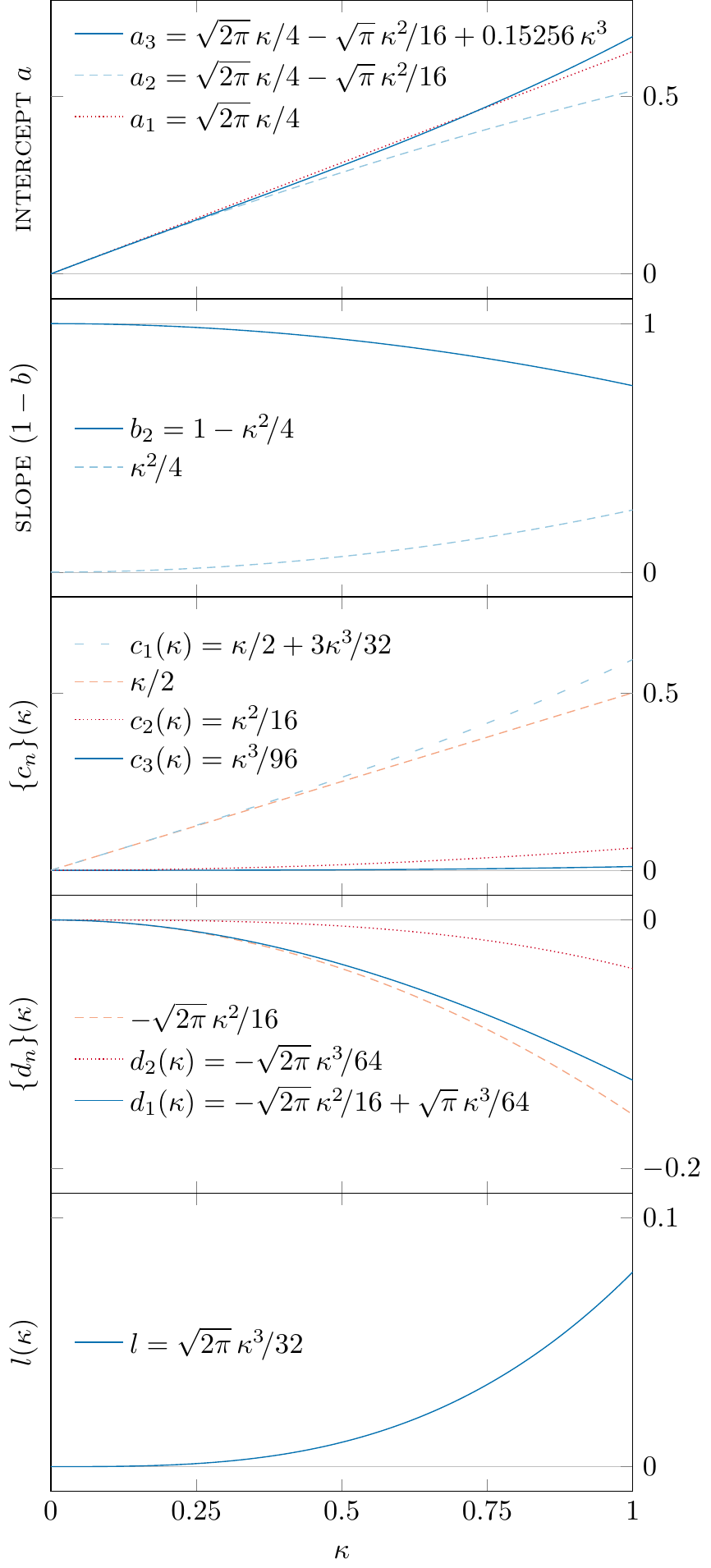}
\caption{Features and coefficients of solutions (\ref{eq:FrequencySolution1}) -- (\ref{eq:FrequencySolution3}) as functions of the mode parameter \(\kappa\).}
\label{fig:frequency_features_single_mode}
\end{figure}
\end{minipage}
\hspace*{\columnsep}
\begin{minipage}[t]{0.45\textwidth}
\begin{figure}[H]
\includegraphics[scale=1]{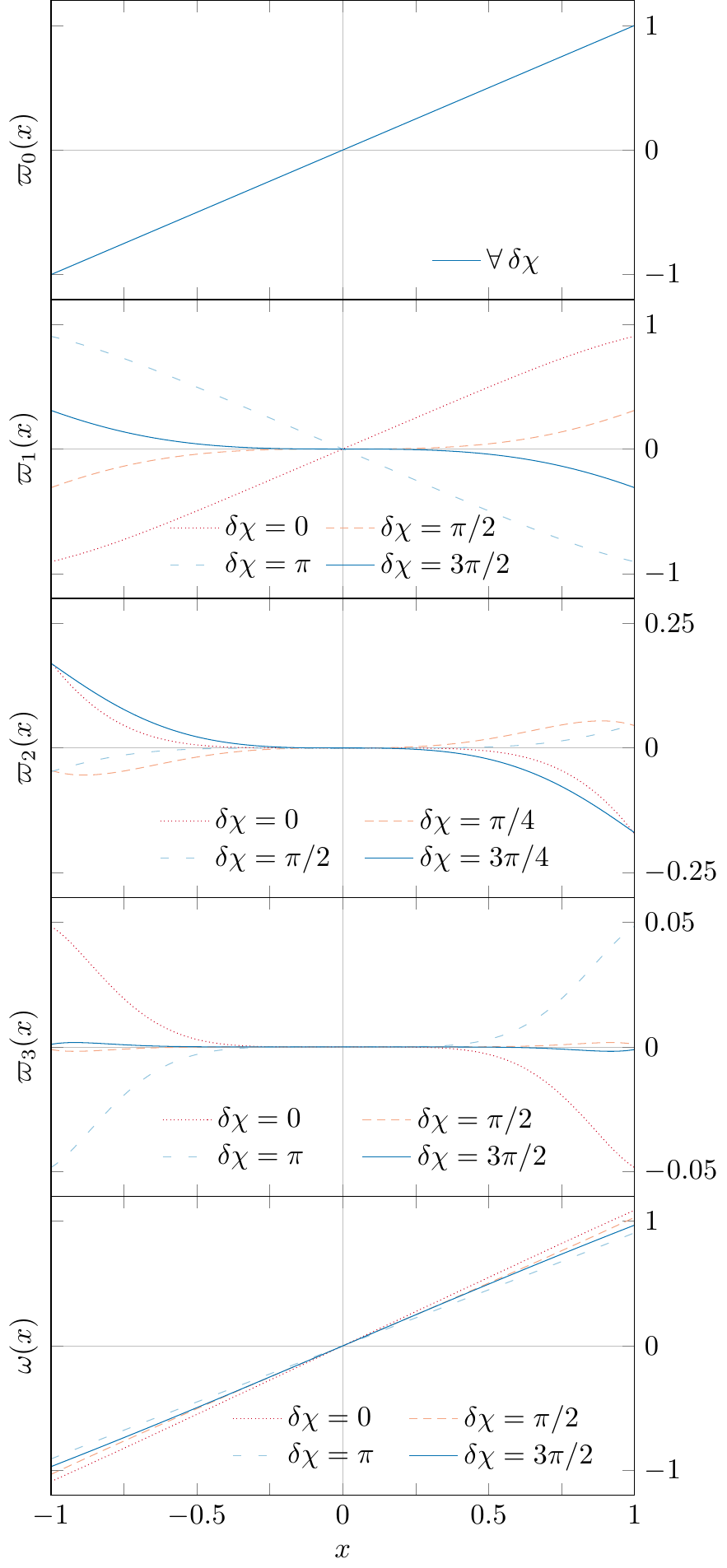}
\caption{Variation of the components of the solution to the frequency resonance equation for different values of the phase angle. The bottom panel uses \(\kappa = 0.1\).}
\label{fig:frequencySolutionSmallXPhasePlot}
\end{figure}
\end{minipage}
\end{minipage}
\newpage
\end{widetext}
and a suitable substitution can be applied to obtain:
\begin{align}
\frac{1}{n^{2\alpha - 1}} \int_{\sfrac{1}{n}}^{1-\sfrac{1}{n}}\!\!\!\!\frac{\dd \tilde{p}}{(1-\tilde{p})^{\alpha - 1}\,\tilde{p}^{\,\alpha + 1}}.
\end{align}
We proceed to split the integral for some \(n > 1\), say 2:
\begin{align}
\int_{\sfrac{1}{n}}^{\sfrac{1}{2}}\!\!\frac{\dd \tilde{p}}{(1-\tilde{p})^{\alpha - 1}\,\tilde{p}^{\,\alpha + 1}} + \int_{\sfrac{1}{2}}^{1-\sfrac{1}{n}}\!\!\!\!\!\!\!\!\frac{\dd \tilde{p}}{(1-\tilde{p})^{\alpha - 1}\,\tilde{p}^{\,\alpha + 1}}
\end{align}
Each of the two integrals can be bounded as follows:
\begin{subequations}
\begin{align}
\int_{\sfrac{1}{n}}^{\sfrac{1}{2}}\!\!\!\frac{\dd \tilde{p}}{(1-\tilde{p})^{\alpha - 1}\,\tilde{p}^{\,\alpha + 1}} < \frac{1}{(1-\sfrac{1}{2})^{\alpha - 1}}\int_{\sfrac{1}{n}}^{\sfrac{1}{2}}\!\!\frac{\dd \tilde{p}}{\tilde{p}^{\,\alpha + 1}} \label{eq:upsilon_bound1}\\
\int_{\sfrac{1}{2}}^{1-\sfrac{1}{n}}\!\!\!\!\!\!\!\!\!\frac{\dd \tilde{p}}{(1-\tilde{p})^{\alpha - 1}\,\tilde{p}^{\,\alpha + 1}} < \frac{1}{(\sfrac{1}{2})^{\alpha + 1}}\int_{\sfrac{1}{2}}^{1-\sfrac{1}{n}}\!\!\!\!\!\!\!\!\!\frac{\dd \tilde{p}}{(1-\tilde{p})^{\alpha - 1}}. \label{eq:upsilon_bound2}
\end{align}
\end{subequations}
Evaluating these and putting everything back together, we conclude that a bound on the term~(\ref{eq:frequency_term_2}) is given by
\begin{widetext}
\begin{align}
\Upsilon_{n} (\kappa_{1}) = \frac{\kappa_{1}^{2}}{8n}\,\sum_{m = 1}^{n-1} \frac{1}{m^{\alpha - 1} (n-m)^{\alpha + 1}} < \frac{\kappa_{1}^{2}}{8n} \Bigg(\!\frac{1}{(n-1)^{\alpha - 1}} + \frac{1}{(n+1)^{\alpha + 1}} + \frac{\upsilon_{1}}{n^{\alpha - 1}} + \frac{\upsilon_{2}}{n^{\alpha + 1}} + \frac{\upsilon_{3}}{n^{2\alpha - 1}}\!\Bigg)\!\sim \frac{\kappa_{1}^{2}}{n^{\alpha}}
\end{align}
\end{widetext}
where \(\upsilon_{i}\) are constants. We have found that the latter two terms in eq.~(\ref{eq:frequency_multimode_c_n_init}) are of the order of \(\kappa_{1}^{2}/n^{\alpha}\). We still need to determine how this compares to the magnitude of the first term in this result, namely \(\kappa_{1} / n^{\alpha + 1}\):
\begin{align}
\frac{\kappa_{1}}{n^{\alpha + 1}} \gtrsim \frac{\kappa_{1}^{2}}{n^{\alpha}} \quad \Leftrightarrow \quad n \lesssim \kappa_{1}^{-1}. \label{eq:frequency_bound}
\end{align}
We can therefore say that for values of \(n\) smaller than \(\kappa_{1}^{-1}\) the second and third terms are much smaller than the first and can therefore be ignored. For values of \(n\) above this limit all terms are of comparable magnitude. However, this \(\bar{n}\) can be related to the order of the parameters \(\{\kappa_{i}\}\):
\begin{align}
\kappa_{\bar{n}} = \frac{\kappa_{1}}{n^{\alpha}} \sim \kappa_{1}^{1 + \alpha} < \kappa_{1}^{3}.
\end{align}
and so ignoring these higher terms leads to an error smaller than \(\kappa_{1}^{3}\). Therefore, we have discovered that we are free to ignore the terms discussed above, and in general ignore the interference between the magnitude of \(\kappa_{1}\) and the large values of the index \(n\), as long as we do not concern ourselves with corrections beyond third order. We may write:
\begin{align}
\bm{c}_{1}^{(n)} \sim \frac{\kappa_{n}}{2n}.
\end{align}
As before, the constant \(\bm{a}\) in eq.~(\ref{eq:multi_mode_ansatz}) can be determined if we compare our multimode ansatz to solution~(\ref{eq:frequencyMMSolutionSmallX}) in the buffer zone where both solutions are valid. This means expanding (\ref{eq:frequencyMMSolutionSmallX}) for large \(x\), and comparing with eq.~(\ref{eq:multi_mode_ansatz}) (see eq.~(\ref{eq:frequencyMMFrstOrdMatching}). This suggests that the value of \(\bm{a}\) is:
\begin{align}
\bm{a} = \pm \sum_{n} \kappa_{n}\,\frac{\spi}{2 \sqrt{2n}}
\end{align}
Our final solution with this ansatz is given by eq.~(\ref{eq:frequencySolutionMM}).

\begin{figure}[H]
\includegraphics[scale=1]{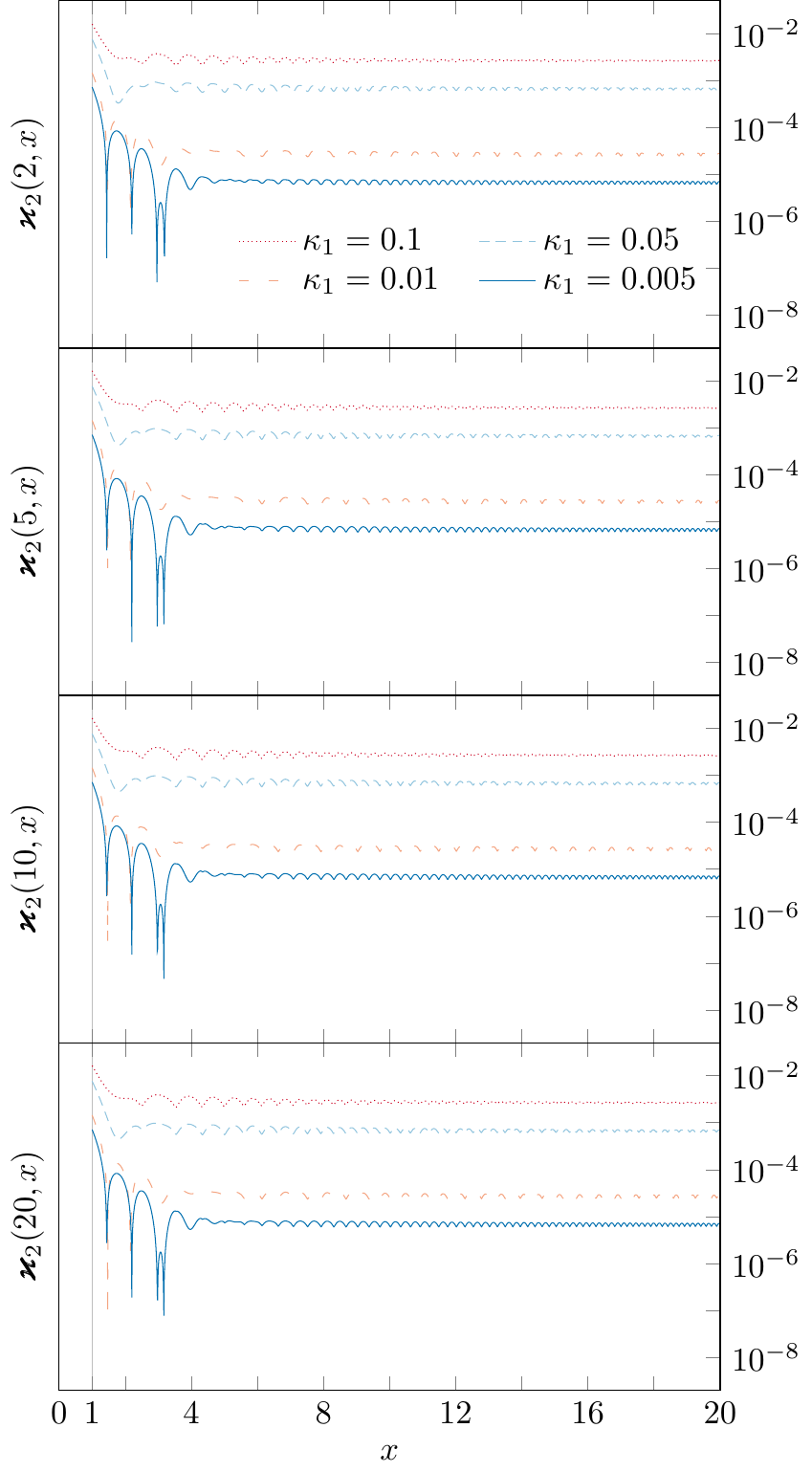}
\caption{Graphical demonstration of the validity of solution~(\ref{eq:frequencySolutionMM}) for different number of modes. Here the coefficients \(\{\kappa_{n}\}\) follow the power law \(\kappa_{1} / n^{2.5}\). The fractional errors scale with the value of the primary resonance modification \(\kappa_{1}\).}
\label{fig:frequencyMMValidityPlot}
\end{figure}

To test the validity of this solution, we devise a test similar to eq.~(\ref{eq:frequencyValidityFormula}), but include the presence of multiple modes in the definition of the quantifier:
\begin{align}
\bm{\varkappa}_{2} (m, x) = \left|\vph\frac{\bm{\omega}_{\text{num}} - \bom^{(1)} (x)}{\bm{\omega}_{\text{num}}}\right|_{m}. \label{eq:frequencyMMValidityFormula}
\end{align}
Here, \(\bm{\omega} (x)\) is given by eq.~(\ref{eq:frequencySolutionMM}), \(\bm{\omega}_{\text{num}}\) is the respective numerical solution, and the subscript \(m\) signifies that all equations and solutions have been limited to the first \(m\) terms only. In Figures~\ref{fig:frequencyMMValidityPlot} and \ref{fig:frequencyMMSmallXValidityPlot} we demonstrate graphically the validity of our solution for \(m = \{2, 5, 10, 20\}\) and in the intervals \(|x| \le 1\) and \(1\le x \le 20\), respectively.

We could try to extend this approach further, by naively including \(S_{2n} (x)\) terms in the ansatz. These terms, however, are redundant, as for \(n\) ranging from 1 to \(+\infty\), the set \(\{2n\}\) is a subset of \(\{n\}\), and therefore the latter terms in the sum can be absorbed into the former one. Thus, eq.~(\ref{eq:multi_mode_ansatz}) is the only ansatz we can use for the equation with an infinite number of modes. Finding higher-order solutions is also complicated by the fact that taking powers of the infinite sum introduces an increasing number of cross-terms.

\begin{widetext}
\begin{align}
x + \sum_{n} \kappa_{n} \frac{\spi}{\sqrt{2n}}\left(\!\pm\,\frac{1}{2} + \frac{\sin\!\big(n x^{2}\big)}{\sqrt{2 \pi n}\,x}\!\right)\!+ \bigo{x^{-2}} \sim \bm{a} + x + \sum_{n} \frac{\kappa_{n}}{2n}\,\frac{1}{x}\,\sin\!\big(n x^{2}\big) \label{eq:frequencyMMFrstOrdMatching}
\end{align}
\begin{m}{align}
&\bom_{1} (x) = \pm \sum_{n} \kappa_{n}\,\frac{\spi}{2 \sqrt{2n}} + x + \sum_{n} \kappa_{n}\,\frac{1}{2n}\,\frac{1}{x}\,\sin\!\big(n x^{2}\big) \label{eq:frequencySolutionMM}
\end{m}
\end{widetext}

\begin{figure*}[t]
\includegraphics[scale=1]{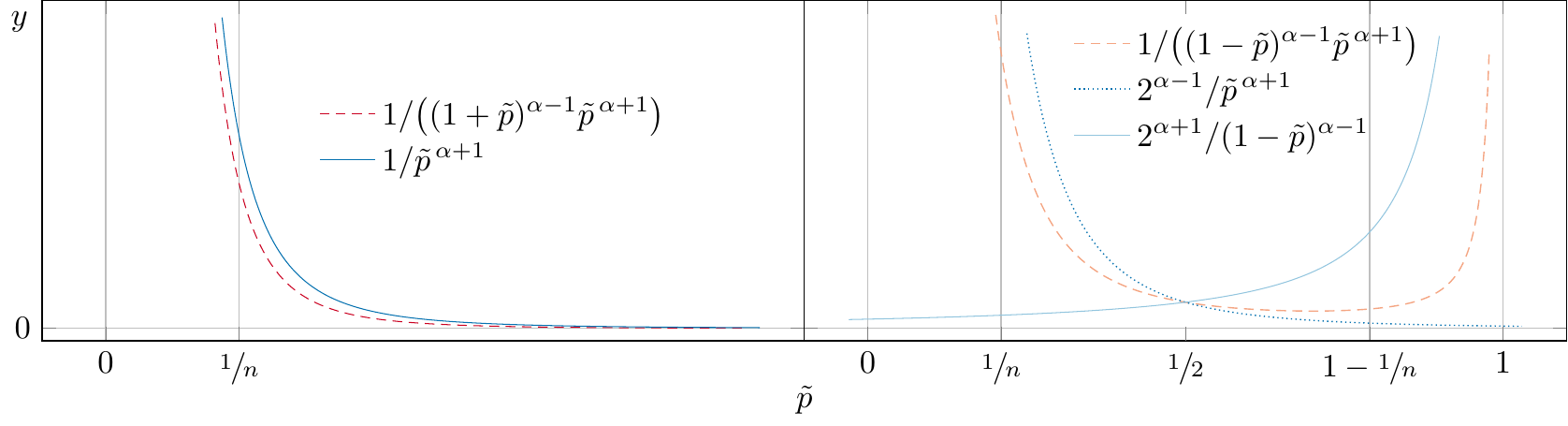}
\caption{Visual representation of the integral bounds we use in the discussion of the multi-mode frequency equation. \emph{Left panel:} Graph of inequality~(\ref{eq:xi_bound}). \emph{Right panel:} Graph of inequalities~(\ref{eq:upsilon_bound1}) and (\ref{eq:upsilon_bound2}). Note that these plots show only the integrands, while the integrals we discuss are given by the area between these curves and the \(\tilde{p}-\)axis.}
\label{fig:integral_bounds_plot}
\vspace*{-10pt}
\end{figure*}

Instead, we return to the single-mode version of the frequency equation and include a non-zero phase \(\delta \chi\):
\begin{align}
\omega^{\prime} (x) = 1 + \kappa \cos\left(x \omega - \delchi\right). \label{eq:FrequencySMEq_phase}
\end{align}
To find the solution for small \(|x|\), we can expand the cosine term on the right-hand side of the equation, and proceed in an analogous way to the zero-phase case:
\begin{align}
\omega^{\prime} (x) = 1 + \kappa \cos\left(\delchi\right) \cos\left(x \omega\right) + \kappa \sin\left(\delchi\right) \sin\left(x \omega\right).
\end{align}
The zeroth-order solution is \(\varpi_{0} (x) = x\), while higher-order solutions will differ from our previous results. The first-order function can be found from
\begin{align}
\varpi_{1}^{\prime} (x) = \cos(\delchi) \cos\!\big(x^{2}\big)\!+ \sin(\delchi) \sin\!\big(x^{2}\big).
\end{align}
After integrating, we arrive at the following result:
\begin{align}
\varpi_{1} (x) = \frac{\spi}{\sqrt{2}}\!\left(\vphbig\!\cos(\delchi)\,C\!\left(\!\frac{\sqrt{2}\,x}{\spi}\!\right)\!+ \sin(\delchi)\,S\!\left(\!\frac{\sqrt{2}\,x}{\spi}\!\right)\!\!\right). \label{eq:frequencySmallXFrstOrderPhase}
\end{align}
The next-order function obeys the differential equation
\begin{align}
\varpi_{2}^{\prime} (x) = -\,x\,\sin\!\big(x^{2} - \delchi\big)\,\varpi_{1} (x),
\end{align}
and its solution is given in eq.~(\ref{eq:frequencySmallXScndOrderPhase}). The third-order function, \(\varpi_{3} (x)\) deserves a little more attention, as was the case for the zero-phase equation. Part of the derivative is again not integrable analytically, but could be approximated for the cases \(|x| \ll 1\) and \(|x| \gg 1\), so we split the solution in two parts, given in eqs.~(\ref{eq:frequencySmallXIntegralAThrdOrdPhase}) and (\ref{eq:frequencySmallXIntegralBThrdOrdPhase}).

\begin{align}
\begin{split}
&\varpi_{3a} (x) = - \frac{\pi}{4} \bigintsss_{0}^{x} \dd y \, y^{2} \cos\!\big(y^{2} - \delchi\big)\,\times \\
&\RepQuad{1} \times\!\left(\!\vphbig\cos^{2}\!\left(\delchi\right) C^{2}\!\!\left(\!\frac{\sqrt{2}\,y}{\spi}\!\right)\!+ \sin^{2}\!\left(\delchi\right) S^{2}\!\!\left(\!\frac{\sqrt{2}\,y}{\spi}\!\right)\!\!\right)
\end{split} \label{eq:frequencySmallXIntegralAThrdOrdPhase}
\end{align}
and
\begin{align}
& \varpi_{3b} (x) = - \frac{\spi}{4 \sqrt{2}} \bigintsss_{0}^{x} \dd y \, y \sin\!\big(2 y^{2} - 2 \delchi\big)\,\times \nonumber\\
& \RepQuad{4} \times\!\left(\!\vphbig\cos\!\left(\delchi\right) C\!\left(\!\frac{\sqrt{2}\,y}{\spi}\!\right)\!+ \sin\!\left(\delchi\right) S\!\left(\!\frac{\sqrt{2}\,y}{\spi}\!\right)\!\!\right) \nonumber\\[5pt]
& \RepQuad{2} + \frac{\spi}{8} \bigintsss_{0}^{x} \dd y \, y \sin\!\big(y^{2} - \delchi\big)\,\times \nonumber\\
& \RepQuad{4} \times\!\left(\!\vphbig\cos\!\left(2 \delchi\right) C\!\left(\!\frac{\sqrt{4}\,y}{\spi}\!\right)\!+ \sin\!\left(2 \delchi\right) S\!\left(\!\frac{\sqrt{4}\,y}{\spi}\!\right)\!\!\right) \nonumber\\[5pt]
& \RepQuad{2} + \frac{1}{4} \int_{0}^{x} \dd y\,y^{2} \sin\!\left(y^{2} - \delchi\right) \label{eq:frequencySmallXIntegralBThrdOrdPhase}
\end{align}
To find a solution for \(\varpi_{3a} (x)\), we proceed in an analogous way to the \(\delchi = 0\) case, and take the appropriate limits of interest of the Fresnel function. For \(|x| \ll 1\), the expansion is a polynomial, finding an analytical approximation is straightforward, and the result is given by eq.~(\ref{eq:frequencySmallXThOrdAproxPhase}). Hence the third-order solution near the origin~is
\begin{align}
\varpi_{3} (x) = \varpi_{3a_{\rm approx}} (x) + \varpi_{3b} (x). \label{eq:frequencySmallXThrdOrderPhase}
\end{align}
We also calculate the asymptotic expression for \(\varpi_{3a} (x)\), by expanding the Fresnel function for \(|x| \gg 1\), the solution is given by eq.~(\ref{eq:frequencyThOrdAAsympPhase}). The additive constant accounting for the value of the integral at the origin is
\begin{align}
\begin{split}
\alpha =& \mp 0.00991 \cos^{3}\!\left(\delchi\right) \pm 0.08018 \cos^{2}\!\left(\delchi\right) \sin\!\left(\delchi\right) \\
& \mp 0.00185 \cos\!\left(\delchi\right) \sin^{2}\!\left(\delchi\right) \pm 0.03110 \sin^{3}\!\left(\delchi\right)
\end{split} \label{eq:smallXAsymptoticAdditiveConstPhase}
\end{align}
The asymptotic expression for \(\varpi_{3} (x)\) can be written as
\begin{align}
\varpi_{3} (x) = \alpha + \varpi_{3a_{\rm asymp}} (x) + \varpi_{3b} (x). \label{eq:frequencySmallXThrdOrdAsympPhase}
\end{align}
The validity of these solutions can be demonstrated in an analogous way to the zero-phase case, by computing:
\begin{align}
\varkappa_{n+1} = \left|\vph\frac{\omega_{\text{num}} - \omega^{(n)} (x)}{\omega_{\text{num}}}\right|, \label{eq:frequencyPhaseValidityFormula}
\end{align}
where \(\omega^{(n)} (x)\) is the solution to eq.~(\ref{eq:FrequencySMEq_phase}) for \(|x| \ll 1\), and \(\omega_{\text{num}}\) is the respective numerical solution. Plots of these functions for \(n = 0, \dots, 3\) are given in Figure~\ref{fig:frequencyPhaseValiditySmallXPlot}.
\begin{figure}[t]
\includegraphics[scale=1]{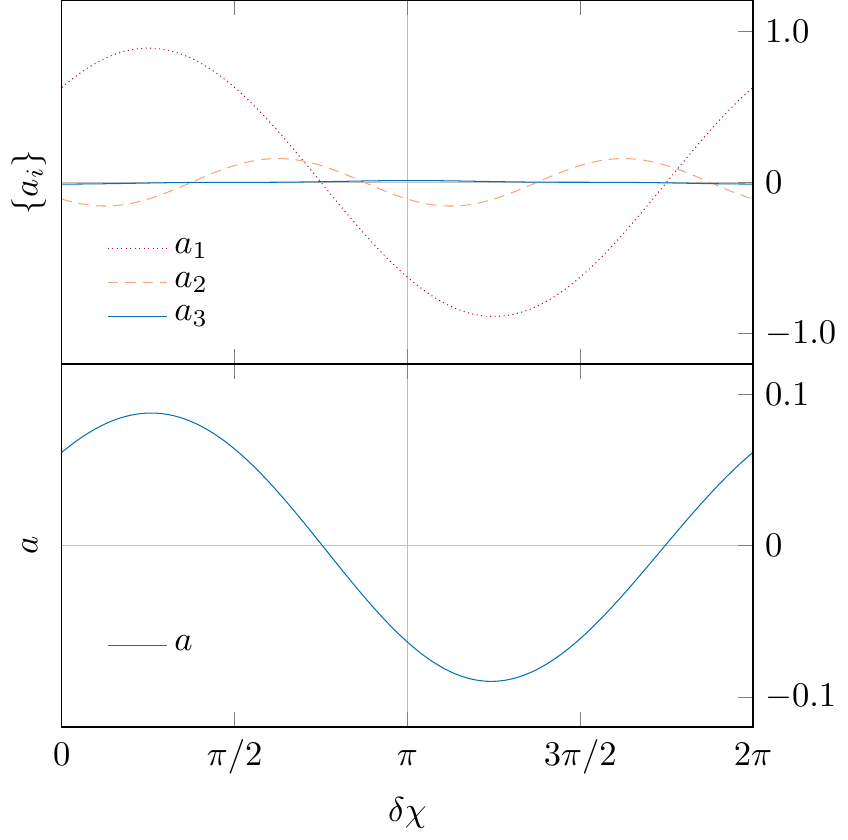}
\caption{Plot of the components of the constant \(a\), and plot of the combined constant for \(\kappa = 0.1\) as functions of \(\delchi\).}
\label{fig:frequencySmallXInitialConstPhasePlot}
\end{figure}
No further calculations are needed to find the constants \(\{b_{i}, c_{1}, c_{2}, c_{3}, d_{1}, d_{2}, l\}\), since we can use modified ans\"{a}tze which include a phase term in their oscillating factors:
\begin{subequations}
\begin{align}
&\mathcal{S}_{n} (x) \equiv \sin\!\left[n\Big(a x + \left(1 + b\right) x^{2} - \delchi\Big)\vphantom{\Big(\Big)\np{2}}\!\right] \label{eq:frequencySnPhase} \\\nonumber\\[-12pt]
&\mathcal{C}_{n} (x) \equiv \cos\!\left[n\Big(a x + \left(1 + b\right) x^{2} - \delchi\Big)\vphantom{\Big(\Big)\np{2}}\!\right].
\end{align}
\label{eq:shorthand_trigonometric_phase}
\end{subequations}
To establish the constants \(\{a_{i}\}\), we match the ansatz against the new (asymptotic) solution:
\begin{align}
\omega (x) &= \varpi_{0} (x) + \kappa \varpi_{1} (x) + \kappa^{2} \varpi_{2} (x) + \kappa^{3} \varpi_{3} (x),
\end{align}
where the last term is the combination of terms (\ref{eq:frequencySmallXThrdOrdAsympPhase}) and \(\alpha\) is given by eq.~(\ref{eq:smallXAsymptoticAdditiveConstPhase}). The three separate orders of the constant \(a\) as a function of the phase difference \(\delchi\) are:

\begin{subequations}
\begin{align}
& a_{1} = \pm\,\frac{\sqrt{2 \pi}}{4} \big(\!\cos\!\left(\delchi\right) + \sin\!\left(\delchi\right)\!\big) \\[5pt]
& a_{2} = \mp\,\frac{\spi}{16} \big(\!\cos\!\left(2 \delchi\right) + \sin\!\left(2 \delchi\right)\!\big) \\[5pt]
& a_{3} = \mp\,0.00991 \cos^{3}\!\left(\delchi\right) \pm 0.08018 \cos^{2}\!\left(\delchi\right) \sin\!\left(\delchi\right) \nonumber\\
& \RepQuad{3} \mp 0.00185 \cos\!\left(\delchi\right) \sin^{2}\!\left(\delchi\right) \pm 0.03110 \sin^{3}\!\left(\delchi\right) \nonumber\\
& \RepQuad{3} \pm \frac{\sqrt{2 \pi}}{256}\big(\!\cos\!\left(3 \delchi\right) - \sin\!\left(3 \delchi\right) \big. \nonumber\\
& \RepQuad{11} \big. + 7 \cos\!\left(\delchi\right) + 7 \sin\!\left(\delchi\right)\!\big) \nonumber\\
& \RepQuad{3} \pm \frac{\sqrt{6 \pi}}{768}\big(3 \cos\!\left(3 \delchi\right) + 3 \sin\!\left(3 \delchi\right) \big. \nonumber\\
& \RepQuad{11} \big. + \cos\!\left(\delchi\right) - \sin\!\left(\delchi\right)\!\big) \nonumber\\
& \RepQuad{3} \pm \frac{\sqrt{2}\,\pi^{\sfrac{3}{2}}}{128} \big(\!\cos\!\left(\delchi\right) - \sin\!\left(\delchi\right)\!\big)
\end{align}
\end{subequations}
The constant \(a\) itself is given by
\begin{align}
a = a_{1} \kappa + a_{2} \kappa^{2} + a_{3} \kappa^{3}
\end{align}
Plots of these functions are given in Figure~\ref{fig:frequencySmallXInitialConstPhasePlot}.

\begin{widetext}
\begin{align}
\begin{split}
&\varpi_{2} (x) = \frac{\spi}{2 \sqrt{2}} \, \cos\!\big(x^{2} - \delchi\big)\!\!\left(\vphbig\!\cos(\delchi)\,C\!\left(\!\frac{\sqrt{2}\,x}{\spi}\!\right)\!+ \sin(\delchi)\,S\!\left(\!\frac{\sqrt{2}\,x}{\spi}\!\right)\!\!\right) \\
& \RepQuad{24} - \frac{\spi}{8}\!\left(\vphbig\!\cos(2 \delchi)\,C\!\left(\!\frac{\sqrt{4}\,x}{\spi}\!\right)\!+ \sin(2 \delchi)\,S\!\left(\!\frac{\sqrt{4}\,x}{\spi}\!\right)\!\!\right)\!- \frac{x}{4}
\end{split} \label{eq:frequencySmallXScndOrderPhase} \\[5pt]
\begin{split}
&\varpi_{3a_{\rm approx}} (x) = \left(\!\vph 33 \cos^{2}\!\left(\delchi\right) - \frac{35}{3} \sin^{2}\!\left(\delchi\right)\!\right)\!\!\left[\vphbigg\frac{\sqrt{2\pi}}{64}\!\left(\!\vphbig\cos\!\left(\delchi\right) C\!\left(\!\frac{\sqrt{2}\,x}{\spi}\!\right)\!+ \sin\!\left(\delchi\right) S\!\left(\!\frac{\sqrt{2}\,x}{\spi}\!\right)\!\!\right)\!- \frac{1}{32}\,x \cos\!\big(x^{2} - \delchi\big)\right. \\
& \RepQuad{8} \left.\vphbigg - \frac{1}{48}\,x^{3} \sin\!\big(x^{2} - \delchi\big)\!\right]\!+ \left(\!\vph \frac{1}{5} \cos^{2}\!\left(\delchi\right) - \frac{1}{9} \sin^{2}\!\left(\delchi\right)\!\right)\!\!\left[\vph\frac{7}{8}\,x^{5}\cos\!\big(x^{2} - \delchi\big) + \frac{1}{4}\,x^{7} \sin\!\big(x^2 - \delchi\big)\!\right]
\end{split} \label{eq:frequencySmallXThOrdAproxPhase} \\[5pt]
\begin{split}
& \varpi_{3a_{\rm asymp}} (x) = \pm\,\frac{\sqrt{2\pi}}{32} \left(\cos^{3}\!\left(\delchi\right) + \sin^{3}\!\left(\delchi\right)\right) \ln\!|x| \mp \frac{\sqrt{2\pi}}{64}\,\sin\!\left(2 \delchi\right)\!\big(\!\cos\!\left(\delchi\right) - \sin\!\left(\delchi\right)\!\big)\!\left(\!x^{2} + \frac{3}{4 x^{2}}\!\right) \\
& \RepQuad{8} - \frac{\sqrt{2\pi}}{64}\!\left(\!\vphbig\cos^{3}\!\left(\delchi\right)C\!\left(\!\frac{\sqrt{2}\,x}{\spi}\!\right)\!+ \sin^{3}\!\left(\delchi\right)S\!\left(\!\frac{\sqrt{2}\,x}{\spi}\!\right)\!\!\right) \\
& \RepQuad{10} - \frac{3 \sqrt{2\pi}}{128}\sin\!\left(2 \delchi\right)\!\left(\!\vphbig\cos\!\left(\delchi\right)S\!\left(\!\frac{\sqrt{2}\,x}{\spi}\!\right)\!+ \sin\!\left(\delchi\right)C\!\left(\!\frac{\sqrt{2}\,x}{\spi}\!\right)\!\!\right) \\
& \RepQuad{12} + \frac{\sqrt{2}\,\pi^{\sfrac{3}{2}}}{64}\!\left(\!\vphbig\cos\!\left(\delchi\right)S\!\left(\!\frac{\sqrt{2}\,x}{\spi}\!\right)\!- \sin\!\left(\delchi\right)C\!\left(\!\frac{\sqrt{2}\,x}{\spi}\!\right)\!\!\right) \\
& \RepQuad{14} + \frac{\sqrt{6 \pi}}{192}\cos\!\left(2 \delchi\right)\!\left(\!\vphbig\cos\!\left(\delchi\right)C\!\left(\!\frac{\sqrt{6}\,x}{\spi}\!\right)\!+ \sin\!\left(\delchi\right)S\!\left(\!\frac{\sqrt{6}\,x}{\spi}\!\right)\!\!\right) \\
& \RepQuad{7} - \frac{\pi}{32}\, x \sin\!\left(x^{2} - \delchi\right) \pm \frac{\sqrt{2\pi}}{64}\!\left(\!1 - \frac{1}{x^{4}}\!\right)\!\left(\cos^{2}\!\left(\delchi\right)\cos\!\left(2 x^{2} - \delchi\right)+ \sin^{2}\!\left(\delchi\right)\sin\!\left(2 x^{2} - \delchi\right)\!\right) \\
& \RepQuad{8} \pm \frac{\sqrt{2\pi}}{128}\,\frac{1}{x^{2}}\!\left(\cos^{2}\!\left(\delchi\right)\sin\!\left(2 x^{2} - \delchi\right)- \sin^{2}\!\left(\delchi\right)\cos\!\left(2 x^{2} - \delchi\right)\!\right) \\
& \RepQuad{10} - \frac{1}{64}\,\frac{1}{x^{3}}\cos\!\left(2 \delchi\right)\!\left(\cos\!\left(x^{2} + \delchi\right) + \frac{1}{3}\cos\!\left(3 x^{2} - \delchi\right)\!\right) \\
& \RepQuad{12} - \frac{5}{128}\sin\!\left(2 \delchi\right)\frac{\cos\!\left(x^{2} + \delchi\right)}{x^{5}} - \frac{3}{256}\,\frac{1}{x^{5}}\cos\!\left(2 \delchi\right)\!\left(\sin\!\left(3 x^{2} - \delchi\right)\!+ \sin\!\left(x^{2} + \delchi\right)\!\right)
\end{split} \label{eq:frequencyThOrdAAsympPhase} \\[5pt]
\begin{split}
& \varpi_{3b} (x) = \frac{\sqrt{2 \pi}}{32} \cos\!\left[2 (x^{2} - \delchi)\right]\!\!\left(\!\vphbig\cos\!\left(\delchi\right)C\!\left(\!\frac{\sqrt{2}\,x}{\spi}\!\right)\!+ \sin\!\left(\delchi\right)S\!\left(\!\frac{\sqrt{2}\,x}{\spi}\!\right)\!\!\right) \\
& \RepQuad{6} - \frac{\spi}{16} \cos\!\left(x^{2} - \delchi\right)\!\!\left(\!\vphbig\cos\!\left(2 \delchi\right)C\!\left(\!\frac{\sqrt{4}\,x}{\spi}\!\right)\!+ \sin\!\left(2 \delchi\right)S\!\left(\!\frac{\sqrt{4}\,x}{\spi}\!\right)\!\!\right) \\
& \RepQuad{8} + \frac{\sqrt{6 \pi}}{192}\!\left(\!\vphbig\cos\!\left(3 \delchi\right)C\!\left(\!\frac{\sqrt{6}\,x}{\spi}\!\right)\!+ \sin\!\left(3 \delchi\right)S\!\left(\!\frac{\sqrt{6}\,x}{\spi}\!\right)\!\!\right) \\
& \RepQuad{10} + \frac{5 \sqrt{2 \pi}}{64}\!\left(\!\vphbig\cos\!\left(\delchi\right)C\!\left(\!\frac{\sqrt{2}\,x}{\spi}\!\right)\!+ \sin\!\left(\delchi\right)S\!\left(\!\frac{\sqrt{2}\,x}{\spi}\!\right)\!\!\right)\!-\frac{1}{8}\,x \cos\!\left(x^{2} - \delchi\right)
\end{split} \label{eq:frequencyThOrdBPhase}
\end{align}
\end{widetext}

\begin{widetext}
\begin{m}{align}
&\omega_{3} (x) = \big(a_{1} \kappa + a_{2} \kappa^{2} + a_{3} \kappa^{3}\big)\!+ \left(\!\vph 1 - \frac{\kappa^{2}}{4}\!\right)\!x + \kappa\,\frac{1}{2}\,\frac{1}{x}\, \sin\!\left[\vphbig\!\left(a_{1} \kappa + a_{2} \kappa^{2}\right)\!x +\!\Bigg(\!1 - \frac{\kappa^{2}}{4}\!\Bigg)x^{2} - \delchi\right] \nonumber\\
& \RepQuad{1} + \kappa^{2}\frac{1}{16}\,\frac{1}{x} \, \sin\!\Big[ 2\!\left(a_{1} \kappa\,x + x^{2} - \delchi\right)\!\Big]\!+ \kappa^{3}\,\frac{1}{x}\!\left(\!\vph\frac{3}{32}\,\sin\!\left(x^{2} - \delchi\right) + \frac{1}{96}\,\sin\!\Big[3 \big(x^{2} - \delchi\big)\Big]\!\right) \label{eq:frequencySMPhaseSolutionThrdOrd}\\
& \RepQuad{2} \mp \kappa^{2}\frac{\sqrt{2 \pi}}{16}\,\frac{1}{x^{2}} \, \sin\!\left(a_{1} \kappa\,x + x^{2} - \delchi\right)\!\pm \kappa^{3}\!\left(\!\vph\frac{\spi}{64}\,\frac{1}{x^{2}}\,\sin\!\left(x^{2} - \delchi\right) - \frac{\sqrt{2 \pi}}{64}\,\frac{1}{x^{2}}\,\sin\!\Big[2\big(x^{2} - \delchi\big)\Big] + \frac{\sqrt{2 \pi}}{32}\,\ln\!|x|\!\right). \nonumber
\end{m}
\end{widetext}

\begin{figure}[H]
\includegraphics[scale=1]{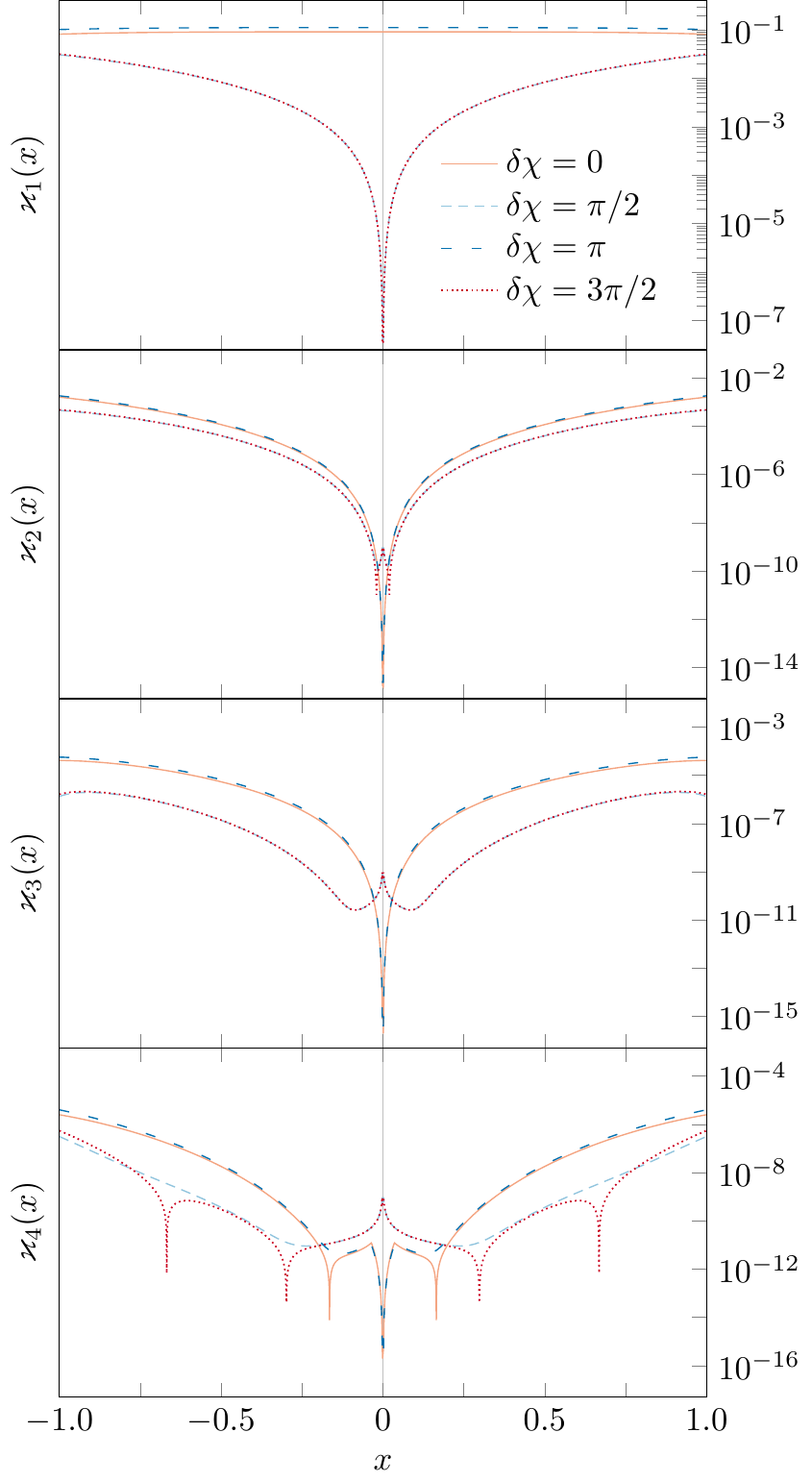}
\caption{Graphical demonstration of the validity of solutions~\(\varpi_{0} (x) = x\), (\ref{eq:frequencySmallXFrstOrderPhase}), (\ref{eq:frequencySmallXScndOrderPhase}), and (\ref{eq:frequencySmallXThrdOrderPhase}). In each plot we have used \(\kappa = 0.1\). It is evident that the fractional errors scale with the value of the resonance modification \(\kappa\), and for each order \(n\) of the solution, the remainder is of order \(\bigo{\kappa^{n+1}}\).}
\label{fig:frequencyPhaseValiditySmallXPlot}
\vspace*{-10pt}
\end{figure}

Solutions \(\varpi_{0} (x)\), \(\varpi_{1} (x)\), \(\varpi_{2} (x)\) and \(\varpi_{3} (x)\) are plotted in Fig.~(\ref{fig:frequencySolutionSmallXPhasePlot}), while a more rigorous derivation of the principle we use here is presented in Appendix~\ref{sec:app_frequency_equation_one_mode_phase}. This paves the way for finding a solution to the general equation with an infinite number of modes.

\begin{figure}[t]
\includegraphics[scale=1]{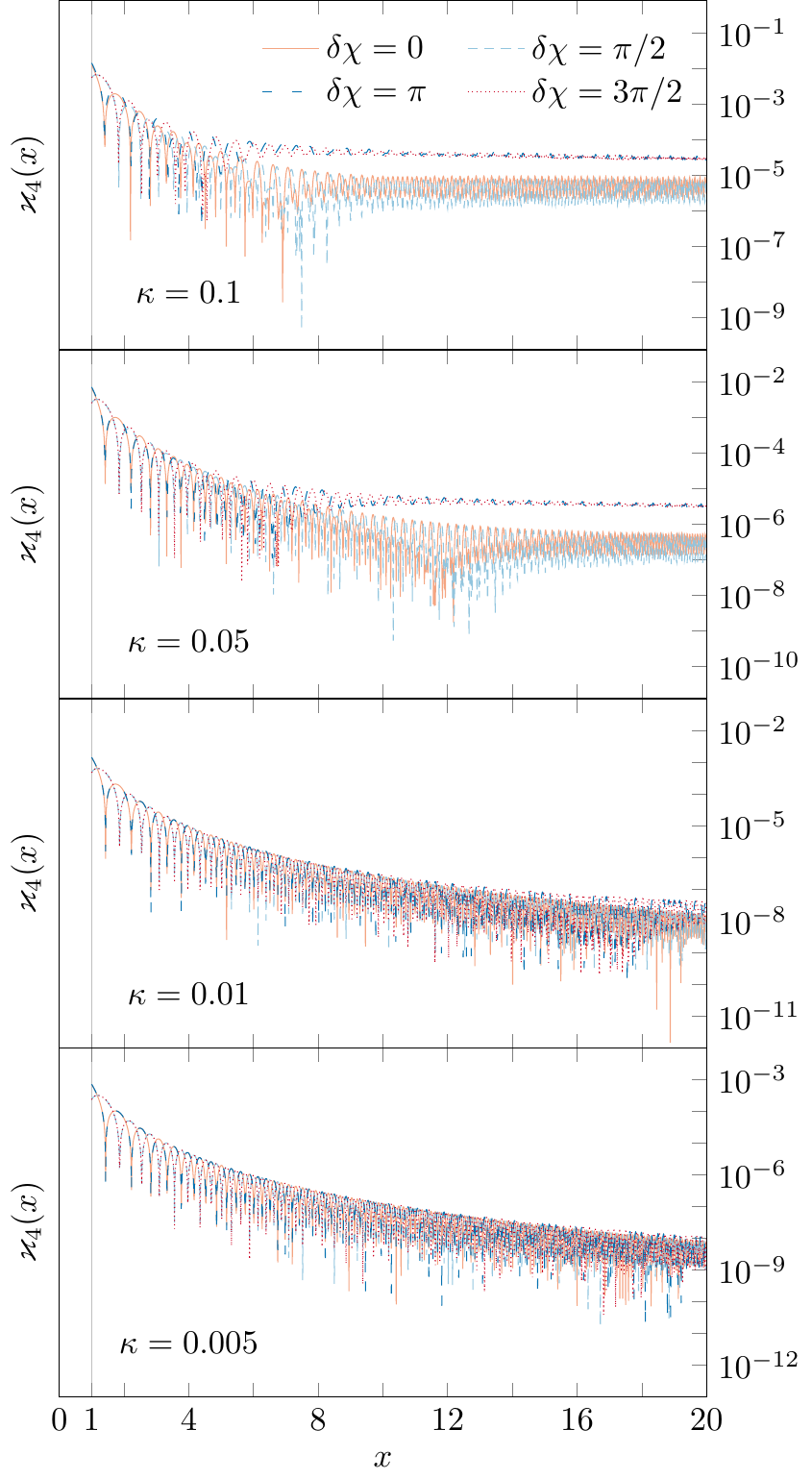}
\caption{Graphical demonstration of the validity of solution~(\ref{eq:frequencySMPhaseSolutionThrdOrd}) for several characteristic values of \(\kappa\) and for sample values of the phase angle \(\delchi\).}
\label{fig:frequencyPhaseValidityPlot}
\vspace*{-10pt}
\end{figure}

Before we proceed with this, we check graphically the validity of solution~(\ref{eq:frequencySMPhaseSolutionThrdOrd}). Fig.~\ref{fig:frequencyPhaseValidityPlot} depicts the fractional errors \(\varkappa_{4} (x)\) for the usual values of \(\kappa\), and for 4 different values of the initial phase offset.

Finally, we consider the multi-mode frequency equation with a non-vanishing initial phase:
\begin{align}
\bom^{\prime}(x) = 1 + \sum_{n} \kappa_{n} \cos\!\left(n x_{\omega}\,\bom - \delchi_{n}\right),
\tag{\ref{eq:frequency_final}}
\end{align}
where the array of constants \(\{\delchi_{n}\}\) is known and can in principle be computed from the constants of motion through eq.~(\ref{eq:parameter_constant_relations}). Before proceeding, we elect to write the equation out in the following more revealing way
\begin{align}
\begin{split}
\bom^{\prime}(x) = 1 + \sum_{n} \kappa_{n} &\cos\!\left(\delchi_{n}\right)\cos\!\left(n x_{\omega}\,\bom\right)\\
&+ \kappa_{n} \sin\!\left(\delchi_{n}\right)\sin\!\left(n x_{\omega}\,\bom\right).
\end{split} \label{eq:frequency_final_modified_form}
\end{align}

For small \(|x|\), we find consecutively:
\begin{align}
\bm{\varpi}_{0} (x) = x
\end{align}
and
\begin{align}
\begin{split}
& \bm{\varpi}_{1, n} (x) = \frac{\spi}{\sqrt{2n}}\!\left(\!\cos(\delchi_{n})\,C\!\left(\!\frac{\sqrt{2n}\,x}{\spi}\!\right) \vphbig \right. \\
& \RepQuad{10} \left. \vphbig + \sin(\delchi_{n})\,S\!\left(\!\frac{\sqrt{2n}\,x}{\spi}\!\right)\!\!\right)
\end{split}
\end{align}
As before, higher-order solutions are not necessary, so we can write the solution for \(|x| \ll 1\) as
\begin{m}{align}
\bom_{1} (x) = \bm{\varpi}_{0} (x) + \sum_{n} \kappa^{n} \bm{\varpi}_{1, n} (x). \label{eq:frequencyMMPhaseSolutionSmallX}
\end{m}

Next we choose an elaborate ansatz of the form:
\begin{align}
\bom_{1} (x) = \bm{a} +\left(1 + \bm{b}\right) x + \sum_{n} \bm{c}_{n} \frac{\mathcal{S}_{n} (x)}{x} \label{eq:multi_mode_phase_ansatz}
\end{align}
where we note that we neglect terms of \(\bigo{x^{-2}}\), which correspond to oscillations which die out quickly with \(|x|\). The derivative of this ansatz is given by
\begin{align}
\bom_{1}^{\prime} (x) = 1 + \bm{b} + 2\left(1 + \bm{b}\right) \sum_{n} n \, \bm{c}_{n} \, \mathcal{C}_{n} (x). \label{eq:multi_mode_phase_ansatz_derivative}
\end{align}
This needs to match the right-hand side of eq.~(\ref{eq:frequency_final}). We perform analysis similar to our earlier discussion on the zero-phase multi-mode version to find the solution~(\ref{eq:frequencySolutionMMPhase}). Note that the slope correction is still zero at leading order, and we discover that the constants \(\{\delchi_{n}\}\) only appear in the argument of the oscillation modes. The constant \(\bm{a}\) is found by comparing the above ansatz with the \(|x| \gg 1\) limit of solution~(\ref{eq:frequencyMMPhaseSolutionSmallX}).

\begin{align}
\bm{a} = \pm \sum_{n} \kappa_{n} \frac{\spi}{2 \sqrt{2 n}}\,\big(\!\cos(\delchi_{n}) + \sin(\delchi_{n})\big)
\end{align}

\begin{figure}[b]
\includegraphics[scale=1]{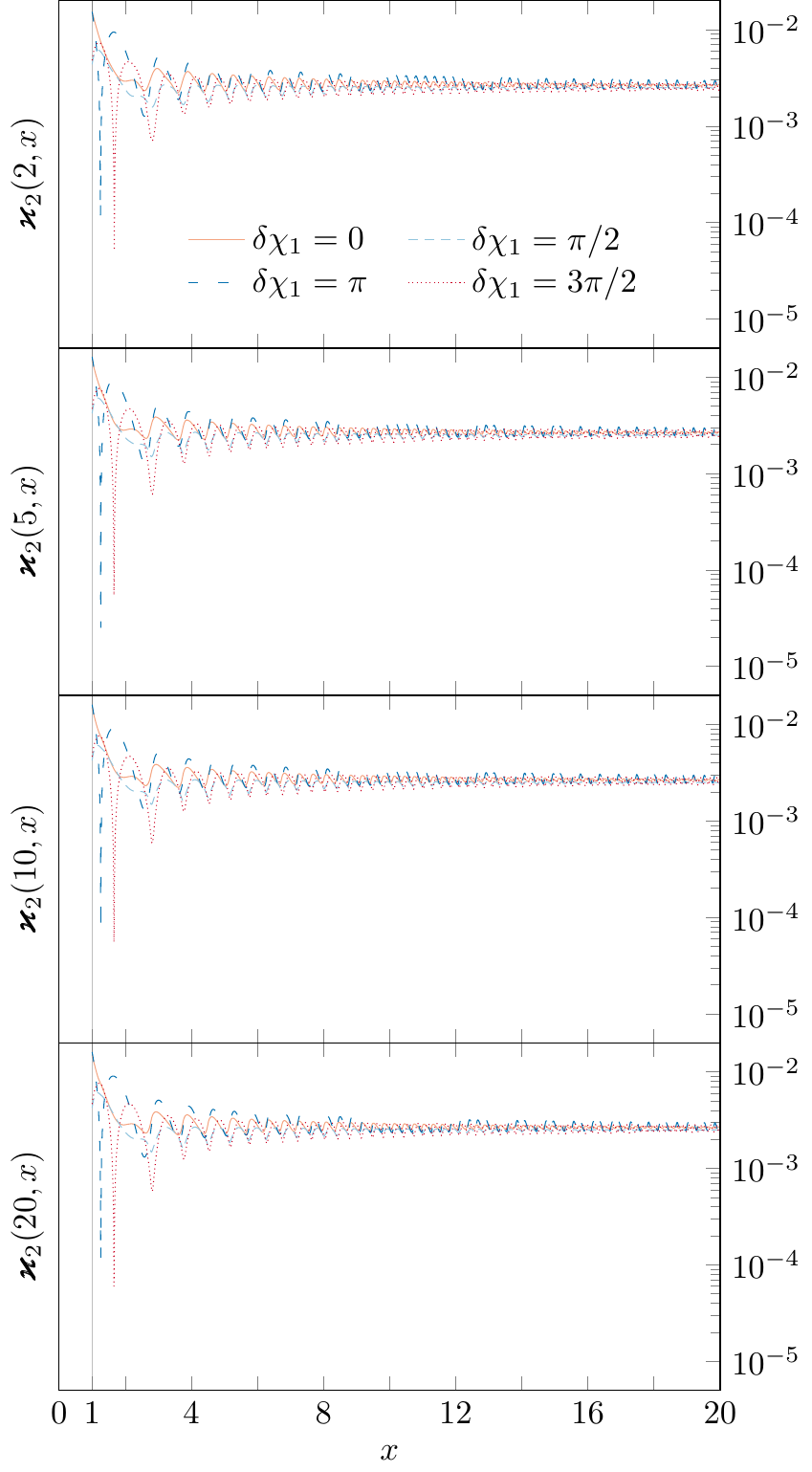}
\caption{Graphical demonstration of the validity of solution~(\ref{eq:frequencySolutionMMPhase}) for different number of terms on the right-hand side and for values of \(\{\delchi_{n}\}\) given by eq.~(\ref{eq:initialPhaseLaw}). In all panels we have used \(\kappa_{1} = 0.1\) with consecutive values given by \(\kappa_{n} = \kappa_{1} / n^{2.5}\).}
\label{fig:frequencyMMPhaseValidityPlot}
\end{figure}

Validating these solutions can be done in the same manner as earlier, by computing the quantities analogous to eq.~(\ref{eq:frequencyValidityFormula}) and eq.~(\ref{eq:frequencyMMValidityFormula}), but this time incorporating the phase variable into the checks. These are depicted in Figures~\ref{fig:frequencyMMPhaseValidityPlot} and \ref{fig:frequencyMMPhaseSmallXValidityPlot}. For these figures, we have assumed that the phases \(\{\delchi_{n}\}\) cycle the set \(\{0, \pi/2, \pi, 3\pi/2\}\),
\begin{align}
\delchi_{n} = \big(\delchi_{1} + (n-1)\,\sfrac{\pi}{2}\big) \mod 2 \pi \label{eq:initialPhaseLaw}
\end{align}

As before, it is very challenging to correctly determine higher-order solutions in terms of the parameters \(\{\kappa_{n}\}\) as higher order corrections include contributions from cross-terms between modes, which significantly complicate the analysis. Instead, we will focus on the more interesting phase resonance equation, where this problem is avoided, and higher-order solutions can still be found in the multi-mode case.
\begin{widetext}
\begin{m}{align}
&\bom_{1} (x) = \pm \sum_{n} \kappa_{n} \frac{\spi}{2 \sqrt{2 n}}\,\big(\!\cos(\delchi_{n}) + \sin(\delchi_{n})\big) + x + \sum_{n} \kappa_{n}\,\frac{1}{2n}\,\frac{1}{x} \, \sin\!\big(n x^{2} - \delchi_{n}\big).\label{eq:frequencySolutionMMPhase}
\end{m}
\vspace*{5pt}
\end{widetext}

This concludes our discussion of the frequency resonance equation. We have produced a series of solutions with increasing generality, and have provided graphical evidence of their validity. We note that some results have been intentionally left to Appendix~\ref{sec:AppendixFrequencyResEq}, since they are simple extensions of less general cases presented within this Section. We can now turn our attention to the phase resonance equation and its solutions.

\subsection{Phase resonance equation}
\label{sec:phase_solution}
\noindent
The phase-resonance equation is derived in Section~\ref{sec:InstantaneousFrequencyApproach} and makes use of the action-angle variables \(\{q_{\mu}\}\), instead of the fundamental frequencies \(\{\omega_{\mu}\}\) to write down the Fourier modes of the self-force function. In the case of a multitude of modes in this expansion, as well as a non-vanishing initial phase offset, the equation takes the form
\begin{align}
\phi^{\prime\prime} (x) = 1 + \sum_{n > 0} \lambda_{n} \cos\!\left(n \phi - \delta \psi_{n}\right). \tag{\ref{eq:phaseEquation}}
\end{align}
Similarly to our approach to the frequency resonance equation, we examine the case in which we ignore all higher-order modes beyond the first one (\(n = 1\)), and assume a zero initial phase, so eq.~(\ref{eq:phaseEquation}) simplifies to
\begin{align}
\phi^{\prime\prime} (x) = 1 + \lambda \cos\!\big(\phi (x)\big). \label{eq:phaseEquationSMZeroPh}
\end{align}
The equation is again non-linear due to the additive constant, and an analytic solution cannot be provided in the general case. Instead we employ an iterative approach to establish a series of solutions in increasing order of the mode parameter \(\lambda \equiv \lambda_{1} \ll 1\). At first, we ignore the term involving \(\lambda\), and seek a solution to the equation:
\begin{align}
\phi_{0}^{\prime\prime} (x) = 1 + \bigo{\lambda}, \label{eq:phaseEqSMZeroPhReduced}
\end{align}
which straightforwardly yields the zeroth-order solution to the phase resonance equation (this and higher order solutions are denote by PH) in the current regime
\begin{ph}{align}
\phi_{0} (x) = \frac{x^{2}}{2}. \label{eq:PhaseSolutionSMZeroOrd}
\end{ph}
Further solutions are obtained by substituting the current solution on the right-hand side of the source equation (\ref{eq:phaseEquationSMZeroPh}), while on the left-hand side we include this solution plus the next-order correction. In the general case, we would need to solve the equation (for \(n \ge 1\))
\begin{align}
\phi_{n}^{\prime\prime} (x) = 1 + \lambda \cos\!\left(\phi_{n-1} (x)\right), \label{eq:generalPhaseEquationIteration}
\end{align}
where \(\phi_{n} (x)\) is the complete \(n^{\text{th}}\)-order solution. We apply this iterative technique to find the first-order correction: we use the function \(\phi_{0} (x)\) as the source for the next order solution \(\phi_{1} (x) = \phi_{0} (x) + \phi_{(1)} (x)\). That means we write \(\phi_{0} (x)\) as the argument on the right-hand side, while on the left-hand side we solve for \(\phi_{1} (x)\), a combination of the zeroth-order \(\phi_{0} (x)\) and the first-order correction \(\phi_{(1)} (x)\). Throughout our current discussion, we shall use \(\phi_{n} (x)\) to denote the complete solution to order \(n\), while \(\phi_{(n)} (x)\) would denote the \(n^{\text{th}}\)-order correction to \(\phi_{0} (x)\), i.e. \(\phi_{n} (x) = \phi_{0} (x) + \phi_{(n)} (x)\). Using this, we write:
\begin{align}
\phi_{1}^{\prime\prime} (x) = \left(\phi_{0} + \phi_{(1)}\right)^{\prime\prime} = 1 + \phi_{(1)}^{\prime\prime} = 1 + \lambda \cos\!\left(\phi_{0}\right).
\end{align}
From this, we find a differential equation for \(\phi_{(1)} (x)\):
\begin{align}
\phi_{(1)}^{\prime\prime} (x) = \lambda \cos\!\left(\!\frac{x^{2}}{2}\!\vph\right). \label{eq:phase_single_mode_second_order}
\end{align}
The details of finding this solution are described in Appendix~\ref{sec:appendixPhaseSolutionSMZeroPh}, here we only quote the final result:
\begin{align}
\phi_{(1)} (x) = \lambda\!\left[\vphbig\sqrt{\pi} \, x \, C\!\left(\!\frac{x}{\sqrt{\pi}}\!\right)\!- \sin\!\Bigg(\!\frac{x^{2}}{2}\!\vph\Bigg)\!\right].
\end{align}
Combining this with \(\phi_{0} (x)\), we find the complete first-order-in-\(\lambda\) solution to eq.~(\ref{eq:phaseEquationSMZeroPh}), given by (\ref{eq:phase_solution_1}). It is worth explaining that the solution has been rearranged into a form which is more useful for our discussion later on. More specifically, we have grouped together terms which oscillate around the value \(\phi (x) = 0\) -- both \((\FrC{\Cdot[2]} \mp \sfrac{1}{2})\) and \(\sin (\Cdot[2])\) are such functions (see Appendix~\ref{sec:AppFresnelFuncs} for details about the former). Moreover, we have collected the terms proportional to \(x^{2}\) and \(x\) into a single quadratic term (by completing the square, which inevitably introduces a further constant, however it is of second order in the mode parameter \(\lambda\), and not relevant here). Hence the solution is comprised of a modified (shifted along the \(x\)-axis) quadratic term and a term \(\rho_{1}\) which oscillates around \(\phi = 0\), and whose frequency of oscillations increases with the value of \(x\).

A number of results in this section contain the symbols ``\(\pm\)'' and ``\(\mp\)''. Similarly to the discussion of the solutions of the frequency resonance equation, these do not signify two alternative equations, but rather that the lower signs are valid for \(x \le 0\), while the upper signs hold for \(x \ge 0\), both part of the same solution. 
\begin{widetext}
\vspace*{-10pt}
\begin{ph}{align}
\phi_{1} (x) = \frac{\pi}{2}\!\LinearTerm[][\pm \frac{1}{2}\,\lambda]\np{2}\!+ \lambda\!\left[\vphbig\pi\!\LinearTerm\!\!\FresnelCTerm\! - \SinTerm\!\right]\!\equiv \frac{\pi}{2}\!\LinearTerm[][\pm \frac{1}{2}\,\lambda]\np{2}\!+ \lambda\,\rho_{1} \!\!\left(\!\frac{x}{\spi}\!\right). \label{eq:phase_solution_1}
\end{ph}
\vspace*{-10pt}
\end{widetext}
It is important to note that for large values of \(x\), the oscillating terms become small, and the approximate slope of \(\phi_{1} (x)\) is given by \(\spi \, \lambda / 2\). We will see later how this slope evolves when we add higher-order-in-\(\lambda\) terms, but for now let us denote \(\sigma_{1} (\lambda) = \lambda / 2\), which is the shift in the position of the leading parabolic term.

\begin{figure}[t]
\vspace*{30pt}
\includegraphics[scale=1.0]{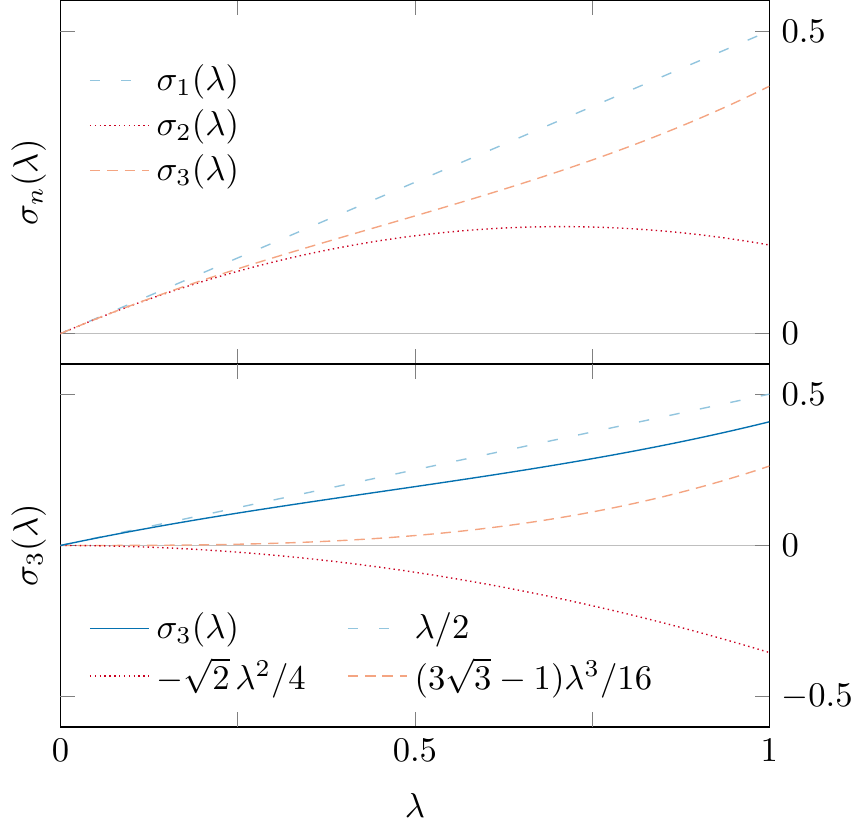}
\caption{Evolution of the slope of the phase resonance solutions. The domain of the plot shows all possible values for the mode parameter: \(0 \le \lambda \le 1\). \emph{Upper panel:} the evolution of the slope for all three solutions which have been calculated explicitly in the paper. \emph{Lower panel:} the individual components which make up the slope function~\(\sigma_{3} (\lambda)\).}
\label{fig:phase_solution_slope}
\end{figure}

Applying the same iterative approach, we use the combined function \(\phi_{1} (x)\) as the source for the second-order asymptotic solution \(\phi_{2} (x) = \phi_{0} (x) + \phi_{(2)} (x)\), where the second-order correction \(\phi_{(2)} (x)\) is comprised both of first- and second- order terms. This notation is convenient since \(\phi_{0}^{\prime\prime} (x) = 1\) cancels with the constant term on the right-hand side,  and upon integrating twice, we obtain an expression for \(\phi_{(2)} (x)\), the complete solution to second order in \(\lambda\), only without the zeroth-order quadratic term.  On the right-hand side, we substitute the complete first-order solution, \(\phi_{1} (x) = \phi_{0} (x) + \phi_{(1)} (x)\),
\begin{align}
\phi_{(2)}^{\prime\prime} (x) = \lambda \, \cos\!\left(\phi_{0} + \phi_{(1)}\right). \label{eq:phase_eq_second_order_equation}
\end{align}
In fact, from eqs.~(\ref{eq:phase_single_mode_second_order}) and (\ref{eq:phase_eq_second_order_equation}) we can infer the general form of this differential equation (for \(n > 1\)):
\begin{align}
\begin{split}
\phi_{(n)}^{\prime\prime} (x) &= \left(\phi_{n} - \phi_{0}\right)^{\prime\prime}\!(x) =\\
&= \lambda \cos\!\left(\phi_{n-1}\right) = \lambda \cos\!\left(\phi_{0} + \phi_{(n-1)}\right).
\end{split} \label{eq:gen_phase_equation_reduced_single_mode}
\end{align}
The details involved in solving eq.~(\ref{eq:phase_eq_second_order_equation}) are presented in Appendix~\ref{sec:appendixPhaseSolutionSMZeroPh}, together with the full-form solution. Here, we omit these details, and simply quote the final combined result \(\phi_{2} (x) = \phi_{0} (x) + \phi_{(2)} (x)\) in the form of a modified quadratic term, plus two terms, both of them \(\bigo{\lambda^{2}}\) that oscillate and remain small for all \(x\).
\begin{widetext}
\vspace*{-10pt}
\begin{ph}{align}
\label{eq:phase_solution_2}
\begin{split}
\phi_{2} (x) &= \frac{\pi}{2}\!\LinearTerm[][\pm \left[\frac{1}{2} \, \lambda - \frac{\sqrt{2}}{4} \, \lambda^{2}\right]]\np{2}\!+ \lambda\,\rho_{1}\!\!\left(\!\frac{x}{\spi}\!\pm \sigma_{1} (\lambda)\!\right) + \lambda^{2} \rho_{2}\!\!\left(\!\frac{x}{\spi}\!\right).
\end{split}
\end{ph}
\vspace*{-10pt}
\end{widetext}
We note some interesting features of this solution. Most notably, the shift in the quadratic \(x\) term has changed from \(\sigma_{1} (\lambda) = \lambda / 2\) to \(\sigma_{2} (\lambda) = \lambda / 2 - \sqrt{2}\,\lambda^{2}/4\) (and the slope for large \(|x|\) has changed from \(\spi \, \sigma_{1} (\lambda)\) to \(\spi \, \sigma_{2} (\lambda)\)). The term which is proportional to \(\lambda\) has the same functional form as the corresponding term in solution~(\ref{eq:phase_solution_1}), however we see that in (\ref{eq:phase_solution_2}) each instance of \(x /\!\spi\) has been modified by exactly \(\pm \sigma_{1} (\lambda)\). We also have a new class of terms, all of which also oscillate around the \(x\)-axis, and they are proportional to \(\lambda^{2}\). We will investigate the numerical performance of these solutions after we explore the next-order solution.

The solutions presented in \cite*{gair2012} do not continue further than this order. Since our goal is to solve the multi-mode equation~(\ref{eq:phaseEquation}), as a form of sanity check, we would like to present the solution to third order in the single-mode parameter \(\lambda\), i.e. the solution to the following generating equation:
\begin{align}
\phi_{3}^{\prime\prime} (x) = 1 + \lambda \cos\!\big(\phi_{2} (x)\big) \label{eq:phase_resonance_equation_third_order}
\end{align}
The solution follows via the same technique as the previous two, and again it is not practical to present the details of the derivation here. Instead, the reader is invited to consult Appendix~\ref{sec:appendixPhaseSolutionSMZeroPh} for the technicalities of the calculation as well as the full solution without abbreviations. With every successive iteration we obtain new terms, and we have grouped them together in a similar fashion to before. Here we have terms which are proportional to \(\lambda^{3}\), as well as modified terms, originating from the lower-order solutions. The result (\ref{eq:phase_solution_3}) is the complete solution to third order, for a single mode with amplitude \(\lambda\), with vanishing initial phase \(\delpsi = 0\).
\begin{widetext}
\begin{ph}{align}
\phi_{3} (x) =& \, \frac{\pi}{2}\!\LinearTerm[][\pm \left[\frac{1}{2} \, \lambda - \frac{\sqrt{2}}{4}\,\lambda^{2} + \frac{3\sqrt{3} - 1}{16}\,\lambda^{3}\right]]\np{2}\!+\lambda\,\rho_{1}\!\!\left(\!\frac{x}{\spi}\!\pm \sigma_{2} (\lambda)\!\right) + \lambda^{2} \rho_{2}\!\!\left(\!\frac{x}{\spi}\!\pm \sigma_{1} (\lambda)\!\right) + \lambda^{3} \rho_{3}\!\!\left(\!\frac{x}{\spi}\!\right) \label{eq:phase_solution_3}
\end{ph}
\end{widetext}
In (\ref{eq:phase_solution_3}), the \(x\)-shift of the quadratic term is now
\begin{align}
\sigma_{3} (\lambda) = \frac{1}{2} \, \lambda - \frac{\sqrt{2}}{4}\,\lambda^{2} + \frac{3\sqrt{3} - 1}{16}\,\lambda^{3}, \label{eq:phaseExprSMThrdOrd}
\end{align}
and consequently, the slope of the graph for large \(x\) is given by \(\spi \, \sigma_{3} (\lambda)\). In fact \cite{gair2012} numerically computed the change in the slope of the phase solution at \(\lambda^{3}\) order:
\begin{align}
\phi_{3}^{\prime} (x) \sim \frac{\spi}{2} \, \lambda - \frac{\spi}{2\sqrt{2}}\,\lambda^{2} + 0.4648 \, \lambda^{3} + \bigo{\lambda^{4}}.
\end{align}
As a form of consistency check, we can directly compare this with the (analytically derived) factor multiplying the \(\lambda^{3}\) term in the coefficient of \(x\) in eq.~(\ref{eq:phase_solution_3}) and confirm that both numerical values agree:
\begin{align}
\spi \times \frac{3\sqrt{3} - 1}{16} \, \lambda^{3} \approx  0.4648 \, \lambda^{3}.
\end{align}

Moreover, we can notice a parallel between the terms proportional to \(\lambda\) in each of (\ref{eq:phase_solution_1}) --- (\ref{eq:phase_solution_3}): all three brackets have the same dependence on \(x\), meaning that for \(\lambda = 0\), they would be equal. However, in each of them, each instance of \(x /\!\spi\) is modified by a different function of the mode parameter \(\lambda\). Compared to (\ref{eq:phase_solution_1}), each \(x /\!\spi\) in (\ref{eq:phase_solution_2}) is shifted by \(\pm\,\sigma_{1} (\lambda)\), and in (\ref{eq:phase_solution_3}), it is shifted by \(\pm\,\sigma_{2} (\lambda)\). Similar evolution can be noticed in the terms proportional to \(\lambda^{2}\) in (\ref{eq:phase_solution_2}) and (\ref{eq:phase_solution_3}). In the latter, each instance of a term involving \(x\) is shifted by \(\sigma_{1} (\lambda)\) compared to the former. While it is hard to predict what terms at higher order would look like, because the \(x\)-dependence cannot be inferred from the first three solutions, we can nevertheless expect that at fourth order, the term proportional to \(\lambda^{3}\) would look similar to its counterpart in (\ref{eq:phase_solution_3}), with each \(x/\!\spi\) term modified by \(\pm\,\sigma_{1} (\lambda)\), and so on. It is evident that the functions \(\sigma_{n} (\lambda)\) play an important role in the phase resonance solutions, and at least in the evolution of the slope of the graph for large \(|x|\). In Figure~\ref{fig:phase_solution_slope}, we plot these functions and present what they look like for different values of \(\lambda\).

\begin{figure}[b]
\includegraphics[scale=1.0]{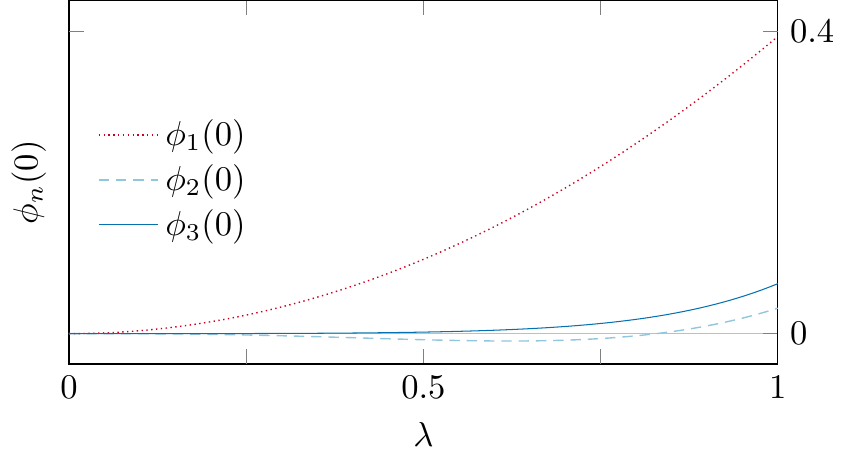}
\caption{Plot of the functions \(\phi_{\{1, 2, 3\}} (0)\) as functions of \(\lambda\).}
\label{fig:phaseAtZeroPlot}
\end{figure}

Let us investigate the properties of solutions (\ref{eq:PhaseSolutionSMZeroOrd})~---~(\ref{eq:phase_solution_3}): each of them contains a quadratic term of the form \(\pi/2\,(x/\!\spi \pm \sigma_{i})^{2}\). Since the magnitudes of the rest of the terms in each solution scale with \(\lambda^{n}\), the functions look similar when plotted together, as can be seen in Figure~\ref{fig:phase_solutions_single_mode}. It is hard to distinguish them in this plot, unless we zoom in closely on a region. For example, if we focus on the region around \(x = 0\) we see that while \(\phi_{0} (x) = x^{2} / 2\) goes through the origin, the rest of the solutions deviate slightly from zero. This is due to the fact that a perturbative solution of order \(n\) is only approximate, and is correct up to a remainder of \(\bigo{\lambda^{n+1}}\). Indeed, if we consider each solution \(\phi_{\{1, 2, 3\}} (x)\) for \(x = 0\):
\begin{widetext}
\begin{align*}
\phi_{1} (0) = \frac{\pi}{8}\,\lambda^{2},  \;
\phi_{2} (0) = \pi\Bigg(\!\frac{1 - \sqrt{2}}{8}\,\lambda^{3} + \frac{1}{16}\,\lambda^{4}\!\Bigg), \;
\phi_{3} (0) = \pi\Bigg(\!\frac{3\sqrt{3} - 4\sqrt{2} + 1}{32}\,\lambda^{4} - \frac{3\sqrt{6} - 4 - \sqrt{2}}{64}\,\lambda^{5} + \frac{28 - 6 \sqrt{3}}{512}\,\lambda^{6}\!\Bigg)
\end{align*}
\end{widetext}
we find that they are indeed of the expected order in each case (see Figure~\ref{fig:phaseAtZeroPlot} for a plot of these functions). These expressions have further higher order corrections coming from sub-leading order terms in the expansion of Fresnel cosine integrals (see eq.~(\ref{eq:fresnelc_expansion}) in Appendix~\ref{sec:AppFresnelFuncs}). Since these solutions are hard to examine graphically, we need to employ other techniques to compare the accuracy of the solutions. Specifically, we compare them with a purely numerical solution of the phase equation.

A simple plot like the lower one in Fig.~\ref{fig:phase_solutions_single_mode} shows that our highest-order solution, eq.~(\ref{eq:phase_solution_3}), does indeed compare well --- within \(\bigo{\lambda^{4}}\) --- to a purely numerical solution \citep{mathematicav10.4}. We need to quantify the comparison between this numerical solution and our analytical solutions as the mode parameter \(\lambda\) varies.

\begin{figure*}[t]
\includegraphics[scale=1.0]{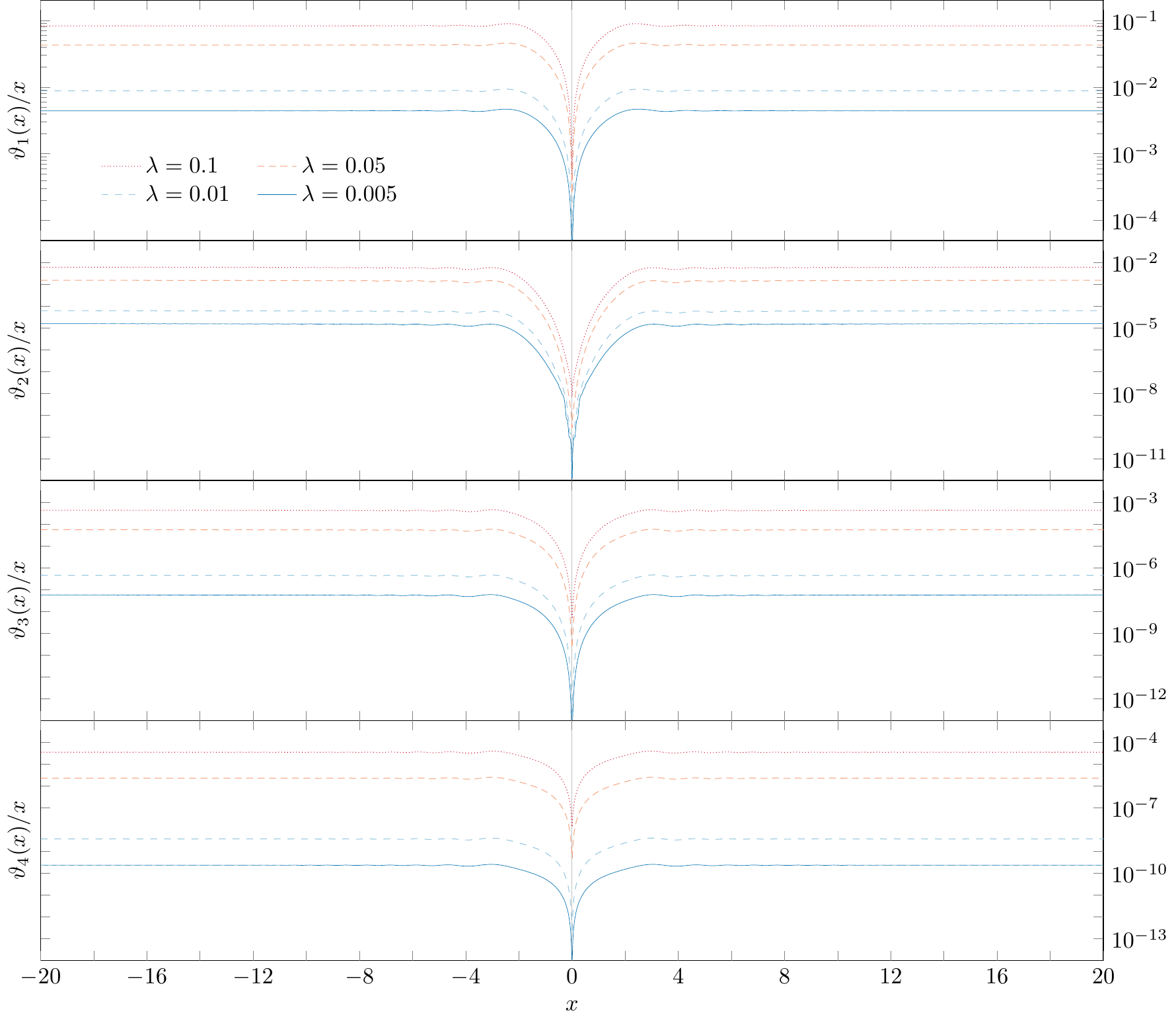}
\caption{Graphical demonstration of the validity of the phase resonance solution for 4 different values of the parameter \(\lambda\). The remainders \(\{\vartheta_{n} (x)\}\) are normalised by \(x\) in order to remove the linear trend. It is evident that the remainders scale with the value of the resonance modification \(\lambda\), and for each order \(n\) of the solution, the remainder is of order \(\bigo{\lambda^{n+1}}\).}
\label{fig:phaseValidityPlot}
\end{figure*}

It is important to demonstrate that each of the corrections \(\phi_{(n)} (x)\) appropriately captures all of the behaviour of the solution up to order \(\bigo{\lambda^{n}}\), i.e., that the difference between the full numerical solution \(\phi_{\text{num}} (x)\) and the correction is of order \(\bigo{\lambda^{n+1}}\) at most. To do this, we employ a technique that allows us to solve numerically for that difference (instead of naively subtracting each analytical solution from the numerical map). Consider eq.~(\ref{eq:phaseEquationSMZeroPh}) which is our initial source equation:
\begin{align}
\phi^{\prime\prime} (x) = 1 + \lambda \cos\!\big(\phi (x)\big). \tag{\ref{eq:phaseEquationSMZeroPh}}
\end{align}
As per our recipe, we would substitute the highest-known-order solution \(\phi_{n} (x)\) on the right-hand side, and solve the equation for the next-order solution \(\phi_{n+1} (x)\). In reality, the solution we are looking for appears on either side of the equation, and furthermore the functions \(\phi (x)\) also contain the orders beyond \(n\) which we have so far been ignoring in eqs.~(\ref{eq:generalPhaseEquationIteration}) and (\ref{eq:gen_phase_equation_reduced_single_mode}):
\begin{align}
\phi (x) = \phi_{0} (x) + \phi_{(n)} (x) + \vartheta_{n+1} (x). \label{eq:complete_solution_function}
\end{align}
Here, \(\phi_{0} (x) = x^{2} / 2\) and \(\phi_{(n)} (x)\) is the correction to \(\phi_{0}\). \(\vartheta_{n+1} (x)\) is the remainder of the complete solution which is not included into \(\phi_{n} (x)\). We can easily find a generic equation satisfied by the functions \(\vartheta_{n+1} (x)\) by using expression~(\ref{eq:complete_solution_function}) in eq.~(\ref{eq:phaseEquationSMZeroPh}) and making use of eq.~(\ref{eq:phaseEquationSMZeroPh}). In general, \(\vartheta_{n+1} (x)\) satisfies the differential equation
\begin{align}
\vartheta_{n+1}^{\prime\prime} (x) = \lambda\cos\!\left(\phi_{n} + \vartheta_{n+1} (x)\right) - \phi_{(n)}^{\prime\prime} (x).
\end{align}
For \(n = 0\) we can find directly from the source equation
\begin{align}
\vartheta_{1}^{\prime\prime} (x) = \lambda \cos\!\left(\phi_{0} (x) + \vartheta_{1} (x)\right).
\end{align}
By solving for \(\vartheta_{n+1} (x)\) and confirming that the result is \(\bigo{\lambda^{n+1}} \forall x\) we verify that \(\phi_{n} (x)\) capture the complete \(\bigo{\lambda^{n}}\) behaviour. The plots for \(\vartheta_{n} (x) / x\) (to remove the expected linear-in-\(x\) general trend) are shown in Figure~\ref{fig:phaseValidityPlot}, where we have performed the computation for 4 distinct values of \(\lambda\): \(0.1, 0.05, 0.01,\) and \(0.005\).

The corrections \(\phi_{(n)} (x)\) themselves are plotted (for \(\lambda^{(1)} = 0.1\)) in Figure~\ref{fig:raw_corrections_comparison}, found in the Appendix. Similarly, plots of the oscillating functions \(\rho_{1} (x)\), \(\rho_{2} (x)\), and \(\rho_{3} (x)\) can be found in Figures~\ref{fig:eta_plots1}, \ref{fig:eta_plots2}, and \ref{fig:eta_plots3}, respectively, with other relevant discussions in Appendix~\ref{sec:app_extra_plots}.

\begin{figure*}[t]
\includegraphics[scale=1.0]{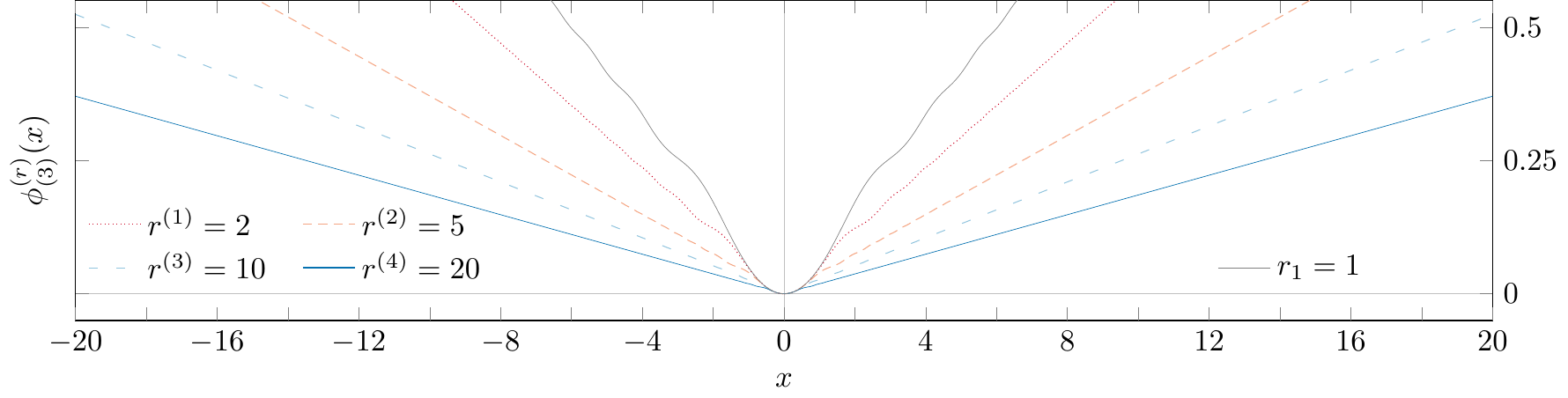}
\caption{The third-order correction to the phase resonance equation for different values of the mode number \(r\). These solutions all have mode parameters \(\lambda_{r} = 0.1\), so the graph only reflects the effect which the number of the mode has on the solution.}
\label{fig:phaseSMrThirdOrdPlot}
\end{figure*}

Before moving on to the case of multiple modes, it is intriguing to look into several other toy models. So far we have only looked at the resonance equation with a single mode \(n = 1\). We now consider the same equation, but with the sole mode \(n = r\) on the right-hand side. The leading-order solution is still (\ref{eq:PhaseSolutionSMZeroOrd}).
\begin{align}
\left(\phi^{(r)} (x)\right)^{\prime\prime} = 1 + \lambda_{r} \cos\!\left(r \phi\right). \label{eq:phase_equation_single_mode_r}
\end{align}
The explicit solution of the above equation is redundant, since it can be obtained from eq.~(\ref{eq:phaseEquationSMZeroPh}) by setting
\begin{subequations}
\begin{minipage}{\textwidth}
\begin{minipage}{.247\textwidth}
\begin{align}
\phi \; \mapsto \; \phi^{(r)} = r \phi, \\[-6pt]\nonumber
\end{align}
\end{minipage}
\begin{minipage}{.23\textwidth}
\begin{align}
x \; \mapsto \; \sqrt{r} \, x. \\[-6pt]\nonumber
\end{align}
\end{minipage}
\end{minipage}
\end{subequations}
and therefore its corrections could be obtained from eqs.~(\ref{eq:phase_solution_1}) -- (\ref{eq:phase_solution_3}) by applying these transformations. The explicit form of the solutions is presented in Appendix~\ref{sec:app_phase_equation_single_mode_r}. In Figure~\ref{fig:phaseSMrThirdOrdPlot} we show these solutions (at third order, i.e. complete) for different values of the mode number \(r\). The modified \(\sigma\)-terms are presented in Figure~\ref{fig:app_slope_r_plot}, while the numerical validation of these solutions can be found in Figure~\ref{fig:phaseRValidityPlot}.

Another toy model which deserves attention is the case of two distinct modes on the right-hand side:
\begin{align}
\left(\phi^{(r, s)} (x)\right)^{\prime\prime} = 1 + \lambda_{r} \cos\!\left(r \phi\right) + \lambda_{s} \cos\!\left(s \phi\right). \label{eq:phase_equation_two_modes}
\end{align}
The leading-order solution is unchanged, while the higher-order corrections are presented in Appendix~\ref{sec:app_phase_equation_two_modes}. We note that at first order, the correction is just the sum of the corrections due to each of the separate oscillating modes on the right-hand side. This, however, is not the case at higher order, where cross-terms appear. It is important to note that these cross-terms are not sub-leading to the other quadratic terms in the solution. This is in contrast to the frequency-resonance equation, where these cross-terms were sub-leading in this respect (see Appendix~\ref{sec:freq_extension_double}). In Figure~\ref{fig:slopeRSPlot} we can examine the \(\sigma\)-terms associated with this equation, while the validity of the solutions is presented in Figure~\ref{fig:phaseRSValidityPlot}.

Our objective in this paper is to extend this approach beyond the scope which has been developed thus far. So far we have explored the solution to third order in the mode parameter. However, we are still limited to the case of a single mode, and we are assuming that the initial phase is \(\phi (0) = 0\). We now relax the first of these restrictions and allow for an infinite number of oscillating modes, each with a distinct parameter \(\lambda_{i}\). This is achieved through allowing for multiple modes in the Fourier expansion, thus lifting the assumption that the contribution derives primarily from the first Fourier mode.

\begin{figure*}[t]
\includegraphics[scale=1.0]{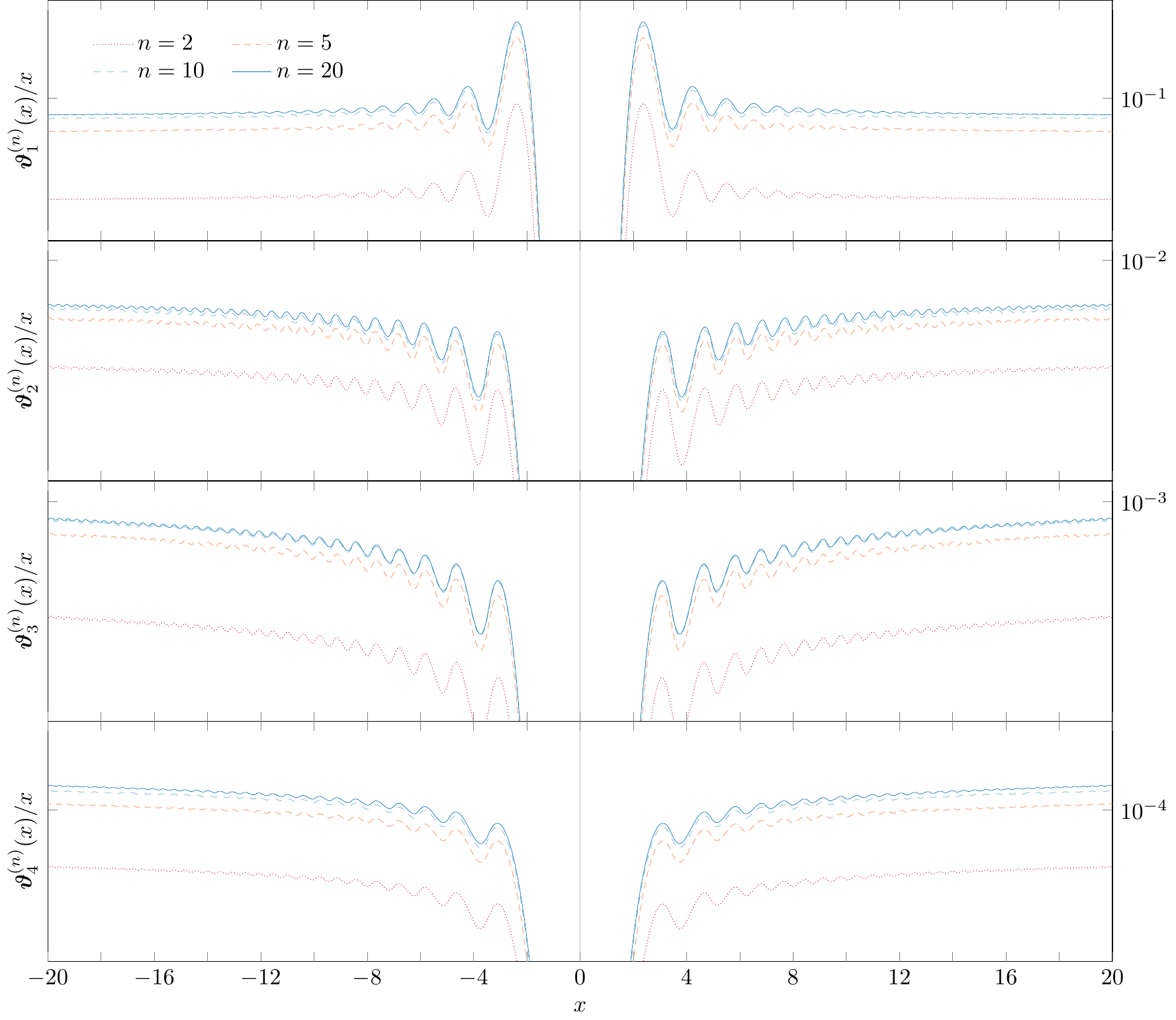}
\caption{Graphical demonstration of the validity of solutions (\ref{eq:phi0_m_solution}), (\ref{eq:phase_equation_multi_mode_first_order_solution}), (\ref{eq:phaseMMSolutionScndOrd}), and (\ref{eq:phase_equation_multi_mode_third_order_solution}). All these solutions are for \(\lambda_{1} = 0.1\), but include numbers of modes. The \(\lambda\) parameters follow the law \(\lambda_{n} = \lambda_{1} / n^{2.5}\).}
\label{fig:phaseMMValidityPlot}
\end{figure*}

We derived the phase resonance equation in Section~\ref{sec:InstantaneousFrequencyApproach}. In the regime of an unrestricted number of oscillating modes, still assuming that the initial phase is 0, the form of the phase resonance equation is given by
\begin{align}
\bphi^{\prime\prime} (x) = 1 + \sum_{n} \lambda_{n} \cos\left(n \bphi\right) \label{eq:phase_equation_multi}
\end{align}
Similar to the previous equation under consideration, this is a second order non-linear differential equation, and therefore finding an exact solution is not possible, and a perturbative solution quickly grows to become heavily involved.  For brevity, here we only present the final solutions, while details about the solutions and relevant discussion are collected in Appendix~\ref{sec:app_phase_solutions}. At zero'th order in the mode parameters \(\{\lambda_{n}\}\), we ignore any oscillating terms on the right-hand side, and the resulting differential equation is identical to that in the single-mode case, eq.~(\ref{eq:phaseEqSMZeroPhReduced}); the solution is obtained by integrating twice:
\begin{ph}{align}
\bphi_{0}^{\prime\prime} = 1 \quad \Leftrightarrow \quad \bphi_{0} (x) = \frac{x^{2}}{2} \label{eq:phi0_m_solution}
\end{ph}
Building onto this result, we construct the equivalent to eq.~(\ref{eq:generalPhaseEquationIteration}), which in the multi-mode case is now given by
\begin{align}
\bphi_{m+1}^{\prime\prime} (x) = 1 + \sum_{n} \lambda_{n} \cos\big(n\,\bphi_{m} (x)\big). \label{eq:gen_phase_equation_multi_first_order} \\ \nonumber
\end{align}
We will solve this consecutively for \(m=0,1\) and 2, and we will obtain the respective solutions to the phase resonance equation for the multi-mode case. At first instance, we use \(\bphi_{0} (x) = x^{2} / 2\) as the argument of the trigonometric function on the right-hand side, and investigate the solution to this equation to first order in \(\{\lambda_{n}\}\), which will yield the first-order correction \(\bphi_{(1)}(x)\):
\begin{align}
\bphi_{(1)}^{\prime\prime} (x) = \sum_{n} \lambda_{n} \cos\left(n \bphi_{0}\right) \label{eq:phase_multimode_first_correction_equation}
\end{align}
The details of finding the first-order correction can be found in Appendix~\ref{sec:app_phase_solution_multimode}. Combining \(\bphi_{(1)} (x)\) with (\ref{eq:phi0_m_solution}), we establish the complete first-order solution.
\begin{widetext}
\vspace*{-10pt}
\begin{ph}{align}
\bphi_{1} (x) = \frac{\pi}{2}\!\left(\!\frac{x}{\spi} \pm \frac{1}{2} \sum_{n} \frac{\lambda_{n}}{\sqrt{n}}\!\right)\np{2} + \sum_{n} \frac{\lambda_{n}}{n}\,\bm{\rho}_{1}^{n}\!\LinearTerm.\label{eq:phase_equation_multi_mode_first_order_solution}
\end{ph}
\end{widetext}
Here \(\bm{\rho}_{1}^{n} (x)\) is an oscillating term proportional to \(|\lambda|\), equivalent to \(\rho_{1} (x)\), and given explicitly in Appendix~\ref{sec:appendixPhaseSolutionSMZeroPh}. Compared to (\ref{eq:phi0_m_solution}), eq.~(\ref{eq:phase_equation_multi_mode_first_order_solution}) has a modified quadratic term and a part which oscillates around 0 for all \(x\).

We apply the same technique to find the second-order solution, namely \(\bphi_{(2)} (x)\), which contains terms that are quadratic in the parameters \(\{\lambda_{n}\}\). Similarly to previous iterations, we use the current highest-order solution \(\bphi_{1} (x)\) to establish a differential equation for \(\bphi_{(2)} (x)\). The differential equation we want to solve is given by
\begin{align}
\bphi_{(2)}^{\prime\prime} (x) = \sum_{n} \lambda_{n} \cos\big(n\,\bphi_{1} (x)\big)
\end{align}
We should remark again that the solution of this equation would yield \(\bphi_{(2)} (x)\), the second-order correction to the quadratic solution. It will contain terms which are first-order in the mode parameters, as well as other terms which are of second order. This is an artefact of the method we use for asymptotically establishing higher-order corrections to our source equation. In practice, we need to solve the equation
\begin{align}
\bphi_{(2)}^{\prime\prime} (x) = \sum_{n} \lambda_{n} \cos\left(n \!\left(\bphi_{0} (x) + \bphi_{(1)} (x)\right)\right) \label{eq:phase_equation_second_order_multi_mode_reworked}
\end{align}
Then, by adding the zero'th order part \(\bphi_{0} (x)\) to the solution of eq.~(\ref{eq:phase_equation_second_order_multi_mode_reworked}) we will obtain the complete solution to second order in \(\{\lambda_{n}\}\). Solving this equation is rather involved, therefore it is better to leave the practicalities of obtaining the solution to Appendix~\ref{sec:app_phase_solution_multimode}. Here we only present the final solution:
\begin{widetext}
\vspace*{-10pt}
\begin{ph}{align}
\begin{split}
&\bphi_{2} (x) = \frac{\pi}{2}\!\LinearTerm[][\pm\!\left[\frac{1}{2} \sum_{n} \frac{\lambda_{n}}{\sqrt{n}} + \sum_{n,\,m} \frac{1}{8}\left(|n-m|^{\sfrac{3}{2}} - (n+m)^{\sfrac{3}{2}}\right)\frac{\lambda_{n}}{n} \frac{\lambda_{m}}{m}\right]]\np{2} \\
&\quad\quad\quad\quad\quad\quad\quad\quad\quad\quad\quad\quad\quad\quad\quad\quad\quad\quad\quad\quad\quad\quad\quad\quad\quad + \sum_{n} \frac{\lambda_{n}}{n}\,\bm{\rho}_{1}^{n}\!\LinearTerm[][\FirstOrderMultiMode] + \sum_{n,\,m} \frac{\lambda_{n}}{\sqrt{n}} \frac{\lambda_{m}}{\sqrt{m}}\,\bm{\rho}_{2}^{nm}\!\LinearTerm \label{eq:phase_equation_multi_mode_second_order_solution}
\end{split}
\end{ph}
\end{widetext}

While the solution to second order is already quite involved, we continued our calculations and derived the solution to third order, so we can draw a comparison with the single-mode result eq.~(\ref{eq:phase_solution_3}). The generating equation involves the new correction \(\bphi_{(3)} (x)\) on the left-hand side, and the current solution, \(\bphi_{2} (x)\) as the argument on the right-hand side:
\begin{align}
\bphi_{(3)}^{\prime\prime} (x) = \sum_{n} \lambda_{n} \cos\big(n\,\bphi_{2}(x)\big).
\end{align}
The solution is rather long, and hence we have placed it, together with explanations, in Appendix~\ref{sec:app_phase_solution_multimode}. The correction at third order is given by eq.~(\ref{eq:phase_equation_multi_mode_third_order_solution}).

After having derived these solutions, it is important to verify their validity numerically, using the same method as in the single-mode case. The results from this verification are presented in Figure~\ref{fig:phaseMMValidityPlot}, where we have shown that the equation is valid for several different values of the total number of modes. To depict the effect a larger number of modes has on the solution, we have used the same initial value of the resonant flux modification \(\lambda\).

So far we have solved the single- and multi-mode equations up to third order in the parameters \(\{\lambda_{n}\}\). It is important to carefully consider these solutions we have obtained thus far, and evaluate the need for further iterations. Let us investigate the governing equation
\begin{align}
\bphi^{\prime\prime} (x) = 1 + \sum_{n} \lambda_{n} \cos\left(n \bphi\right) \tag{\ref{eq:phase_equation_multi}}
\end{align}
and consider those terms for which the small-angle approximation for the trigonometric functions breaks down:
\begin{align}
\cos \!\Bigg[n \!\left(\frac{x^{2}}{2} + \bphi_{(i)} (x)\right)\!\Bigg] \, , \; \bphi_{(i)} = \sum_{m} \frac{\lambda_{m}}{m} \, (\bullet) + \ldots\,.
\end{align}
This occurs when \(n \, \lambda_{1} \sim 1\). Therefore, we need to impose a restriction on the upper bound of \(n\):
\begin{align}
\bar{n} \, \lambda_{1} \ll 1 \quad \Leftrightarrow \quad \bar{n} \ll \lambda_{1}^{-1}.
\end{align}
Although the cross-term corrections here are of the same magnitude as the single-mode solutions, the bound is comparable to that in the frequency equation (cf. eq.~(\ref{eq:frequency_bound})). We invoke the coefficient scaling law which we discussed in Section~\ref{sec:RegimeDomainOfVariables} to link this to values of \(\lambda_{1}\):
\begin{align}
\lambda_{\bar{n}} = \frac{\lambda_{1}}{\bar{n}^{\alpha}} \sim \lambda_{1}^{1 + \alpha} < \lambda_{1}^{3}
\end{align}
for the suggested values of \(2 < \alpha < 3\) \citep{drasco2006}. We can therefore argue that beyond that order, we would not be able to distinguish our solutions, valid in the regime where the small-angle approximation can be applied, from effects which arise from the violation of this assumption. For this reason, higher-order corrections cannot be easily calculated without restricting the total number of modes \(n\). Finally, proceeding beyond third order is pointless since for the range of values we are investigating, \(\lambda^{3} \sim \epsilon^{\sfrac{1}{2}} \sim \bigo{10^{-3}}\), hence in calculating further corrections we need to account for terms arising from higher-order mass-ratio corrections. In this article, we only investigate solutions up to third order inclusive, and disregard higher-order corrections.

\begin{figure}[t]
\includegraphics[scale=1.0]{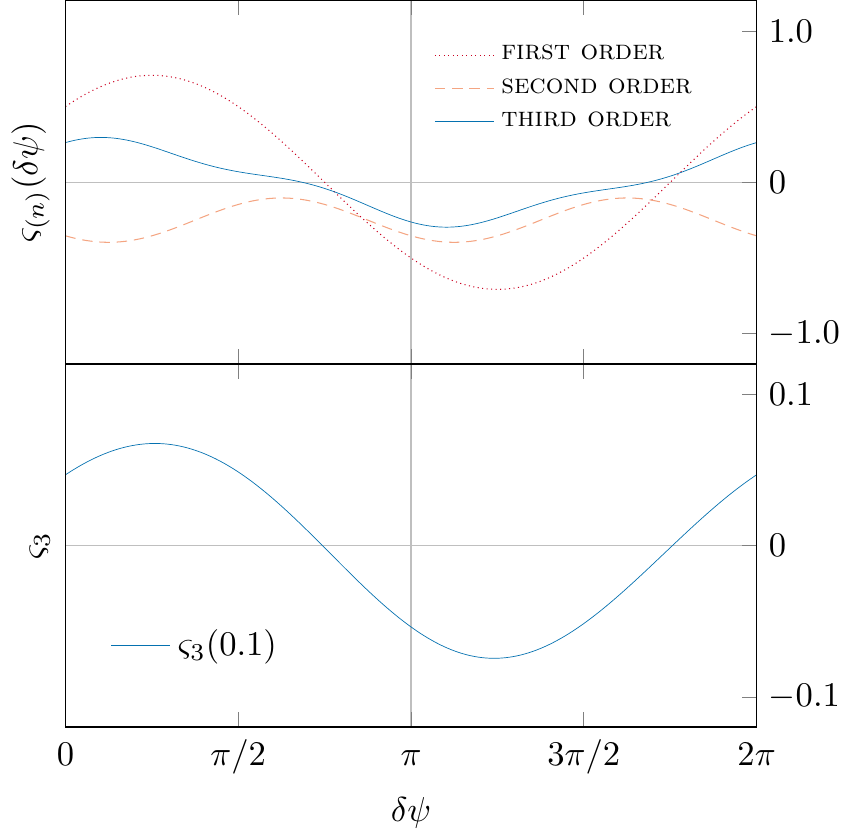}
\caption{\emph{Upper panel:} plots of the functions (of \(\delpsi\)) proportional to \(\lambda, \lambda^{2}\) and \(\lambda^{3}\) in the slope of solution~(\ref{eq:phaseEqSMPhaseThirdOrderSol}). \emph{Lower panel:} plot of the slope \(\varsigma_{3} (\lambda)\) as a function of \(\delpsi\) for \(\lambda = 0.1\).}
\label{fig:phasePhasePlot}
\end{figure}

\begin{figure*}[t]
\includegraphics[scale=1.0]{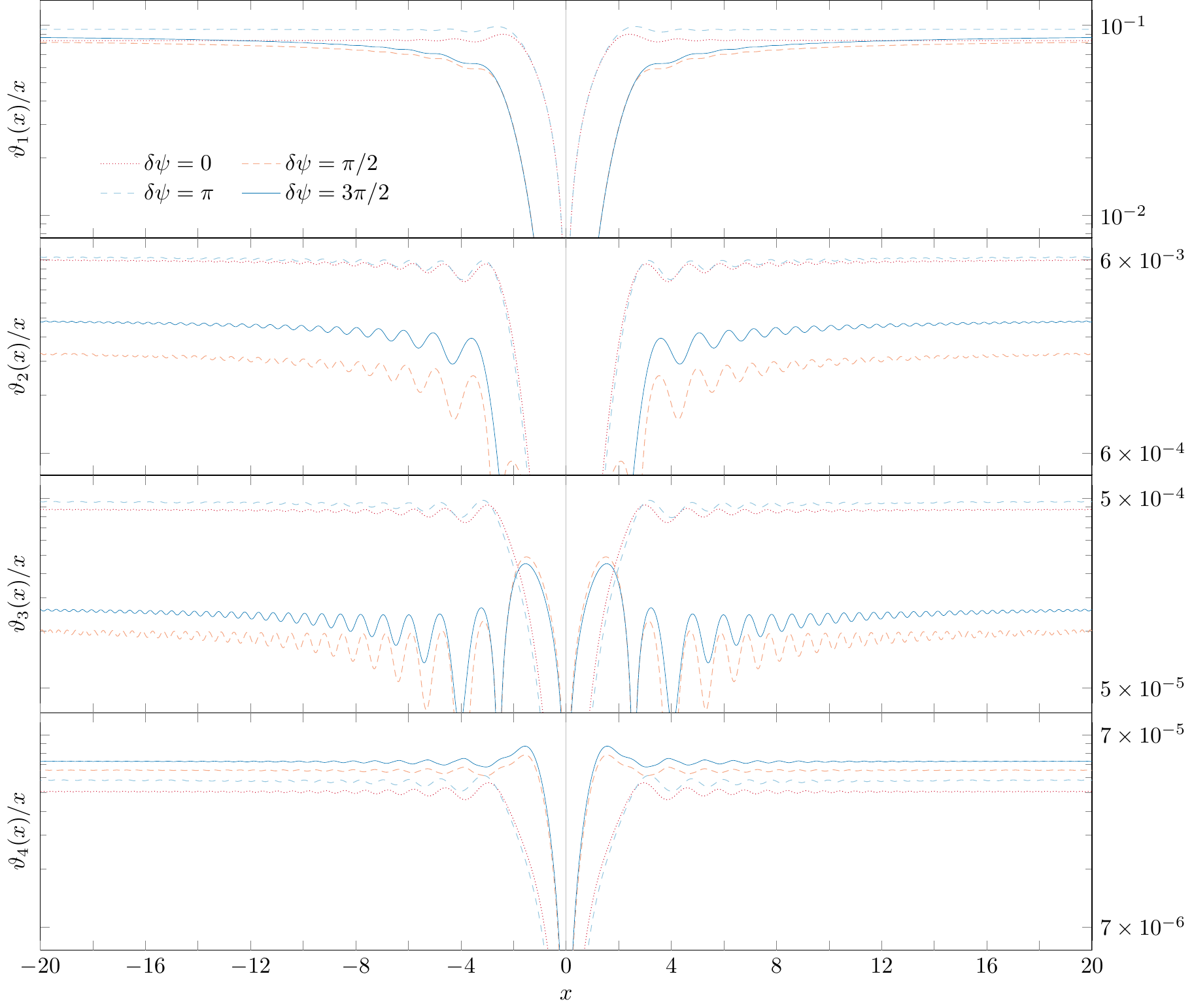}
\caption{Graphical demonstration of the validity of the solutions to the phase resonance equation with a single mode and a non-zero initial phase, eq.~(\ref{eq:phaseSMPhaseEquation}). All these solutions involve the same \(\lambda = 0.1\), but have different initial phases \(\delpsi\).}
\label{fig:phasePhaseValidityPlot}
\end{figure*}

We now return to the original differential equation (\ref{eq:phaseEquation}), and again consider a single mode, however this time we allow for a non-vanishing initial phase \(\delta \psi \neq 0\)
\begin{align}
\varphi^{\prime\prime} (x) = 1 + \lambda \cos\big(\varphi (x) - \delpsi\big). \label{eq:phaseSMPhaseEquation}
\end{align}
The zero'th-order solution, when we ignore the term proportional to \(\lambda\) on the right-hand side, is again given by \(\varphi_{0} (x) = x^{2} /2\). To establish first- and higher-order corrections, we proceed according to the same procedure as before, solving equations of the form:
\begin{align}
\varphi_{(n+1)}^{\;\prime\prime} (x) = \lambda \cos\big(\varphi_{n} (x) - \delpsi\big). \label{eq:phaseSMPhEq}
\end{align}
Using this, we find the first-order solution below.
\begin{widetext}
\vspace*{-20pt}
\begin{ph}{align}
\begin{split}
\varphi_{1} (x) = \frac{\pi}{2} \!\LinearTerm[][\SlopeFrstOrdSMP]\np{2}\!+ \lambda\,\varrho_{1}\!\!\LinearTerm \label{eq:phase_equation_nonzero_phase_first_order_solution}
\end{split}
\end{ph}
\vspace*{-10pt}
\end{widetext}
We notice that the new slope \(\varsigma_{1} (\lambda)\) involves the constant \(\delpsi\), and that the oscillating term \(\varrho_{1}\) has a different form, now involving a Fresnel sine term too. They reduce to their zero-phase counterparts for \(\delpsi = 0\). Following our established algorithm, we can easily derive the second- and third-order corrections to \(\varphi_{0} (x)\). Details of these derivations are left to Appendix~\ref{sec:app_phase_solution_multimode}, here we only present the final results.
\begin{widetext}
\vspace*{-20pt}
\begin{ph}{align}
\varphi_{2} (x) &= \frac{\pi}{2}\!\LinearTerm[][\pm \left[\vphbig\frac{1}{2}\big(\!\cos(\delpsi) + \sin(\delpsi)\big)\lambda - \left(\!\frac{\sqrt{2}}{4} \big(\!\cos(2\delpsi) + \sin(2\delpsi)\big) - \frac{1}{2} \sin(\delpsi) \big(\!\cos(\delpsi) - \sin(\delpsi)\big)\!\right)\!\lambda^{2}\right]]\np{2} \nonumber\\
& \RepQuad{30} + \lambda\,\varrho_{1}\!\!\LinearTerm[][\pm \varsigma_{1} (\lambda)]\!+ \lambda^{2}\varrho_{2}\!\!\LinearTerm \label{eq:phase_equation_nonzero_phase_second_order_solution} \\
\varphi_{3} (x) &= \frac{\pi}{2}\!\left(\vphbigggg\!\frac{x}{\spi}\pm\left[\vphbigg\frac{1}{2}\big(\!\cos(\delpsi) + \sin(\delpsi)\big)\lambda -\!\left(\!\frac{\sqrt{2}}{4} \big(\!\cos(2\delpsi) + \sin(2\delpsi)\big) - \frac{1}{2} \sin(\delpsi) \big(\!\cos(\delpsi) - \sin(\delpsi)\big)\!\right)\!\lambda^{2}\right.\right. \nonumber\\[5pt]
&\RepQuad{1} + \left(\!\frac{3\sqrt{3}}{16}\big(\!\cos(3\delpsi) + \sin(3\delpsi)\big) - \frac{\sqrt{2}}{2}\,\sin(\delpsi)\big(\!\cos(2\delpsi) - \sin(2\delpsi)\big) - \frac{3}{16}\big(\!\cos(\delpsi) + \sin(\delpsi)\big) \right. \nonumber\\
&\RepQuad{2} \left.\vphbigggg \left.\vphbigg \left.\vph + \frac{1}{8}\,\cos(2\delpsi)\big(\!\cos(\delpsi) + \sin(\delpsi)\big)\!\right)\!\lambda^{3}\right]\!\right)\np{2}\!+ \lambda\,\varrho_{1}\!\!\LinearTerm[][\pm \varsigma_{2} (\lambda)]\!+ \lambda^{2}\varrho_{2}\!\!\LinearTerm[][\pm \varsigma_{1} (\lambda)]\!+ \lambda^{3}\varrho_{3}\!\!\LinearTerm \label{eq:phaseEqSMPhaseThirdOrderSol}
\end{ph}
\end{widetext}
It is important to investigate how the slope of these solutions evolves with \(\delpsi\). At each order we encounter a different function of the initial phase. Figure~\ref{fig:phasePhasePlot} shows how these functions behave over the range of \(\delpsi\). Furthermore, we need to verify that these solutions are valid for different values of the initial phase. We computed the quantities \(\vartheta_{n} (x)\) for different values of the initial phase (keeping \(\lambda\) constant) using the same method as before, and show the results in Figure~\ref{fig:phasePhaseValidityPlot}.

Finally, we approach the phase resonance equation in the most general regime: including an infinite number of modes each with a different non-zero phase:
\begin{align}
\bm{\varphi}^{\prime\prime} (x) = 1 + \sum_{n > 1} \lambda_{n} \cos\!\big(n \bm{\varphi}(x) - \delta \psi_{n}\big). \tag{\ref{eq:phaseEquation}}
\end{align}
The details of the solution are too lengthy to be presented in enough detail here and can be found in Appendix~\ref{sec:app_phase_solution_multi_mode_nonzero_phase}: the first-order correction is given by eq.~(\ref{eq:appMMPhFrstOrdSol}) and the second-order correction by eq.~(\ref{eq:appMMPhScndOrdSol}). Upon closer examination these bear the same features of our previously establishes solutions. Finally, we have established the third-order solution to this equation, \(\bm{\varphi}_{3} (x)\). It is important to note that solution~(\ref{eq:phaseSolFinal}) is the most general solution to the phase resonance equation that we could find. It allows for multiple modes and a non-vanishing initial phase, and treats perturbations up to and including third order. The third-order slope function \(\bm{\varsigma}_{3}\) is given by eq.~(\ref{eq:appMMPThrdOrdSol}) in Appendix~\ref{sec:app_phase_solution_multi_mode_nonzero_phase}. Furthermore, each of the three corrections are given by eqs.~(\ref{eq:correctionFirstOrder}), (\ref{eq:correctionSecondOrder}), and (\ref{eq:correctionThirdOrder}).
\begin{widetext}
\begin{ph}{align}
\bm{\varphi}_{3} (x) = \frac{\pi}{2}\!\LinearTerm[][\SlopeThrdOrdMMP]\np{2}\!+ \sum_{n} \frac{\lambda_{n}}{n} \, \bm{\varrho}_{1}^{n}\!\LinearTerm[][\SlopeScndOrdMMP]\!+ \sum_{n,\,m} \frac{\lambda_{n}}{\sqrt{n}} \frac{\lambda_{m}}{\sqrt{m}} \, \bm{\varrho}_{2}^{nm}\!\LinearTerm[][\SlopeFrstOrdMMP]\!+\!\!\sum_{n,\,m,\,p}\!\!\lambda_{n} \frac{\lambda_{m}}{\sqrt{m}}  \frac{\lambda_{p}}{\sqrt{p}} \, \bm{\varrho}_{3}^{nmp}\!\LinearTerm \label{eq:phaseSolFinal}
\end{ph}
\end{widetext}

This is the final and most important result of this article. While it is not easily usable in theoretical calculations, it could find applications in modelling resonances in new kludge waveforms, as this is not a feature currently implemented. We have provided a solution to the highest possible order of perturbation, which is important for meaningful numerical simulations. Furthermore, we have also provided solutions to simplified versions of this problem, which could be employed in various theoretical calculations regarding \textsc{emri}s.

\section{Discussion and conclusions}
\label{sec:discussion}
\noindent
In this paper we have described in Sections~\ref{sec:InstantaneousFrequencyApproach} and \ref{sec:two} the assumptions and conditions which form the basis of two different methods for analysis of resonances in \textsc{emri}s. The two-timescale approach, described in \cite{meent2014} presumes that the functional dependence of the orbital characteristics can be expanded in power series of the mass-ratio and time from resonance. The instantaneous frequency model, as per \cite{gair2012}, relies on the instant values of the on-resonance frequencies as a basis for asymptotic analysis, and this is a more complicated function of the (rescaled) time.

After describing the basic approach, Section~\ref{sec:SolutionsResonanceEqs} described in detail the solution to the frequency and phase resonance equations derived in the course of describing the instantaneous frequency approach. Some of these solutions were previously published in \cite{gair2012}. The difference between the frequency and phase resonances lies in the assumption about the phasing of the self-force. It is more natural that the phase will be determined by the instantaneous orbital phase (as in the phase treatment) than the orbital phase at some fiducial \(\tau = 0\) (as in the frequency investigation). Fully-detailed self-force calculations would be needed to make an informed choice between the two, but it is likely that the phase approach is physically relevant for the resonance problem. The frequency resonance solutions exhibits interesting effects, in particular the ``memory" effect on the slope of the graph, which would be interesting if it really occurred in any physical situation. Here we extended these results, by including higher-order terms and a generalised form of the solutions. We have derived solutions for the case of multiple oscillating modes, given by eq.~(\ref{eq:frequencySolutionMM}). Working further, we have relaxed the assumption of a vanishing initial phase, and have presented the single-mode solution in eq.~(\ref{eq:frequencySMPhaseSolutionThrdOrd}). Finally, combining these results we presented the complete frequency solution, eq.~(\ref{eq:frequencySolutionMMPhase}).

As mentioned above, the phase resonance equation is probably more relevant physically, as it describes the evolution of the phase coordinates near resonance, and its solutions can be related to the two-timescale method results. It is also interesting from a mathematical view point, since it is a second-order non-linear differential equation which can be approached using asymptotic methods. We extended the previously established sequence of results by deriving the third-order solution for the case of a single oscillating mode, eq.~(\ref{eq:phase_solution_3}). We presented further toy models, discussed in detail in Appendices~\ref{sec:app_phase_equation_single_mode_r} and \ref{sec:app_phase_equation_two_modes} before demonstrating an order-by-order solution to the phase resonance equation with an arbitrary (possibly infinite) number of modes, eqs.~(\ref{eq:phase_equation_multi_mode_first_order_solution}), (\ref{eq:phase_equation_multi_mode_second_order_solution}), and (\ref{eq:phase_equation_multi_mode_third_order_solution}). Following this we presented the solution to the phase equation when the alternative initial assumption is relaxed: namely, the starting phase is allowed to vary. The first- and second-order solutions are given by eqs.~(\ref{eq:phase_equation_nonzero_phase_first_order_solution}) and (\ref{eq:phase_equation_nonzero_phase_second_order_solution}). Finally, we have derived the solution in the most general case, allowing for an arbitrary (possibly infinite) number of modes, each with an individual initial phase. This solution was given in eq.~(\ref{eq:phaseSolFinal}). In the case that there is only a single mode, this calculation agrees with the result given in \cite{gair2012}, to leading order in the parameter \(\lambda\), which is equivalent to the order of \(G_1(\mathbf{J})\) in the two-timescale analysis.

Both the instantaneous frequency approach and the two-timescale analysis are valid treatments of the \textsc{emri} problem --- though their foundational assumptions appear different, they both have the same computational complexity and arrive at equivalent conclusions. Previous work on the two-timescale method has only been presented to the lowest order in the mass ratio \(\epsilon\) and the resonance flux change parameters \(\{\kappa, \lambda\}\). As has been discovered, in certain scenarios higher-order corrections in the latter can be more important than those to the former, hence it is vital to investigate the form of these corrections. This is what the current paper accomplishes.

Further work on solving the phase-resonance equation would use these equations in numerical waveform models. These could be utilised in implementing a computationally cheap approximate (kludge) model of \textsc{emri} waveforms which include resonance effects \citep{barakcutler2004, babakal2006}. At the moment such models lack the necessary features to predict the change in frequency or phase on resonance, which is precisely what the results in this article offer. Therefore, this should be pursued in a future publication.

\section*{Electronic resources}
\label{sec:electronicResources}
\noindent
Many of the results in this article are available in a Mathematica notebook, which can be found at:
\begin{center}
\texttt{http://resonances.deyanmihaylov.net/}
\end{center}

\section*{Acknowledgements}
\noindent
Deyan P Mihaylov is supported by the \textsc{stfc}. He is thankful to Dr Christopher J Moore and to Alvin Chua for the engaging conversations about general relativity, cosmology, and gravitational wave astronomy.

\bibliographystyle{apsrev4-1}
\bibliography{main}

\begin{thebibliography}{53}%
\makeatletter
\providecommand \@ifxundefined [1]{%
 \@ifx{#1\undefined}
}%
\providecommand \@ifnum [1]{%
 \ifnum #1\expandafter \@firstoftwo
 \else \expandafter \@secondoftwo
 \fi
}%
\providecommand \@ifx [1]{%
 \ifx #1\expandafter \@firstoftwo
 \else \expandafter \@secondoftwo
 \fi
}%
\providecommand \natexlab [1]{#1}%
\providecommand \enquote  [1]{``#1''}%
\providecommand \bibnamefont  [1]{#1}%
\providecommand \bibfnamefont [1]{#1}%
\providecommand \citenamefont [1]{#1}%
\providecommand \href@noop [0]{\@secondoftwo}%
\providecommand \href [0]{\begingroup \@sanitize@url \@href}%
\providecommand \@href[1]{\@@startlink{#1}\@@href}%
\providecommand \@@href[1]{\endgroup#1\@@endlink}%
\providecommand \@sanitize@url [0]{\catcode `\\12\catcode `\$12\catcode
  `\&12\catcode `\#12\catcode `\^12\catcode `\_12\catcode `\%12\relax}%
\providecommand \@@startlink[1]{}%
\providecommand \@@endlink[0]{}%
\providecommand \url  [0]{\begingroup\@sanitize@url \@url }%
\providecommand \@url [1]{\endgroup\@href {#1}{\urlprefix }}%
\providecommand \urlprefix  [0]{URL }%
\providecommand \Eprint [0]{\href }%
\providecommand \doibase [0]{http://dx.doi.org/}%
\providecommand \selectlanguage [0]{\@gobble}%
\providecommand \bibinfo  [0]{\@secondoftwo}%
\providecommand \bibfield  [0]{\@secondoftwo}%
\providecommand \translation [1]{[#1]}%
\providecommand \BibitemOpen [0]{}%
\providecommand \bibitemStop [0]{}%
\providecommand \bibitemNoStop [0]{.\EOS\space}%
\providecommand \EOS [0]{\spacefactor3000\relax}%
\providecommand \BibitemShut  [1]{\csname bibitem#1\endcsname}%
\let\auto@bib@innerbib\@empty
\bibitem [{\citenamefont {{Einstein}}(1918)}]{einstein1918}%
  \BibitemOpen
  \bibfield  {author} {\bibinfo {author} {\bibfnamefont {A.}~\bibnamefont
  {{Einstein}}},\ }\href@noop {} {\bibfield  {journal} {\bibinfo  {journal}
  {Sitzungsberichte der K{\"o}niglich Preu{\ss}ischen Akademie der
  Wissenschaften (Berlin), Seite 154-167.}\ ,\ \bibinfo {pages} {154}}
  (\bibinfo {year} {1918})}\BibitemShut {NoStop}%
\bibitem [{\citenamefont {{Hulse}}\ and\ \citenamefont
  {{Taylor}}(1975)}]{hulsetaylor}%
  \BibitemOpen
  \bibfield  {author} {\bibinfo {author} {\bibfnamefont {R.~A.}\ \bibnamefont
  {{Hulse}}}\ and\ \bibinfo {author} {\bibfnamefont {J.~H.}\ \bibnamefont
  {{Taylor}}},\ }\href {\doibase 10.1086/181708} {\bibfield  {journal}
  {\bibinfo  {journal} {\apjl}\ }\textbf {\bibinfo {volume} {195}},\ \bibinfo
  {pages} {L51} (\bibinfo {year} {1975})}\BibitemShut {NoStop}%
\bibitem [{\citenamefont {{Taylor}}\ and\ \citenamefont
  {{Weisberg}}(1982)}]{taylorweis1982}%
  \BibitemOpen
  \bibfield  {author} {\bibinfo {author} {\bibfnamefont {J.~H.}\ \bibnamefont
  {{Taylor}}}\ and\ \bibinfo {author} {\bibfnamefont {J.~M.}\ \bibnamefont
  {{Weisberg}}},\ }\href {\doibase 10.1086/159690} {\bibfield  {journal}
  {\bibinfo  {journal} {\apjl}\ }\textbf {\bibinfo {volume} {253}},\ \bibinfo
  {pages} {908} (\bibinfo {year} {1982})}\BibitemShut {NoStop}%
\bibitem [{\citenamefont {{Abbott}}\ \emph
  {et~al.}(2016{\natexlab{a}})\citenamefont {{Abbott}}, \citenamefont
  {{Abbott}}, \citenamefont {{Abbott}}, \citenamefont {{Abernathy}},
  \citenamefont {{Acernese}}, \citenamefont {{Ackley}}, \citenamefont
  {{Adams}}, \citenamefont {{Adams}}, \citenamefont {{Addesso}}, \citenamefont
  {{Adhikari}},\ and\ \citenamefont {et~al.}}]{ligo2016detect}%
  \BibitemOpen
  \bibfield  {author} {\bibinfo {author} {\bibfnamefont {B.~P.}\ \bibnamefont
  {{Abbott}}}, \bibinfo {author} {\bibfnamefont {R.}~\bibnamefont {{Abbott}}},
  \bibinfo {author} {\bibfnamefont {T.~D.}\ \bibnamefont {{Abbott}}}, \bibinfo
  {author} {\bibfnamefont {M.~R.}\ \bibnamefont {{Abernathy}}}, \bibinfo
  {author} {\bibfnamefont {F.}~\bibnamefont {{Acernese}}}, \bibinfo {author}
  {\bibfnamefont {K.}~\bibnamefont {{Ackley}}}, \bibinfo {author}
  {\bibfnamefont {C.}~\bibnamefont {{Adams}}}, \bibinfo {author} {\bibfnamefont
  {T.}~\bibnamefont {{Adams}}}, \bibinfo {author} {\bibfnamefont
  {P.}~\bibnamefont {{Addesso}}}, \bibinfo {author} {\bibfnamefont {R.~X.}\
  \bibnamefont {{Adhikari}}}, \ and\ \bibinfo {author} {\bibnamefont
  {et~al.}},\ }\href {\doibase 10.1103/PhysRevLett.116.061102} {\bibfield
  {journal} {\bibinfo  {journal} {Physical Review Letters}\ }\textbf {\bibinfo
  {volume} {116}},\ \bibinfo {eid} {061102} (\bibinfo {year}
  {2016}{\natexlab{a}})},\ \Eprint {http://arxiv.org/abs/1602.03837}
  {arXiv:1602.03837 [gr-qc]} \BibitemShut {NoStop}%
\bibitem [{\citenamefont {{The \textsc{LIGO} Scientific Collaboration}}\ \emph
  {et~al.}(2015)\citenamefont {{The \textsc{LIGO} Scientific Collaboration}},
  \citenamefont {{Aasi}}, \citenamefont {{Abbott}}, \citenamefont {{Abbott}},
  \citenamefont {{Abbott}}, \citenamefont {{Abernathy}}, \citenamefont
  {{Ackley}}, \citenamefont {{Adams}}, \citenamefont {{Adams}}, \citenamefont
  {{Addesso}},\ and\ \citenamefont {et~al.}}]{LIGO2015}%
  \BibitemOpen
  \bibfield  {author} {\bibinfo {author} {\bibnamefont {{The \textsc{LIGO}
  Scientific Collaboration}}}, \bibinfo {author} {\bibfnamefont
  {J.}~\bibnamefont {{Aasi}}}, \bibinfo {author} {\bibfnamefont {B.~P.}\
  \bibnamefont {{Abbott}}}, \bibinfo {author} {\bibfnamefont {R.}~\bibnamefont
  {{Abbott}}}, \bibinfo {author} {\bibfnamefont {T.}~\bibnamefont {{Abbott}}},
  \bibinfo {author} {\bibfnamefont {M.~R.}\ \bibnamefont {{Abernathy}}},
  \bibinfo {author} {\bibfnamefont {K.}~\bibnamefont {{Ackley}}}, \bibinfo
  {author} {\bibfnamefont {C.}~\bibnamefont {{Adams}}}, \bibinfo {author}
  {\bibfnamefont {T.}~\bibnamefont {{Adams}}}, \bibinfo {author} {\bibfnamefont
  {P.}~\bibnamefont {{Addesso}}}, \ and\ \bibinfo {author} {\bibnamefont
  {et~al.}},\ }\href {\doibase 10.1088/0264-9381/32/7/074001} {\bibfield
  {journal} {\bibinfo  {journal} {Classical and Quantum Gravity}\ }\textbf
  {\bibinfo {volume} {32}},\ \bibinfo {eid} {074001} (\bibinfo {year}
  {2015})},\ \Eprint {http://arxiv.org/abs/1411.4547} {arXiv:1411.4547 [gr-qc]}
  \BibitemShut {NoStop}%
\bibitem [{\citenamefont {{Acernese}}\ \emph {et~al.}(2015)\citenamefont
  {{Acernese}}, \citenamefont {{Agathos}}, \citenamefont {{Agatsuma}},
  \citenamefont {{Aisa}}, \citenamefont {{Allemandou}}, \citenamefont
  {{Allocca}}, \citenamefont {{Amarni}}, \citenamefont {{Astone}},
  \citenamefont {{Balestri}}, \citenamefont {{Ballardin}},\ and\ \citenamefont
  {et~al.}}]{virgo2015}%
  \BibitemOpen
  \bibfield  {author} {\bibinfo {author} {\bibfnamefont {F.}~\bibnamefont
  {{Acernese}}}, \bibinfo {author} {\bibfnamefont {M.}~\bibnamefont
  {{Agathos}}}, \bibinfo {author} {\bibfnamefont {K.}~\bibnamefont
  {{Agatsuma}}}, \bibinfo {author} {\bibfnamefont {D.}~\bibnamefont {{Aisa}}},
  \bibinfo {author} {\bibfnamefont {N.}~\bibnamefont {{Allemandou}}}, \bibinfo
  {author} {\bibfnamefont {A.}~\bibnamefont {{Allocca}}}, \bibinfo {author}
  {\bibfnamefont {J.}~\bibnamefont {{Amarni}}}, \bibinfo {author}
  {\bibfnamefont {P.}~\bibnamefont {{Astone}}}, \bibinfo {author}
  {\bibfnamefont {G.}~\bibnamefont {{Balestri}}}, \bibinfo {author}
  {\bibfnamefont {G.}~\bibnamefont {{Ballardin}}}, \ and\ \bibinfo {author}
  {\bibnamefont {et~al.}},\ }\href {\doibase 10.1088/0264-9381/32/2/024001}
  {\bibfield  {journal} {\bibinfo  {journal} {Classical and Quantum Gravity}\
  }\textbf {\bibinfo {volume} {32}},\ \bibinfo {eid} {024001} (\bibinfo {year}
  {2015})},\ \Eprint {http://arxiv.org/abs/1408.3978} {arXiv:1408.3978 [gr-qc]}
  \BibitemShut {NoStop}%
\bibitem [{\citenamefont {{Abbott}}\ \emph
  {et~al.}(2016{\natexlab{b}})\citenamefont {{Abbott}}, \citenamefont
  {{Abbott}}, \citenamefont {{Abbott}}, \citenamefont {{Abernathy}},
  \citenamefont {{Acernese}}, \citenamefont {{Ackley}}, \citenamefont
  {{Adams}}, \citenamefont {{Adams}}, \citenamefont {{Addesso}}, \citenamefont
  {{Adhikari}},\ and\ \citenamefont {et~al.}}]{abbott2016}%
  \BibitemOpen
  \bibfield  {author} {\bibinfo {author} {\bibfnamefont {B.~P.}\ \bibnamefont
  {{Abbott}}}, \bibinfo {author} {\bibfnamefont {R.}~\bibnamefont {{Abbott}}},
  \bibinfo {author} {\bibfnamefont {T.~D.}\ \bibnamefont {{Abbott}}}, \bibinfo
  {author} {\bibfnamefont {M.~R.}\ \bibnamefont {{Abernathy}}}, \bibinfo
  {author} {\bibfnamefont {F.}~\bibnamefont {{Acernese}}}, \bibinfo {author}
  {\bibfnamefont {K.}~\bibnamefont {{Ackley}}}, \bibinfo {author}
  {\bibfnamefont {C.}~\bibnamefont {{Adams}}}, \bibinfo {author} {\bibfnamefont
  {T.}~\bibnamefont {{Adams}}}, \bibinfo {author} {\bibfnamefont
  {P.}~\bibnamefont {{Addesso}}}, \bibinfo {author} {\bibfnamefont {R.~X.}\
  \bibnamefont {{Adhikari}}}, \ and\ \bibinfo {author} {\bibnamefont
  {et~al.}},\ }\href {\doibase 10.1007/lrr-2016-1} {\bibfield  {journal}
  {\bibinfo  {journal} {Living Reviews in Relativity}\ }\textbf {\bibinfo
  {volume} {19}} (\bibinfo {year} {2016}{\natexlab{b}}),\ 10.1007/lrr-2016-1},\
  \Eprint {http://arxiv.org/abs/1304.0670} {arXiv:1304.0670 [gr-qc]}
  \BibitemShut {NoStop}%
\bibitem [{\citenamefont {{Abadie}}\ \emph {et~al.}(2010)\citenamefont
  {{Abadie}}, \citenamefont {{Abbott}}, \citenamefont {{Abbott}}, \citenamefont
  {{Abernathy}}, \citenamefont {{Accadia}}, \citenamefont {{Acernese}},
  \citenamefont {{Adams}}, \citenamefont {{Adhikari}}, \citenamefont {{Ajith}},
  \citenamefont {{Allen}},\ and\ \citenamefont {et~al.}}]{rate2010}%
  \BibitemOpen
  \bibfield  {author} {\bibinfo {author} {\bibfnamefont {J.}~\bibnamefont
  {{Abadie}}}, \bibinfo {author} {\bibfnamefont {B.~P.}\ \bibnamefont
  {{Abbott}}}, \bibinfo {author} {\bibfnamefont {R.}~\bibnamefont {{Abbott}}},
  \bibinfo {author} {\bibfnamefont {M.}~\bibnamefont {{Abernathy}}}, \bibinfo
  {author} {\bibfnamefont {T.}~\bibnamefont {{Accadia}}}, \bibinfo {author}
  {\bibfnamefont {F.}~\bibnamefont {{Acernese}}}, \bibinfo {author}
  {\bibfnamefont {C.}~\bibnamefont {{Adams}}}, \bibinfo {author} {\bibfnamefont
  {R.}~\bibnamefont {{Adhikari}}}, \bibinfo {author} {\bibfnamefont
  {P.}~\bibnamefont {{Ajith}}}, \bibinfo {author} {\bibfnamefont
  {B.}~\bibnamefont {{Allen}}}, \ and\ \bibinfo {author} {\bibnamefont
  {et~al.}},\ }\href {\doibase 10.1088/0264-9381/27/17/173001} {\bibfield
  {journal} {\bibinfo  {journal} {Classical and Quantum Gravity}\ }\textbf
  {\bibinfo {volume} {27}},\ \bibinfo {eid} {173001} (\bibinfo {year}
  {2010})},\ \Eprint {http://arxiv.org/abs/1003.2480} {arXiv:1003.2480
  [astro-ph.HE]} \BibitemShut {NoStop}%
\bibitem [{\citenamefont {{The LIGO Scientific Collaboration}}\ \emph
  {et~al.}(2016)\citenamefont {{The LIGO Scientific Collaboration}},
  \citenamefont {{the Virgo Collaboration}}, \citenamefont {{Abbott}},
  \citenamefont {{Abbott}}, \citenamefont {{Abbott}}, \citenamefont
  {{Abernathy}}, \citenamefont {{Acernese}}, \citenamefont {{Ackley}},
  \citenamefont {{Adams}}, \citenamefont {{Adams}},\ and\ \citenamefont
  {et~al.}}]{ligovirgo2016}%
  \BibitemOpen
  \bibfield  {author} {\bibinfo {author} {\bibnamefont {{The LIGO Scientific
  Collaboration}}}, \bibinfo {author} {\bibnamefont {{the Virgo
  Collaboration}}}, \bibinfo {author} {\bibfnamefont {B.~P.}\ \bibnamefont
  {{Abbott}}}, \bibinfo {author} {\bibfnamefont {R.}~\bibnamefont {{Abbott}}},
  \bibinfo {author} {\bibfnamefont {T.~D.}\ \bibnamefont {{Abbott}}}, \bibinfo
  {author} {\bibfnamefont {M.~R.}\ \bibnamefont {{Abernathy}}}, \bibinfo
  {author} {\bibfnamefont {F.}~\bibnamefont {{Acernese}}}, \bibinfo {author}
  {\bibfnamefont {K.}~\bibnamefont {{Ackley}}}, \bibinfo {author}
  {\bibfnamefont {C.}~\bibnamefont {{Adams}}}, \bibinfo {author} {\bibfnamefont
  {T.}~\bibnamefont {{Adams}}}, \ and\ \bibinfo {author} {\bibnamefont
  {et~al.}},\ }\href {\doibase 10.1103/PhysRevX.6.041015} {\bibfield  {journal}
  {\bibinfo  {journal} {Phys. Rev.}\ }\textbf {\bibinfo {volume} {X6}},\
  \bibinfo {pages} {041015} (\bibinfo {year} {2016})},\ \Eprint
  {http://arxiv.org/abs/1606.04856} {arXiv:1606.04856 [gr-qc]} \BibitemShut
  {NoStop}%
\bibitem [{\citenamefont {{Amaro-Seoane}}\ \emph {et~al.}(2012)\citenamefont
  {{Amaro-Seoane}}, \citenamefont {{Aoudia}}, \citenamefont {{Babak}},
  \citenamefont {{Bin{\'e}truy}}, \citenamefont {{Berti}}, \citenamefont
  {{Boh{\'e}}}, \citenamefont {{Caprini}}, \citenamefont {{Colpi}},
  \citenamefont {{Cornish}}, \citenamefont {{Danzmann}}, \citenamefont
  {{Dufaux}}, \citenamefont {{Gair}}, \citenamefont {{Jennrich}}, \citenamefont
  {{Jetzer}}, \citenamefont {{Klein}}, \citenamefont {{Lang}}, \citenamefont
  {{Lobo}}, \citenamefont {{Littenberg}}, \citenamefont {{McWilliams}},
  \citenamefont {{Nelemans}}, \citenamefont {{Petiteau}}, \citenamefont
  {{Porter}}, \citenamefont {{Schutz}}, \citenamefont {{Sesana}}, \citenamefont
  {{Stebbins}}, \citenamefont {{Sumner}}, \citenamefont {{Vallisneri}},
  \citenamefont {{Vitale}}, \citenamefont {{Volonteri}},\ and\ \citenamefont
  {{Ward}}}]{eLISA2012}%
  \BibitemOpen
  \bibfield  {author} {\bibinfo {author} {\bibfnamefont {P.}~\bibnamefont
  {{Amaro-Seoane}}}, \bibinfo {author} {\bibfnamefont {S.}~\bibnamefont
  {{Aoudia}}}, \bibinfo {author} {\bibfnamefont {S.}~\bibnamefont {{Babak}}},
  \bibinfo {author} {\bibfnamefont {P.}~\bibnamefont {{Bin{\'e}truy}}},
  \bibinfo {author} {\bibfnamefont {E.}~\bibnamefont {{Berti}}}, \bibinfo
  {author} {\bibfnamefont {A.}~\bibnamefont {{Boh{\'e}}}}, \bibinfo {author}
  {\bibfnamefont {C.}~\bibnamefont {{Caprini}}}, \bibinfo {author}
  {\bibfnamefont {M.}~\bibnamefont {{Colpi}}}, \bibinfo {author} {\bibfnamefont
  {N.~J.}\ \bibnamefont {{Cornish}}}, \bibinfo {author} {\bibfnamefont
  {K.}~\bibnamefont {{Danzmann}}}, \bibinfo {author} {\bibfnamefont {J.-F.}\
  \bibnamefont {{Dufaux}}}, \bibinfo {author} {\bibfnamefont {J.}~\bibnamefont
  {{Gair}}}, \bibinfo {author} {\bibfnamefont {O.}~\bibnamefont {{Jennrich}}},
  \bibinfo {author} {\bibfnamefont {P.}~\bibnamefont {{Jetzer}}}, \bibinfo
  {author} {\bibfnamefont {A.}~\bibnamefont {{Klein}}}, \bibinfo {author}
  {\bibfnamefont {R.~N.}\ \bibnamefont {{Lang}}}, \bibinfo {author}
  {\bibfnamefont {A.}~\bibnamefont {{Lobo}}}, \bibinfo {author} {\bibfnamefont
  {T.}~\bibnamefont {{Littenberg}}}, \bibinfo {author} {\bibfnamefont {S.~T.}\
  \bibnamefont {{McWilliams}}}, \bibinfo {author} {\bibfnamefont
  {G.}~\bibnamefont {{Nelemans}}}, \bibinfo {author} {\bibfnamefont
  {A.}~\bibnamefont {{Petiteau}}}, \bibinfo {author} {\bibfnamefont {E.~K.}\
  \bibnamefont {{Porter}}}, \bibinfo {author} {\bibfnamefont {B.~F.}\
  \bibnamefont {{Schutz}}}, \bibinfo {author} {\bibfnamefont {A.}~\bibnamefont
  {{Sesana}}}, \bibinfo {author} {\bibfnamefont {R.}~\bibnamefont
  {{Stebbins}}}, \bibinfo {author} {\bibfnamefont {T.}~\bibnamefont
  {{Sumner}}}, \bibinfo {author} {\bibfnamefont {M.}~\bibnamefont
  {{Vallisneri}}}, \bibinfo {author} {\bibfnamefont {S.}~\bibnamefont
  {{Vitale}}}, \bibinfo {author} {\bibfnamefont {M.}~\bibnamefont
  {{Volonteri}}}, \ and\ \bibinfo {author} {\bibfnamefont {H.}~\bibnamefont
  {{Ward}}},\ }\href {\doibase 10.1088/0264-9381/29/12/124016} {\bibfield
  {journal} {\bibinfo  {journal} {Classical and Quantum Gravity}\ }\textbf
  {\bibinfo {volume} {29}},\ \bibinfo {eid} {124016} (\bibinfo {year}
  {2012})},\ \Eprint {http://arxiv.org/abs/1202.0839} {arXiv:1202.0839 [gr-qc]}
  \BibitemShut {NoStop}%
\bibitem [{\citenamefont {Moore}\ \emph {et~al.}(2015)\citenamefont {Moore},
  \citenamefont {Cole},\ and\ \citenamefont {Berry}}]{moore2015}%
  \BibitemOpen
  \bibfield  {author} {\bibinfo {author} {\bibfnamefont {C.~J.}\ \bibnamefont
  {Moore}}, \bibinfo {author} {\bibfnamefont {R.~H.}\ \bibnamefont {Cole}}, \
  and\ \bibinfo {author} {\bibfnamefont {C.~P.~L.}\ \bibnamefont {Berry}},\
  }\href {\doibase 10.1088/0264-9381/32/1/015014} {\bibfield  {journal}
  {\bibinfo  {journal} {Class. Quant. Grav.}\ }\textbf {\bibinfo {volume}
  {32}},\ \bibinfo {pages} {015014} (\bibinfo {year} {2015})},\ \Eprint
  {http://arxiv.org/abs/1408.0740} {arXiv:1408.0740 [gr-qc]} \BibitemShut
  {NoStop}%
\bibitem [{\citenamefont {Thorne}(1995)}]{thorne1995}%
  \BibitemOpen
  \bibfield  {author} {\bibinfo {author} {\bibfnamefont {K.~S.}\ \bibnamefont
  {Thorne}},\ }in\ \href@noop {} {\emph {\bibinfo {booktitle} {{Particle and
  nuclear astrophysics and cosmology in the next millennium. Proceedings,
  Summer Study, Snowmass, USA, June 29-July 14, 1994}}}}\ (\bibinfo {year}
  {1995})\ pp.\ \bibinfo {pages} {0160--184},\ \Eprint
  {http://arxiv.org/abs/gr-qc/9506086} {arXiv:gr-qc/9506086 [gr-qc]}
  \BibitemShut {NoStop}%
\bibitem [{\citenamefont {{Aasi}}\ \emph {et~al.}(2014)\citenamefont {{Aasi}},
  \citenamefont {{Abbott}}, \citenamefont {{Abbott}}, \citenamefont {{Abbott}},
  \citenamefont {{Abernathy}}, \citenamefont {{Accadia}}, \citenamefont
  {{Acernese}}, \citenamefont {{Ackley}}, \citenamefont {{Adams}},
  \citenamefont {{Adams}},\ and\ \citenamefont {et~al.}}]{abbott2014}%
  \BibitemOpen
  \bibfield  {author} {\bibinfo {author} {\bibfnamefont {J.}~\bibnamefont
  {{Aasi}}}, \bibinfo {author} {\bibfnamefont {B.~P.}\ \bibnamefont
  {{Abbott}}}, \bibinfo {author} {\bibfnamefont {R.}~\bibnamefont {{Abbott}}},
  \bibinfo {author} {\bibfnamefont {T.}~\bibnamefont {{Abbott}}}, \bibinfo
  {author} {\bibfnamefont {M.~R.}\ \bibnamefont {{Abernathy}}}, \bibinfo
  {author} {\bibfnamefont {T.}~\bibnamefont {{Accadia}}}, \bibinfo {author}
  {\bibfnamefont {F.}~\bibnamefont {{Acernese}}}, \bibinfo {author}
  {\bibfnamefont {K.}~\bibnamefont {{Ackley}}}, \bibinfo {author}
  {\bibfnamefont {C.}~\bibnamefont {{Adams}}}, \bibinfo {author} {\bibfnamefont
  {T.}~\bibnamefont {{Adams}}}, \ and\ \bibinfo {author} {\bibnamefont
  {et~al.}},\ }\href {\doibase 10.1103/PhysRevD.90.062010} {\bibfield
  {journal} {\bibinfo  {journal} {\prd}\ }\textbf {\bibinfo {volume} {90}},\
  \bibinfo {eid} {062010} (\bibinfo {year} {2014})},\ \Eprint
  {http://arxiv.org/abs/1405.7904} {arXiv:1405.7904 [gr-qc]} \BibitemShut
  {NoStop}%
\bibitem [{\citenamefont {Gossan}\ \emph {et~al.}(2016)\citenamefont {Gossan},
  \citenamefont {Sutton}, \citenamefont {Stuver}, \citenamefont {Zanolin},
  \citenamefont {Gill},\ and\ \citenamefont {Ott}}]{gossan2015}%
  \BibitemOpen
  \bibfield  {author} {\bibinfo {author} {\bibfnamefont {S.~E.}\ \bibnamefont
  {Gossan}}, \bibinfo {author} {\bibfnamefont {P.}~\bibnamefont {Sutton}},
  \bibinfo {author} {\bibfnamefont {A.}~\bibnamefont {Stuver}}, \bibinfo
  {author} {\bibfnamefont {M.}~\bibnamefont {Zanolin}}, \bibinfo {author}
  {\bibfnamefont {K.}~\bibnamefont {Gill}}, \ and\ \bibinfo {author}
  {\bibfnamefont {C.~D.}\ \bibnamefont {Ott}},\ }\href {\doibase
  10.1103/PhysRevD.93.042002} {\bibfield  {journal} {\bibinfo  {journal} {Phys.
  Rev.}\ }\textbf {\bibinfo {volume} {D93}},\ \bibinfo {pages} {042002}
  (\bibinfo {year} {2016})},\ \Eprint {http://arxiv.org/abs/1511.02836}
  {arXiv:1511.02836 [astro-ph.HE]} \BibitemShut {NoStop}%
\bibitem [{\citenamefont {Babak}\ \emph {et~al.}(2015)\citenamefont {Babak},
  \citenamefont {Gair},\ and\ \citenamefont {Cole}}]{babak2014}%
  \BibitemOpen
  \bibfield  {author} {\bibinfo {author} {\bibfnamefont {S.}~\bibnamefont
  {Babak}}, \bibinfo {author} {\bibfnamefont {J.~R.}\ \bibnamefont {Gair}}, \
  and\ \bibinfo {author} {\bibfnamefont {R.~H.}\ \bibnamefont {Cole}},\
  }\bibfield  {booktitle} {\emph {\bibinfo {booktitle} {{Proceedings, 524th
  WE-Heraeus-Seminar: Equations of Motion in Relativistic Gravity (EOM 2013):
  Bad Honnef, Germany, February 17-23, 2013}}},\ }\href {\doibase
  10.1007/978-3-319-18335-0_23} {\bibfield  {journal} {\bibinfo  {journal}
  {Fund. Theor. Phys.}\ }\textbf {\bibinfo {volume} {179}},\ \bibinfo {pages}
  {783} (\bibinfo {year} {2015})},\ \Eprint {http://arxiv.org/abs/1411.5253}
  {arXiv:1411.5253 [gr-qc]} \BibitemShut {NoStop}%
\bibitem [{\citenamefont {{Kormendy}}\ and\ \citenamefont
  {{Richstone}}(1995)}]{kormendyrichstone1995}%
  \BibitemOpen
  \bibfield  {author} {\bibinfo {author} {\bibfnamefont {J.}~\bibnamefont
  {{Kormendy}}}\ and\ \bibinfo {author} {\bibfnamefont {D.}~\bibnamefont
  {{Richstone}}},\ }\href {\doibase 10.1146/annurev.aa.33.090195.003053}
  {\bibfield  {journal} {\bibinfo  {journal} {\araa}\ }\textbf {\bibinfo
  {volume} {33}},\ \bibinfo {pages} {581} (\bibinfo {year} {1995})}\BibitemShut
  {NoStop}%
\bibitem [{\citenamefont {{Barack}}\ and\ \citenamefont
  {{Cutler}}(2004)}]{barakcutler2004}%
  \BibitemOpen
  \bibfield  {author} {\bibinfo {author} {\bibfnamefont {L.}~\bibnamefont
  {{Barack}}}\ and\ \bibinfo {author} {\bibfnamefont {C.}~\bibnamefont
  {{Cutler}}},\ }\href {\doibase 10.1103/PhysRevD.69.082005} {\bibfield
  {journal} {\bibinfo  {journal} {\prd}\ }\textbf {\bibinfo {volume} {69}},\
  \bibinfo {eid} {082005} (\bibinfo {year} {2004})},\ \Eprint
  {http://arxiv.org/abs/gr-qc/0310125} {gr-qc/0310125} \BibitemShut {NoStop}%
\bibitem [{\citenamefont {{Gair}}\ and\ \citenamefont
  {{Glampedakis}}(2006)}]{gairglamp2006}%
  \BibitemOpen
  \bibfield  {author} {\bibinfo {author} {\bibfnamefont {J.~R.}\ \bibnamefont
  {{Gair}}}\ and\ \bibinfo {author} {\bibfnamefont {K.}~\bibnamefont
  {{Glampedakis}}},\ }\href {\doibase 10.1103/PhysRevD.73.064037} {\bibfield
  {journal} {\bibinfo  {journal} {\prd}\ }\textbf {\bibinfo {volume} {73}},\
  \bibinfo {eid} {064037} (\bibinfo {year} {2006})},\ \Eprint
  {http://arxiv.org/abs/gr-qc/0510129} {gr-qc/0510129} \BibitemShut {NoStop}%
\bibitem [{\citenamefont {{Drasco}}\ and\ \citenamefont
  {{Hughes}}(2006)}]{drasco2006}%
  \BibitemOpen
  \bibfield  {author} {\bibinfo {author} {\bibfnamefont {S.}~\bibnamefont
  {{Drasco}}}\ and\ \bibinfo {author} {\bibfnamefont {S.~A.}\ \bibnamefont
  {{Hughes}}},\ }\href {\doibase 10.1103/PhysRevD.73.024027} {\bibfield
  {journal} {\bibinfo  {journal} {\prd}\ }\textbf {\bibinfo {volume} {73}},\
  \bibinfo {eid} {024027} (\bibinfo {year} {2006})},\ \Eprint
  {http://arxiv.org/abs/gr-qc/0509101} {gr-qc/0509101} \BibitemShut {NoStop}%
\bibitem [{\citenamefont {{Chua}}\ and\ \citenamefont
  {{Gair}}(2015)}]{chua2015}%
  \BibitemOpen
  \bibfield  {author} {\bibinfo {author} {\bibfnamefont {A.~J.~K.}\
  \bibnamefont {{Chua}}}\ and\ \bibinfo {author} {\bibfnamefont {J.~R.}\
  \bibnamefont {{Gair}}},\ }\href {\doibase 10.1088/0264-9381/32/23/232002}
  {\bibfield  {journal} {\bibinfo  {journal} {Classical and Quantum Gravity}\
  }\textbf {\bibinfo {volume} {32}},\ \bibinfo {eid} {232002} (\bibinfo {year}
  {2015})},\ \Eprint {http://arxiv.org/abs/1510.06245} {arXiv:1510.06245
  [gr-qc]} \BibitemShut {NoStop}%
\bibitem [{\citenamefont {{Babak}}\ \emph {et~al.}(2007)\citenamefont
  {{Babak}}, \citenamefont {{Fang}}, \citenamefont {{Gair}}, \citenamefont
  {{Glampedakis}},\ and\ \citenamefont {{Hughes}}}]{babakal2006}%
  \BibitemOpen
  \bibfield  {author} {\bibinfo {author} {\bibfnamefont {S.}~\bibnamefont
  {{Babak}}}, \bibinfo {author} {\bibfnamefont {H.}~\bibnamefont {{Fang}}},
  \bibinfo {author} {\bibfnamefont {J.~R.}\ \bibnamefont {{Gair}}}, \bibinfo
  {author} {\bibfnamefont {K.}~\bibnamefont {{Glampedakis}}}, \ and\ \bibinfo
  {author} {\bibfnamefont {S.~A.}\ \bibnamefont {{Hughes}}},\ }\href {\doibase
  10.1103/PhysRevD.75.024005} {\bibfield  {journal} {\bibinfo  {journal}
  {\prd}\ }\textbf {\bibinfo {volume} {75}},\ \bibinfo {eid} {024005} (\bibinfo
  {year} {2007})},\ \Eprint {http://arxiv.org/abs/gr-qc/0607007}
  {gr-qc/0607007} \BibitemShut {NoStop}%
\bibitem [{\citenamefont {{Glampedakis}}\ \emph {et~al.}(2002)\citenamefont
  {{Glampedakis}}, \citenamefont {{Hughes}},\ and\ \citenamefont
  {{Kennefick}}}]{glampedakiskennefick2002}%
  \BibitemOpen
  \bibfield  {author} {\bibinfo {author} {\bibfnamefont {K.}~\bibnamefont
  {{Glampedakis}}}, \bibinfo {author} {\bibfnamefont {S.~A.}\ \bibnamefont
  {{Hughes}}}, \ and\ \bibinfo {author} {\bibfnamefont {D.}~\bibnamefont
  {{Kennefick}}},\ }\href {\doibase 10.1103/PhysRevD.66.064005} {\bibfield
  {journal} {\bibinfo  {journal} {\prd}\ }\textbf {\bibinfo {volume} {66}},\
  \bibinfo {eid} {064005} (\bibinfo {year} {2002})},\ \Eprint
  {http://arxiv.org/abs/gr-qc/0205033} {gr-qc/0205033} \BibitemShut {NoStop}%
\bibitem [{\citenamefont {{Hopman}}(2009)}]{hopman2009}%
  \BibitemOpen
  \bibfield  {author} {\bibinfo {author} {\bibfnamefont {C.}~\bibnamefont
  {{Hopman}}},\ }\href {\doibase 10.1088/0264-9381/26/9/094028} {\bibfield
  {journal} {\bibinfo  {journal} {Classical and Quantum Gravity}\ }\textbf
  {\bibinfo {volume} {26}},\ \bibinfo {eid} {094028} (\bibinfo {year}
  {2009})},\ \Eprint {http://arxiv.org/abs/0901.1667} {arXiv:0901.1667
  [astro-ph.GA]} \BibitemShut {NoStop}%
\bibitem [{\citenamefont {{Arun}}\ \emph {et~al.}(2009)\citenamefont {{Arun}},
  \citenamefont {{Babak}}, \citenamefont {{Berti}}, \citenamefont {{Cornish}},
  \citenamefont {{Cutler}}, \citenamefont {{Gair}}, \citenamefont {{Hughes}},
  \citenamefont {{Iyer}}, \citenamefont {{Lang}}, \citenamefont {{Mandel}},
  \citenamefont {{Porter}}, \citenamefont {{Sathyaprakash}}, \citenamefont
  {{Sinha}}, \citenamefont {{Sintes}}, \citenamefont {{Trias}}, \citenamefont
  {{Van Den Broeck}},\ and\ \citenamefont {{Volonteri}}}]{arun2009}%
  \BibitemOpen
  \bibfield  {author} {\bibinfo {author} {\bibfnamefont {K.~G.}\ \bibnamefont
  {{Arun}}}, \bibinfo {author} {\bibfnamefont {S.}~\bibnamefont {{Babak}}},
  \bibinfo {author} {\bibfnamefont {E.}~\bibnamefont {{Berti}}}, \bibinfo
  {author} {\bibfnamefont {N.}~\bibnamefont {{Cornish}}}, \bibinfo {author}
  {\bibfnamefont {C.}~\bibnamefont {{Cutler}}}, \bibinfo {author}
  {\bibfnamefont {J.}~\bibnamefont {{Gair}}}, \bibinfo {author} {\bibfnamefont
  {S.~A.}\ \bibnamefont {{Hughes}}}, \bibinfo {author} {\bibfnamefont {B.~R.}\
  \bibnamefont {{Iyer}}}, \bibinfo {author} {\bibfnamefont {R.~N.}\
  \bibnamefont {{Lang}}}, \bibinfo {author} {\bibfnamefont {I.}~\bibnamefont
  {{Mandel}}}, \bibinfo {author} {\bibfnamefont {E.~K.}\ \bibnamefont
  {{Porter}}}, \bibinfo {author} {\bibfnamefont {B.~S.}\ \bibnamefont
  {{Sathyaprakash}}}, \bibinfo {author} {\bibfnamefont {S.}~\bibnamefont
  {{Sinha}}}, \bibinfo {author} {\bibfnamefont {A.~M.}\ \bibnamefont
  {{Sintes}}}, \bibinfo {author} {\bibfnamefont {M.}~\bibnamefont {{Trias}}},
  \bibinfo {author} {\bibfnamefont {C.}~\bibnamefont {{Van Den Broeck}}}, \
  and\ \bibinfo {author} {\bibfnamefont {M.}~\bibnamefont {{Volonteri}}},\
  }\href {\doibase 10.1088/0264-9381/26/9/094027} {\bibfield  {journal}
  {\bibinfo  {journal} {Classical and Quantum Gravity}\ }\textbf {\bibinfo
  {volume} {26}},\ \bibinfo {eid} {094027} (\bibinfo {year} {2009})},\ \Eprint
  {http://arxiv.org/abs/0811.1011} {arXiv:0811.1011 [gr-qc]} \BibitemShut
  {NoStop}%
\bibitem [{\citenamefont {{Gair}}\ \emph {et~al.}(2010)\citenamefont {{Gair}},
  \citenamefont {{Tang}},\ and\ \citenamefont {{Volonteri}}}]{gair2010}%
  \BibitemOpen
  \bibfield  {author} {\bibinfo {author} {\bibfnamefont {J.~R.}\ \bibnamefont
  {{Gair}}}, \bibinfo {author} {\bibfnamefont {C.}~\bibnamefont {{Tang}}}, \
  and\ \bibinfo {author} {\bibfnamefont {M.}~\bibnamefont {{Volonteri}}},\
  }\href {\doibase 10.1103/PhysRevD.81.104014} {\bibfield  {journal} {\bibinfo
  {journal} {\prd}\ }\textbf {\bibinfo {volume} {81}},\ \bibinfo {eid} {104014}
  (\bibinfo {year} {2010})},\ \Eprint {http://arxiv.org/abs/1004.1921}
  {arXiv:1004.1921 [astro-ph.GA]} \BibitemShut {NoStop}%
\bibitem [{\citenamefont {{MacLeod}}\ and\ \citenamefont
  {{Hogan}}(2008)}]{macleodhogan2008}%
  \BibitemOpen
  \bibfield  {author} {\bibinfo {author} {\bibfnamefont {C.~L.}\ \bibnamefont
  {{MacLeod}}}\ and\ \bibinfo {author} {\bibfnamefont {C.~J.}\ \bibnamefont
  {{Hogan}}},\ }\href {\doibase 10.1103/PhysRevD.77.043512} {\bibfield
  {journal} {\bibinfo  {journal} {\prd}\ }\textbf {\bibinfo {volume} {77}},\
  \bibinfo {eid} {043512} (\bibinfo {year} {2008})},\ \Eprint
  {http://arxiv.org/abs/0712.0618} {arXiv:0712.0618} \BibitemShut {NoStop}%
\bibitem [{\citenamefont {{Gair}}\ \emph {et~al.}(2013)\citenamefont {{Gair}},
  \citenamefont {{Vallisneri}}, \citenamefont {{Larson}},\ and\ \citenamefont
  {{Baker}}}]{TestGRLivRev}%
  \BibitemOpen
  \bibfield  {author} {\bibinfo {author} {\bibfnamefont {J.~R.}\ \bibnamefont
  {{Gair}}}, \bibinfo {author} {\bibfnamefont {M.}~\bibnamefont
  {{Vallisneri}}}, \bibinfo {author} {\bibfnamefont {S.~L.}\ \bibnamefont
  {{Larson}}}, \ and\ \bibinfo {author} {\bibfnamefont {J.~G.}\ \bibnamefont
  {{Baker}}},\ }\href {\doibase 10.12942/lrr-2013-7} {\bibfield  {journal}
  {\bibinfo  {journal} {Living Reviews in Relativity}\ }\textbf {\bibinfo
  {volume} {16}},\ \bibinfo {eid} {7} (\bibinfo {year} {2013})},\ \Eprint
  {http://arxiv.org/abs/1212.5575} {arXiv:1212.5575 [gr-qc]} \BibitemShut
  {NoStop}%
\bibitem [{\citenamefont {{Carter}}(1968)}]{carter1968}%
  \BibitemOpen
  \bibfield  {author} {\bibinfo {author} {\bibfnamefont {B.}~\bibnamefont
  {{Carter}}},\ }\href {\doibase 10.1103/PhysRev.174.1559} {\bibfield
  {journal} {\bibinfo  {journal} {Physical Review}\ }\textbf {\bibinfo {volume}
  {174}},\ \bibinfo {pages} {1559} (\bibinfo {year} {1968})}\BibitemShut
  {NoStop}%
\bibitem [{\citenamefont {Nakamura}\ and\ \citenamefont
  {Sasaki}(1982)}]{nakamura1982}%
  \BibitemOpen
  \bibfield  {author} {\bibinfo {author} {\bibfnamefont {T.}~\bibnamefont
  {Nakamura}}\ and\ \bibinfo {author} {\bibfnamefont {M.}~\bibnamefont
  {Sasaki}},\ }\href {\doibase http://dx.doi.org/10.1016/0375-9601(82)90204-3}
  {\bibfield  {journal} {\bibinfo  {journal} {Physics Letters A}\ }\textbf
  {\bibinfo {volume} {89}},\ \bibinfo {pages} {185 } (\bibinfo {year}
  {1982})}\BibitemShut {NoStop}%
\bibitem [{\citenamefont {{Wardell}}\ \emph {et~al.}(2012)\citenamefont
  {{Wardell}}, \citenamefont {{Vega}}, \citenamefont {{Thornburg}},\ and\
  \citenamefont {{Diener}}}]{wardell2012}%
  \BibitemOpen
  \bibfield  {author} {\bibinfo {author} {\bibfnamefont {B.}~\bibnamefont
  {{Wardell}}}, \bibinfo {author} {\bibfnamefont {I.}~\bibnamefont {{Vega}}},
  \bibinfo {author} {\bibfnamefont {J.}~\bibnamefont {{Thornburg}}}, \ and\
  \bibinfo {author} {\bibfnamefont {P.}~\bibnamefont {{Diener}}},\ }\href
  {\doibase 10.1103/PhysRevD.85.104044} {\bibfield  {journal} {\bibinfo
  {journal} {\prd}\ }\textbf {\bibinfo {volume} {85}},\ \bibinfo {eid} {104044}
  (\bibinfo {year} {2012})},\ \Eprint {http://arxiv.org/abs/1112.6355}
  {arXiv:1112.6355 [gr-qc]} \BibitemShut {NoStop}%
\bibitem [{\citenamefont {{Pound}}\ and\ \citenamefont
  {{Poisson}}(2008)}]{poundpoisson2008}%
  \BibitemOpen
  \bibfield  {author} {\bibinfo {author} {\bibfnamefont {A.}~\bibnamefont
  {{Pound}}}\ and\ \bibinfo {author} {\bibfnamefont {E.}~\bibnamefont
  {{Poisson}}},\ }\href {\doibase 10.1103/PhysRevD.77.044013} {\bibfield
  {journal} {\bibinfo  {journal} {\prd}\ }\textbf {\bibinfo {volume} {77}},\
  \bibinfo {eid} {044013} (\bibinfo {year} {2008})},\ \Eprint
  {http://arxiv.org/abs/0708.3033} {arXiv:0708.3033 [gr-qc]} \BibitemShut
  {NoStop}%
\bibitem [{\citenamefont {{Gair}}\ \emph {et~al.}(2011)\citenamefont {{Gair}},
  \citenamefont {{Flanagan}}, \citenamefont {{Drasco}}, \citenamefont
  {{Hinderer}},\ and\ \citenamefont {{Babak}}}]{gairflanagan2011}%
  \BibitemOpen
  \bibfield  {author} {\bibinfo {author} {\bibfnamefont {J.~R.}\ \bibnamefont
  {{Gair}}}, \bibinfo {author} {\bibfnamefont {{\'E}.~{\'E}.}\ \bibnamefont
  {{Flanagan}}}, \bibinfo {author} {\bibfnamefont {S.}~\bibnamefont
  {{Drasco}}}, \bibinfo {author} {\bibfnamefont {T.}~\bibnamefont
  {{Hinderer}}}, \ and\ \bibinfo {author} {\bibfnamefont {S.}~\bibnamefont
  {{Babak}}},\ }\href {\doibase 10.1103/PhysRevD.83.044037} {\bibfield
  {journal} {\bibinfo  {journal} {\prd}\ }\textbf {\bibinfo {volume} {83}},\
  \bibinfo {eid} {044037} (\bibinfo {year} {2011})},\ \Eprint
  {http://arxiv.org/abs/1012.5111} {arXiv:1012.5111 [gr-qc]} \BibitemShut
  {NoStop}%
\bibitem [{\citenamefont {{Sopuerta}}\ and\ \citenamefont
  {{Yunes}}(2012)}]{sopuerta2012}%
  \BibitemOpen
  \bibfield  {author} {\bibinfo {author} {\bibfnamefont {C.~F.}\ \bibnamefont
  {{Sopuerta}}}\ and\ \bibinfo {author} {\bibfnamefont {N.}~\bibnamefont
  {{Yunes}}},\ }\href {\doibase 10.1088/1742-6596/363/1/012021} {\bibfield
  {journal} {\bibinfo  {journal} {Journal of Physics Conference Series}\
  }\textbf {\bibinfo {volume} {363}},\ \bibinfo {eid} {012021} (\bibinfo {year}
  {2012})},\ \Eprint {http://arxiv.org/abs/1201.5715} {arXiv:1201.5715 [gr-qc]}
  \BibitemShut {NoStop}%
\bibitem [{\citenamefont {Flanagan}\ and\ \citenamefont
  {Hinderer}(2012)}]{flanagan2012}%
  \BibitemOpen
  \bibfield  {author} {\bibinfo {author} {\bibfnamefont {E.~E.}\ \bibnamefont
  {Flanagan}}\ and\ \bibinfo {author} {\bibfnamefont {T.}~\bibnamefont
  {Hinderer}},\ }\href {\doibase 10.1103/PhysRevLett.109.071102} {\bibfield
  {journal} {\bibinfo  {journal} {Phys. Rev. Lett.}\ }\textbf {\bibinfo
  {volume} {109}},\ \bibinfo {pages} {071102} (\bibinfo {year}
  {2012})}\BibitemShut {NoStop}%
\bibitem [{\citenamefont {{Barack}}(2009)}]{barack2009}%
  \BibitemOpen
  \bibfield  {author} {\bibinfo {author} {\bibfnamefont {L.}~\bibnamefont
  {{Barack}}},\ }\href {\doibase 10.1088/0264-9381/26/21/213001} {\bibfield
  {journal} {\bibinfo  {journal} {Classical and Quantum Gravity}\ }\textbf
  {\bibinfo {volume} {26}},\ \bibinfo {eid} {213001} (\bibinfo {year}
  {2009})},\ \Eprint {http://arxiv.org/abs/0908.1664} {arXiv:0908.1664 [gr-qc]}
  \BibitemShut {NoStop}%
\bibitem [{\citenamefont {Berry}\ \emph {et~al.}(2016)\citenamefont {Berry},
  \citenamefont {Cole}, \citenamefont {Cañizares},\ and\ \citenamefont
  {Gair}}]{berry2016}%
  \BibitemOpen
  \bibfield  {author} {\bibinfo {author} {\bibfnamefont {C.~P.~L.}\
  \bibnamefont {Berry}}, \bibinfo {author} {\bibfnamefont {R.~H.}\ \bibnamefont
  {Cole}}, \bibinfo {author} {\bibfnamefont {P.}~\bibnamefont {Cañizares}}, \
  and\ \bibinfo {author} {\bibfnamefont {J.~R.}\ \bibnamefont {Gair}},\ }\href
  {\doibase 10.1103/PhysRevD.94.124042} {\bibfield  {journal} {\bibinfo
  {journal} {Phys. Rev.}\ }\textbf {\bibinfo {volume} {D94}},\ \bibinfo {pages}
  {124042} (\bibinfo {year} {2016})},\ \Eprint
  {http://arxiv.org/abs/1608.08951} {arXiv:1608.08951 [gr-qc]} \BibitemShut
  {NoStop}%
\bibitem [{\citenamefont {{Gralla}}\ and\ \citenamefont
  {{Wald}}(2008)}]{gralla2008}%
  \BibitemOpen
  \bibfield  {author} {\bibinfo {author} {\bibfnamefont {S.~E.}\ \bibnamefont
  {{Gralla}}}\ and\ \bibinfo {author} {\bibfnamefont {R.~M.}\ \bibnamefont
  {{Wald}}},\ }\href {\doibase 10.1088/0264-9381/25/20/205009} {\bibfield
  {journal} {\bibinfo  {journal} {Classical and Quantum Gravity}\ }\textbf
  {\bibinfo {volume} {25}},\ \bibinfo {eid} {205009} (\bibinfo {year}
  {2008})},\ \Eprint {http://arxiv.org/abs/0806.3293} {arXiv:0806.3293 [gr-qc]}
  \BibitemShut {NoStop}%
\bibitem [{\citenamefont {{Poisson}}\ \emph {et~al.}(2011)\citenamefont
  {{Poisson}}, \citenamefont {{Pound}},\ and\ \citenamefont
  {{Vega}}}]{poisson2011}%
  \BibitemOpen
  \bibfield  {author} {\bibinfo {author} {\bibfnamefont {E.}~\bibnamefont
  {{Poisson}}}, \bibinfo {author} {\bibfnamefont {A.}~\bibnamefont {{Pound}}},
  \ and\ \bibinfo {author} {\bibfnamefont {I.}~\bibnamefont {{Vega}}},\ }\href
  {\doibase 10.12942/lrr-2011-7} {\bibfield  {journal} {\bibinfo  {journal}
  {Living Reviews in Relativity}\ }\textbf {\bibinfo {volume} {14}},\ \bibinfo
  {pages} {7} (\bibinfo {year} {2011})},\ \Eprint
  {http://arxiv.org/abs/1102.0529} {arXiv:1102.0529 [gr-qc]} \BibitemShut
  {NoStop}%
\bibitem [{\citenamefont {Poisson}(2011)}]{poisson2009}%
  \BibitemOpen
  \bibfield  {author} {\bibinfo {author} {\bibfnamefont {E.}~\bibnamefont
  {Poisson}},\ }\bibfield  {booktitle} {\emph {\bibinfo {booktitle} {{Mass and
  motion in general relativity. Proceedings, School on Mass, Orleans, France,
  June 23-25, 2008}}},\ }\href@noop {} {\bibfield  {journal} {\bibinfo
  {journal} {Fundam. Theor. Phys.}\ }\textbf {\bibinfo {volume} {162}},\
  \bibinfo {pages} {309} (\bibinfo {year} {2011})},\ \bibinfo {note}
  {[,309(2009)]},\ \Eprint {http://arxiv.org/abs/0909.2994} {arXiv:0909.2994
  [gr-qc]} \BibitemShut {NoStop}%
\bibitem [{\citenamefont {Wald}(2011)}]{wald2009}%
  \BibitemOpen
  \bibfield  {author} {\bibinfo {author} {\bibfnamefont {R.~M.}\ \bibnamefont
  {Wald}},\ }\bibfield  {booktitle} {\emph {\bibinfo {booktitle} {{Mass and
  motion in general relativity. Proceedings, School on Mass, Orleans, France,
  June 23-25, 2008}}},\ }\href@noop {} {\bibfield  {journal} {\bibinfo
  {journal} {Fundam. Theor. Phys.}\ }\textbf {\bibinfo {volume} {162}},\
  \bibinfo {pages} {253} (\bibinfo {year} {2011})},\ \bibinfo {note}
  {[,253(2009)]},\ \Eprint {http://arxiv.org/abs/0907.0412} {arXiv:0907.0412
  [gr-qc]} \BibitemShut {NoStop}%
\bibitem [{\citenamefont {{Gair}}\ \emph {et~al.}(2012)\citenamefont {{Gair}},
  \citenamefont {{Yunes}},\ and\ \citenamefont {{Bender}}}]{gair2012}%
  \BibitemOpen
  \bibfield  {author} {\bibinfo {author} {\bibfnamefont {J.}~\bibnamefont
  {{Gair}}}, \bibinfo {author} {\bibfnamefont {N.}~\bibnamefont {{Yunes}}}, \
  and\ \bibinfo {author} {\bibfnamefont {C.~M.}\ \bibnamefont {{Bender}}},\
  }\href {\doibase 10.1063/1.3691226} {\bibfield  {journal} {\bibinfo
  {journal} {Journal of Mathematical Physics}\ }\textbf {\bibinfo {volume}
  {53}},\ \bibinfo {pages} {032503} (\bibinfo {year} {2012})},\ \Eprint
  {http://arxiv.org/abs/1111.3605} {arXiv:1111.3605 [gr-qc]} \BibitemShut
  {NoStop}%
\bibitem [{\citenamefont {{Schmidt}}(2002)}]{schmidt2002}%
  \BibitemOpen
  \bibfield  {author} {\bibinfo {author} {\bibfnamefont {W.}~\bibnamefont
  {{Schmidt}}},\ }\href {\doibase 10.1088/0264-9381/19/10/314} {\bibfield
  {journal} {\bibinfo  {journal} {Classical and Quantum Gravity}\ }\textbf
  {\bibinfo {volume} {19}},\ \bibinfo {pages} {2743} (\bibinfo {year}
  {2002})},\ \Eprint {http://arxiv.org/abs/gr-qc/0202090} {gr-qc/0202090}
  \BibitemShut {NoStop}%
\bibitem [{\citenamefont {{Warburton}}\ \emph {et~al.}(2013)\citenamefont
  {{Warburton}}, \citenamefont {{Barack}},\ and\ \citenamefont
  {{Sago}}}]{warburton2013}%
  \BibitemOpen
  \bibfield  {author} {\bibinfo {author} {\bibfnamefont {N.}~\bibnamefont
  {{Warburton}}}, \bibinfo {author} {\bibfnamefont {L.}~\bibnamefont
  {{Barack}}}, \ and\ \bibinfo {author} {\bibfnamefont {N.}~\bibnamefont
  {{Sago}}},\ }\href {\doibase 10.1103/PhysRevD.87.084012} {\bibfield
  {journal} {\bibinfo  {journal} {\prd}\ }\textbf {\bibinfo {volume} {87}},\
  \bibinfo {eid} {084012} (\bibinfo {year} {2013})},\ \Eprint
  {http://arxiv.org/abs/1301.3918} {arXiv:1301.3918 [gr-qc]} \BibitemShut
  {NoStop}%
\bibitem [{\citenamefont {{Ruangsri}}\ and\ \citenamefont
  {{Hughes}}(2014)}]{ruangsri2014}%
  \BibitemOpen
  \bibfield  {author} {\bibinfo {author} {\bibfnamefont {U.}~\bibnamefont
  {{Ruangsri}}}\ and\ \bibinfo {author} {\bibfnamefont {S.~A.}\ \bibnamefont
  {{Hughes}}},\ }\href {\doibase 10.1103/PhysRevD.89.084036} {\bibfield
  {journal} {\bibinfo  {journal} {\prd}\ }\textbf {\bibinfo {volume} {89}},\
  \bibinfo {eid} {084036} (\bibinfo {year} {2014})},\ \Eprint
  {http://arxiv.org/abs/1307.6483} {arXiv:1307.6483 [gr-qc]} \BibitemShut
  {NoStop}%
\bibitem [{\citenamefont {{Hinderer}}\ and\ \citenamefont
  {{Flanagan}}(2008)}]{hinderer2008}%
  \BibitemOpen
  \bibfield  {author} {\bibinfo {author} {\bibfnamefont {T.}~\bibnamefont
  {{Hinderer}}}\ and\ \bibinfo {author} {\bibfnamefont {{\'E}.~{\'E}.}\
  \bibnamefont {{Flanagan}}},\ }\href {\doibase 10.1103/PhysRevD.78.064028}
  {\bibfield  {journal} {\bibinfo  {journal} {\prd}\ }\textbf {\bibinfo
  {volume} {78}},\ \bibinfo {eid} {064028} (\bibinfo {year} {2008})},\ \Eprint
  {http://arxiv.org/abs/0805.3337} {arXiv:0805.3337 [gr-qc]} \BibitemShut
  {NoStop}%
\bibitem [{\citenamefont {Kevorkian}(1987)}]{kevorkian1987}%
  \BibitemOpen
  \bibfield  {author} {\bibinfo {author} {\bibfnamefont {J.}~\bibnamefont
  {Kevorkian}},\ }\href {http://www.jstor.org/stable/2031238} {\bibfield
  {journal} {\bibinfo  {journal} {SIAM Review}\ }\textbf {\bibinfo {volume}
  {29}},\ \bibinfo {pages} {391} (\bibinfo {year} {1987})}\BibitemShut
  {NoStop}%
\bibitem [{\citenamefont {{van de Meent}}(2014)}]{meent2014}%
  \BibitemOpen
  \bibfield  {author} {\bibinfo {author} {\bibfnamefont {M.}~\bibnamefont {{van
  de Meent}}},\ }\href {\doibase 10.1103/PhysRevD.89.084033} {\bibfield
  {journal} {\bibinfo  {journal} {\prd}\ }\textbf {\bibinfo {volume} {89}},\
  \bibinfo {eid} {084033} (\bibinfo {year} {2014})},\ \Eprint
  {http://arxiv.org/abs/1311.4457} {arXiv:1311.4457 [gr-qc]} \BibitemShut
  {NoStop}%
\bibitem [{\citenamefont {Berry}(2013)}]{berry2013d}%
  \BibitemOpen
  \bibfield  {author} {\bibinfo {author} {\bibfnamefont {C.~P.~L.}\
  \bibnamefont {Berry}},\ }\emph {\bibinfo {title} {{Exploring Gravity}}},\
  \href {http://www.repository.cam.ac.uk/handle/1810/245139} {\bibinfo {type}
  {Ph.d. dissertation}},\ \bibinfo  {school} {University of Cambridge}
  (\bibinfo {year} {2013})\BibitemShut {NoStop}%
\bibitem [{\citenamefont {{Mino}}(2003)}]{mino2003p}%
  \BibitemOpen
  \bibfield  {author} {\bibinfo {author} {\bibfnamefont {Y.}~\bibnamefont
  {{Mino}}},\ }\href {\doibase 10.1103/PhysRevD.67.084027} {\bibfield
  {journal} {\bibinfo  {journal} {\prd}\ }\textbf {\bibinfo {volume} {67}},\
  \bibinfo {eid} {084027} (\bibinfo {year} {2003})},\ \Eprint
  {http://arxiv.org/abs/gr-qc/0302075} {gr-qc/0302075} \BibitemShut {NoStop}%
\bibitem [{\citenamefont {{Wolfram Research, Inc.}}()}]{mathematicav10.4}%
  \BibitemOpen
  \bibfield  {author} {\bibinfo {author} {\bibnamefont {{Wolfram Research,
  Inc.}}},\ }\href {https://www.wolfram.com} {\enquote {\bibinfo {title}
  {Mathematica 10.4},}\ }\BibitemShut {NoStop}%
\bibitem [{\citenamefont {{Boyer}}\ and\ \citenamefont
  {{Lindquist}}(1982)}]{boyerlindquist1982}%
  \BibitemOpen
  \bibfield  {author} {\bibinfo {author} {\bibfnamefont {R.~H.}\ \bibnamefont
  {{Boyer}}}\ and\ \bibinfo {author} {\bibfnamefont {R.~W.}\ \bibnamefont
  {{Lindquist}}},\ }\enquote {\bibinfo {title} {{Maximal Analytic Extension of
  the Kerr Metric}},}\ in\ \href@noop {} {\emph {\bibinfo {booktitle} {Black
  Holes: Selected Reprints}}},\ \bibinfo {editor} {edited by\ \bibinfo {editor}
  {\bibfnamefont {S.}~\bibnamefont {{Detweiler}}}}\ (\bibinfo  {publisher} {the
  American Association of Physics Teachers},\ \bibinfo {year} {1982})\
  p.~\bibinfo {pages} {61}\BibitemShut {NoStop}%
\bibitem [{\citenamefont {Visser}(2007)}]{visser2007}%
  \BibitemOpen
  \bibfield  {author} {\bibinfo {author} {\bibfnamefont {M.}~\bibnamefont
  {Visser}},\ }in\ \href
  {https://inspirehep.net/record/752316/files/arXiv:0706.0622.pdf} {\emph
  {\bibinfo {booktitle} {{Kerr Fest: Black Holes in Astrophysics, General
  Relativity and Quantum Gravity Christchurch, New Zealand, August 26-28,
  2004}}}}\ (\bibinfo {year} {2007})\ \Eprint {http://arxiv.org/abs/0706.0622}
  {arXiv:0706.0622 [gr-qc]} \BibitemShut {NoStop}%
\bibitem [{\citenamefont {{Misner}}\ \emph {et~al.}(1973)\citenamefont
  {{Misner}}, \citenamefont {{Thorne}},\ and\ \citenamefont
  {{Wheeler}}}]{gravitation}%
  \BibitemOpen
  \bibfield  {author} {\bibinfo {author} {\bibfnamefont {C.~W.}\ \bibnamefont
  {{Misner}}}, \bibinfo {author} {\bibfnamefont {K.~S.}\ \bibnamefont
  {{Thorne}}}, \ and\ \bibinfo {author} {\bibfnamefont {J.~A.}\ \bibnamefont
  {{Wheeler}}},\ }\href@noop {} {\emph {\bibinfo {title} {San Francisco:
  W.H.~Freeman and Co., 1973}}}\ (\bibinfo {year} {1973})\BibitemShut {NoStop}%
\end{thebibliography}%

\onecolumngrid
\newpage
\renewcommand{\thesubsection}{\Roman{subsection}}

\begin{appendices}

\section{Kerr space-time}
\label{sec:KerrSpacetime}
\noindent
Since the background we consider in this article is exclusively the Kerr space-time, it is worth providing a summary of the metric for a rotating non-charged black hole. Alternatively, a charged and spinning black hole is described by the Kerr-Newman metric, while a non-rotating charged black hole solution is given by the Reissner-Nordst\"{o}m background. The Kerr metric, in Boyer-Lindquist coordinates \citep{boyerlindquist1982} \((t, r, \theta, \phi)\), is given by:\\
\begin{subequations}
\begin{minipage}{\textwidth}
\begin{minipage}{\textwidth}
\begin{align}
\dd s^{2} = - \dfrac{\Delta - a^{2} \sin^{2} \theta}{\Sigma} \dd t^{2} - 2 a \sin^{2} \theta \dfrac{r^{2} + a^{2} - \Delta}{\Sigma} \dd t \dd \phi + \dfrac{\left(r^{2} + a^{2}\right)^{2} - \Delta a^{2} \sin^{2} \theta}{\Sigma} \sin^{2} \theta \dd \phi^{2} + \dfrac{\Sigma}{\Delta} \dd r^{2} + \Sigma \dd \theta^{2} \label{eq:_metric}
\end{align}
\begin{flalign*}
& \text{where} &
\end{flalign*}
\end{minipage}
\begin{minipage}{.5\textwidth}
\begin{align}
\Delta (r) = r^{2} - 2 M r + a^{2}\;,
\end{align}
\end{minipage}
\begin{minipage}{.5\textwidth}
\begin{align}
\Sigma (r, \theta) = r^{2} + a^{2} \cos^{2} \theta\;,
\end{align}
\end{minipage}
\vspace{1em}
\end{minipage}
\end{subequations}
\begin{minipage}{\textwidth}
\begin{minipage}{.35\textwidth}
and \(a = \mathcal{L}_{z} / M\) is the specific black hole spin parameter. This metric is singular at the points where \(\theta = \{0, \pi\}\), however these are the usual coordinate singularities associated with the spherical polar system, remedied by the introduction of a second coordinate system. Further, we encounter coordinate singularities at the points where \(\Delta (r) = 0\), i.e., for values of the radial coordinate \(r_{\pm} = M \pm \sqrt{M^{2} + a^{2}}\). To demonstrate that these are merely coordinate singularities (and not an inherent curvature singularity of the metric), it is customary to make the change to Kerr coordinates \((v, r, \theta, \chi)\), given by \citep{visser2007}:\\
\begin{subequations}
\begin{align}
\dd v &= \dd t + \frac{r^{2} + a^{2}}{\Delta} \, \dd r\;,\\[5pt]
\dd \chi &= \dd \phi + \frac{a}{\Delta} \, \dd r\;.
\end{align}
\end{subequations}
\vspace{0.1em}
\end{minipage}
\begin{minipage}{.65\textwidth}
\begin{figure}[H]
\includegraphics[scale=1]{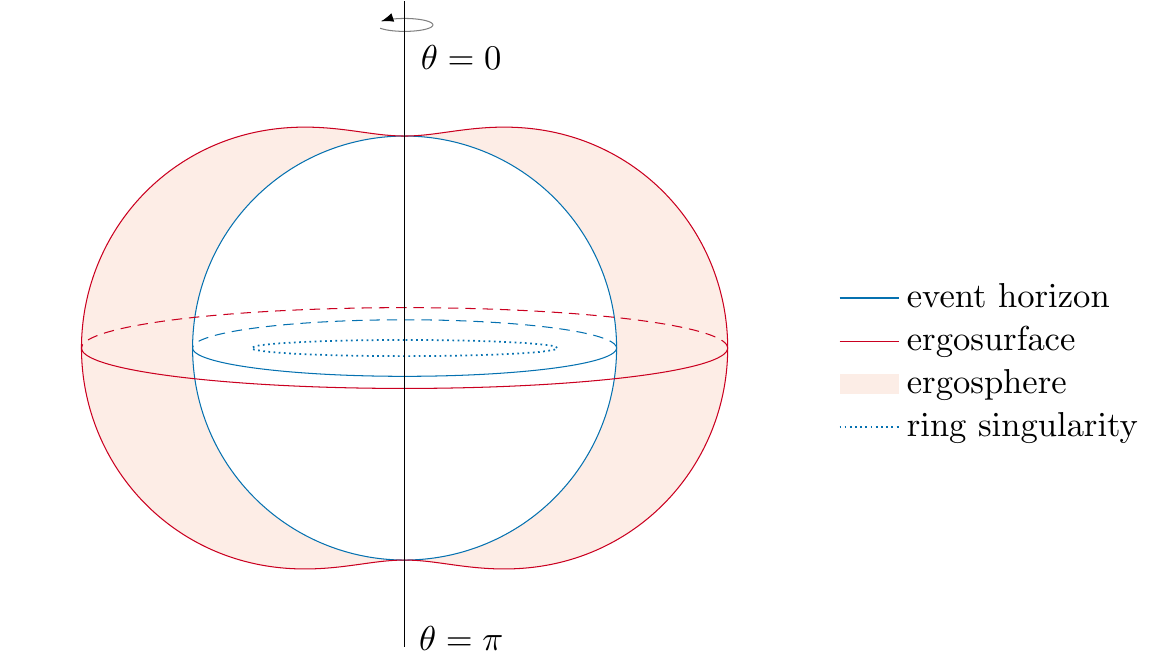}
\caption{Visualisation of the Kerr space-time for \(a/M = 0.95\).}
\label{fig:kerr}
\end{figure}
\vspace{1em}
\end{minipage}
\end{minipage}
This shows that the metric could be analytically continued through the surface \(r_{\pm}\) to the region \(0 < r < r_{+}\):
\begin{equation}
\begin{split}
\dd s^{2} = -\frac{\Delta - a^{2} \sin^{2} \theta}{\Sigma} \dd v^{2} + 2 \dd v \dd r &- 2 a \sin^{2} \theta \, \frac{r^{2} + a^{2} -\Delta}{\Sigma} \dd v \dd \chi \\
&-2 a \sin^{2} \theta \dd \chi \dd r + \frac{\left(r^{2} + a^{2}\right)^{2} - \Delta \, a^{2} \sin^{2} \theta}{\Sigma} \, \sin^{2} \theta \dd\chi^{2} + \Sigma \dd\theta^{2}.
\end{split}
\end{equation}
The surface \(r = r_{+} = M + \sqrt{M^{2} - a^{2}}\) is a null hypersurface with normal 4-vector given by
\begin{align}
\xi^{a} = \left(\frac{\partial}{\partial v}\right)\np{a} + \frac{a}{a^{2} + r_{+}^{2}} \left(\frac{\partial}{\partial \chi}\right)\np{a} = \left(\frac{\partial}{\partial v}\right)\np{a} + \Omega_{\mathrm{Kerr}} \left(\frac{\partial}{\partial \chi}\right)\np{a}
\end{align}
and the region \(r \le r_{+}\) is (part of) the black hole region. \(\Omega_{\mathrm{}}\) is the angular velocity of the black hole with respect to an observer at infinity. There are several important surfaces in the Kerr space-time. Firstly, the curvature singularity \(r = 0\) is in fact a ring singularity, which can be explained as follows. A rotating black hole would not be spherically symmetric, as matter would bulge around the equator, the effect quantified by the spin parameter \(a\). Since a point singularity cannot support angular momentum, the lowest-dimensional shape required is a ring. This is evidenced by substituting \(t \equiv const.\), \(r = 0\), \(\theta=\sfrac{\pi}{2}\) (i.e., the singularity associated with \(\Sigma (r, \theta) = 0\)) in eq.~(\ref{eq:_metric}), obtaining \(\dd s^{2} = a^{2} \dd\phi^{2}\), i.e., a ring of radius \(a\). Another important surface is the event horizon \(r_{+} = M + \sqrt{M^{2} - a^{2}}\). To understand the final feature of interest, we consider the norm of the Killing vector field \((\partial / \partial t)^{a}\) in Boyer-Lindquist coordinates:
\begin{align}
\left(\frac{\partial}{\partial t}\right)\np{2} = -\frac{\Delta - a^{2} \sin^{2} \theta}{\Sigma} = -1 + \frac{2Mr}{r^{2} + a^{2} \sin^{2} \theta}
\end{align}
and notice that it is space-like (\((\partial / \partial t)^{2} < 0\)) in the following region outside the event horizon \(r = r_{+}\):
\begin{align}
M + \sqrt{M^{2} - a^{2}} < r < M + \sqrt{M^{2} - a^{2} \cos^{2} \theta}.
\end{align}
This region is called the ergosphere, and its outside surface is named the ergosurface, which intersects the event horizon at the poles \(\theta = 0, \pi\). These surfaces are illustrated on the Kerr black-hole schematic on Fig.~\ref{fig:kerr}.

\section{Geodesic constants}
\label{sec:GeodesicConstants}
\noindent
One immediately notices that the Kerr metric does not depend on the time coordinate, \(t\), or the azimuthal angle, \(\phi\). Therefore, using the Euler-Lagrange equations, it can be shown that the conjugate momenta \(p_{\mu} = \partial \mathcal{L} / \partial \dot{q}^{\mu}\) associated to those coordinates are constant with respect to the affine parameter of the geodesic:
\begin{align}
\frac{\dd}{\dd \tau}\!\left(\frac{\partial \mathcal{L}}{\partial \dot{q}^{\mu}}\right) = \frac{\partial \mathcal{L}}{\partial q^{\mu}} \quad \Leftrightarrow \quad \frac{\dd p_{\mu}}{\dd \tau} = \frac{\partial \mathcal{L}}{\partial q^{\mu}} =0 \quad\Leftrightarrow \quad \begin{cases}
p_{t} \equiv const. \\
p_{\phi} \equiv const.\;,
\end{cases}
\end{align}
where in general relativity, the Lagrangian is formally given by \(\mathcal{L}(q, \dot{q}) = \sfrac{1}{2} \, g_{\mu \nu} \, \dot{q}^{\mu} \dot{q}^{\nu}\). Thus we have identified 2 quantities which are constant along geodesics of the Kerr space-time. These are related to the energy of the orbit \(\mathcal{E}\), and the axial component of the angular momentum \(\mathcal{L}_{z}\), respectively:
\begin{align}
\mathcal{E} &= -p_{t} = -g_{t \nu} \dot{q}^{\nu} = \dfrac{\Delta - a^{2} \sin^{2} \theta}{\Sigma} \, \dot{t} \, + \, a \sin^{2} \theta \, \dfrac{r^{2} + a^{2} - \Delta}{\Sigma} \,\dot{\phi}\\
\mathcal{L}_{3} &= p_{\phi} = g_{\phi \nu} \dot{q}^{\nu} = - a \sin^{2} \theta \, \dfrac{r^{2} + a^{2} - \Delta}{\Sigma} \,\dot{t} \, + \, \dfrac{\left(r^{2} + a^{2}\right)^{2} - \Delta a^{2} \sin^{2} \theta}{\Sigma} \, \sin^{2} \theta \, \dot{\phi}
\end{align}
For equatorial geodesics (\(\theta = \sfrac{\pi}{2}\)), these two constants are sufficient to uniquely determine a given geodesic. However, for general geodesics with 3 spatial degrees of freedom, we can identify a third constant of the motion, which is not associated with a linear symmetry of the background space-time.

Apart from the two Killing vectors associated with the Kerr metric, there exists a second-order Killing tensor for this background, \(Q_{\mu \nu}\). This Killing tensor gives rise to another conserved quantity, which was originally discovered in 1968 as a consequence of the separable Hamilton-Jacobi equation. This quantity is called the Carter's constant, and can be written as \citep{carter1968}:
\begin{align}
\mathcal{Q} = Q^{\mu \nu} p_{\mu} p_{\nu} = p_{\theta}^{2} + \cos^{2} \theta \left(\!a^{2}\big(m^{2} - \mathcal{E}^{2}\big) + \left(\!\frac{\mathcal{L}_{z}}{\sin \theta}\!\right)\np{\,2}\!\right).
\end{align}

\section{Geodesics of the Kerr metric}
\label{sec:KerrGeodesics}
\noindent
\begin{figure}[t!]
	\centering
	\subfloat[(a) Projections of the geodesic on the principal planes.]{\rule{0.5\textwidth}{0pt}}\\
	\subfloat{\includegraphics[scale=1.0]{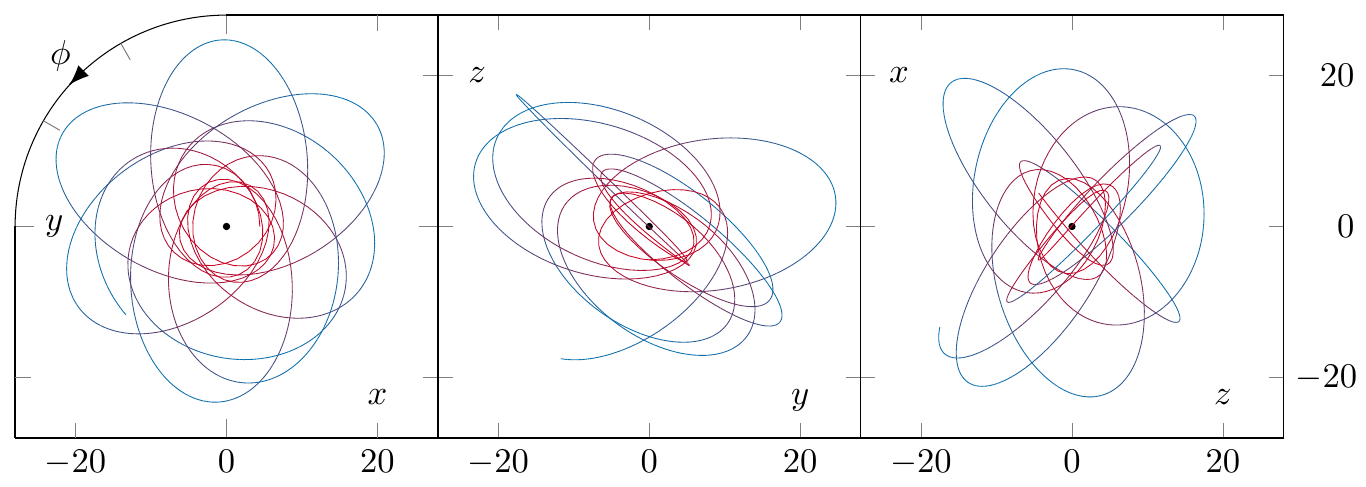}\label{fig:kerr_c}} \\
	\subfloat[(b) Evolution of the coordinates \(x\), \(y\) and \(z\) with physical time \(t\).]{\includegraphics[scale=1.0]{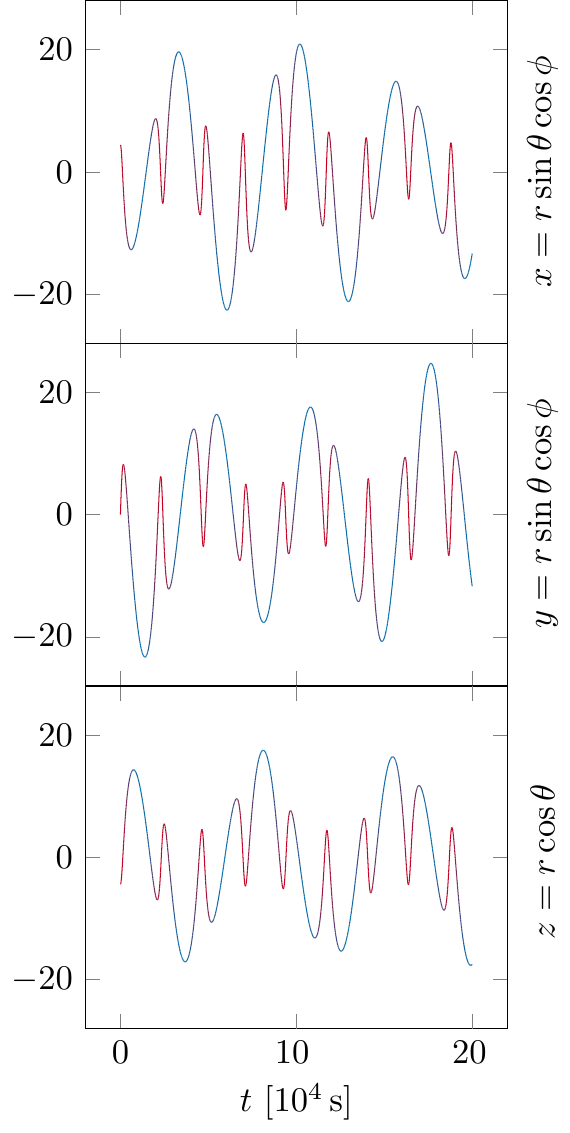}\label{fig:kerr_a}}
	\subfloat[(c) A tridimensional view of the geodesic in Cartesian and cylindrical coordinate systems.]{\includegraphics[scale=1.0]{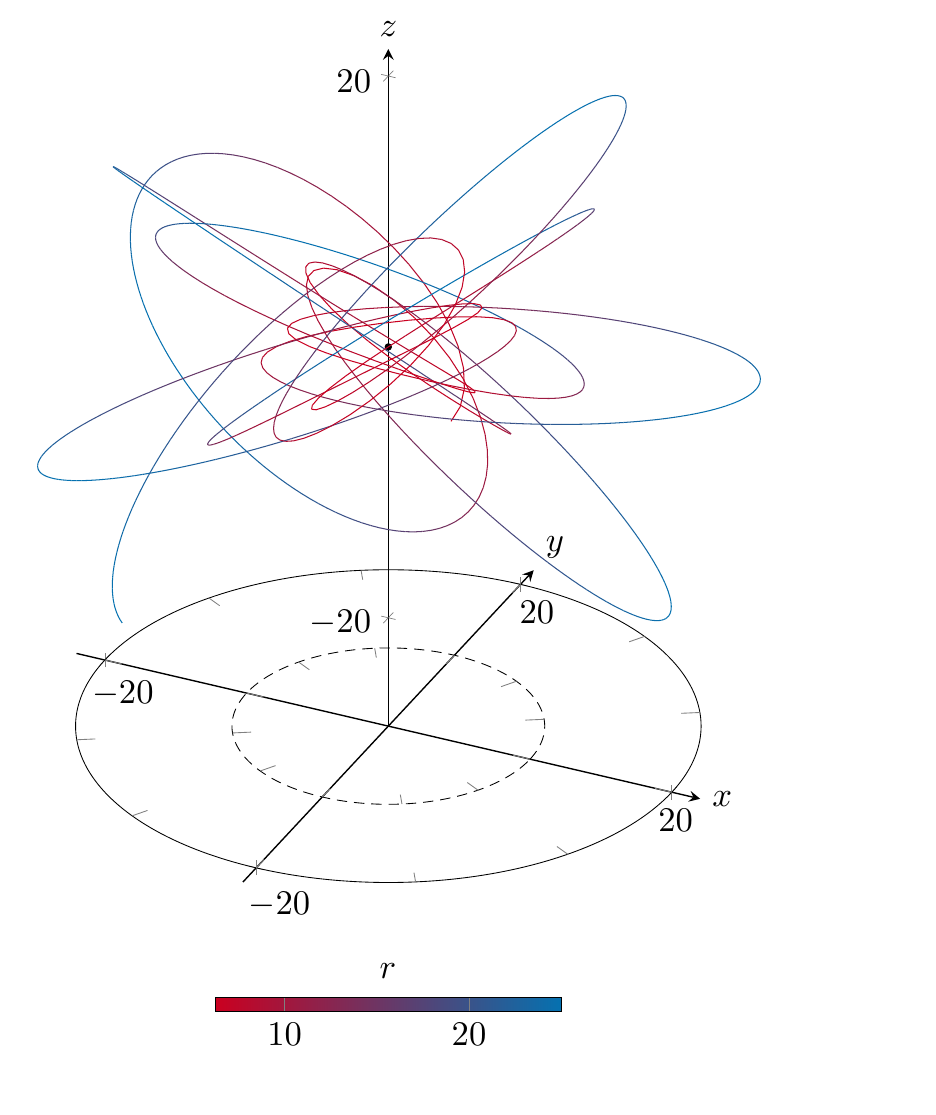}\label{fig:kerr_b}}
	\caption{A geodesics of the Kerr metric, and several projections, evolved for a period of \(2 \times 10^{5}\) s \(\approx 5.56\) hours. The orbit has eccentricity \(e=0.6\) and semi-latus rectum \(10 M\), while the central black hole has mass \(M = 10^{6} M_{\odot}\) and spin \(0.9 M^{2}\). All distances are in units of the black-hole mass \(M\). The angle of inclination is \(\iota = 45^{\circ}\), and the plane of inclination precesses.}
	\label{fig:kerr_geodesic}
\end{figure}

The equatorial geodesics of the Kerr metric can be characterised by the constants \(\mathcal{E}\) and \(\mathcal{L}_{z}\), combined with the Hamiltonian, which is an additional integral of motion. General 3-dimensional geodesics, however, require the introduction of a further integral of motion, Carter's constant, discussed in Appendix~\ref{sec:GeodesicConstants}. The equations of motion for a massless particle, as per \cite*{gravitation}, are given by\\
\begin{subequations}
\begin{minipage}{\textwidth}
\begin{minipage}{.44\textwidth}
\begin{align}
\frac{\dd t}{\dd \tau} = \frac{1}{\Delta}\!\left(\!\!\vphbig\left(\!r^{2} + a^{2}\,\frac{2Ma^{2}}{r}\!\vph\right)\!\mathcal{E} - \frac{2Ma}{r} \, \mathcal{L}_{z}^{2}\!\right) \, \\[-23pt] \nonumber
\end{align}
\end{minipage}
\begin{minipage}{.57\textwidth}
\begin{align}
\frac{\dd r}{\dd \tau} = \pm \frac{1}{\Sigma} \, \sqrt{\Big(\mathcal{E}\big(r^{2} + a^{2}\big) - a \, \mathcal{L}_{z}\Big)^{\!2}\!- \Delta\!\left(\!\mathcal{Q} + \left(\mathcal{L}_{z} - a \mathcal{E}\right)^{2}\right)} \quad\quad\quad \nonumber\\[-23pt]
\end{align}
\end{minipage}
\begin{minipage}{.44\textwidth}
\begin{align}
\frac{\dd \theta}{\dd \tau} = \pm \frac{1}{\Sigma} \, \sqrt{\mathcal{Q} + \cos^{2} \theta\!\left(\!a^{2} \mathcal{E}^{2} - \frac{\mathcal{L}_{z}}{\sin^{2} \theta}\!\vph\right)} \quad
\end{align}
\end{minipage}
\begin{minipage}{.57\textwidth}
\begin{align}
\frac{\dd \phi}{\dd \tau} &= \frac{1}{\Delta}\!\left(\!\!\vphbig\left(\!1 - \frac{2M}{r}\!\vph\right)\!\mathcal{L}_{z} + \frac{2Ma}{r} \, \mathcal{E}\!\right) \quad\quad\quad\quad\quad\quad\quad\quad\quad\,\, 
\end{align}
\end{minipage}
\end{minipage}
\end{subequations}%
In the general case, these equation cannot be solved analytically to find a closed form of the shape of the geodesic. However, it is possible to integrate the equations numerically to find an approximate shape for a Kerr geodesic, and one possible solutions is shown in Fig.~\ref{fig:kerr_geodesic}. The Kerr geodesics are not closed, but rather fill the entire parameter space \(r_{\text{\textsc{min}}} < r < r_{\text{\textsc{max}}}\), \(\theta_{\text{\textsc{min}}} < \theta < \theta_{\text{\textsc{max}}}\), \(0 < \phi < 2 \pi\) (which has the shape of a torus) before returning to the starting \((r, \theta, \phi)\). Moreover, the plane of inclination of the orbit is not stationary, but precesses around the black hole. In the case of a resonance, when two of the frequencies become commensurate, this configuration breaks down and the geodesic becomes a (simpler) closed curve around the black hole, defined only by 2 parameters.

\section{Fresnel integrals}
\label{sec:AppFresnelFuncs}
\noindent
\begin{figure}[b]
\includegraphics[width=\textwidth]{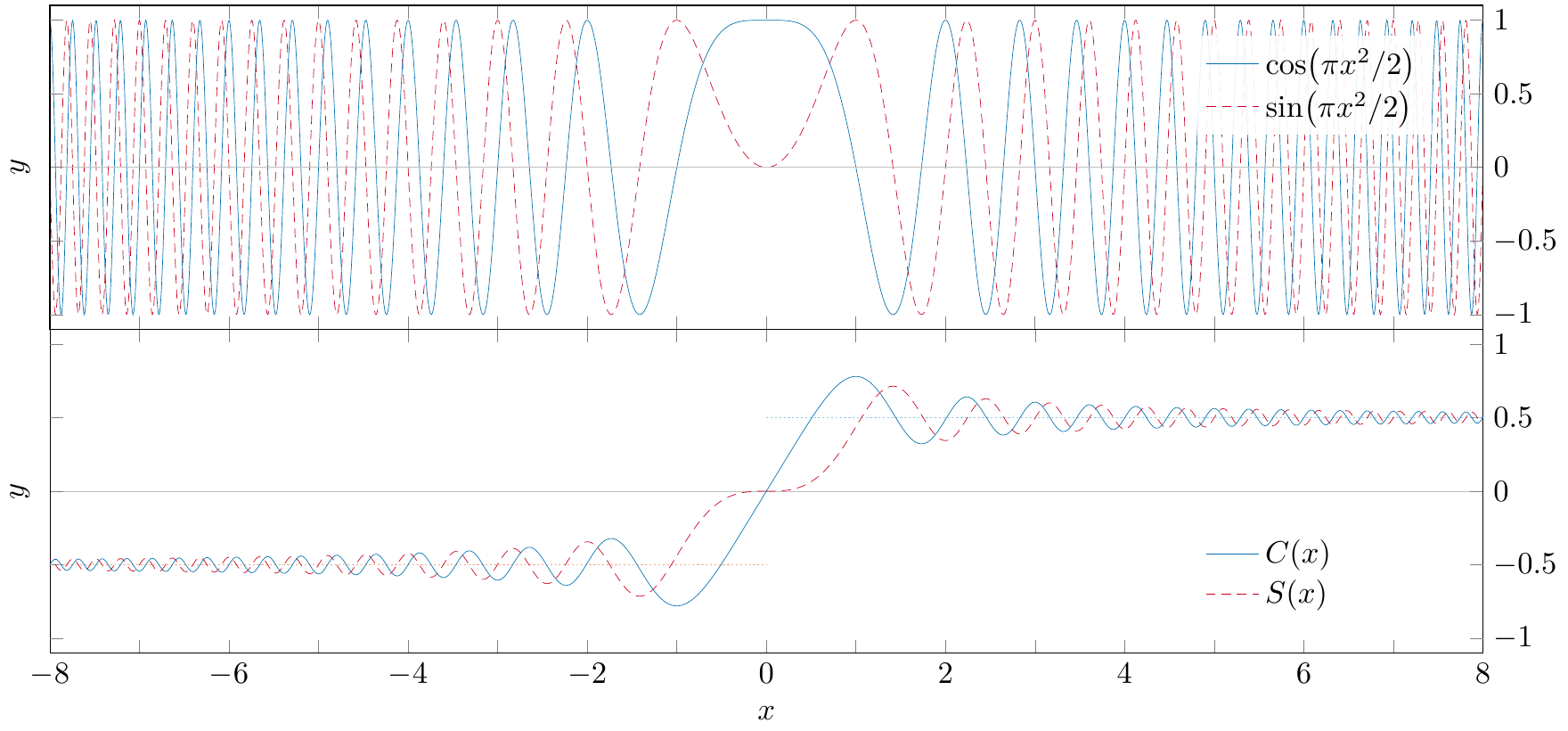}
\caption{\emph{Upper panel:} the integrands of the two Fresnel integrals, the functions \(\cos\left(\pi x^{2}/2\right)\) and \(\sin\left(\pi x^{2}/2\right)\), and the axis \(y = 0\) around which they oscillate. The frequency of oscillation increases with the absolute value of \(x\). \emph{Lower panel:} numerical integration of the functions above gives the desired graphs of \(C(x)\) and \(S(x)\). They converge towards the line \(y = \pm\sfrac{1}{2}\).}
\label{fig:fresnel}
\end{figure}

In the course of solving the phase resonance equations, as well as in the presentation of the frequency resonance solution, we encounter integrals of trigonometric functions with quadratic arguments:
\begin{align}
\bigintsss_{0}^{x}\!\!\!\dd y\,\sin\!\left(\!\vph\frac{\pi y^{2}}{2}\!\right) \quad \text{and} \quad \bigintsss_{0}^{x}\!\!\!\dd y\,\cos\!\left(\!\vph\frac{\pi y^{2}}{2}\!\right).
\label{eq:fresnel_integrals}
\end{align}
These are known as the Fresnel integrals, after the French engineer and physicist Augustin-Jean Fresnel, who used them in optics. Here, we have adopted the notation where the argument of the integrand is premultiplied by \(\sfrac{\pi}{2}\). Please note that this definition is equivalent to the one used in \cite{gair2010}:
\begin{align}
C(x) = \frac{\sqrt{2}}{\sqrt{\pi}} \int_{0}^{x} \dd y \, \cos\!\left(y^{2}\right).
\label{eq:fresnel_integral_alternative_definition}
\end{align}
These integrals are not directly solvable analytically, and therefore other methods for integration must be used. One approach to finding the two integrals in (\ref{eq:fresnel_integrals}) is by expanding the integrands in converging Taylor series:
\begin{subequations}
\label{eq:fresnel_int_solutions}
\begin{align}
S (x) &= \bigintsss_{0}^{x}\!\!\!\dd y \sin\!\left(\!\vph\frac{\pi y^{2}}{2}\!\right) = \sum_{n = 0}^{\infty} \frac{(-1)^{n}}{(2n + 1)!} \bigintsss_{0}^{x}\!\!\!\dd y \!\left(\!\vph\frac{\pi y^{2}}{2}\!\right)\np{2n + 1} \!\!\!\! = \sum_{n = 0}^{\infty} \frac{(-1)^{n}\,\pi^{2n+1}}{(2n + 1)!\,(4n + 3)\,2^{2n+1}}\, x^{4n + 3}, \\
C (x) &= \bigintsss_{0}^{x}\!\!\!\dd y \cos\!\left(\!\vph\frac{\pi y^{2}}{2}\!\right) = \sum_{n = 0}^{\infty} \frac{(-1)^{n}}{(2n)!} \bigintsss_{0}^{x}\!\!\!\dd y \!\left(\!\vph\frac{\pi y^{2}}{2}\!\right)\np{2n} \!\!\!\! = \sum_{n = 0}^{\infty} \frac{(-1)^{n}\,\pi^{2n}}{(2n)!\,(4n + 1)\,2^{2n}}\, x^{4n + 1}.
\end{align}
\end{subequations}%
Often, a symbol like the ones appearing on the left-hand sides of eqs.~(\ref{eq:fresnel_int_solutions}) are used in lieu of the actual series solutions. Hence, it is useful to know the first few terms of the series for each of the integrals:

\begin{subequations}
\begin{minipage}{\textwidth}
\begin{minipage}{.475\textwidth}
\begin{align}
S (x) &= \frac{\pi}{6}\,x^{3} - \frac{\pi^{3}}{336}\,x^{7} + \bigo{x^{11}}, \label{eq:fresnels_expansion}
\end{align}
\end{minipage}
\begin{minipage}{.5\textwidth}
\begin{align}
C (x) &= x - \frac{\pi^{2}}{40}\,x^{5} + \bigo{x^{9}}. \label{eq:fresnelc_expansion}
\end{align}
\end{minipage}
\end{minipage}
\end{subequations}%

\vspace{1em}

The integrands and the integral functions are plotted in Fig.~\ref{fig:fresnel}. We make note of several features: at \(x = 0\), \(S(x)\) starts flat, and progresses as \(\sim x^{3}\), while \(C (x)\) is linear. For positive or negative infinity, both functions tend to \(\pm\sfrac{1}{2}\), and can be represented by an asymptotic expansion for large absolute values of \(x\):
\begin{subequations}
\begin{align}
&S (x) = \frac{\Sgn(x)}{2} - \frac{1}{\pi x}\,\cos\!\Bigg(\!\frac{\pi x^{2}}{2}\!\Bigg)\!- \frac{1}{\pi^{2} x^{3}}\,\sin\!\Bigg(\!\frac{\pi x^{2}}{2}\!\Bigg)\!+\frac{3}{\pi^{3} x^{5}}\,\cos\!\Bigg(\!\frac{\pi x^{2}}{2}\!\Bigg)\!+ \bigo{x^{-6}}, \\
& C (x) = \frac{\Sgn(x)}{2} + \frac{1}{\pi x}\,\sin\!\Bigg(\!\frac{\pi x^{2}}{2}\!\Bigg)\!- \frac{1}{\pi^{2} x^{3}}\,\cos\!\Bigg(\!\frac{\pi x^{2}}{2}\!\Bigg)\!- \frac{3}{\pi^{3} x^{5}}\,\sin\!\Bigg(\!\frac{\pi x^{2}}{2}\!\Bigg)\!+ \bigo{x^{-6}}.
\end{align}
\end{subequations}%

\section{Shorthand trigonometric notation}
\label{sec:ShorthandTrig}
\noindent
In eqs.~(\ref{eq:shorthand_trigonometric}) we define the following to make the discussion regarding the frequency mode equation easier to read:\\
\begin{minipage}{\textwidth}
\begin{minipage}{.5\textwidth}
\begin{align*}
S_{n} (x) \equiv \sin\!\left[n\Big(a x + \left(1 + b\right) x^{2}\Big)\vphantom{\Big(\Big)\np{2}}\!\right] \tag{\ref{eq:def_snx}}
\end{align*}
\end{minipage}
\begin{minipage}{.5\textwidth}
\begin{align*}
C_{n} (x) \equiv \cos\!\left[n\Big(a x + \left(1 + b\right) x^{2}\Big)\vphantom{\Big(\Big)\np{2}}\!\right]. \tag{\ref{eq:def_cnx}}
\end{align*}
\end{minipage}
\vspace{1em}
\end{minipage}
for some arbitrary constants \(a\) and \(b\). 

\begin{figure}
\includegraphics[width=\textwidth]{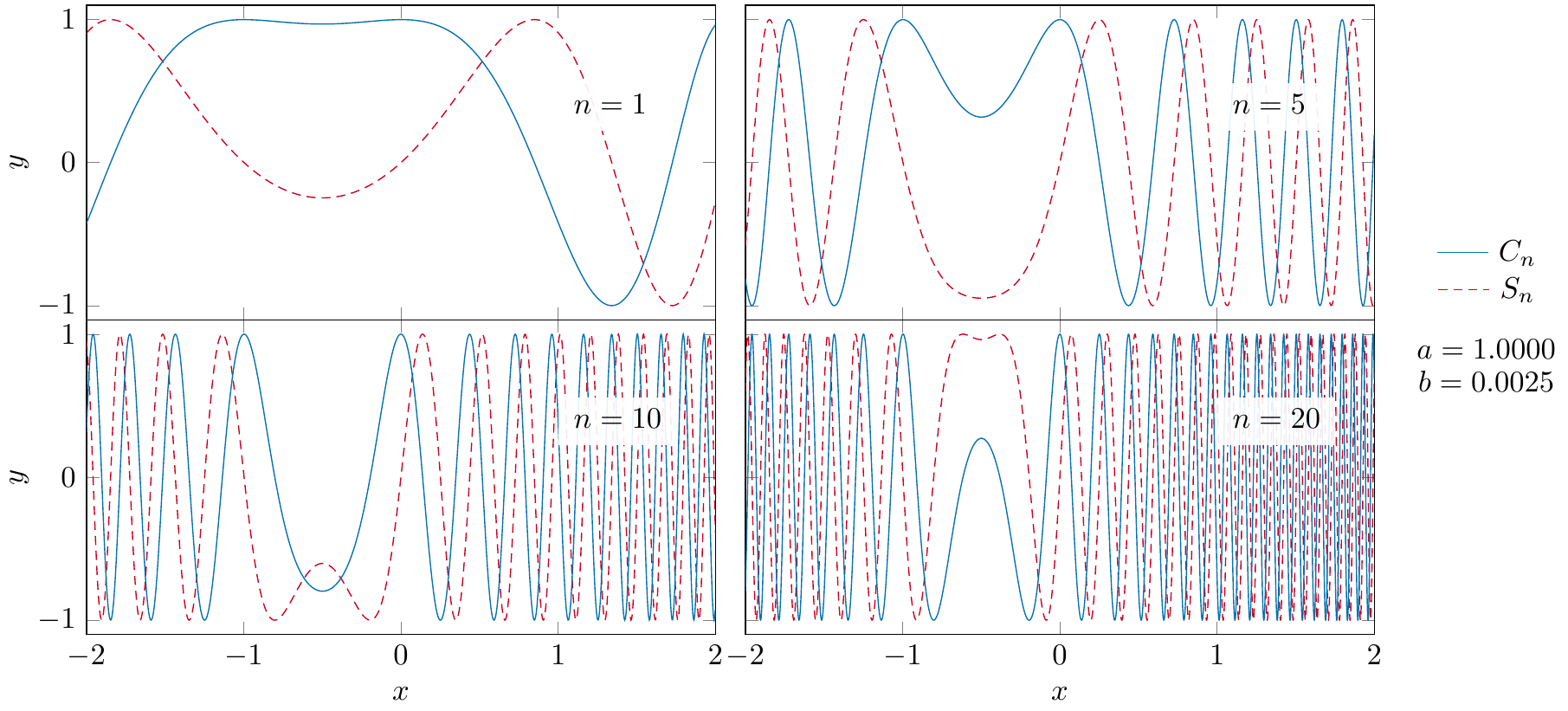}
\caption{The functions \(S_{n} (x)\) and \(C_{n} (x)\) for several values of \(n\). As we can see, the functions oscillate (away from the region around \(x=0\)) more rapidly as \(n\) increases. We have used values of \(a = 1\) and \(b = 0.0025\) for the constants.}
\label{fig:sncn}
\end{figure}

Please note that these are different from the Fresnel integrals, denoted by the same letters, but without subscripts, discussed in Appendix \ref{sec:AppFresnelFuncs}. These functions are plotted for several representative values of \(n\) in Fig.~\ref{fig:sncn}. It is worth noting several identities which we make use of in the article, but do not derive explicitly. These follow directly from the product-to-sum formulae and the double angle formulae in trigonometry.
\begin{subequations}
\noindent
\begin{minipage}{\textwidth}
\begin{minipage}{.44\textwidth}
\begin{align}
S_{n} (x) S_{m} (x) = \frac{1}{2}\!\left(C_{\left(n - m\right)} (x) - C_{\left(n + m\right)} (x)\right)
\end{align}
\end{minipage}
\begin{minipage}{.56\textwidth}
\begin{align}
S_{n} (x) C_{m} (x) = \frac{1}{2}\!\left(S_{\left(n + m\right)} (x) - S_{\left(n - m\right)} (x)\right)
\end{align}
\end{minipage}
\begin{minipage}{.442\textwidth}
\begin{align}
C_{n} (x) C_{m} (x) = \frac{1}{2}\!\left(C_{\left(n - m\right)} (x) + C_{\left(n + m\right)} (x)\right)
\end{align}
\end{minipage}
\begin{minipage}{.277\textwidth}
\begin{align}
S_{n}^{2} (x) = \frac{1}{2} \left(1 - C_{2n} (x)\right)
\end{align}
\end{minipage}
\begin{minipage}{.277\textwidth}
\begin{align}
C_{n}^{2} (x) = \frac{1}{2} \left(1 + C_{2n} (x)\right)
\end{align}
\end{minipage}
\end{minipage}
\end{subequations}

\vspace{1em}
We only consider positive values of the subscripts, with the following rules for negative values: \(S_{-n} (x) = - S_{n} (x)\) and \(C_{-n} (x) = C_{n} (x)\). It is important to note that these functions are linearly independent, i.e., they form a basis in functional space. This is most easily verified by considering the Wronskian for the sequence of functions
\begin{align}
f_{0} (x) = 1\;, \quad f_{1} (x) = S_{1} (x)\;, \quad f_{2} (x) = C_{1} (x)\;, \quad f_{3} (x) = S_{2} (x)\;, \quad f_{4} (x) = C_{2} (x) \; \text{ and so on.}
\end{align}
For \(\{f_{0} (x), f_{1} (x), f_{2} (x)\}\), the Wronskian is \(-(a+2(1+b)x)^{3}\), while for the first 5 functions in the series, it is \(72(a+2(1+b)x)^{10}\), both of which do not vanish in the general case. Since their Wronskian does not vanish, we can conclude that these functions are linearly independent over \(\mathbb{R} \, \backslash \{0\}\):
\begin{align}
W (f_{0}, f_{1}, f_{2}, \ldots) (x) = \det \left|
\begin{matrix}
f_{0} (x) & f_{1} (x) & f_{2} (x) & f_{3} (x) & f_{4} (x) & \\[2pt]
f_{0}^{\prime} (x) & f_{1}^{\prime} (x) & f_{2}^{\prime} (x) & f_{3}^{\prime} (x) & f_{4}^{\prime} (x) & \cdots \;\;\;\; \\[2pt]
f_{0}^{\prime\prime} (x) & f_{1}^{\prime\prime} (x) & f_{2}^{\prime\prime} (x) & f_{3}^{\prime\prime} (x) & f_{4}^{\prime\prime} (x) & \\[2pt]
& & \vdots & & & \ddots \;\;\;\; \\
\end{matrix}
\right| \neq 0.
\end{align}
It should be noted that these functions are not orthogonal. First, we show how to integrate a single factor of \(S_{n} (x)\):
\begin{align*}
\int_{-\infty}^{\infty} \dd x \, S_{n} (x) &= \lim_{\substack{L_{1} \rightarrow -\infty \\ L_{2} \rightarrow +\infty}} \int_{L_{1}}^{L_{2}} \dd x\,\sin\!\left[\frac{\pi}{2}\!\left(\!\sqrt{\frac{2 n (1+b)}{\pi}} \left(\!x + \frac{a}{2 (1+b)}\!\right)\!\right)^{2}\!\!- \frac{n a^{2}}{4 (1+b)}\right] = \\
&= \lim_{\substack{L_{1} \rightarrow -\infty \\ L_{2} \rightarrow +\infty}} \sqrt{\frac{\pi}{2 n (1+b)}} \left(\cos\!\left[\vph\frac{n a^{2}}{4 (1+b)}\right]\!\big(S\left(L_{2}\right) - S\left(L_{1}\right)\!\big) - \sin\!\left[\vph\frac{n a^{2}}{4 (1+b)}\right]\!\big(C\left(L_{2}\right) - C\left(L_{1}\right)\!\big)\!\vphbig\right) = \\
&= \sqrt{\frac{\pi}{n (1+b)}} \; \cos\!\left[\vph\frac{n a^{2}}{4 (1+b)} + \frac{\pi}{4}\right]
\end{align*}
or similarly, a single factor of \(C_{n} (x)\):
\begin{align*}
\int_{-\infty}^{\infty} \dd x \, C_{n} (x) &= \lim_{\substack{L_{1} \rightarrow -\infty \\ L_{2} \rightarrow +\infty}} \sqrt{\frac{\pi}{2 n (1+b)}} \left(\cos\!\left[\vph\frac{n a^{2}}{4 (1+b)}\right]\!\big(C\left(L_{2}\right) - C\left(L_{1}\right)\!\big) - \sin\!\left[\vph\frac{n a^{2}}{4 (1+b)}\right]\!\big(S\left(L_{2}\right) - S\left(L_{1}\right)\!\big)\!\vphbig\right) = \\
&= \sqrt{\frac{\pi}{n (1+b)}} \; \sin\!\left[\vph\frac{n a^{2}}{4 (1+b)} + \frac{\pi}{4}\right].
\end{align*}
Here, \(S (x)\) and \(C (x)\) are the Fresnel integrals, whose properties and limits are discussed in Appendix~\ref{sec:AppFresnelFuncs}. Using these results, we can easily demonstrate that these functions do not form an orthogonal basis:
\begin{align}
\begin{split}
\int_{-\infty}^{\infty} \dd x \, S_{n} (x) \, S_{m} (x) &= \frac{1}{2} \int_{-\infty}^{\infty} \dd x \, \left(C_{\left(n - m\right)} (x) - C_{\left(n + m\right)} (x)\right) = \\
&= \frac{1}{2} \, \sqrt{\frac{\pi}{1+b}} \left(\!\vphbig\frac{1}{\sqrt{\left|n - m\right|}} \; \sin\!\left[\vph\frac{\left|n-m\right| a^{2}}{4 (1+b)} + \frac{\pi}{4}\right]\!- \frac{1}{\sqrtbig{n+m}} \; \sin\!\left[\vph\frac{(n+m) a^{2}}{4 (1+b)} + \frac{\pi}{4}\right]\!\right),
\end{split}
\end{align}
which does not vanish in general. Similar expressions can be obtained for integrals of \(S_{n} (x) \, C_{m} (x)\) and \(C_{n} (x) \, C_{m} (x)\).

\section{Solutions to the Frequency resonance equation}
\label{sec:AppendixFrequencyResEq}
\noindent
\subsection{Third-order solution}
\label{sec:AppFrequencyThirdOrderSMNP}
\noindent
In the body of the article we derived in detail the solution to the single-mode frequency resonance equation (\ref{eq:FrequencySMEq}) up to second order in the parameter \(\kappa\). Here we extend the same solution method to derive the behaviour of \(\omega(x)\) to \(\bigo{\kappa^{3}}\). Building onto our previous ansätze, eqs.~(\ref{eq:ansatz_zero}) and (\ref{eq:freq_ansatz_order2}), we suggest the third-order function
\begin{align}
\omega_{3} (x) =  a &+ \left(1 + b_{3}\right)x + c_{1} \frac{S_{1}(x)}{x} + c_{2} \frac{S_{2}(x)}{x} + c_{3} \frac{S_{3}(x)}{x} +  d_{1} \frac{S_{1}(x)}{x^{2}} + d_{2} \frac{S_{2}(x)}{x^{2}} + l \ln\!|x|. \tag{\ref{eq:FreqThirdOrderAnsatz}}
\end{align}
where \(S_{3} (x) = \sin\!\left[3\!\left(ax + (1+b)\,x^{2}\right)\right]\) and we assume that \(c_{3} \sim \bigo{\kappa^{3}}\), since it needs to match terms of \(\bigo{\kappa^{3}}\) in eq.~(\ref{eq:freq_order2_rhs}). We substitute this proposed solution into our governing equation~(\ref{eq:FrequencySMEq}), and following rearrangement of both sides using familiar techniques, but this time keeping terms \(\bigo{x^{-1}}\), we find
\begin{align}
\begin{split}
&b + 2 (1+b_{3}) \, c_{1} \, C_{1} (x) + 4 (1+b_{3}) \, c_{2} \, C_{2} (x) + 6 (1+b_{3}) \, c_{3} \, C_{3} (x) + \frac{l}{x} \\
& \RepQuad{1} + \big(c_{1} a_{3} + 2 (1+b_{3}) \, d_{1}\big) \frac{C_{1}(x)}{x} + \big(2 c_{2} a_{3} + 4 (1+b_{3}) \, d_{2}\big) \frac{C_{2}(x)}{x} + \big(3 c_{3} a_{3} + 6 (1+b_{3}) \, d_{3}\big) \frac{C_{3}(x)}{x} + \bigo{x^{-2}}\sim \\
& \RepQuad{2} \sim -\frac{\kappa c_{1}}{2} + \left(\!\kappa - \frac{\kappa c_{1}^{2}}{8} - \frac{\kappa c_{2}}{2}\!\right)\!C_{1} (x) + \frac{\kappa c_{1}}{2}\,C_{2} (x) + \left(\!\frac{\kappa c_{1}^{2}}{8} + \frac{\kappa c_{2}}{2}\!\right)\!C_{3} (x) - \frac{\kappa d_{1}}{2} \, \frac{1}{x} + \\
& \RepQuad{4} \left(\!- \frac{\kappa c_{1} d_{1}}{2} - \frac{\kappa d_{2}}{2}\!\right)\!\frac{C_{1}(x)}{x} + \frac{\kappa d_{1}}{2} \, \frac{C_{2}(x)}{x} + \frac{\kappa d_{2}}{2} \, \frac{C_{3}(x)}{x} + \bigo{\kappa^{4}}.
\end{split}
\end{align}
Matching terms on both sides yields a system of coupled equations, which determines the coefficients:
\begin{align}
\begin{split}
b \sim -\frac{\kappa c_{1}}{2} \;, \;\; 2 (1+b) \, c_{1} \sim \kappa - \frac{\kappa c_{1}^{2}}{8} - \frac{\kappa c_{2}}{2} \;, &\;\; 4 (1+b) \, c_{2} \sim \frac{\kappa c_{1}}{2} \;, \;\; 6 (1+b) \, c_{3} \sim \frac{\kappa c_{1}^{2}}{8} + \frac{\kappa c_{2}}{2} \;, \\[2pt]
l \sim - \frac{\kappa d_{1}}{2} \;, \;\; c_{1} a_{3} + 2 (1+b_{3}) \, d_{1} \sim - \frac{\kappa c_{1} d_{1}}{2} - \frac{\kappa d_{2}}{2} \;, &\;\; 2 c_{2} a_{3} + 4 (1+b_{3}) \, d_{2} \sim \frac{\kappa d_{1}}{2} \;, \;\; 3 c_{3} a_{3} + 6 (1+b_{3}) \, d_{3} \sim \frac{\kappa d_{2}}{2}.
\end{split}
\end{align}
These yield the following expressions for the undetermined coefficients of the anzats, eq.~(\ref{eq:FreqThirdOrderAnsatz}):
\begin{align*}
b \sim -\frac{1}{4}\,\kappa^{2}\,, \;\, c_{1} \sim \frac{1}{2}\,\kappa + \frac{3}{32}\,\kappa^{3}\,, \;\, c_{2} \sim \frac{1}{16}\,\kappa^{2}\,, \;\, c_{3} \sim \frac{1}{96}\,\kappa^{3}\,, \;\, d_{1} \sim \mp \frac{\sqrt{2 \pi}}{16}\,\kappa^{2} \pm \frac{\spi}{64}\,\kappa^{3}\,, \;\, d_{2} \sim \mp \frac{\sqrt{2 \pi}}{64}\,\kappa^{3}\,, \;\, l \sim \pm \frac{\sqrt{2 \pi}}{32}\,\kappa^{3}
\end{align*}
and the solution to third order in \(\kappa\) can be written as
\begin{m}{align}
&\omega_{3} (x) = a +\!\left(\!\vph 1 - \frac{\kappa^{2}}{4}\!\right)\!x + \kappa\,\frac{1}{2}\,\frac{1}{x}\, \sin\!\left[\vphbig\!\left(\!\pm\frac{\sqrt{2\pi}}{4}\,\kappa \mp \frac{\spi}{16}\,\kappa^{2}\!\right)\!x +\!\Bigg(\!1 - \frac{\kappa^{2}}{4}\!\Bigg)x^{2}\right] \nonumber\\
& \RepQuad{4} + \kappa^{2}\frac{1}{16}\,\frac{1}{x} \, \sin\!\left[\vphbigg 2\!\left(\!\!\vphbig \left(\!\pm\,\frac{\sqrt{2 \pi}}{4}\,\kappa\!\right)\!x + x^{2}\!\right)\!\right]\!+ \kappa^{3}\,\frac{1}{x}\!\left(\!\vph\frac{3}{32}\,\sin\!\left(x^{2}\right) + \frac{1}{96}\,\sin\!\left(3 x^{2}\right)\!\right) \tag{\ref{eq:FrequencySolution3}}\\
& \RepQuad{8} \mp \kappa^{2}\frac{\sqrt{2 \pi}}{16}\,\frac{1}{x^{2}} \, \sin\!\left[\vphbig\!\!\vphbig \left(\!\pm\,\frac{\sqrt{2 \pi}}{4}\,\kappa\!\right)\!x + x^{2}\right]\!\pm \kappa^{3}\!\left(\!\vph\frac{\spi}{64}\,\frac{1}{x^{2}}\,\sin\!\left(x^{2}\right) - \frac{\sqrt{2 \pi}}{64}\,\frac{1}{x^{2}}\,\sin\!\left(2 x^{2}\right) + \frac{\sqrt{2 \pi}}{32}\,\ln\!|x|\!\right) \nonumber
\end{m}
The value of the initial value parameter \(a_{3}\) is determined in the main body of the article.

\subsection{Frequency resonance equation with \(\mathbf{n \neq 1}\)}
\label{sec:freq_extension_simple}
\noindent
In Section~\ref{sec:frequency_solution_ch} we solved the single-mode frequency resonance equation to obtain the solution (\ref{eq:FrequencySolution1}). Instead of including the first oscillatory mode \(n=1\), we now solve this equation for \(n=r\):
\begin{align}
\omega^{\prime} (x) = 1 + \kappa_{r} \cos \left(r x \omega\right).
\end{align}
We propose a version of the ansatz (\ref{eq:ansatz_zero}), with a modified oscillating term, reflecting the modified governing equation:
\begin{align}
\omega_{r} (x) = a + (1 + b)\,x + c_{1}^{(r)}\,\frac{\sin \left(r\left(a x + (1 + b) x^{2}\right)\right)}{x} + \bigo{x^{-1}} = a + (1 + b)\,x + c_{1}^{(r)} \frac{S_{r} (x)}{x} + \bigo{x^{-1}}.
\end{align}
Upon substitution and carrying out the derivative and the Taylor expansion (valid since \(r \kappa_{r} \sim 1/r^{\alpha - 1}\)), we obtain:
\begin{align}
b + 2 r (1 + b)\, c_{1}^{(r)} C_{r} (x) \sim -\frac{1}{2} \, r \kappa_{r} c_{1}^{(r)} + \kappa_{r} C_{r} (x) +\,\text{\textsc{term involving} \(C_{2r} (x)\)}
\end{align}
Matching the respective coefficients in the manner already familiar from above, we obtain the following expressions:\\
\begin{subequations}
\begin{minipage}{\textwidth}
\begin{minipage}{.5\textwidth}
\begin{align}
b \sim - \frac{\kappa_{r}^{2}}{4},
\end{align}
\end{minipage}
\begin{minipage}{.5\textwidth}
\begin{align}
c_{1}^{(r)} \sim \frac{\kappa_{r}}{2r}.
\end{align}
\end{minipage}
\vspace{1em}
\end{minipage}
\end{subequations}

We have shown that \(b \sim \bigo{\kappa_{r}^{2}}\) and \({c_{1}}^{(r)} \sim \bigo{\kappa_{r} / r}\) for the \(r^{\mathrm{th}}\) oscillating mode. Note that since \citep{drasco2006} propose a value of \(\alpha \gtrsim 2\), the series \({c_{1}}^{(n)}\) converge, and our use of the Taylor expansion is valid.

\begin{minipage}{\textwidth}
\begin{minipage}[t]{0.45\textwidth}
\begin{figure}[H]
\includegraphics[scale=1]{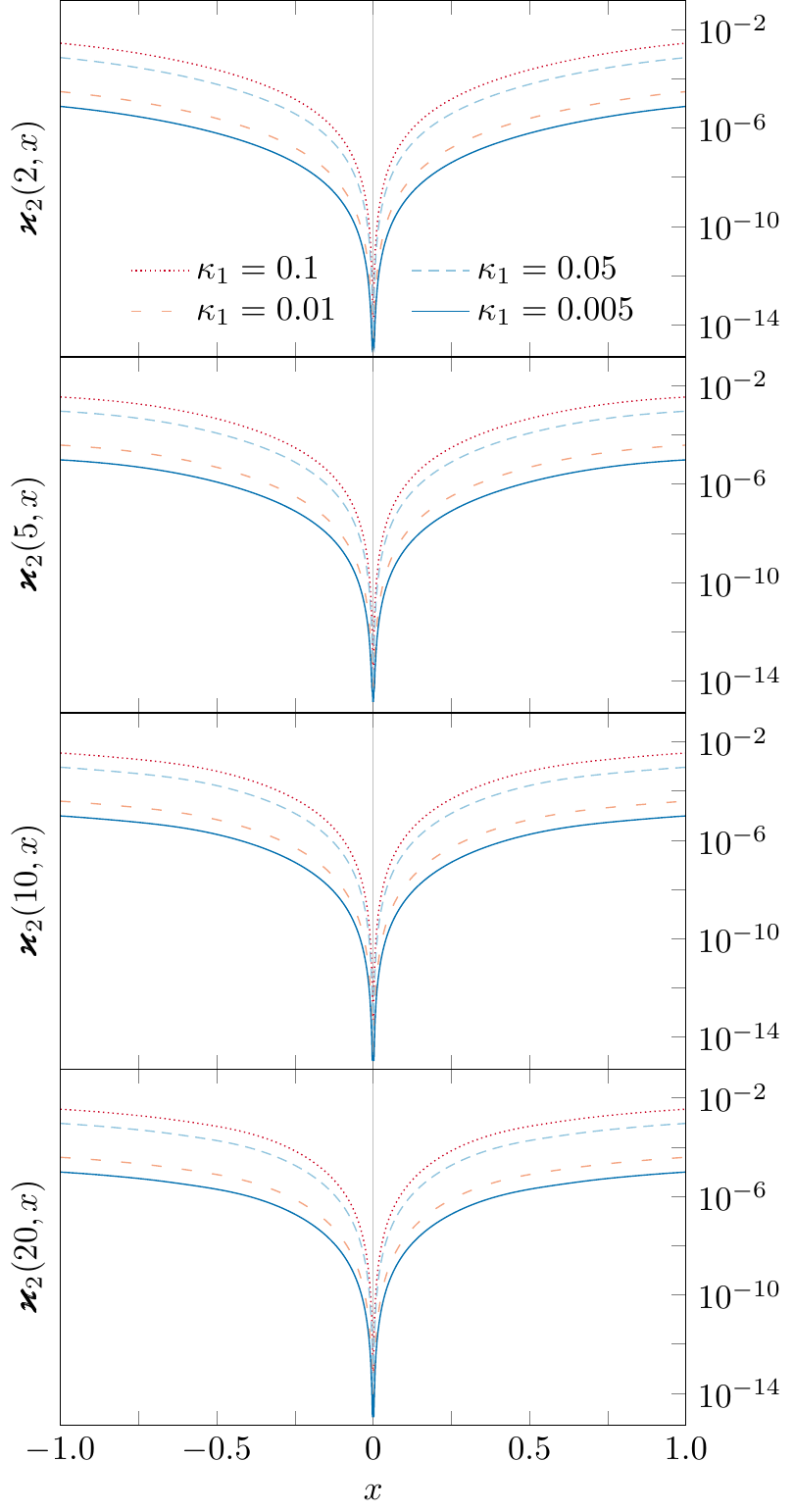}
\caption{Graphical demonstration of the validity of solution~(\ref{eq:frequencyMMSolutionSmallX}) for different number of terms on the right-hand side. The coefficients \(\{\kappa_{n}\}\) follow the power law \(\kappa_{1} / n^{2.5}\). It is evident that the fractional errors scale with the value of the primary resonance modification \(\kappa_{1}\).}
\label{fig:frequencyMMSmallXValidityPlot}
\end{figure}
\end{minipage}
\hspace*{\columnsep}
\begin{minipage}[t]{0.45\textwidth}
\begin{figure}[H]
\includegraphics[scale=1]{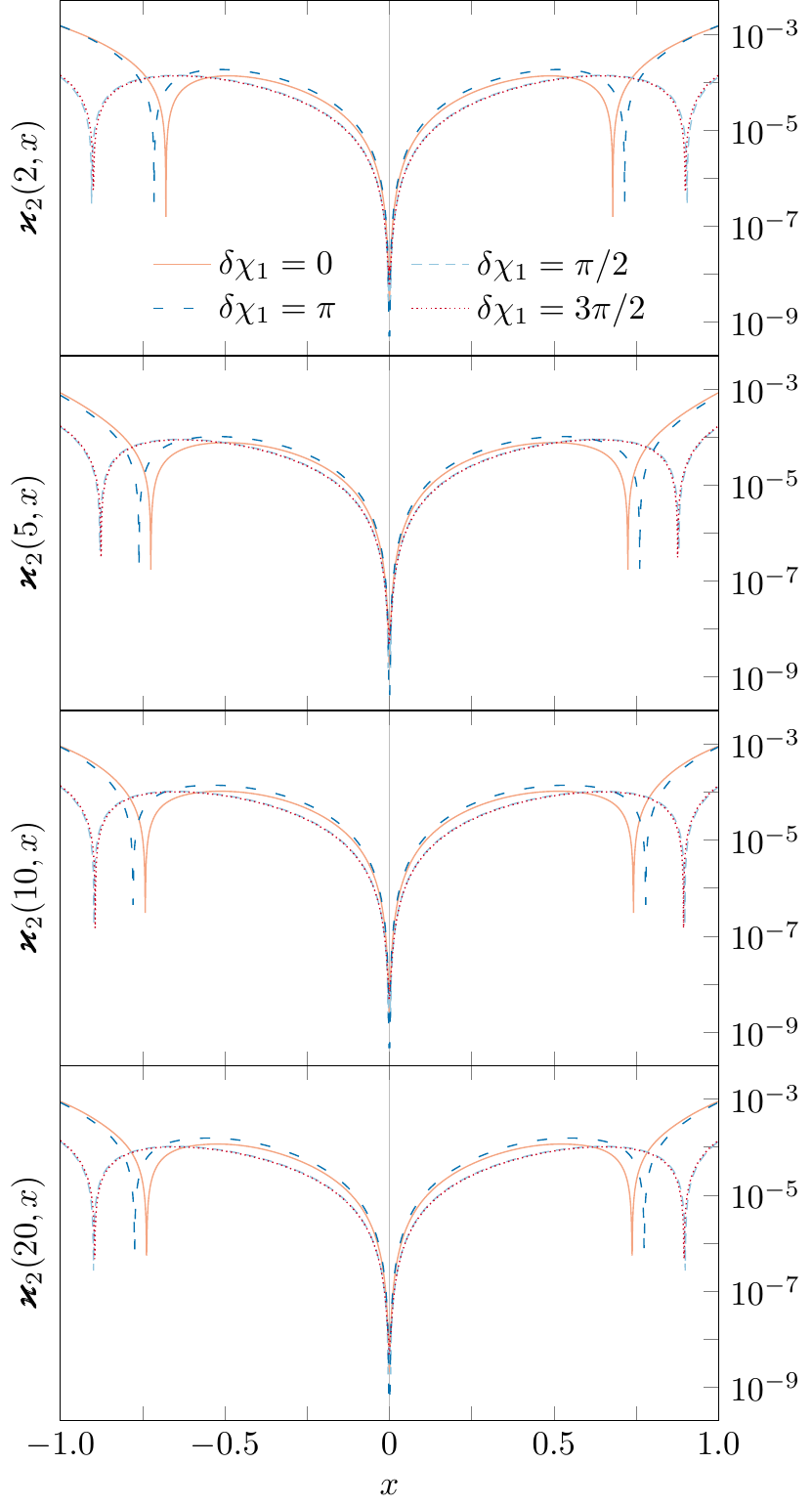}
\caption{Graphical demonstration of the validity of solution~(\ref{eq:frequencyMMPhaseSolutionSmallX}) for different number of oscillating terms and for values of \(\{\delchi_{n}\}\) given by eq.~(\ref{eq:initialPhaseLaw}). In all panels we have used \(\kappa_{1} = 0.1\) with consecutive values given by \(\kappa_{n} = \kappa_{1} / n^{2.5}\).}
\label{fig:frequencyMMPhaseSmallXValidityPlot}
\end{figure}
\end{minipage}
\end{minipage}

\subsection{Frequency resonance equation with two distinct modes}
\label{sec:freq_extension_double}
\noindent
If we investigate the same equation with two distinct modes on the right-hand side, say \(n = \{r, s\}\), with \(r < s\):
\begin{align}
\omega^{\prime} = 1 + \kappa_{r} \cos \left(r x \omega\right) + \kappa_{s} \cos \left(s x \omega\right),
\end{align}
then the solution would include two leading-order terms associated with each of the two separate oscillating terms, with corrections at sub-leading order due to cross-terms between these. Specifically, we find \({c_{1}}^{(r)} \sim \kappa_{r} / 2 r (1+b)\) and \({c_{1}}^{(s)} \sim \kappa_{s} / 2 s (1+b)\), which implies that the coefficient \(b\) is given by the expression
\begin{align}
b \sim -\frac{1}{2} \, r \, \kappa_{r} \, c_{1}^{(r)} -\frac{1}{2} \, s \, \kappa_{s} \, c_{1}^{(s)} \sim - \frac{\kappa_{r}^{2} + \kappa_{s}^{2}}{4} \, ,
\end{align}
and consequently\\
\begin{subequations}
\begin{minipage}{\textwidth}
\begin{minipage}{.496\textwidth}
\begin{align}
c_{1}^{(r)} \sim \frac{\kappa_{r}}{2r},
\end{align}
\end{minipage}
\begin{minipage}{.496\textwidth}
\begin{align}
c_{1}^{(s)} \sim \frac{\kappa_{s}}{2s},
\end{align}
\end{minipage}
\vspace{1em}
\end{minipage}
\end{subequations}
where each expression also includes sub-leading terms of order \(\bigo{\kappa_{r} \kappa_{s}}\). This proves the important result that for the frequency equation with 2 distinct modes, all mixed terms (which depend on both \(\kappa_{r}\) and \(\kappa_{s}\)) are of sub-leading order compared to the quadratic contributions originating from the individual modes as found earlier.

\subsection{Frequency equation with a non-vanishing phase term}
\label{sec:app_frequency_equation_one_mode_phase}
\noindent
Let us consider the frequency equation with a single oscillating mode but with a non-vanishing phase term \(\delchi\):
\begin{align}
\omega^{\prime} (x) = 1 + \kappa \cos\!\left(x \omega - \delchi\right),
\tag{\ref{eq:FrequencySMEq_phase}}
\end{align}
and let us expand the cosine term on the right-hand side to obtain the following differential equation:
\begin{align}
\omega^{\prime} (x) = 1 + \kappa \cos\!\left(\delchi\right) \cos\!\left(x \omega\right) + \kappa \sin\!\left(\delchi\right) \sin\!\left(x \omega\right).
\end{align}
We may naively try the ansatz which previously proved suitable for the zero-phase version of the equation, given by eq.~(\ref{eq:ansatz_zero}), however it will not work this time, as there will be unmatched terms left on the right-hand side. Instead, we suggest the following alternative ansatz, including an extra term proportional to \(C_{1} (x)\)
\begin{align}
\omega_{1} (x) = a + \left(1 + b\right) x + c_{1} \frac{S_{1} (x)}{x} + d_{1} \frac{C_{1} (x)}{x}.
\end{align}
Upon differentiating and substituting on both sides of eq.~(\ref{eq:FrequencySMEq_phase}), we obtain a more involved relation than before:
\begin{align}
b + 2 (1+b) c_{1} C_{1} (x) - 2 (1+b) d_{1} S_{1} (x) \sim -\frac{1}{2}\,\kappa\,c_{1} \cos\!\left(\delchi\right) - \frac{1}{2}\,\kappa\,d_{1} \sin\!\left(\delchi\right) + \kappa \cos\!\left(\delchi\right) C_{1} (x) - \kappa \sin\!\left(\delchi\right) S_{1} (x).
\end{align}
Making the informed assumption that \(b \sim \bigo{\kappa}\), we find the constants \(c_{1}\) and \(d_{1}\) by matching terms on both sides:
\begin{subequations}
\begin{minipage}{\textwidth}
\begin{minipage}{.496\textwidth}
\begin{align}
c_{1} \sim \frac{\kappa}{2} \cos\!\left(\delchi\right),
\end{align}
\end{minipage}
\begin{minipage}{.496\textwidth}
\begin{align}
d_{1} \sim - \frac{\kappa}{2} \sin\!\left(\delchi\right),
\end{align}
\end{minipage}
\vspace{1em}
\end{minipage}
\end{subequations}
which is different from the previously found value of \(c_{1}\). These yield the same value of the slope correction \(b\) as before:
\begin{align}
b \sim - \frac{1}{2}\,\kappa\,c_{1} \cos\!\left(\delchi\right) + \frac{1}{2}\,\kappa\,d_{1} \sin\!\left(\delchi\right) = - \frac{\kappa^{2}}{2}\left(\cos^{2}\!\left(\delchi\right) + \sin^{2}\!\left(\delchi\right)\right) = - \frac{\kappa^{2}}{2}.
\end{align}
If we substitute these into our proposed ansatz, we see that we recover the proposed alternative ansatz:
\begin{align}
\omega_{1} (x) = a + x + \frac{\kappa}{2}\,\frac{1}{x} \left(S_{1} (x) \cos\!\left(\delchi\right) - C_{1} (x) \sin\!\left(\delchi\right)\right) \equiv a + x + \frac{\kappa}{2}\,\frac{1}{x} \sin\!\left[\vphbig ax +\!\Bigg(\!1 - \frac{\kappa^{2}}{2}\!\Bigg)x^{2} - \delchi\right],
\end{align}
where the sine factor in the last term is identical to \(\mathcal{S}_{1} (x)\), as defined in eq.~(\ref{eq:frequencySnPhase}).

\section{Solutions to the phase resonance equation}
\label{sec:app_phase_solutions}
\noindent
In Section \ref{sec:phase_solution}, we explain our algorithm for finding iterative solutions to the phase resonance equation, and present the solutions to third order in the parameter \(\lambda\). However, we omit the details of finding these solutions, since these are often too tedious and irrelevant to our discussion. Here, we present these solutions in more detail. The zeroth-order equation (\ref{eq:phaseEqSMZeroPhReduced}) is straightforward to solve, by integrating both sides twice.

\subsection{Phase equation with a single mode and zero phase}
\label{sec:appendixPhaseSolutionSMZeroPh}

\begin{figure}[b]
\centering
\begin{minipage}[t]{0.43\textwidth}
	\includegraphics[scale=1]{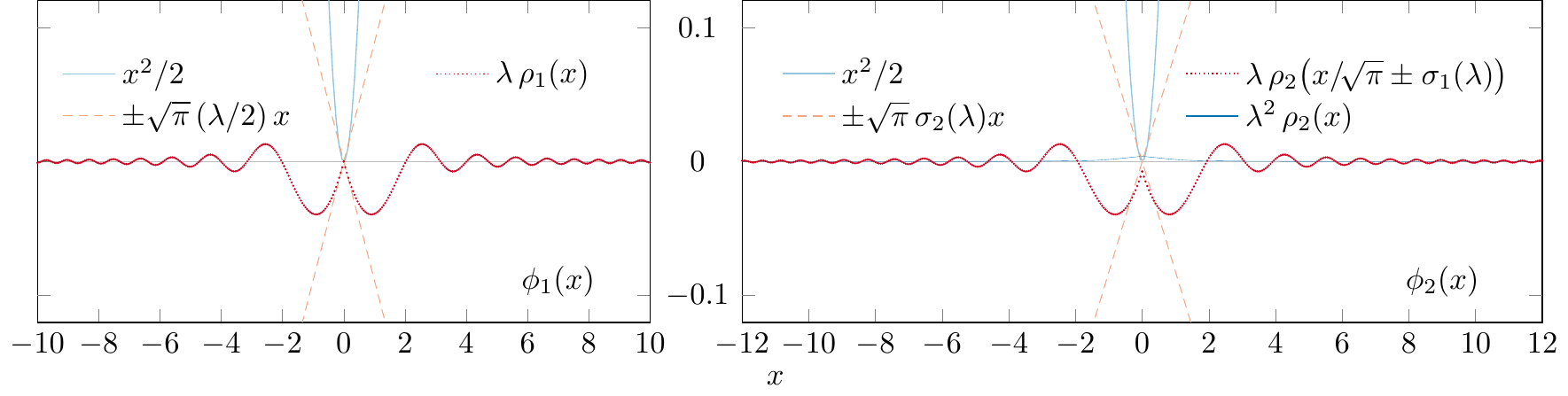}
	\caption{Graphical comparison of the terms in the argument on the right-hand side of eq.~(\ref{eq:appendix_second_order_initial}). Evidently the term \(\lambda \, \rho_{1} \) stays close to \(0\) for all positive \(x\).}
	\label{fig:perturbation_first_order}
\end{minipage}~\hspace{0.01\textwidth}~
\begin{minipage}[t]{0.559\textwidth}
	\vspace{4pt}
	\caption{Graphical comparison of the terms in the argument on the right-hand side of eq.~(\ref{eq:appendix_third_order_initial}). We can see that both terms, \(\lambda \, \rho_{1}\) and \(\lambda^{2} \rho_{2}\), are small compared to the other terms in the argument.}
	\label{fig:perturbation_second_order}
\end{minipage}
\end{figure}
\noindent
We source the first-order correction from the zeroth-order solution, and arrive at eq.~(\ref{eq:phase_single_mode_second_order}). Integrating both sides between \(0\) and \(x\), making use of the Fresnel integrals, with initial condition \(\phi^{\prime} (0) = 0\), we find:
\begin{align}
\phi_{(1)}^{\prime} (x) = \lambda \int_{0}^{x} \dd y \, \cos\!\left(\vph\!\frac{y^{2}}{2}\!\right) = \sqrt{\pi} \, \lambda \, C\!\left(\!\frac{x}{\sqrt{\pi}}\!\right).
\end{align}
Integrating once again, using the method of integration by parts, we obtain an initial form of the first-order correction:
\begin{align}
\phi_{(1)} (x) = \sqrt{\pi} \, \lambda\!\int_{0}^{x}\!\dd y \; C\!\left(\!\frac{y}{\sqrt{\pi}}\!\right)\!= \sqrt{\pi}\,\lambda \left[\vphbig x\,C\!\left(\!\frac{x}{\sqrt{\pi}}\!\right)\!- \frac{1}{\sqrt{\pi}} \,\sin\!\left(\vph\!\frac{x^{2}}{2}\!\right)\!\right].
\label{eq:phi_1_app_single_mode_first_order_raw}
\end{align}
While this is indeed a valid result, we prefer to re-arrange it in a form which is more useful for our tasks. If we look at the graph of this result for \(\lambda = 0.1\) (Fig.~\ref{fig:raw_corrections_comparison}), we can notice several features. First of all, since there are no constants dependant on \(\lambda\) in eq.~(\ref{eq:phi_1_app_single_mode_first_order_raw}), it vanishes at \(x = 0\). Furthermore, we immediately notice that the solution is a linear function which is modified by a decaying oscillating function. Therefore, we split the solution into two parts: a linear part, whose coefficient is a function of \(\lambda\), and another part, a collection of periodic and quasi-periodic functions (trigonometric functions and Fresnel integrals) that collectively oscillates around \(\phi = 0\), and whose magnitude is small compared to the linear term. Since the graph is symmetric relative to the vertical axis, it is necessary to write down solutions for both cases \(x > 0\) and \(x < 0\). The Fresnel integrals oscillate around the value \(\pm\sfrac{1}{2}\) (instead of 0) for large \(\pm\,x\), so we subtract this value from each of them in the rapidly oscillating terms. Finally, we note of an emerging pattern: the argument of the Fresnel integral is \(x / \spi\), therefore we express the linear term and the remaining factors of \(x\) in terms of this scaled variable. The upper sign is for positive \(x\), while the lower sign is for \(x < 0\).
\begin{align}
\phi_{(1)} (x) = \underbrace{\pm \, \pi \! \left[\vph\frac{1}{2} \, \lambda \right]\!\!\LinearTerm}_{\text{\textsc{linear term}}} + \underbrace{\lambda\!\left[\vphbig\pi\!\LinearTerm\!\!\FresnelCTerm\!- \SinTerm\!\right]}_{\text{\textsc{oscillating term proportional to \(\lambda\)}}}. \label{eq:appendix_phase_sol1}
\end{align}
We note that this result is strictly first order in the mode parameter \(\lambda\). To construct the complete solution of the phase resonance equation to first order in \(\lambda\), eq.~(\ref{eq:phase_solution_1}), we combine eq.~(\ref{eq:appendix_phase_sol1}) with the zeroth-order function (\ref{eq:PhaseSolutionSMZeroOrd}):
\begin{align}
\phi_{1} (x) = \frac{\pi}{2} \! \LinearTerm[][\SlopeFrstOrdSM]\np{2}\!+ \lambda\!\left[\vphbig\pi \! \LinearTerm\!\!\FresnelCTerm\!- \SinTerm\!\right]\!+ \bigo{\lambda^{2}}. \tag{\ref{eq:phase_solution_1}}
\end{align}
We have combined the quadratic and linear terms into a single quadratic term, by implicitly adding the constant term \((\pi \lambda^{2} / 8)\), which is nevertheless ignored at linear order in the mode parameter, thus obtaining the form (\ref{eq:phase_solution_1}):
\begin{align}
\frac{\pi}{2}\!\left(\!\frac{x}{\spi}\!\right)\np{2} \pm \frac{\pi}{2}\,\lambda\!\left(\!\frac{x}{\spi}\!\right)\!+ \bigo{\lambda^{2}} = \frac{\pi}{2}\!\left(\!\frac{x}{\spi} \pm \frac{1}{2}\,\lambda\!\right)\np{2}\!+ \bigo{\lambda^{2}}.
\end{align}
Finally, we make note of the functional form of the oscillating term, which will be investigated in detail later together with the functions incurred at higher order (let \(X = x /\!\spi\) and see Appendix~\ref{sec:app_extra_plots} for plots and discussion):
\begin{align}
\rho_{1} (X) = \pi \, X \bigg(\!C\!\left(X\right) \mp \frac{1}{2}\!\bigg)\!- \sin\!\bigg(\!\frac{\pi}{2} \, X^{2}\!\bigg)
\label{eq:eta_1_function}
\end{align}
The above result allows us to solve the equation at the next higher order, which is discussed below.

Finding the solution up to second order in \(\lambda\) involves substituting \(\phi_{1} (x) = \phi_{0} (x) + \phi_{(1)} (x)\) on the right-hand side of the source equation~(\ref{eq:phaseEquationSMZeroPh}), and solving for \(\phi_{(2)} (x)\). Explicitly, eq.~(\ref{eq:phase_eq_second_order_equation}) is (we use the variant (\ref{eq:phi_1_app_single_mode_first_order_raw}) of \(\phi_{(1)} (x)\)):
\begin{align}
\phi_{(2)}^{\prime\prime}(x) = \lambda \cos\!\left[\vphbigg\frac{x^{2}}{2} \pm \frac{\spi}{2} \, \lambda \, x + \lambda\!\left[\vphbig\spi \, x \FresnelCTerm\!- \sin\!\left(\frac{x^{2}}{2}\right)\!\right]\!\right]. \label{eq:appendix_second_order_initial}
\end{align}
We write the last term in the argument of the cosine as \(\lambda \, \rho_{1} \big(x/\!\spi\big)\) and invoke our knowledge of trigonometric identities to write the right-hand side as:
\begin{align}
\phi_{(2)}^{\prime\prime}(x) = \lambda \cos\!\left(\vph\!\frac{x^{2}}{2} \pm \frac{\spi}{2} \, \lambda \, x\!\right) \cos\!\left(\vph\!\lambda \, \rho_{1}\!\!\left(\!\frac{x}{\spi}\!\right)\!\!\right)\!- \lambda \sin\!\left(\vph\!\frac{x^{2}}{2} \pm \frac{\spi}{2} \, \lambda \, x\!\right) \sin\!\left(\vph\!\lambda \, \rho_{1}\!\!\left(\!\frac{x}{\spi}\!\right)\!\!\right)
\end{align}
The function \(\rho_{1} \big(x/\!\spi\big)\) oscillates around 0, and further, is pre-multiplied by the coefficient \(\lambda \ll 1\), hence we can assume that it is always small (compared to the quadratic and linear \(x\)-terms, see Figure~\ref{fig:perturbation_first_order} for comparison of the relative magnitude of the terms), and use this fact to substitute the trigonometric functions involving \(\rho_{1}\) for Taylor series to leading order in \(\lambda\). We ignore all terms of \(\bigo{\lambda^{3}}\) and higher, since they are not included at the current order.
\begin{align}
\phi_{(2)}^{\prime\prime}(x) = \lambda \cos\!\left(\vph\!\frac{x^{2}}{2} \pm \frac{\spi}{2} \, \lambda \, x\!\right)\!\left(1 + \bigo{\lambda^{2}}\right)\!- \lambda \sin\!\left(\vph\!\frac{x^{2}}{2} \pm \frac{\spi}{2} \, \lambda \, x\!\right)\!\left(\vph\!\lambda \, \rho_{1}\!\!\left(\!\frac{x}{\spi}\!\right)\!+ \bigo{\lambda^{3}}\!\right)
\end{align}
We carry out the same procedure with the term multiplied by \(\lambda^{2}\), and neglect resulting terms proportional to \(\lambda^{3}\):
\begin{align}
\lambda^{2} \sin\!\left(\vph\!\frac{x^{2}}{2} \pm \frac{\spi}{2} \, \lambda \, x\!\right)\rho_{1}\!\!\left(\!\frac{x}{\spi}\!\right)  = \lambda^{2} \sin\!\left(\vph\!\frac{x^{2}}{2}\!\right)\rho_{1}\!\!\left(\!\frac{x}{\spi}\!\right)\!+ \bigo{\lambda^{3}}.
\end{align}
Thus, we are left with the integral below, which can be easily performed using standard techniques, integration by parts and utilising the Fresnel integrals notation. We keep in mind to throw out terms of \(\bigo{\lambda^{3}}\) and higher.
\begin{align}
\phi_{(2)}^{\prime}(x) &= \bigintss_{0}^{x}\!\!\!\!\dd y \left[\vphbig\lambda \cos\!\left(\vph\!\frac{y^{2}}{2} \pm \frac{\spi}{2} \, \lambda \, y\!\right)\!- \spi \, \lambda^{2} \sin\!\left(\vph\!\frac{y^{2}}{2}\!\right) y \left(\vph\!C\!\left(\!\frac{y}{\spi}\!\right)\!\mp \frac{1}{2}\!\right)\!- \sin\!\left(\vph\!\frac{y^{2}}{2}\!\right)\right] = \\[5pt]
&= \sqrt{\pi} \, \lambda \Bigg[C\!\left(\!\frac{x}{\spi} \pm \frac{\lambda}{2}\!\right)\!\mp C\!\left(\frac{\lambda}{2}\right)\!\Bigg]\!+ \spi \, \lambda^{2} \!\left[\vphbig\cos\!\left(\vph\!\frac{x^{2}}{2}\!\right) C\!\left(\!\frac{x}{\spi}\!\right)\!- \frac{1}{\sqrt{2}} \, C\!\left(\!\frac{\sqrt{2}\,x}{\spi}\!\right)\!\mp \frac{1}{2} \cos\!\left(\vph\!\frac{x^{2}}{2}\!\right)\!\pm \frac{1}{2}\right].
\end{align}
We need to integrate this further once again, keeping track of the initial condition, \(\phi (0) = 0\). We obtain the following expression for the correction, where again the upper sign is valid for \(x > 0\) and the lower for \(x < 0\):
\begin{align}
\begin{split}
\phi_{(2)} (x) &= \spi \, \lambda \!\left[\vphbigg x \, C\!\left(\!\frac{x}{\spi} \pm \frac{\lambda}{2}\!\right) - \frac{1}{\spi} \, \SinTerm[][\SlopeFrstOrdSM]\!\mp C\!\left(\frac{\lambda}{2}\right)x\right] \\
&+ \spi \, \lambda^{2} \! \left[\vphbigg \frac{\spi}{2} \, C^{2}\!\left(\!\frac{x}{\spi}\!\right) \mp \frac{\spi}{2} \, C\!\left(\!\frac{x}{\spi}\!\right)\!\pm \frac{\spi}{2} \, C\!\left(\!\frac{x}{\spi} \pm \frac{\lambda}{2}\!\right)\!- \frac{1}{\sqrt{2}} \, x \, C\!\left(\!\frac{\sqrt{2}\,x}{\spi}\!\right) + \frac{1}{2 \spi} \, \sin \!\left(x^{2}\right) \pm \frac{x}{2}\right]
\end{split}
\end{align}
We have obtained an expression which gives the correction to the quadratic term to second order in \(\lambda\), which can be seen plotted in Figure~\ref{fig:raw_corrections_comparison}. Again, we seek to express this in a modified form where the linear term is explicit, and all the oscillating terms are grouped by order of the parameter \(\lambda\), which allows for better understanding of the solution.

\begin{figure*}[t]
\includegraphics[scale=1.0]{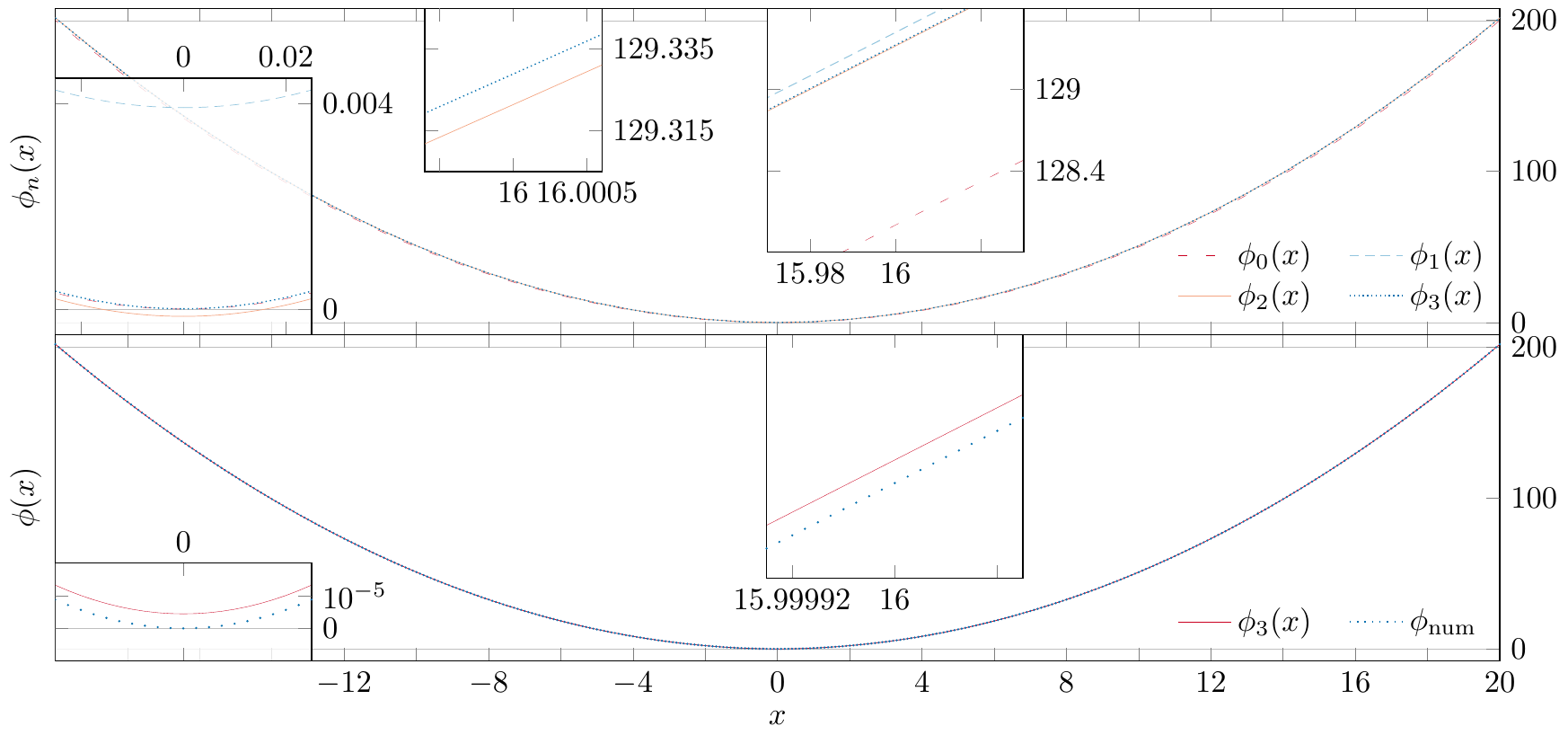}
\caption{\emph{Upper panel:} All four solutions (\ref{eq:PhaseSolutionSMZeroOrd}) -- (\ref{eq:phase_solution_3}) represented on the same plot. Evidently, they are all very close to each other, and cannot be easily distinguished in this manner. The three insets show features of the plot. At the origin, the solutions vanish up to a small deviation of magnitude comparable with the next respective order. Upon zooming in closer, one can distinguish the graphs from each other.  \emph{Lower panel:} Visual comparison of the highest-order solution, \(\phi_{3} (x)\), and the numerical solution of the phase resonance equation. In all calculations and simulations we have used value of \(\lambda = 0.1\).}
\label{fig:phase_solutions_single_mode}
\end{figure*}

\begin{ph}{align}
\begin{split}
\phi_{(2)} (x) &= \underbrace{\pm \sqrt{\pi} \left[\frac{\lambda}{2} - \lambda \, C\!\left(\frac{\lambda}{2}\right) + \frac{\sqrt{2} - 1}{2 \sqrt{2}} \, \lambda^{2}\right] x}_{\text{\textsc{linear term}}} \;\; + \underbrace{\vphantom{\Bigg(\Bigg)} \frac{\pi}{8} \, \lambda^{2}}_{\text{\textsc{constant}}} \\[5pt]
& \phantom{\quad\quad} + \underbrace{\lambda\!\left[\vphbigg \pi \! \LinearTerm[][\SlopeFrstOrdSM]\!\!\left(\vphbig\!C\!\LinearTerm[][\SlopeFrstOrdSM]\!\mp \frac{1}{2}\!\right)\!- \SinTerm[][\SlopeFrstOrdSM]\!\right]}_{\text{\textsc{oscillating term proportional to \(\lambda\)}}} \\[5pt]
& \phantom{\quad\quad\quad\quad} + \underbrace{\lambda^{2}\!\left[\vphbigg\frac{\pi}{2}\!\FresnelCTerm\np{2}\!- \frac{\pi}{2}\!\LinearTerm[2]\!\!\FresnelCTerm[2]\!+ \frac{1}{2}\,\SinTerm[2]\!\right]}_{\text{\textsc{oscillating term proportional to \(\lambda^2\)}}}.
\end{split}\label{eq:app_phase_solution_second_order}
\end{ph}
We note that some of these terms, if expanded, would be \(\bigo{\lambda^{3}}\) or higher, but we leave them in for presentational purposes. We simplify the coefficient of the linear term (the slope of the derivative \(\phi_{2}^{\;\prime} (x)\)), by writing the Fresnel integral in power series, and retaining only its linear term (the next higher-order term is of fifth order):
\begin{align}
\frac{\lambda}{2} - \lambda \, C\!\left(\frac{\lambda}{2}\right) + \frac{\sqrt{2} - 1}{2 \sqrt{2}} \, \lambda^{2} = \frac{\lambda}{2} - \frac{\lambda^{2}}{2} + \frac{\sqrt{2} - 1}{2 \sqrt{2}} \, \lambda^{2} + \bigo{\lambda^{6}} \approx \frac{1}{2} \, \lambda - \frac{\sqrt{2}}{4} \, \lambda^{2}.
\end{align}
When we combine this with the zeroth-order function \(\phi_{0} (x) = x^{2} / 2\), we find the complete second-order solution, given by eq.~(\ref{eq:phase_solution_2}). Specifically, we can complete the square with the quadratic, linear, and constant terms, to obtain
\begin{align}
\frac{x^2}{2} \pm \spi \left[\frac{1}{2} \, \lambda - \frac{\sqrt{2}}{4} \, \lambda^{2}\right] x + \frac{\pi}{8} \, \lambda^{2} = \frac{\pi}{2}\!\LinearTerm[][\pm \!\left[\frac{1}{2} \, \lambda - \frac{\sqrt{2}}{4} \, \lambda^{2}\right]]\np{2}\!+ \bigo{\lambda^{3}}.
\end{align}
In deriving this solution, we have used familiar integration techniques, and the definition of the Fresnel integrals, however it is useful to quote one of the ``tricks" for the benefit of the reader who wishes to replicate these results:
\begin{align}
\bigintsss_{\pm f(\lambda)}^{x / \!\spi \, \pm f(\lambda)} \!\!\!\!\!\!\!\! \dd y \cos\!\bigg(\!\frac{\pi}{2}\,y^{2}\!\bigg)\!= C\!\left(\!\frac{x}{\spi} \pm f(\lambda)\!\right)\!\mp C\big(f(\lambda)\big).
\end{align}
It is important to pay attention to the two oscillating terms in eq.~(\ref{eq:app_phase_solution_second_order}): the first one, has the same functional form as the oscillating term in eq.~(\ref{eq:appendix_phase_sol1}), and can in fact be related to \(\rho_{1} (X)\) in the following way:
\begin{align}
\text{\footnotesize{\textsc{oscillating term proportional to \(\lambda\)}}} = \rho_{1}\!\!\left(\!\frac{x}{\spi} \pm \frac{1}{2}\,\lambda\!\right).
\end{align}
The second oscillating term has not been encountered before, and will be investigated in detail in Appendix~\ref{sec:app_extra_plots}:
\begin{align}
\rho_{2} (X) = \frac{\pi}{2} \bigg(\!C\!\left(X\right) \mp \frac{1}{2}\bigg)\np{\,2}\!- \frac{\pi}{2}\!\left(\!\sqrt{2}\,X\right)\!\left(\!C\!\left(\sqrt{2}\,X\right)\!\mp \frac{1}{2}\right)\!+ \frac{1}{2} \,\sin\!\left[\frac{\pi}{2}\!\left(\!\sqrt{2}\,X\right)\np{\, 2}\right]. \label{eq:eta_2_function}
\end{align}
As we can see, the second-order solution \(\phi_{2} (x) = \phi_{0} (x) + \phi_{(2)} (x)\) is already heavily involved. Extending the results in \citep{gair2010}, we use this function to source the third-order solution, see below for details.

\begin{figure}[t]
\includegraphics[scale=1.0]{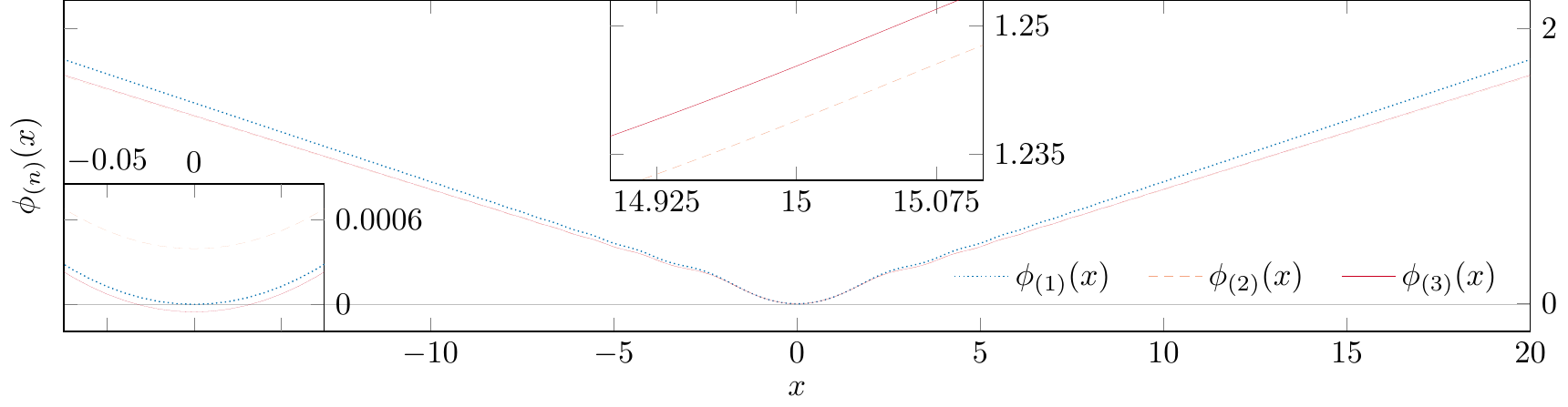}
\caption{Graphical comparison of the three corrections \(\phi_{(1)} (x)\), \(\phi_{(2)} (x)\), and \(\phi_{(3)} (x)\) for \(\lambda=0.1\). At the origin, \(\phi_{(1)} (0) = 0\) as it does not involve a constant term, while \(\phi_{(2)} (0)\) and \(\phi_{(3)} (0)\) do not vanish, but rather obey the condition to \(\bigo{\lambda^{3}}\) and \(\bigo{\lambda^{4}}\), respectively, which is within their errors: \(|\phi_{(2)} (0)| < \lambda^{3}\) and \(|\phi_{(3)} (0)| < \lambda^{4}\).}
\label{fig:raw_corrections_comparison}
\end{figure}

Substituting the expression (\ref{eq:phase_solution_3}) for \(\phi_{2} (x)\) into eq.~(\ref{eq:phase_resonance_equation_third_order}) gives the equation we need to solve at third order:
\begin{align}
\phi_{(3)}^{\; \prime\prime} (x) &= \lambda\,\cos\!\left[\vphbig\frac{x^{2}}{2} \pm \spi \left(\!\frac{1}{2} \, \lambda - \frac{\sqrt{2}}{4} \, \lambda^{2}\!\right)\!x + \frac{\pi}{8} \, \lambda^{2} + \lambda\,\rho_{1}\!\!\left(\!\frac{x}{\spi} \pm \frac{1}{2}\,\lambda\!\right) + \lambda^{2}\,\rho_{2}\!\!\left(\!\frac{x}{\spi}\!\right)\!\right], \label{eq:appendix_third_order_initial} \\
\text{where} & \quad \rho_{1}\!\!\left(\!\frac{x}{\spi} \pm \frac{1}{2}\,\lambda\!\right) = \pi \! \LinearTerm[][\SlopeFrstOrdSM]\!\!\left(\vphbig\!C\!\LinearTerm[][\SlopeFrstOrdSM]\!\mp \frac{1}{2}\!\right)\!- \SinTerm[][\SlopeFrstOrdSM] \\
\text{and} & \quad \rho_{2}\!\!\left(\!\frac{x}{\spi}\!\right) = \frac{\pi}{2}\!\FresnelCTerm\np{2}\!- \frac{\pi}{2}\!\LinearTerm[2]\!\!\FresnelCTerm[2]\!+ \frac{1}{2}\,\SinTerm[2].
\end{align}
It is important to realise that \(\lambda \, \rho_{1} (\bullet)\) and \(\lambda^{2} \rho_{2} (\bullet)\) are small in magnitude compared to the other terms, as is obvious when these are plotted on the same graph, see Figure~\ref{fig:perturbation_second_order}. Hence we can treat these as a perturbation to the rest of the argument. This becomes obvious when the functions are plotted on the same graph, see Figure~\ref{fig:perturbation_second_order}. This prompts us to expand the cosine term using the sum-to-product formulae, and expand the terms containing the perturbation:
\begin{align}
& \lambda\,\cos\!\left[\vphbig\frac{x^{2}}{2} \pm \spi \left(\!\frac{1}{2} \, \lambda - \frac{\sqrt{2}}{4} \, \lambda^{2}\!\right)\!x + \frac{\pi}{8} \, \lambda^{2} + \lambda\,\rho_{1}\!\!\left(\!\frac{x}{\spi} \pm \frac{1}{2}\,\lambda\!\right) + \lambda^{2}\,\rho_{2}\!\!\left(\!\frac{x}{\spi}\!\right)\!\right] = \\[5pt]
\begin{split}
&= \lambda\,\cos\!\left[\vphbig\frac{x^{2}}{2} \pm \spi \left(\!\frac{1}{2} \, \lambda - \frac{\sqrt{2}}{4} \, \lambda^{2}\!\right)\!x + \frac{\pi}{8} \, \lambda^{2}\right] \! \times \cos\!\left(\vph\!\lambda\,\rho_{1}\!\!\left(\!\frac{x}{\spi} \pm \frac{1}{2}\,\lambda\!\right) + \lambda^{2}\,\rho_{2}\!\!\left(\!\frac{x}{\spi}\!\right)\!\right) \\
& \RepQuad{10} - \lambda \, \sin\!\left[\vphbig\frac{x^{2}}{2} \pm \spi \left(\!\frac{1}{2} \, \lambda - \frac{\sqrt{2}}{4} \, \lambda^{2}\!\right)\!x + \frac{\pi}{8} \, \lambda^{2}\right] \! \times \sin\!\left(\vph\!\lambda\,\rho_{1}\!\!\left(\!\frac{x}{\spi} \pm \frac{1}{2}\,\lambda\!\right) + \lambda^{2}\,\rho_{2}\!\!\left(\!\frac{x}{\spi}\!\right)\!\right) =
\end{split}\\[5pt]
\begin{split}
&= \lambda\,\cos\!\left[\vphbig\frac{x^{2}}{2} \pm \spi \left(\!\frac{1}{2} \, \lambda - \frac{\sqrt{2}}{4} \, \lambda^{2}\!\right)\!x + \frac{\pi}{8} \, \lambda^{2}\right]\!- \lambda^{2} \sin\!\left(\vph\!\frac{x^{2}}{2} \pm \frac{\spi}{2} \, \lambda \, x\!\right) \rho_{1}\!\!\left(\!\frac{x}{\spi} \pm \frac{1}{2}\,\lambda\!\right) \\
& \RepQuad{10} - \frac{1}{2}\,\lambda^{3}\left(\vph\!\cos\!\left(\vph\!\frac{x^{2}}{2}\!\right) \rho_{1}^{2}\!\!\left(\!\frac{x}{\spi} \pm \frac{1}{2}\,\lambda\!\right)  + 2 \, \sin\!\left(\vph\!\frac{x^{2}}{2}\!\right) \rho_{2}\!\!\left(\!\frac{x}{\spi}\!\right)\!\right)\! + \bigo{\lambda^{4}}.
\end{split} \label{eq:app_third_order_expression}
\end{align}
We have combined both third-order terms together, since it makes the integration more straightforward. Integrating the expression (\ref{eq:app_third_order_expression}), subject to the boundary condition \(\phi^{\prime} (0) = 0\) gives the derivative of the third-order correction:
\begin{align}
\phi_{(3)}^{\prime} (x) &= \spi\,\lambda\!\left[\vphbigg C\!\left(\vphbig\!\frac{x}{\spi} \pm\! \left[\frac{1}{2} \, \lambda - \frac{\sqrt{2}}{4} \, \lambda^{2}\right]\!\right)\!\mp C\!\left(\!\frac{1}{2} \, \lambda - \frac{\sqrt{2}}{4} \, \lambda^{2}\!\right)\!\right] \nonumber\\[5pt]
&+ \spi \, \lambda^{2}\left[\vphbig \cos\!\left(\vph\!\frac{x^{2}}{2} \pm \frac{\spi}{2}\,\lambda \, x\!\right) C\!\left(\!\frac{x}{\spi} \pm \frac{\lambda}{2}\!\right)\!\mp \frac{1}{2}\,\cos\!\left(\vph\!\frac{x^{2}}{2} \pm \frac{\spi}{2}\,\lambda \, x\!\right) \right. \nonumber\\
& \RepQuad{8} \left. \vphbig - \frac{1}{\sqrt{2}} \, C\!\left(\!\frac{\sqrt{2}\,x}{\spi} \pm \frac{\lambda}{\sqrt{2}}\!\right)\!\pm \frac{1}{\sqrt{2}} \, C\!\left(\!\frac{\lambda}{\sqrt{2}}\!\right)\!\mp C\!\left(\frac{\lambda}{2}\right)\!\pm \frac{1}{2}\right] \\[5pt]
&- \spi \, \lambda^{3} \left[\vphbigg\frac{\spi}{2} \, x \, \sin\!\left(\vph\!\frac{x^{2}}{2}\!\right)\!\left[\vph C^{\,2}\!\left(\!\frac{x}{\spi}\!\right)\!\mp C\!\left(\!\frac{x}{\spi}\!\right)\!+ \frac{1}{4}\right]\!- \frac{3}{8} \, C\!\left(\!\frac{x}{\spi}\!\right)\!- \frac{3\sqrt{3}}{8} \, C\!\left(\!\frac{\sqrt{3}\,x}{\spi}\!\right) \right. \nonumber\\
& \RepQuad{8} \left. \vphbigg +\,\frac{1}{2} \cos\!\left(x^{2}\right)\!\left[\vph C\!\left(\!\frac{x}{\spi}\!\right)\!\mp \frac{1}{2}\right]\!+ \frac{1}{\sqrt{2}}\cos\!\left(\vph\!\frac{x^{2}}{2}\!\right)\!\left[\vphbig C\!\left(\!\frac{\sqrt{2}\,x}{\spi}\!\right)\!\mp \frac{1}{2}\right]\!\pm \frac{\sqrt{2}}{4} \pm \frac{1}{4}\right]. \nonumber
\end{align}
Integrating this again, using a combination of integration by parts, substitution, double-angle formulae and other trigonometric identities, finally applying the current boundary condition \(\phi(0) = 0\), we arrive at an expression for the third-order correction to the solution (\ref{eq:PhaseSolutionSMZeroOrd}). Below is given an initial form of the solution, and as before the upper sign holds for positive \(x\), while the lower sign is valid for negative \(x\) (see Figure~\ref{fig:raw_corrections_comparison} for a plot of this function):
\begin{align}
\phi_{(3)} (x) &= \spi \,\lambda\!\left[\vphbigggg x \, C\!\left(\vphbig\!\frac{x}{\spi} \pm\! \left[\frac{1}{2} \, \lambda - \frac{\sqrt{2}}{4} \, \lambda^{2}\right]\!\right)\!- \frac{1}{\spi}\,\sin\!\left[\frac{\pi}{2}\!\left(\vphbig\!\frac{x}{\spi} \pm\! \left[\frac{1}{2} \, \lambda - \frac{\sqrt{2}}{4} \, \lambda^{2}\right]\!\right)\np{2}\right]\!+ \frac{\spi}{2} \!\left(\!\frac{1}{2} \, \lambda - \frac{\sqrt{2}}{4} \, \lambda^{2}\!\right)\np{2} \right. \nonumber\\
& \RepQuad{2} \left. \vphbigggg \pm \spi \left(\!\frac{1}{2} \, \lambda - \frac{\sqrt{2}}{4} \, \lambda^{2}\!\right) C\!\left(\vphbig\!\frac{x}{\spi} \pm\! \left[\frac{1}{2} \, \lambda - \frac{\sqrt{2}}{4} \, \lambda^{2}\right]\!\right)\!- \frac{\spi}{2} \, \lambda \, C\!\left(\!\frac{1}{2} \, \lambda - \frac{\sqrt{2}}{4} \, \lambda^{2}\!\right) \mp C\!\left(\!\frac{1}{2} \, \lambda - \frac{\sqrt{2}}{4} \, \lambda^{2}\!\right) x \right] \nonumber\\[5pt]
\begin{split}
& + \spi \, \lambda^{2} \!\left[\frac{\spi}{2} \, C^{2}\!\left(\!\frac{x}{\sqrt{\pi}} \pm \frac{1}{2}\,\lambda\!\right)\!\mp \frac{\spi}{2} \, C\!\left(\!\frac{x}{\sqrt{\pi}} \pm \frac{1}{2}\,\lambda\!\right)\!+ \frac{1}{2 \spi}\,\sin\!\left[\frac{\pi}{2}\!\left(\vph\!\sqrt{2}\!\left[\frac{x}{\sqrt{\pi}} \pm \frac{1}{2}\,\lambda\right]\!\right)\np{2}\right]\!\mp C\!\left(\frac{\lambda}{2}\right) x \vphbigg \right. \\
& \RepQuad{6} \left. \vphbigg\;\;\; - \frac{1}{\sqrt{2}} \, x \, C\!\left(\vph\!\sqrt{2}\left[\frac{x}{\sqrt{\pi}} \pm \frac{1}{2}\,\lambda\right]\!\right)\!\mp \frac{\spi}{2 \sqrt{2}} \, \lambda \, C\!\left(\!\frac{\sqrt{2} \, x}{\sqrt{\pi}}\!\right)\!\pm \frac{1}{2} \, x + \frac{\spi}{2} \, C\!\left(\frac{\lambda}{2}\right)\!\pm \frac{1}{\sqrt{2}}\,C\!\left(\!\frac{\lambda}{\sqrt{2}}\!\right) x\right]
\end{split} \\[5pt]
& + \spi \, \lambda^{3}\! \left[\frac{\spi}{2} \cos\!\left(\vph\!\frac{x^{2}}{2}\!\right)\!\left[\vph C^{\,2}\!\left(\!\frac{x}{\spi}\!\right)\!\mp C\!\left(\!\frac{x}{\spi}\!\right)\!+ \frac{1}{4}\right]\!- \frac{\spi}{\sqrt{2}} \, C\!\left(\!\frac{\sqrt{2} x}{\spi}\!\right) C\!\left(\!\frac{x}{\spi}\!\right)\!+ \frac{3 \sqrt{3}}{8} \, x \, C\!\left(\!\frac{\sqrt{3} x}{\spi}\!\right) \vphbigg \right. \nonumber\\
& \RepQuad{2} \left. \vphbigg - \frac{x}{8} \, C\!\left(\!\frac{x}{\spi}\!\right)\!\pm \frac{\spi}{2\sqrt{2}} \, C\!\left(\!\frac{\sqrt{2} \, x}{\spi}\!\right)\!\pm \frac{\spi}{2\sqrt{2}}\,C\!\left(\frac{x}{\spi}\right)\!- \frac{3}{8 \spi}\, \sin\!\left(\vph\!\frac{3x^{2}}{2}\!\right)\!+ \frac{1}{8 \spi}\, \sin\!\left(\vph\!\frac{x^{2}}{2}\!\right)\!\mp \frac{x}{2\sqrt{2}} - \frac{\spi}{8}\right]. \nonumber
\end{align}
\begin{figure}[b]
\includegraphics[scale=1.0]{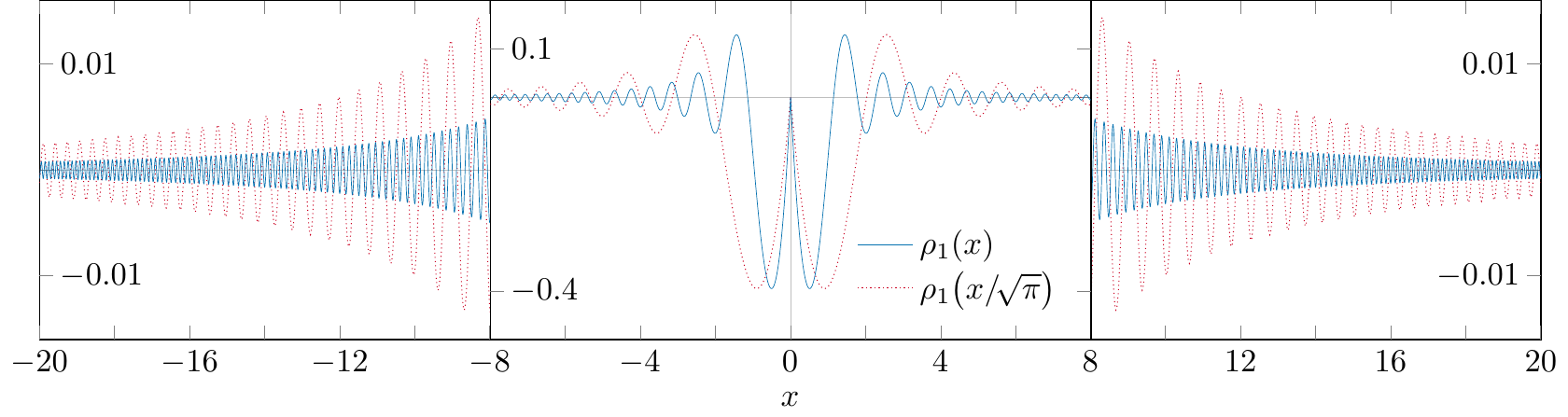}
\caption{Plots of the two functions \(\rho_{(1)} (x)\) and \(\rho_{1} \big(x/\!\spi\big)\). We see that both branches (for positive and negative \(x\)) are comprised of oscillations which peter out as we increase distance from the origin.}
\label{fig:eta_plots1}
\end{figure}
As before, we re-arrange this form of the solution to obtain groups of terms which oscillate around \(0\) (both for negative and positive \(x\)) and beside them, terms which are either linear or constant with \(x\), but are otherwise functions of \(\lambda\).
\begin{align}
&\phi_{(3)} (x) = \phi_{3} (x) - \phi_{0} (x) = \nonumber \\
& = \underbrace{\pm \spi \left[\frac{\lambda}{2} - \lambda \, C\!\left(\!\frac{\lambda}{2} - \frac{\lambda^{2}}{2\sqrt{2}}\!\right)\!+ \frac{\lambda^{2}}{2} - \lambda^{2} \, C\!\left(\frac{\lambda}{2}\right)\!- \frac{\lambda^{2}}{2\sqrt{2}} + \frac{\lambda^{2}}{\sqrt{2}} \, C\!\left(\!\frac{\lambda}{\sqrt{2}}\!\right)\!- \frac{\lambda^{3}}{16} - \frac{\lambda^{3}}{2\sqrt{2}} + \frac{3\sqrt{3}}{16}\,\lambda^{3}\right]\!x}_{\text{\textsc{linear term}}} \nonumber\\[5pt]
&+ \underbrace{\lambda\,\sin\!\left[\frac{\pi}{2} \! \left(\!\frac{\lambda}{2} - \frac{\lambda^{2}}{2\sqrt{2}}\!\right)\np{2}\right]\!+ \frac{\pi}{2}\,\lambda \left(\!\frac{\lambda}{2} - \frac{\lambda^{2}}{2\sqrt{2}}\!\right)\!- \pi\,\lambda\,\frac{\lambda}{2}\,C\!\left(\frac{\lambda}{2}\right)\!- \frac{\pi}{8}\, \lambda^{2} + \frac{\pi}{2} \, \lambda^{2} \, C\!\left(\frac{\lambda}{2}\right) - \frac{\pi}{4\sqrt{2}}\,\lambda^{3} + \frac{\pi}{4\sqrt{2}}\,\lambda^{3} - \frac{\pi}{8}\,\lambda^{3}}_{\text{\textsc{constant term}}} \nonumber\\[5pt]
&+ \underbrace{\lambda \! \left[\vphbigggg \pi \! \LinearTerm[][\SlopeScndOrdSM]\!\!\left(\vphbigg\!C\!\LinearTerm[][\SlopeScndOrdSM]\!\mp \frac{1}{2}\!\right)\!- \SinTerm[][\SlopeScndOrdSM]\!\right]}_{\text{\textsc{oscillating term proportional to \(\lambda\)}}} \nonumber\\[5pt]
&+ \underbrace{\lambda^{2} \!\! \left[\vphbigg\frac{\pi}{2}\!\left(\vphbig\!C\!\LinearTerm[][\SlopeFrstOrdSM]\!\mp \frac{1}{2}\!\right)\np{2}\!- \frac{\pi}{2}\!\LinearTerm[2][\SlopeFrstOrdSM]\!\!\FresnelCTerm[2][\SlopeFrstOrdSM]\!+ \frac{1}{2}\,\SinTerm[2][\SlopeFrstOrdSM]\!\right]}_{\text{\textsc{oscillating term proportional to \(\lambda^2\)}}} \nonumber\\[5pt]
&+ \lambda^{3} \!\!\left[\frac{\pi}{2} \CosTerm\!\!\FresnelCTerm\np{2}\!- \frac{\pi}{\sqrt{2}}\!\FresnelCTerm[2]\!\!\FresnelCTerm\!-\frac{\pi}{8}\!\LinearTerm\!\!\FresnelCTerm \right. \nonumber\\
& \RepQuad{1} \underbrace{\RepQuad{3} \left. + \frac{3\pi}{8}\!\LinearTerm[3]\!\!\FresnelCTerm[3]\!+ \frac{1}{8}\,\SinTerm - \frac{3}{8}\,\SinTerm[3]\!\right] + \bigo{\lambda^{4}}\quad\quad}_{\text{\textsc{oscillating term proportional to \(\lambda^{3}\)}}}
\label{eq:app_third_order_final_expression}
\end{align}
A useful trick that we use in evaluating these integrals is the following: if \(f \sim \bigo{\lambda}\) is a polynomial function of the mode parameter, we can establish which terms are relevant at each order of the solution:
\begin{align}
\begin{split}
C\!\left(\!\frac{x}{\spi} \pm f(\lambda)\!\right) &= \int_{0}^{x/\!\spi} \!\!\!\! \dd y \, \cos\!\left(\frac{\pi}{2}\,y^{2}\right) + \int_{x/\!\spi}^{x/\!\spi \, \pm f(\lambda)} \!\!\!\! \dd y \, \cos\!\left(\frac{\pi}{2}\,y^{2}\right) = \\
&= C\!\left(\!\frac{x}{\spi}\!\right)\!\pm f(\lambda) \times \CosTerm = C\!\left(\!\frac{x}{\spi}\!\right)\!+ \bigo{\lambda}.
\end{split}
\end{align}
The linear and constant terms can be simplified by expanding the sine function and the Fresnel integral (see eq.~(\ref{eq:fresnelc_expansion})):
\begin{align}
&\pm\spi \left[\vphbig\frac{\lambda}{2} - \lambda \, C\!\left(\vph\!\frac{\lambda}{2} - \frac{\lambda^{2}}{2\sqrt{2}}\!\right)\!+ \frac{\lambda^{2}}{2} - \lambda^{2} \, C\!\left(\frac{\lambda}{2}\right)\!- \frac{\lambda^{2}}{2\sqrt{2}} + \frac{\lambda^{2}}{\sqrt{2}} \, C\!\left(\!\frac{\lambda}{\sqrt{2}}\!\right)\!- \frac{\lambda^{3}}{16} - \frac{\lambda^{3}}{2\sqrt{2}} + \frac{3\sqrt{3}}{16}\,\lambda^{3}\right]\!x \nonumber \\
& + \lambda\,\sin\!\left[\frac{\pi}{2}\!\left(\vph\!\frac{\lambda}{2} - \frac{\lambda^{2}}{2\sqrt{2}}\right)\np{2}\!\right]\!+ \frac{\pi}{2}\,\lambda\left(\vph\!\frac{\lambda}{2} - \frac{\lambda^{2}}{2\sqrt{2}}\right)\!- \pi\,\lambda\,\frac{\lambda}{2}\,C\!\left(\frac{\lambda}{2}\right)\!- \frac{\pi}{8}\, \lambda^{2} + \frac{\pi}{2} \, \lambda^{2} \, C\!\left(\frac{\lambda}{2}\right)\!- \frac{\pi}{4\sqrt{2}}\,\lambda^{3} + \frac{\pi}{4\sqrt{2}}\,\lambda^{3} - \frac{\pi}{8}\,\lambda^{3} = \nonumber \\
&= \pm\,\pi \!\left[\frac{1}{2}\,\lambda - \frac{\sqrt{2}}{4}\,\lambda^{2} + \frac{3\sqrt{3}-1}{16}\,\lambda^{3}\right]\!\!\LinearTerm + \pi \!\left[\frac{1}{8}\,\lambda^{2} - \frac{\sqrt{2}}{8}\,\lambda^{3}\right] + \bigo{\lambda^{4}}
\end{align}
Combining eq.~(\ref{eq:app_third_order_final_expression}) with \(\phi_{(0)} (x) = x^{2} / 2\), we obtain the complete solution to third order in \(\lambda\), eq. (\ref{eq:phase_solution_3}). In particular, we have combined the quadratic, linear, and constant terms into a single quadratic term:
\begin{align}
\frac{x^{2}}{2} \pm\,\pi \!\left[\frac{1}{2}\,\lambda - \frac{\sqrt{2}}{4}\,\lambda^{2} + \frac{3\sqrt{3}-1}{16}\,\lambda^{3}\right]\!\!\LinearTerm\!+ \pi \Bigg[\frac{\lambda^{2}}{8} - \frac{\lambda^{3}}{4\sqrt{2}}\Bigg]\!= \frac{\pi}{2}\!\LinearTerm[][\pm\!\left[\frac{1}{2}\,\lambda - \frac{\sqrt{2}}{4}\,\lambda^{2} + \frac{3\sqrt{3}-1}{16}\,\lambda^{3}\right]]\np{2}\!+ \bigo{\lambda^{4}}.
\end{align}
\begin{figure}[b]
\includegraphics[scale=1.0]{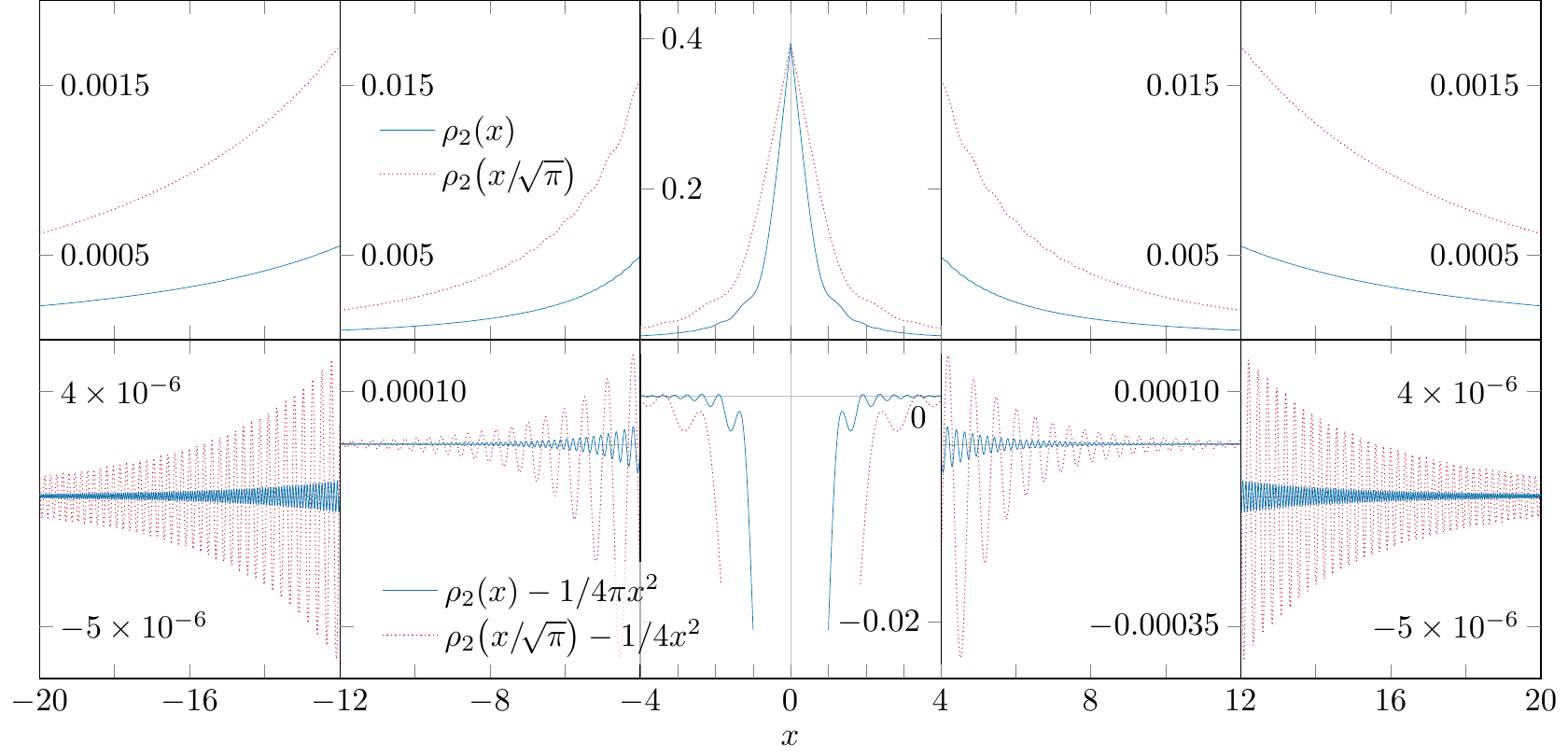}
\caption{\emph{Upper pannel}: Graphical representation of the two functions \(\rho_{2} (x)\) and \(\rho_{2} \big(x/\!\spi\big)\). \emph{Lower panel}: The same functions, however here the highest-order guide function is removed to reveal the oscillations. It is evident from the deviation from \(y = 0\), especially around the origin, that there are lower-order contributions to the guide function. From both panels, however, we see that the function \(\rho_{2} (X)\) oscillates more rapidly and decreases in magnitude away from the origin.}
\label{fig:eta_plots2}
\end{figure}
Finally, we pay attention to the form of the oscillating terms. If we compare eq.~(\ref{eq:app_third_order_final_expression}) to eqs.~(\ref{eq:appendix_phase_sol1}) and (\ref{eq:app_phase_solution_second_order}), we can immediately notice how the oscillating terms proportional to \(\lambda\) and \(\lambda^{2}\) are related to previous results:\\%
\begin{subequations}
\begin{minipage}{\textwidth}
\begin{minipage}{.545\textwidth}
\begin{align}
\parbox{8em}{\footnotesize{\textsc{oscillating term \\ proportional to \(\lambda\)}}} \!\!= \rho_{1}\!\left(\vphbig\!\frac{x}{\sqrt{\pi}} \pm \! \left[\frac{1}{2} \, \lambda - \frac{\sqrt{2}}{4}\,\lambda^{2}\right]\!\right),
\end{align}
\end{minipage}
\begin{minipage}{.45\textwidth}
\begin{align}
\parbox{8em}{\footnotesize{\textsc{oscillating term \\ proportional to \(\lambda^{2}\)}}} \!= \rho_{2}\!\left(\vph\!\frac{x}{\sqrt{\pi}} \pm \frac{1}{2}\,\lambda\!\right).
\end{align}
\end{minipage}
\end{minipage}
\end{subequations}
We are in a position to establish the functional form of the third-order oscillating term, \(\rho_{(3)} (X)\):
\begin{align}
\begin{split}
\rho_{3} (X) &= \frac{\pi}{2} \cos\!\bigg[\frac{\pi}{2}\,X^{2}\bigg]\!\left(\!C\!\left(X\right) \mp \frac{1}{2}\!\right)\np{2}\!- \frac{\pi}{\sqrt{2}}\!\left(\!C\!\left(\sqrt{2}\,X\right) \mp \frac{1}{2}\!\right)\!\left(\!C\!\left(X\right) \mp \frac{1}{2}\!\right)\!- \frac{\pi}{8}\,X \!\left(\!C\!\left(X\right) \mp \frac{1}{2}\!\right) \\[1pt]
& + \frac{3\pi}{8}\!\left(\!\sqrt{3}\,X\!\right)\!\left(\!C\!\left(\!\sqrt{3}\,X\right) \mp \frac{1}{2}\!\right)\!+ \frac{1}{8} \, \sin\!\bigg[\frac{\pi}{2}\,X^{2}\bigg]\!- \frac{3}{8} \, \sin\!\left[\frac{\pi}{2}\!\left(\!\sqrt{3}\,X\right)^{\!2}\right].
\end{split} \label{eq:eta_3_function}
\end{align}

\subsection{Supplementary plots and remarks}
\label{sec:app_extra_plots}
\noindent
The corrections have been plotted in Figure~\ref{fig:raw_corrections_comparison} for value of \(\lambda = 0.1\). Interesting is their behaviour around the origin: even though we have imposed the condition \(\phi(0) = 0\) at each order of the solution, it is important to note that they have corrections of the next-higher order of \(\lambda\): \(\phi_{(1)} (x)\) is accurate up to \(\bigo{\lambda^{2}}\), \(\phi_{(2)} (x)\) to \(\bigo{\lambda^{3}}\), and so on. Therefore, at the origin \(x = 0\), \(\phi_{(1)} (x)\) satisfies the initial condition and goes through (0,0), since it does not contain any constant terms in \(\lambda\) (cf. eq.~(\ref{eq:appendix_phase_sol1})). The next-order correction, \(\phi_{(2)} (x)\), however, does not go through the origin: we can see this both from eq.~(\ref{eq:app_phase_solution_second_order}) and from Figure~\ref{fig:raw_corrections_comparison}. Note that this is not only due to the constant term \(\pi \lambda^{2} / 8\), but also because the oscillating term proportional to \(\lambda\) in eq.~(\ref{eq:app_phase_solution_second_order}) does not go through the origin. Similarly, \(|\phi_{(3)} (0)| \sim \bigo{\lambda^{4}}\), and the third-order correction is also valid.

\begin{figure}[b]
\includegraphics[scale=1.0]{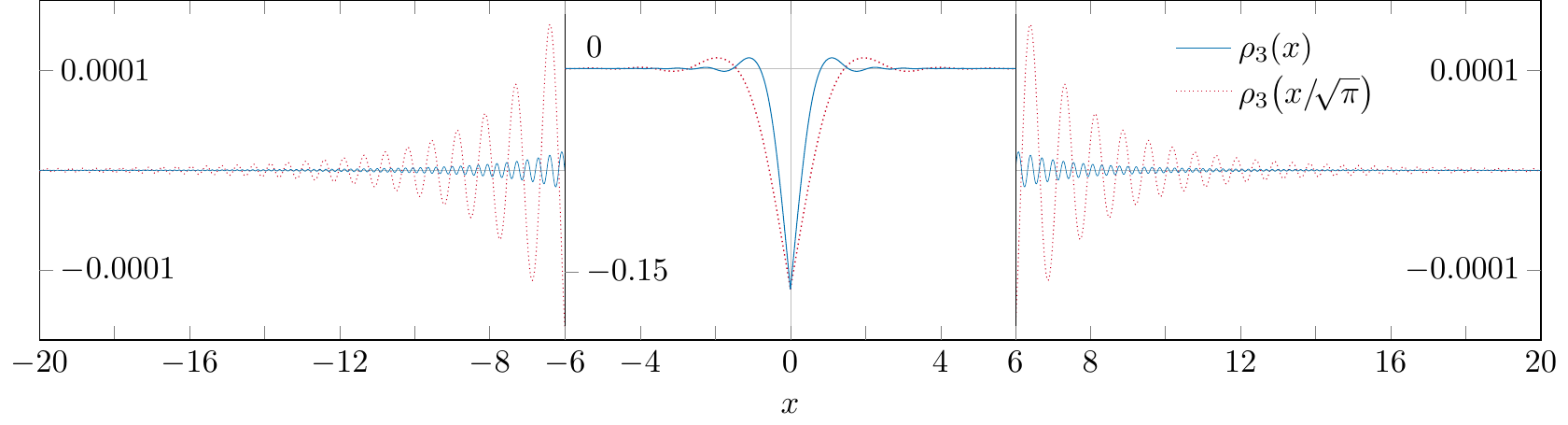}
\caption{Plots of the two functions \(\rho_{3} (x)\) and \(\rho_{3} \big(x/\!\spi\big)\). We see that both branches (for positive and negative \(x\)) are comprised of oscillations which peter out as the distance from the origin increases.}
\label{fig:eta_plots3}
\end{figure}

As we can see, certain function emerge in the arguments of the phase functions: the coefficients of the linear \(x\) term in each of the phase corrections can be written as \(\pm \spi \, \sigma_{\{1,2,3\}} (\lambda)\), where the functions are given by:
\begin{minipage}{\textwidth}
\begin{minipage}{.24\textwidth}
\begin{align}
\sigma_{1} (\lambda) = \frac{1}{2}\,\lambda,\\[-6pt]\nonumber
\end{align}
\end{minipage}
\begin{minipage}{.32\textwidth}
\begin{align}
\sigma_{2} (\lambda) = \frac{1}{2}\,\lambda - \frac{\sqrt{2}}{4}\,\lambda^{2},\\[-6pt]\nonumber
\end{align}
\end{minipage}
\begin{minipage}{.43\textwidth}
\begin{align}
\sigma_{3} (\lambda) = \frac{1}{2}\,\lambda - \frac{\sqrt{2}}{4}\,\lambda^{2}+\frac{3\sqrt{3} - 1}{16}\,\lambda^{3}. \tag{\ref{eq:phaseExprSMThrdOrd}} \\[-6pt]\nonumber
\end{align}
\end{minipage}
\end{minipage}
These have been plotted (for \(\lambda=0.1\)) in Fig.~\ref{fig:phase_solution_slope}, and discussed in more detail in Section~\ref{sec:phase_solution}.

The last issue we will discuss in this section is the functional form of the oscillating terms \(\rho_{\{1,2,3\}} (X)\), given by eqs.~(\ref{eq:eta_1_function}), (\ref{eq:eta_2_function}), and (\ref{eq:eta_3_function}). These represent the functions which arise as the oscillating terms at first, second, and third orders, respectively. Two of them, \(\rho_{1} (X)\) and \(\rho_{3} (X)\), oscillate around the abscissa, and are depicted in Figure~\ref{fig:eta_plots1} and Figure~\ref{fig:eta_plots3}. The magnitude of the oscillations decreases with \(|X|\), while their frequency increases, which means that they can safely be treated as perturbations to terms which grow quadratically or linearly with \(X\). The function \(\rho_{2} (X)\) differs slightly from the others in that it does not oscillate around the abscissa, but rather around the curve \(1/4 \pi X^{2}\), as can be seen from analytically expanding the function for large values of \(|X|\) (i.e., \(|X| > 1\)):
\begin{align}
\rho_{2} (X) = \frac{\pi}{2} \bigg(C\!\left(X\right) \mp \frac{1}{2}\bigg)\np{2} - \frac{\pi}{\sqrt{2}}\,X \!\left(C\!\left(\sqrt{2}\,X\right) \mp \frac{1}{2}\right) + \frac{1}{2} \,\sin\!\left(\pi X^{2}\right) \sim \frac{1}{4 \pi X^{2}} \left(1 - \cos\!\left(\pi X^{2}\right)\right) + \bigo{X^{-3}}
\end{align}
The details about \(\rho_{2} (X)\) are plotted on Figure~\ref{fig:eta_plots2}, and we can see that despite the dissimilarities to the other two functions, it still decreases in magnitude with increasing \(|X|\), and can be treated as a perturbation to the other terms.

\subsection{Phase resonance equation with \(\mathbf{n \neq 1}\)}
\label{sec:app_phase_equation_single_mode_r}
\noindent
If we consider the source equation (\ref{eq:phaseEquation}) in the single-mode version again, however this time instead of keeping the mode with \(n = 1\), we include only the \(r^{\text{th}}\) mode, then we need to solve the following differential equation:
\begin{figure}[t]
\includegraphics[scale=1.0]{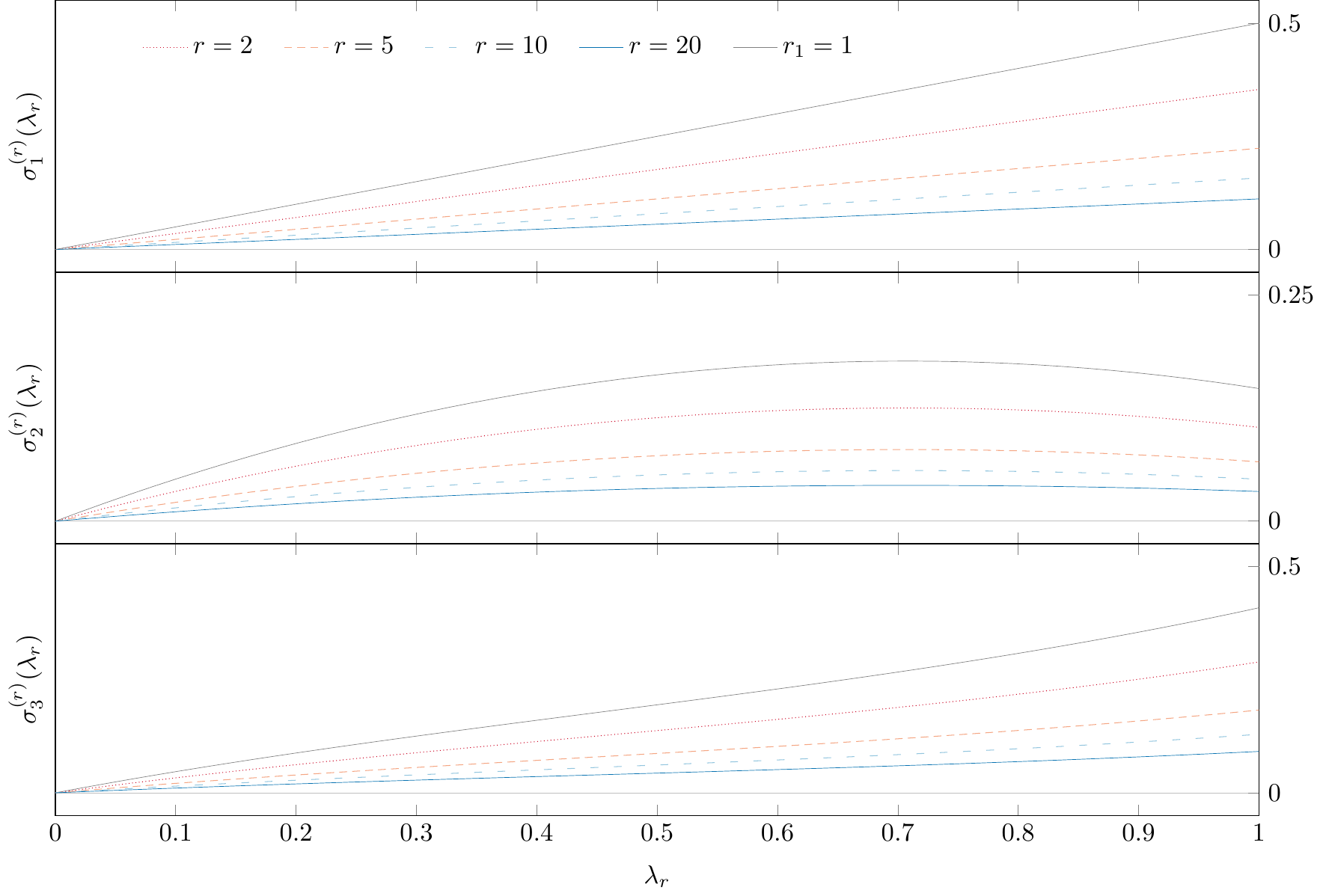}
\caption{Evolution of the slope of the phase resonance solution (for large \(|x|\)) with changing number of the mode \(n = r\).}
\label{fig:app_slope_r_plot}
\end{figure}
\begin{align}
\left(\phi^{(r)} (x)\right)^{\prime\prime}\!= 1 + \lambda_{r} \cos\!\left(r \phi^{(r)}\right). \label{eq:app_phase_equation_moder}
\end{align}
The algorithm for solution is identical to the one presented above for the case with \(r = 1\), hence we only quote the results at each order. The zeroth-order correction is left unchanged, given by eq.~(\ref{eq:PhaseSolutionSMZeroOrd}). The first-order correction is:
\begin{ph}{align}
\phi_{(1)}^{(r)} (x) = \pm \, \pi\!\left[\vph\frac{1}{2} \, \frac{\lambda_{r}}{\sqrt{r}}\right]\!\!\LinearTerm\!+ \frac{\lambda_{r}}{r} \! \left[\vphbigg\pi\!\LinearTerm[r]\!\!\FresnelCTerm[r]\!- \SinTerm[r]\!\right]\!+ \bigo{\lambda_{r}^{2}}. \label{eq:appendix_phase_sol_first_r}
\end{ph}
In the second-order correction we begin to notice the evolution of the terms from the original solution eq.~(\ref{eq:app_phase_solution_second_order}):
\begin{ph}{align}
\phi_{(2)}^{(r)} (x) &= \pm \, \pi \! \left[\frac{1}{2} \, \frac{\lambda_{r}}{\sqrt{r}} - \frac{\sqrt{2}}{4} \, \frac{\lambda_{r}^{2}}{\sqrt{r}}\right] \!\! \LinearTerm\!+ \pi\!\left[\vph\frac{1}{8} \, \frac{\lambda_{r}^{2}}{r}\right] \nonumber \\
& \RepQuad{1} + \frac{\lambda_{r}}{r} \! \left[\vphbigg\pi \! \LinearTerm[r][\SlopeFrstOrdR] \!\! \FresnelCTerm[r][\SlopeFrstOrdR]\!- \SinTerm[r][\SlopeFrstOrdR]\!\right] \label{eq:appendix_phase_sol_second_r}\\
& \RepQuad{1} + \frac{\lambda_{r}^{2}}{r} \! \left[\frac{\pi}{2} \! \FresnelCTerm[r]\np{2}\!- \frac{\pi}{2} \! \LinearTerm[2r] \!\! \FresnelCTerm[2r]\!+ \frac{1}{2} \, \SinTerm[2r]\!\right]\!+ \bigo{\lambda_{r}^{3}}. \nonumber
\end{ph}
We can already notice some ways in which this result differs from the second-order correction with \(r = 1\), for instance the change of variables from \(x /\!\spi\) to \(\sqrt{r}\,x /\!\spi\), and the factor of \(1 / r\) added to the multipliers of the oscillating terms. Before we explain how this arises from the governing equation, we present the third-order result:
\begin{ph}{align}
&\phi_{(3)}^{(r)} (x) = \pm \, \pi \! \left[\frac{1}{2} \, \frac{\lambda_{r}}{\sqrt{r}} - \frac{\sqrt{2}}{4} \, \frac{\lambda_{r}^{2}}{\sqrt{r}} + \frac{3\sqrt{3} - 1}{16} \, \frac{\lambda_{r}^{3}}{\sqrt{r}}\right] \!\! \LinearTerm\!+ \pi \! \left[\frac{1}{8} \,\frac{\lambda_{r}^{2}}{r} - \frac{\sqrt{2}}{8} \, \frac{\lambda_{r}^{3}}{r}\right] \nonumber \\[5pt]
&\RepQuad{3} + \frac{\lambda_{r}}{r} \! \left[\vphbigg\pi \! \LinearTerm[r][\SlopeScndOrdR] \!\! \FresnelCTerm[r][\SlopeScndOrdR]\!- \SinTerm[r][\SlopeScndOrdR] \!\right] \nonumber\\[5pt]
&\RepQuad{3} + \frac{\lambda_{r}^{2}}{r} \! \left[\frac{\pi}{2} \! \FresnelCTerm[r][\SlopeFrstOrdR]\np{2}\! + \frac{1}{2}\,\SinTerm[2r][\SlopeFrstOrdR] \vphbigg\right. \nonumber\\
&\RepQuad{20} \left.\vphbigg - \frac{\pi}{2} \! \LinearTerm[2r][\SlopeFrstOrdR] \!\! \FresnelCTerm[2r][\SlopeFrstOrdR]\!\right] \label{eq:app_third_order_final_expressionR}\\[5pt]
&\RepQuad{3} + \frac{\lambda_{r}^{3}}{r} \! \left[\frac{\pi}{2} \, \CosTerm[r] \!\! \FresnelCTerm[r]\np{2}\!\!- \frac{\pi}{\sqrt{2}} \! \FresnelCTerm[2r] \!\! \FresnelCTerm[r] \vphbigggg\right. \nonumber\\
&\RepQuad{7} + \frac{3\pi}{8} \! \LinearTerm[3r] \!\! \FresnelCTerm[3r]\!- \frac{\pi}{8} \! \LinearTerm[r] \!\! \FresnelCTerm[r] \nonumber\\
&\RepQuad{11} \left. \vphbigggg- \frac{3}{8} \, \SinTerm[3r]\!+ \frac{1}{8} \, \SinTerm[r]\!\right]\!+ \bigo{\lambda_{r}^{4}}. \nonumber
\end{ph}
These results can be obtained from the ones involving \(\lambda_{1} \equiv \lambda\) by transforming eq.~(\ref{eq:phaseEquationSMZeroPh}) with a change of variables:
\begin{subequations}
\begin{minipage}{\textwidth}
\begin{minipage}{.495\textwidth}
\begin{align}
\phi \quad \mapsto \quad \phi^{(r)} = r \phi \\[-6pt]\nonumber
\end{align}
\end{minipage}
\begin{minipage}{.5\textwidth}
\begin{align}
x \quad \mapsto \quad \sqrt{r} \, x, \\[-6pt]\nonumber
\end{align}
\end{minipage}
\end{minipage}
\end{subequations}
which yields eq.~(\ref{eq:app_phase_equation_moder}), hence the solutions to this equation can be found from eq.~(\ref{eq:phaseEquationSMZeroPh}) by applying a change of variables. It is interesting to see how the number of the mode \(r\) affects the graphical behaviour of the solution. In Figure~\ref{fig:phaseSMrThirdOrdPlot} we have plotted the effect of the mode number \(r\) on the correction \({\phi_{3}}^{(r)} (x)\), and in Figure~\ref{fig:app_slope_r_plot} we see the effect of \(r\) on the slope of the solution. The validity of solutions (\ref{eq:appendix_phase_sol_first_r}) -- (\ref{eq:app_third_order_final_expressionR}) is confirmed graphically in Figure~\ref{fig:phaseRValidityPlot}.

\begin{figure}[t]
\includegraphics[scale=1.0]{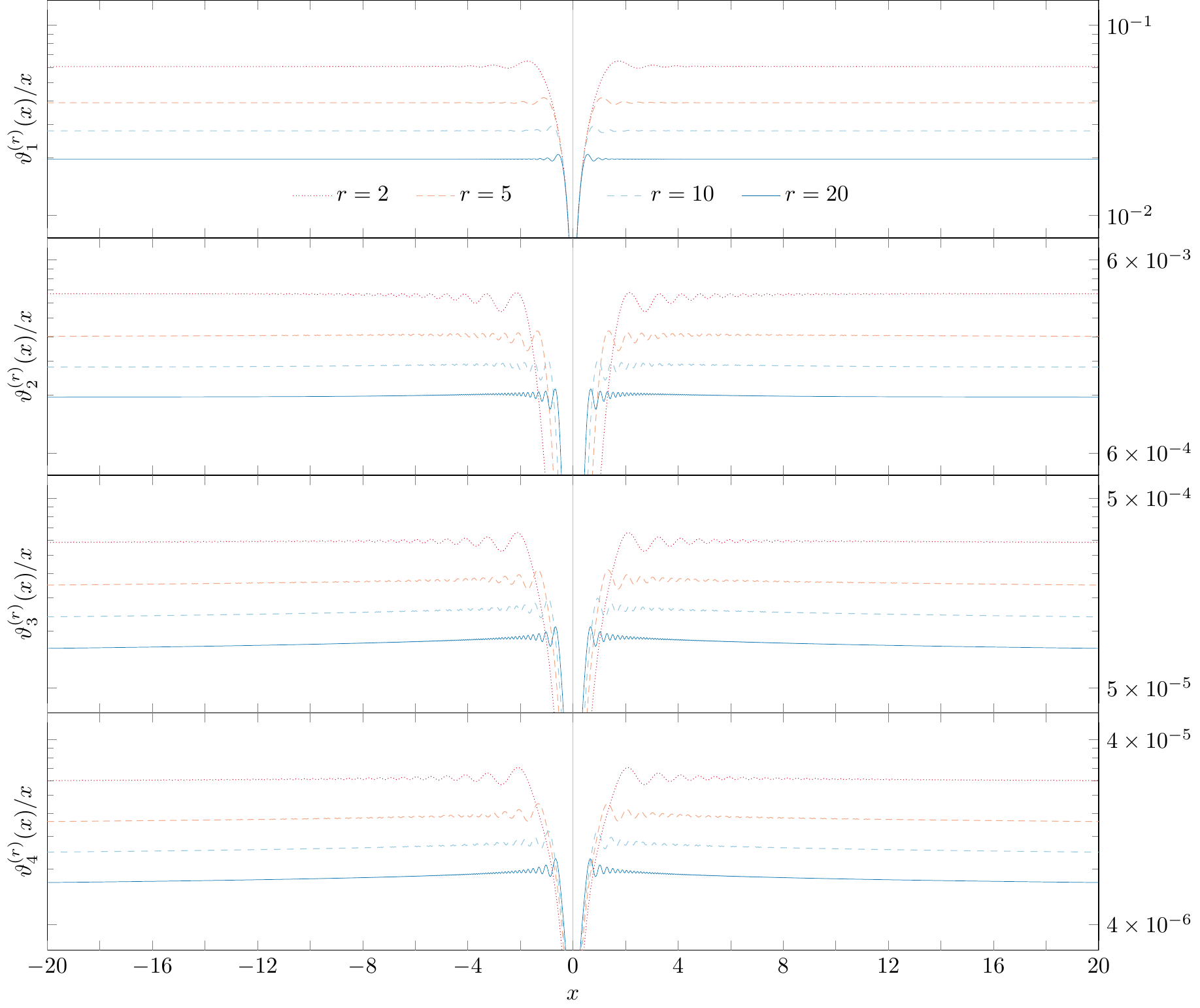}
\caption{Graphical validation of the phase resonance equation with a single \(n \neq 1\) mode. solutions (\ref{eq:appendix_phase_sol_first_r}) -- (\ref{eq:app_third_order_final_expressionR})}
\label{fig:phaseRValidityPlot}
\end{figure}

\subsection{Phase resonance equation with two distinct modes}
\label{sec:app_phase_equation_two_modes}
\noindent
As an interlude to solving the multi-mode version of the phase resonance equation, we present the solution for two modes on the right-hand side, labelled by integers \(r\) and \(s\). Without loss of generality, we can let let \(r < s\).
\begin{align}
\left(\phi^{(r,s)} (x)\right)^{\prime\prime} = 1 + \lambda_{r} \cos\left(r \phi\right) + \lambda_{s} \cos\left(s \phi\right). \tag{\ref{eq:phase_equation_two_modes}}
\end{align}
The zeroth-order solution remains unchanged, \(\phi_{0}^{(r,s)} = x^{2}/2\). The first-order correction can be written in the form:
\begin{ph}{align}
\phi_{(1)}^{(r,s)} (x) = \sum_{l\in\{r,s\}}\!\left\{\vphbigggg\!\!\pm \frac{\pi}{2}\frac{\lambda_{l}}{\sqrt{l}}\!\LinearTerm\!+ \frac{\lambda_{l}}{l}\!\left[\vphbigg\pi\!\LinearTerm[l]\!\!\FresnelCTerm[l]\!- \SinTerm[l]\!\right]\!\!\right\}, \label{eq:appendix_phase_sol_first_rs}
\end{ph}
i.e., the two separate solutions do not mix at first order, and effectively \(\phi_{(1)}^{(r, s)} (x) = \phi_{(1)}^{(r)} (x) + \phi_{(1)}^{(s)} (x)\).

At next order, however, mixing occurs, since we need to substitute the above expression (\ref{eq:appendix_phase_sol_first_rs}) in both terms on the right-hand side of eq.~(\ref{eq:phase_equation_two_modes}). The second-order correction to the original solution can be found to be a combination of (modified) single-mode corrections plus a new mixing term at second order.
\begin{ph}{align} \label{eq:app_phase_solution_two_modes_second_order}
&\phi_{(2)}^{(r,s)} (x) = \pm\,\pi\!\left[\vphbig\frac{1}{2}\!\left(\!\frac{\lambda_{r}}{\sqrt{r}} + \frac{\lambda_{s}}{\sqrt{s}}\!\right)\!- \frac{\sqrt{2}}{4}\!\left(\vph\!\frac{\lambda_{r}^{2}}{\sqrt{r}} + \frac{\lambda_{s}^{2}}{\sqrt{s}}\!\right) + \frac{1}{4} \!\left(\left(s-r\right)^{\sfrac{3}{2}} - \left(r+s\right)^{\sfrac{3}{2}}\right)\!\frac{\lambda_{r}}{r} \frac{\lambda_{s}}{s}\right]\!\!\LinearTerm + \frac{\pi}{8}\!\left(\!\frac{\lambda_{r}}{\sqrt{r}} + \frac{\lambda_{s}}{\sqrt{s}}\!\right)\np{2} \nonumber\\
&+ \frac{\lambda_{r}}{r}\!\left[\vphbigggg\pi\!\LinearTerm[r][\SlopeFrstOrdRS]\!\!\FresnelCTerm[r][\SlopeFrstOrdRS]\!- \SinTerm[r][\SlopeFrstOrdRS]\!\right] \nonumber\\
&+ \frac{\lambda_{s}}{s}\!\left[\vphbigggg\pi\!\LinearTerm[s][\SlopeFrstOrdRS]\!\!\FresnelCTerm[s][\SlopeFrstOrdRS]\!- \SinTerm[s][\SlopeFrstOrdRS]\!\right] \nonumber\\
&+ \frac{\pi}{2} \!\left(\vphbigg\!\frac{\lambda_{r}}{\sqrt{r}}\!\FresnelCTerm[r]\!+ \frac{\lambda_{s}}{\sqrt{s}}\!\FresnelCTerm[s]\!\!\right)\np{2} \nonumber\\
&\RepQuad{2} + \frac{\pi}{2}\,\frac{\lambda_{r}}{r}\frac{\lambda_{s}}{s}\!\left(\vphbigg\!\!\left(s-r\right)\!\LinearTerm[s - r]\!\!\FresnelCTerm[s - r]\!- \left(r+s\right)\!\LinearTerm[r + s]\!\!\FresnelCTerm[r + s]\!\!\right) \nonumber\\
&\RepQuad{6} - \frac{\pi}{2}\!\left(\vphbigg\!\frac{\lambda_{r}^{2}}{r}\!\LinearTerm[2r]\!\!\FresnelCTerm[2r]\!+ \frac{\lambda_{s}^{2}}{s}\!\LinearTerm[2s]\!\!\FresnelCTerm[2s]\!\!\right) \nonumber\\
&\RepQuad{10} - \frac{1}{2}\,\frac{\lambda_{r}}{r}\frac{\lambda_{s}}{s}\!\left(\vphbigg\!\left(s-r\right)\,\SinTerm[s - r]\!- \left(r+s\right)\,\SinTerm[r + s]\!\right) \nonumber\\
&\RepQuad{14} + \frac{1}{2}\!\left(\vphbigg\!\frac{\lambda_{r}^{2}}{r}\,\SinTerm[2r]\!+ \frac{\lambda_{s}^{2}}{s}\,\SinTerm[2s]\!\right) = \\[-6pt]
&=\!\sum_{l\in\{r,s\}}\! \left\{\vphbigggg\!\pm\,\pi\!\left[\frac{1}{2} \, \frac{\lambda_{l}}{l} - \frac{\sqrt{2}}{4} \, \frac{\lambda_{l}^{2}}{l}\right]\!\!\LinearTerm[l]\!+ \pi\!\left[\vph\frac{1}{8} \, \frac{\lambda_{l}^{2}}{l}\right] \right. \nonumber\\
&\RepQuad{1} + \frac{\lambda_{l}}{l}\!\left[\vphbigggg\pi\!\LinearTerm[l][\SlopeFrstOrdL]\!\!\FresnelCTerm[l][\SlopeFrstOrdL]\!- \SinTerm[l][\SlopeFrstOrdL]\!\right] \nonumber\\[-5pt]
\begin{split}
&\RepQuad{4} \left. +\,\frac{\lambda_{l}^{2}}{l}\!\left[\frac{\pi}{2}\!\FresnelCTerm[l]\np{2}\!\!- \frac{\pi}{2}\!\LinearTerm[2l]\!\!\FresnelCTerm[2l]\!+ \frac{1}{2}\,\SinTerm[2l]\!\right]\vphbigggg\!\!\right\} \\[-4pt]
&\RepQuad{1} \pm\frac{\pi}{4}\,\frac{\lambda_{r}}{r}\frac{\lambda_{s}}{s}\!\left[\vphbig\!\left(s-r\right)\!\LinearTerm[s - r]\!- \left(r+s\right)\!\LinearTerm[r+s]\!\right]\!+ \frac{\pi}{4}\,\frac{\lambda_{r}}{\sqrt{r}}\frac{\lambda_{s}}{\sqrt{s}}
\end{split} \\
&\RepQuad{1} + \frac{\lambda_{r}}{\sqrt{r}}\frac{\lambda_{s}}{\sqrt{s}}\!\left[\pi\!\FresnelCTerm[r]\!\!\FresnelCTerm[s] \vphbiggg\right. \nonumber\\
&\RepQuad{4} + \frac{\pi}{2}\,\frac{s-r}{\sqrt{rs}}\!\LinearTerm[s - r]\!\FresnelCTerm[s - r]\!-\,\frac{\pi}{2}\,\frac{r+s}{\sqrt{rs}}\!\LinearTerm[r+s]\!\!\FresnelCTerm[r+s] \nonumber\\
&\RepQuad{7} \left.\vphbiggg + \frac{1}{2}\,\frac{r+s}{\sqrt{rs}}\,\SinTerm[r + s]\!- \frac{1}{2}\,\frac{s-r}{\sqrt{rs}}\,\SinTerm[s - r]\!\right] \nonumber
\end{ph}
Here, \(\bar{l}\) denotes the element of the set \(\{r, s\}\) which is currently not \(l\). It is best to interpret these results graphically.

\begin{figure}[H]
\includegraphics[scale=1.0]{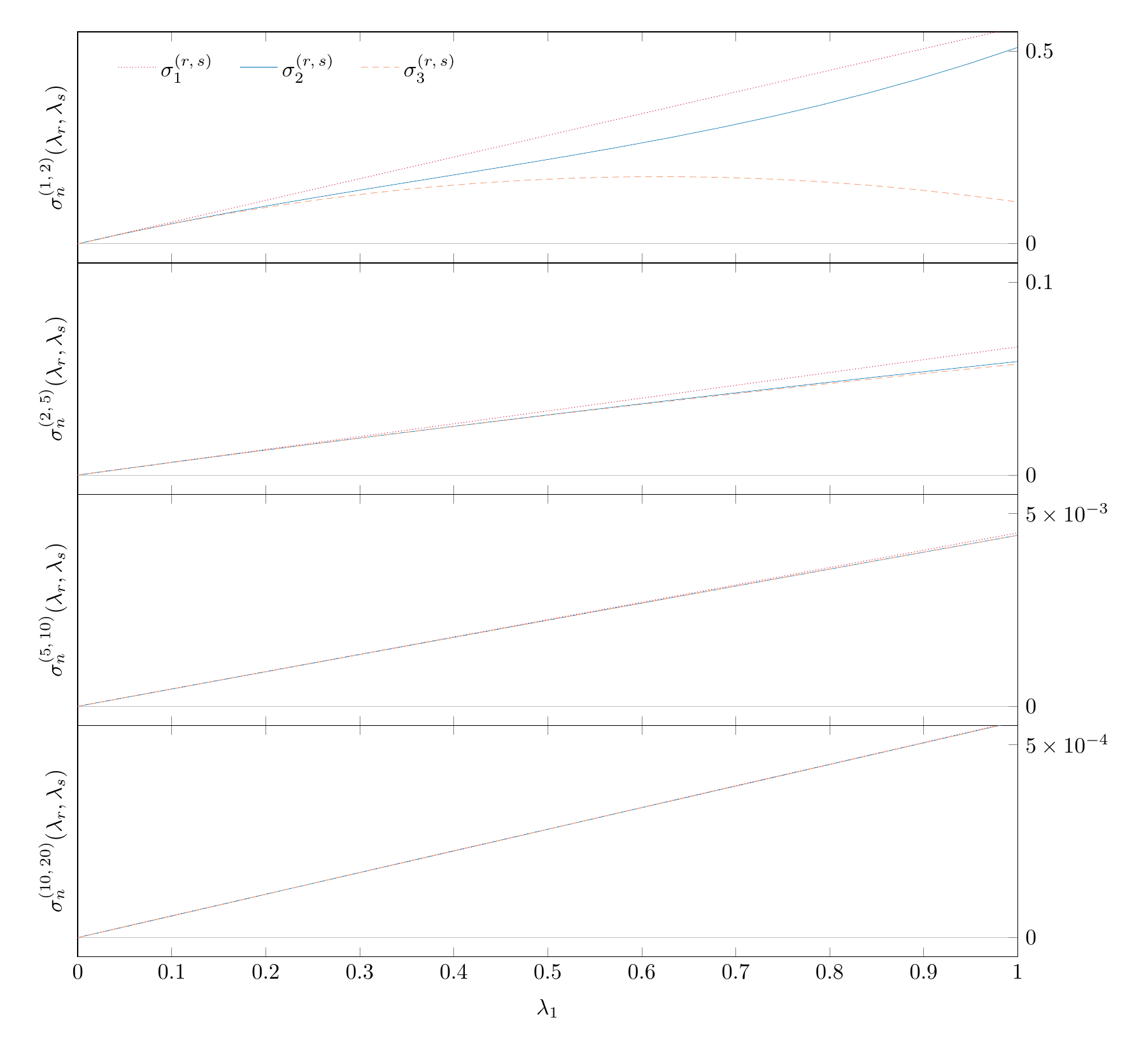}
\caption{Slope of the phase solution with two different oscillating modes, \(r\) and \(s\), for several different combinations of the mode numbers, and for \(0 \le \lambda_{r}, \lambda_{s} \le 1\).}
\label{fig:slopeRSPlot}
\end{figure}

Finally, we present the third-order solution for a source equation with two distinct modes.
\begin{ph}{align}
&\phi_{(3)}^{(r,s)} (x) = \pm\,\pi\!\left[\vphbig\frac{1}{2}\!\left(\!\frac{\lambda_{r}}{\sqrt{r}} + \frac{\lambda_{s}}{\sqrt{s}}\!\right)\!- \frac{\sqrt{2}}{4}\!\left(\vph\!\frac{\lambda_{r}^{2}}{\sqrt{r}} + \frac{\lambda_{s}^{2}}{\sqrt{s}}\!\right)\!+ \frac{1}{4}\!\left(\left(s-r\right)^{\sfrac{3}{2}} - \left(r+s\right)^{\sfrac{3}{2}}\right)\!\frac{\lambda_{r}}{r} \frac{\lambda_{s}}{s} + \frac{3\sqrt{3} - 1}{16}\!\left(\vph\!\frac{\lambda_{r}^{3}}{\sqrt{r}} + \frac{\lambda_{s}^{3}}{\sqrt{s}}\!\right) \right. \nonumber\\[-2.5pt]
&\RepQuad{2} \left.\vphbig + \frac{1}{16}\!\left(\left(2r + s\right)^{\sfrac{5}{2}}\!- \Sgn(2r - s)\!\left|2r - s\right|^{\sfrac{5}{2}}\!- 2s^{\sfrac{5}{2}}\right)\!\frac{\lambda_{r}^{2}}{r^{2}} \frac{\lambda_{s}}{s} + \frac{1}{16}\!\left(\left(2s + r\right)^{\sfrac{5}{2}}\!- \left(2s - r\right)^{\sfrac{5}{2}}\!- 2r^{\sfrac{5}{2}}\right)\!\frac{\lambda_{r}}{r} \frac{\lambda_{s}^{2}}{s^{2}} \right]\!\!\LinearTerm \nonumber \\[5pt]
&\RepQuad{7} + \pi\!\left[\vphbig\frac{1}{8}\!\left(\!\frac{\lambda_{r}}{\sqrt{r}} + \frac{\lambda_{s}}{\sqrt{s}}\!\right)\np{2}\!- \frac{\sqrt{2}}{8}\!\left(\!\frac{\lambda_{r}}{\sqrt{r}} + \frac{\lambda_{s}}{\sqrt{s}}\!\right)\!\!\left(\vph\!\frac{\lambda_{r}^{2}}{\sqrt{r}} + \frac{\lambda_{s}^{2}}{\sqrt{s}}\!\right)\!+ \frac{1}{8}\!\left(\!\frac{\lambda_{r}}{\sqrt{r}} + \frac{\lambda_{s}}{\sqrt{s}}\!\right)\!\!\left(\left(s-r\right)^{\sfrac{3}{2}} - \left(r+s\right)^{\sfrac{3}{2}}\right)\!\frac{\lambda_{r}}{r} \frac{\lambda_{s}}{s}\right] \nonumber\\[5pt]
&\RepQuad{2} + \pi\!\left(\vphbigg\frac{\lambda_{r}}{r}\!\LinearTerm[r][\SlopeScndOrdRSSymb]\!\!\FresnelCTerm[r][\SlopeScndOrdRSSymb] \right. \nonumber\\
&\RepQuad{25}\;\; \left. +\,\frac{\lambda_{s}}{s} \LinearTerm[s][\SlopeScndOrdRSSymb]\!\!\FresnelCTerm[s][\SlopeScndOrdRSSymb]\!\!\right) \nonumber\\
&\RepQuad{6} - \frac{\lambda_{r}}{r}\,\SinTerm[r][\SlopeScndOrdRSSymb]\!- \frac{\lambda_{s}}{s}\,\SinTerm[s][\SlopeScndOrdRSSymb] \nonumber\\[5pt]
&\RepQuad{2} + \frac{\pi}{2}\!\left(\vphbigg\!\frac{\lambda_{r}}{\sqrt{r}}\!\FresnelCTerm[r][\SlopeFrstOrdRSSymb]\!+ \frac{\lambda_{s}}{\sqrt{s}}\!\FresnelCTerm[s][\SlopeFrstOrdRSSymb]\!\!\right)\np{2} \nonumber\\
&\RepQuad{2} + \frac{\pi}{2}\,\frac{\lambda_{r}}{r} \frac{\lambda_{s}}{s}\!\left(\!\left(s-r\right)\!\LinearTerm[s - r][\SlopeFrstOrdRSSymb]\!\!\FresnelCTerm[s - r][\SlopeFrstOrdRSSymb] \vphbigg\right. \nonumber\\[-2.5pt]
&\RepQuad{20} \left.\vphbigg - \left(r+s\right)\!\LinearTerm[r + s][\SlopeFrstOrdRSSymb]\!\!\FresnelCTerm[r + s][\SlopeFrstOrdRSSymb]\!\!\right) \nonumber\\
&\RepQuad{2} - \frac{\pi}{2}\!\left(\vphbigg\!\frac{\lambda_{r}^{2}}{r}\LinearTerm[2r][\SlopeFrstOrdRSSymb]\!\!\FresnelCTerm[2r][\SlopeFrstOrdRSSymb] \right. \nonumber\\
&\RepQuad{24} \left. + \frac{\lambda_{s}^{2}}{s}\LinearTerm[2s][\SlopeFrstOrdRSSymb]\!\!\FresnelCTerm[2s][\SlopeFrstOrdRSSymb]\!\!\right) \nonumber\\
&\RepQuad{6} - \frac{1}{2}\,\frac{\lambda_{r}}{r}\frac{\lambda_{s}}{s}\!\left(\vphbigg\!\!\left(s-r\right)\,\SinTerm[s - r][\SlopeFrstOrdRSSymb]\!- \left(r+s\right)\,\SinTerm[r + s][\SlopeFrstOrdRSSymb]\!\right) \nonumber\\
&\RepQuad{10} + \frac{1}{2}\!\left(\vphbigg\!\frac{\lambda_{r}^{2}}{r}\,\SinTerm[2r][\SlopeFrstOrdRSSymb]\!+ \frac{\lambda_{s}^{2}}{s}\,\SinTerm[2s][\SlopeFrstOrdRSSymb]\!\right) \nonumber\\[5pt]
&\RepQuad{2} + \frac{\pi}{2}\!\left(\vphbigg\!\lambda_{r}\,\CosTerm[r]\!+ \lambda_{s}\,\CosTerm[s]\!\right)\!\!\left(\vphbigg\!\frac{\lambda_{r}}{\sqrt{r}}\!\FresnelCTerm[r]\!+ \frac{\lambda_{s}}{\sqrt{s}}\!\FresnelCTerm[s]\!\!\right)\np{2} \nonumber\\
&\RepQuad{2} + \frac{\pi}{2}\,\frac{\lambda_{r}}{r}\frac{\lambda_{s}}{s}\!\left(\vphbigg\!\!\left(s-r\right)^{\sfrac{3}{2}}\!\FresnelCTerm[s - r]\!- \left(r+s\right)^{\sfrac{3}{2}}\!\FresnelCTerm[r + s]\!\!\right)\!\times \nonumber\\[-2.5pt]
&\RepQuad{25} \times\!\left(\vphbigg\!\frac{\lambda_{r}}{\sqrt{r}}\!\FresnelCTerm[r]\! + \frac{\lambda_{s}}{\sqrt{s}}\!\FresnelCTerm[s]\!\!\right) \nonumber\\
&\RepQuad{2} - \frac{\pi}{\sqrt{2}}\!\left(\vphbigg\!\frac{\lambda_{r}^{2}}{\sqrt{r}}\!\FresnelCTerm[2r]\! + \frac{\lambda_{s}^{2}}{\sqrt{s}}\!\FresnelCTerm[2s]\!\!\right)\!\times \nonumber\\[-2.5pt]
&\RepQuad{25} \times\!\left(\vphbigg\!\frac{\lambda_{r}}{\sqrt{r}}\!\FresnelCTerm[r]\! + \frac{\lambda_{s}}{\sqrt{s}}\!\FresnelCTerm[s]\!\!\right)\!\! \nonumber\\
&\RepQuad{2} + \frac{3 \pi}{8}\!\left(\vphbigg\!\frac{\lambda_{r}^{3}}{r}\!\LinearTerm[3r]\!\!\FresnelCTerm[3r]\!+ \frac{\lambda_{s}^{3}}{s}\!\LinearTerm[3s]\!\!\FresnelCTerm[3s]\!\!\right) \nonumber\\
&\RepQuad{2} + \frac{\pi}{8}\,\frac{\lambda_{r}}{r}\frac{\lambda_{s}}{s}\!\left(\!\frac{\lambda_{r}}{r}\left(2r + s\right)^{2}\!\LinearTerm[2r + s]\!\!\FresnelCTerm[2r + s] \vphbigg\right. \nonumber\\[-2.5pt]
&\RepQuad{23} \left.\vphbigg + \frac{\lambda_{s}}{s} \left(2s + r\right)^{2}\!\LinearTerm[2s + r]\!\!\FresnelCTerm[2s + r]\!\!\right) \nonumber\\
&\RepQuad{2} - \frac{\pi}{8}\,\frac{\lambda_{r}}{r} \frac{\lambda_{s}}{s}\!\left(\!\frac{\lambda_{r}}{r} \Sgn(2r - s) \left(2r - s\right)^{2}\!\LinearTerm[|2r - s|]\!\!\FresnelCTerm[|2r - s|] \vphbigg\right. \nonumber\\[-2.5pt]
&\RepQuad{23} \left.\vphbigg +\,\frac{\lambda_{s}}{s} \left(2s - r\right)^{2}\!\LinearTerm[2s - r]\!\!\FresnelCTerm[2s - r]\!\!\right) \nonumber\\
&\RepQuad{2} - \frac{\pi}{8}\!\left(\vphbigg\!\frac{\lambda_{r}^{3}}{r}\!\LinearTerm[r]\!\!\FresnelCTerm[r]\! + \frac{\lambda_{s}^{3}}{s}\!\LinearTerm[s]\!\!\FresnelCTerm[s]\!\!\right) \nonumber\\
&\RepQuad{6} - \frac{\pi}{4}\,\frac{\lambda_{r}}{r} \frac{\lambda_{s}}{s}\!\left(\vphbigg\!\frac{\lambda_{r}}{r}\,s^{2}\!\LinearTerm[s]\!\!\FresnelCTerm[s]\!+ \frac{\lambda_{s}}{s}\,r^{2}\!\LinearTerm[r]\!\!\FresnelCTerm[r]\!\!\right) \nonumber\\
&\RepQuad{2} - \frac{3}{8}\!\left(\vphbigg\!\frac{\lambda_{r}^{3}}{r}\,\SinTerm[3r]\!+ \frac{\lambda_{s}^{3}}{s}\,\SinTerm[3s]\!\right) \nonumber\\
&\RepQuad{6} - \frac{1}{8}\,\frac{\lambda_{r}}{r} \frac{\lambda_{s}}{s}\!\left(\vphbigg\!\frac{\lambda_{r}}{r}\left(2r + s\right)^{2}\SinTerm[2r + s]\!+ \frac{\lambda_{s}}{s} \left(2s + r\right)^{2}\SinTerm[2s + r]\!\right) \nonumber\\
&\RepQuad{6} + \frac{1}{8}\,\frac{\lambda_{r}}{r} \frac{\lambda_{s}}{s}\!\left(\vphbigg\!\frac{\lambda_{r}}{r}\Sgn(2r - s)\left(2r - s\right)^{2}\SinTerm[|2r - s|]\!+ \frac{\lambda_{s}}{s}\left(2s - r\right)^{2}\SinTerm[2s - r]\!\right) \nonumber\\
&\RepQuad{10} + \frac{1}{8}\!\left(\vphbigg\!\frac{\lambda_{r}^{3}}{r}\,\SinTerm[r]\!+ \frac{\lambda_{s}^{3}}{s}\,\SinTerm[s]\!\right) \nonumber\\
&\RepQuad{14} + \frac{1}{4}\,\frac{\lambda_{r}}{r}\frac{\lambda_{s}}{s}\!\left(\vphbigg\!\frac{\lambda_{r}}{s}\,r^{2}\,\SinTerm[r]\!+ \frac{\lambda_{s}}{r}\,s^{2}\,\SinTerm[s]\!\right) \label{eq:phase_solution_two_modes_third_order}
\end{ph}

\begin{figure}[t]
\includegraphics[scale=1.0]{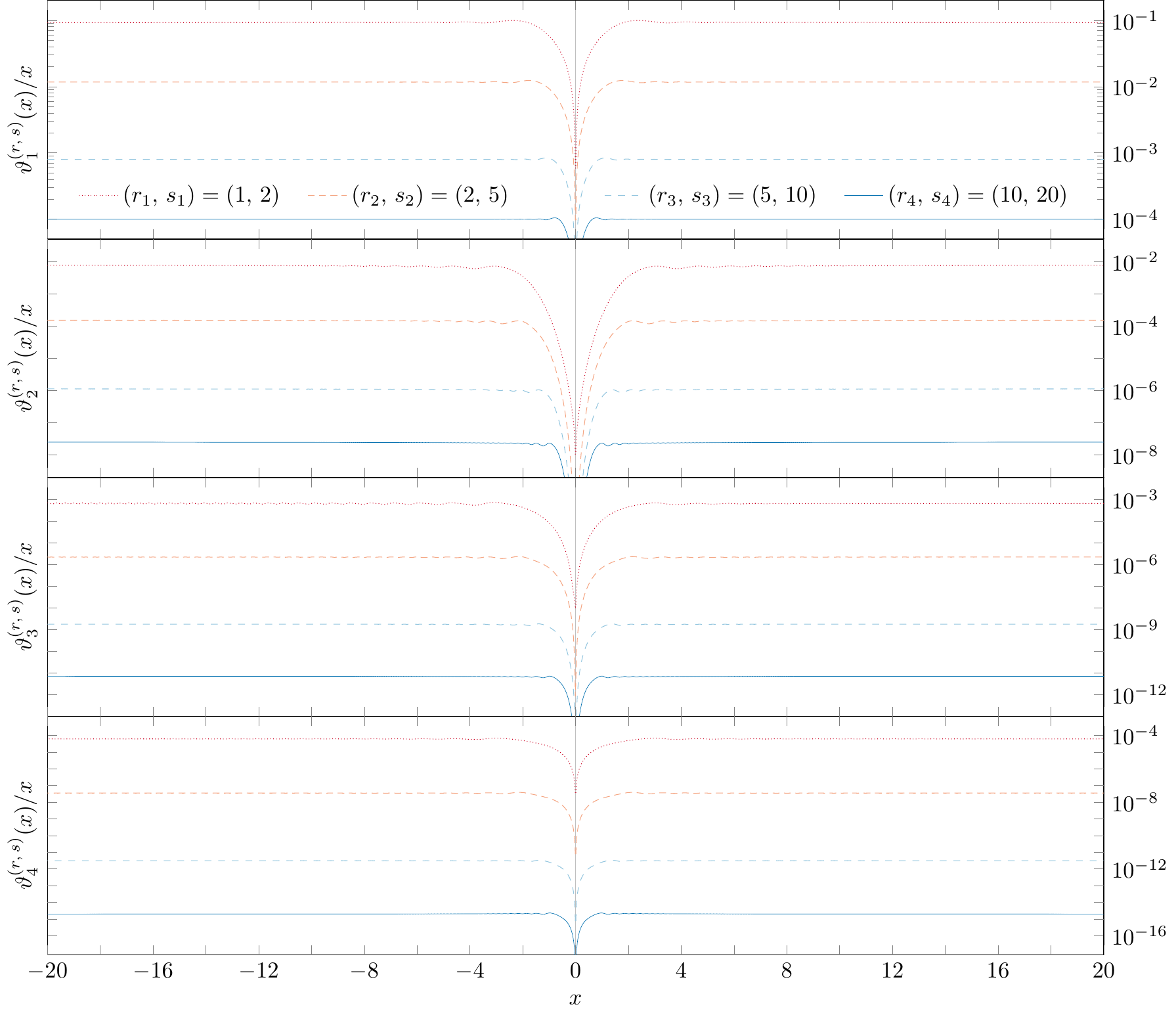}
\caption{Graphical demonstration of the validity of the solutions to the phase resonance equation with two distinct modes \(r\) and \(s\), for several different number combinations. In all these cases, we use \(\lambda_{1} = 0.1\), and then consecutive coefficients are calculated by \(\lambda_{n} = \lambda_{1} / n^{2.5}\). We can see that in all cases, the remainder is an order of magnitude smaller than the larger coefficient \(\lambda_{r}\).}
\label{fig:phaseRSValidityPlot}
\end{figure}

\subsection{Multi-mode phase equation}
\label{sec:app_phase_solution_multimode}
\noindent
We know by designation that the correction \(\bphi_{(1)} (x)\) will be of first order in the \(\{\lambda_{n}\}\). Therefore, we ignore any terms of second or higher order when solving the differential equation (\ref{eq:phase_multimode_first_correction_equation}), as they belong to higher-order corrections:
\begin{align}
\bphi_{(1)}^{\prime\prime} (x) = \sum_{n} \lambda_{n} \cos\left(n\,\bphi_{0}\right). \tag{\ref{eq:phase_multimode_first_correction_equation}}
\end{align}
Substituting for \(\bphi_{0}(x)\) from eq.~(\ref{eq:phi0_m_solution}), we obtain the differential equation whose solution is the first-order correction:
\begin{align}
\bphi_{(1)}^{\prime\prime} (x) = \sum_{n} \lambda_{n} \cos\!\Bigg(\!\frac{n x^{2}}{2}\!\Bigg).
\end{align}
We integrate once to find the first derivative (we have to apply the appropriate boundary condition \(\bphi^{\prime} (0) = 0\)), keeping in mind that the integral of a sum is simply a sum of the integrals. We obtain the following result:
\begin{align}
\bphi_{(1)}^{\prime} (x) = \spi \, \sum_{n} \frac{\lambda_{n}}{\sqrt{n}} \, C\Bigg(\!\frac{\sqrt{n}\,x}{\spi}\!\Bigg).
\end{align}
Integrating this once again, and keeping track of the boundary condition \(\bphi (0) = 0\) yields the first-order correction:
\begin{ph}{align}
\bphi_{(1)} (x) = \sum_{n} \left\{\vphbigggg\!\!\pm \, \frac{\pi}{2} \frac{\lambda_{n}}{n}\LinearTerm[n]\!+ \frac{\lambda_{n}}{n}\!\left[\!\vphbigg\pi\!\LinearTerm[n]\!\!\FresnelCTerm[n]\!- \SinTerm[n]\!\right]\!\!\right\} = \sum_{n} \phi_{(1)}^{(n)}. \label{eq:phase_equation_multi_mode_first_order_raw_solution}
\end{ph}
As we can see, in exactly the same way as solution~(\ref{eq:appendix_phase_sol_first_rs}), this first-order result exhibits no mixing between the solutions for different values of \(n\), which means it can be expressed as a sum over many single-mode solutions of the form~(\ref{eq:appendix_phase_sol_first_r}). Combining this with the zeroth-order solution, eq.~(\ref{eq:phi0_m_solution}), we can write down the complete solution of the phase equation to first order in the case of an infinite number of Fourier-like coefficients \(\{\lambda_{n}\}\). As before, it is beneficial to combine the terms proportional to \(x^{2}\) and \(x\) into a single term, again retaining only terms linear in \(\{\lambda_{n}\}\):
\begin{ph}{align}
\bphi_{1} (x) = \frac{\pi}{2}\!\LinearTerm[][\pm \ConstMultiMode{n}]\np{2}\!+ \sum_{n} \frac{\lambda_{n}}{n}\!\left[\vphbigg\pi\!\LinearTerm[n][]\!\!\FresnelCTerm[n][]\!- \SinTerm[n]\!\right].\tag{\ref{eq:phase_equation_multi_mode_first_order_solution}}
\end{ph}

The second-order correction to eq.~(\ref{eq:phase_equation_multi}) requires us to solve eq.~(\ref{eq:phase_equation_second_order_multi_mode_reworked}). The argument of the trigonometric function on the right-hand side is given by the first-order correction, eq.~(\ref{eq:phase_equation_multi_mode_first_order_raw_solution}) combined with the zeroth-order solution:
\begin{align}
\bphi_{(2)}^{\prime\prime} (x) = \sum_{n} \lambda_{n} \cos\!\left[\vphbigggg n \!\left(\!\vphantom{\left[\Bigg(\Bigg)\np{2}\right]\np{2}}\frac{x^{2}}{2} \pm \sum_{m} \frac{\pi}{2} \frac{\lambda_{m}}{m}\Bigg(\!\frac{\sqrt{m} \, x}{\spi}\!\Bigg)\!+ \sum_{m} \frac{\lambda_{m}}{m} \!\left(\!\vphantom{\left[\Bigg(\Bigg)\np{2}\right]\np{2}}\pi \Bigg(\!\frac{\sqrt{m} \, x}{\spi}\!\Bigg)\!\!\left(\!\vphantom{\left[\Bigg(\Bigg)\np{2}\right]}C\Bigg(\!\frac{\sqrt{m} \, x}{\spi}\!\Bigg)\!\mp \frac{1}{2}\! \right)\!- \sin\!\left[\frac{\pi}{2}\!\left(\frac{\sqrt{m} \, x}{\spi}\right)\np{2}\right]\!\right)\!\!\right)\!\right].
\end{align}
Even though the process is more involved than before, the method of solution here is the same: consecutively integrating the right-hand side twice, while observing the boundary conditions. After the first integration, we obtain
\begin{align}
\begin{split}
\bphi_{(2)}^{\prime} (x) &= \spi \sum_{n} \frac{\lambda_{n}}{\sqrt{n}}\!\left(\vphantom{\left[\Bigg(\Bigg)\np{2}\right]}C\!\left(\!\frac{\sqrt{n} \, x}{\sqrt{\pi}} \pm \frac{\sqrt{n}}{2} \sum_{m} \frac{\lambda_{m}}{\sqrt{m}}\!\right)\!\mp C\!\left(\!\frac{\sqrt{n}}{2} \sum_{m} \frac{\lambda_{m}}{\sqrt{m}}\right)\!\!\right) \\[5pt]
& + \spi \sum_{n,\,m} \lambda_{n} \lambda_{m} \!\left[\vphantom{\left[\Bigg(\Bigg)\np{2}\right]}\frac{1}{\sqrt{m}} \cos\!\Bigg(\!\frac{n x^{2}}{2}\!\Bigg)\,C\Bigg(\!\frac{\sqrt{m}x}{\spi}\!\Bigg)\!\mp \frac{1}{2\sqrt{m}}\cos\!\Bigg(\!\frac{n x^{2}}{2}\!\Bigg)\!\pm \frac{1}{2\sqrt{m}} \right.\\
&\quad\quad\quad\quad\quad\quad\quad\quad\quad\quad\quad\quad \left.\vphantom{\left[\Bigg(\Bigg)\np{2}\right]} + \frac{1}{2m}\frac{n-m}{\sqrt{|n-m|}}\, C\!\left(\!\frac{\sqrt{|n-m|}\,x}{\spi}\!\right) - \frac{1}{2m}\, \sqrt{n+m}\, C\Bigg(\!\frac{\sqrt{n+m}\,x}{\spi}\!\Bigg)\right].
\end{split} \label{eq:app_second_order_multi_mode_intermediate}
\end{align}
While performing this integration, we encounter some unusual integrals, hence we quote the results here:
\begin{align}
\begin{split}
&-\!\!\bigintss_{0}^{x} \!\!\! \dd y\,\cos\!\Bigg(\!\frac{n y^{2}}{2}\!\Bigg) \cos\!\Bigg(\!\frac{m y^{2}}{2}\!\Bigg)\!+ \frac{n}{m}\!\bigintss_{0}^{x} \!\!\! \dd y\,\sin\!\Bigg(\!\frac{n y^{2}}{2}\!\Bigg) \sin\!\Bigg(\!\frac{m y^{2}}{2}\!\Bigg) =\\
&\RepQuad{8} = \frac{\spi}{2} \!\left(\!\!\vphbig\Sgn(n-m)\,\frac{\sqrt{|n-m|}}{m}\,C\!\left(\!\frac{\sqrt{|n-m|}\,x}{\spi}\!\right)\!- \frac{\sqrtbig{n+m}}{m}\,C\Bigg(\!\frac{\sqrtbig{n+m}\,x}{\spi}\!\Bigg)\!\!\right).
\end{split}
\end{align}
Similarly, when we integrate the expression (\ref{eq:app_second_order_multi_mode_intermediate}), we encounter the following integral, whose solution is made possible by the fact that we are summing over both indices \(n\) and \(m\), and hence we can re-label them interchangeably.
\begin{align}
\spi\,\sum_{n,\,m} \lambda_{n}\,\frac{\lambda_{m}}{\sqrt{m}}\!\bigintss_{0}^{x} \!\!\! \dd y\,\cos\!\Bigg(\!\frac{n y^{2}}{2}\!\Bigg)\,C\Bigg(\!\frac{\sqrt{m}\,x}{\spi}\!\Bigg)\!= \frac{\pi}{2} \sum_{n,\,m} \frac{\lambda_{n}}{\sqrt{n}} \frac{\lambda_{m}}{\sqrt{m}}\,C\Bigg(\!\frac{\sqrt{n}\,x}{\spi}\!\Bigg)\,C\Bigg(\!\frac{\sqrt{m}\,x}{\spi}\!\Bigg)
\end{align}
Using this, the second-order correction is found to be as follows (remember that we are using \(\bphi^{\prime} (0) = \bphi (0) = 0\)).
\begin{ph}{align}
&\bphi_{(2)} (x) = \pm\,\pi \Bigg[\frac{1}{2} \sum_{n} \frac{\lambda_{n}}{\sqrt{n}} + \sum_{n,\,m} \frac{1}{8}\left(|n-m|^{\sfrac{3}{2}} - (n+m)^{\sfrac{3}{2}}\right)\frac{\lambda_{n}}{n} \frac{\lambda_{m}}{m}\Bigg]\!\!\LinearTerm + \frac{\pi}{8} \sum_{n,\,m} \frac{\lambda_{n}}{\sqrt{n}} \frac{\lambda_{m}}{\sqrt{m}} \nonumber\\[5pt]
&\!+ \sum_{n} \frac{\lambda_{n}}{n} \!\left[\vphbigggg\pi\!\LinearTerm[n][\pm \ConstMultiMode{m}]\!\!\FresnelCTerm[n][\pm \ConstMultiMode{m}]\!- \SinTerm[n][\pm \ConstMultiMode{m}]\!\right] \nonumber\\[5pt]
&\!+ \sum_{n,\,m} \frac{\lambda_{n}}{\sqrtbig{n}} \frac{\lambda_{m}}{\sqrtbig{m}} \!\left[\frac{\pi}{2}\!\FresnelCTerm[n]\!\!\FresnelCTerm[m]\!- \frac{\pi}{2} \frac{\sqrt{n}}{\sqrt{m}}\!\LinearTerm[n + m]\!\!\FresnelCTerm[n + m] \vphbigggg \right. \nonumber\\
&\RepQuad{4} + \frac{\pi}{2} \frac{\sqrtbig{n}}{\sqrtbig{m}}\,\Sgn(n - m)\!\LinearTerm[|n-m|]\!\!\FresnelCTerm[|n-m|] \nonumber\\
&\RepQuad{8} \left.\vphbigggg - \frac{1}{2} \frac{\sqrtbig{n}}{\sqrtbig{m}}\!\left(\vphbigg\!\Sgn(n - m) \SinTerm[|n - m|]\!- \SinTerm[n + m]\!\right)\!\right]
\label{eq:phaseMMSolutionScndOrd}
\end{ph}
It is important to note that this is consistent with our previous results. For \(n = m = 1\), we retrieve the second-order single-mode solution, eq.~(\ref{eq:app_phase_solution_second_order}) while if we choose \(n, m = \{r,s\}\), we find the second-order solution for an equation with 2 distinct modes, namely the voluminous eq.~(\ref{eq:app_phase_solution_two_modes_second_order}). Moreover, we recognise that the oscillating terms here have similar form to the ones in the single-mode solution, hence we can introduce the notation \(\{\bm{\sigma}_{n}\}\) and \(\{\bm{\rho}_{n}^{m} (x)\}\).

To obtain the third-order correction, we need to solve the following multi-mode second-order differential equation:
\begin{align}
\begin{split}
\bphi_{(3)}^{\prime\prime} (x) &= \sum_{n} \lambda_{n} \CosTerm[n][\SecondOrderMultiMode] - \sum_{n,\,m} \frac{n}{m}\,\lambda_{n} \lambda_{m} \SinTerm[n][\FirstOrderMultiMode]\!\bm{\rho}_{1}^{m} (x)\\
&\RepQuad{1} -\!\sum_{n,\,m,\,p}\!\lambda_{n} \lambda_{m} \lambda_{p}\!\left(\vphbigg\!\frac{1}{2}\,\frac{n^{2}}{m p}\,\CosTerm[n]\!\bm{\rho}_{1}^{m} (x)\,\bm{\rho}_{1}^{p} (x) + \frac{n}{\sqrtbig{m}\sqrtbig{p}}\,\SinTerm[n]\!\bm{\rho}_{2}^{mp} (x)\!\right).
\end{split}
\end{align}

Following through with two rounds of integration yields the following solution for the third-order correction:
\begin{ph}{align}
&\bphi_{(3)} (x) = \pm\,\pi\!\left[\,\sum_{n} \frac{1}{2}\,\frac{\lambda_{n}}{\sqrtbig{n}} + \sum_{n,\,m} \frac{1}{8}\left(|n-m|^{\sfrac{3}{2}} - (n+m)^{\sfrac{3}{2}}\right)\frac{\lambda_{n}}{n} \frac{\lambda_{m}}{m} \vphbig \right. \nonumber\\
&\RepQuad{8} \left. \vphbigg +\!\sum_{n,\,m,\,p} \frac{1}{16}\!\left(\!(n+m+p)^{\sfrac{3}{2}} - |n-m+p|^{\sfrac{3}{2}} - |n+m-p|^{\sfrac{3}{2}} + |n-m-p|^{\sfrac{3}{2}}\right)\lambda_{n}\,\frac{\lambda_{m}}{m}\frac{\lambda_{p}}{p}\right]\!\LinearTerm \nonumber\\[5pt]
& + \pi\!\left[\sum_{n,\,m}\frac{1}{8}\,\frac{\lambda_{n}}{\sqrt{n}}\frac{\lambda_{m}}{\sqrt{m}} +\!\sum_{n,\,m,\,p} \frac{1}{16}\!\left(\!\Sgn(n-m)\frac{\sqrtbig{|n-m|}}{\sqrtbig{m}} + \Sgn(n-p)\frac{\sqrtbig{|n-p|}}{\sqrtbig{p}} - \frac{\sqrtbig{n+m}}{\sqrt{m}} - \frac{\sqrtbig{n+p}}{\sqrtbig{p}}\right) \lambda_{n}\,\frac{\lambda_{m}}{\sqrtbig{m}}\frac{\lambda_{p}}{\sqrtbig{p}} \vphbig\right] \nonumber\\[5pt]
& + \sum_{n} \frac{\lambda_{n}}{n} \!\left[\vphbigg\pi\!\LinearTerm[n][\SecondOrderMultiMode]\!\!\FresnelCTerm[n][\SecondOrderMultiMode]\!- \SinTerm[n][\SecondOrderMultiMode]\!\right] \nonumber\\[5pt]
& + \sum_{n,\,m} \frac{\lambda_{n}}{\sqrt{n}} \frac{\lambda_{m}}{\sqrtbig{m}}\!\left[\vphbigggg\frac{\pi}{2}\!\FresnelCTerm[n][\FirstOrderMultiMode]\!\!\FresnelCTerm[m][\FirstOrderMultiMode] \right. \nonumber\\
&\RepQuad{4} + \frac{\pi}{2}\,\frac{\sqrt{n}}{\sqrt{m}}\,\Sgn{(n - m)}\!\LinearTerm[|n - m|][\FirstOrderMultiMode]\!\!\FresnelCTerm[|n - m|][\FirstOrderMultiMode] \nonumber\\
&\RepQuad{7} - \frac{\pi}{2}\,\frac{\sqrt{n}}{\sqrt{m}}\!\LinearTerm[n+m][\FirstOrderMultiMode]\!\!\FresnelCTerm[n+m][\FirstOrderMultiMode] \nonumber\\
&\RepQuad{11} \left.\vphbigggg \!\!\!\! - \frac{1}{2}\,\frac{\sqrtbig{n}}{\sqrtbig{m}}\!\left(\vphbigg\! \SinTerm[|n - m|][\FirstOrderMultiMode][][\Sgn(n - m)]\!- \SinTerm[n + m][\FirstOrderMultiMode]\!\right)\!\right] \nonumber\\[5pt]
& +\!\sum_{n,\,m,\,p} \lambda_{n}\,\frac{\lambda_{m}}{\sqrtbig{m}} \frac{\lambda_{p}}{\sqrtbig{p}}\!\left[\vphbigggg\frac{\pi}{2}\,\CosTerm[n]\!\!\FresnelCTerm[m]\!\!\FresnelCTerm[p] \right. \nonumber\\
& + \frac{\pi}{4}\!\left(\vphbigg\!\!\Sgn(n-m)\,\frac{\sqrtbig{|n-m|}}{\sqrt{m}}\!\FresnelCTerm[|n - m|]\!- \frac{\sqrtbig{n+m}}{\sqrt{m}}\!\FresnelCTerm[n+m]\!\!\!\right)\!\!\FresnelCTerm[p] \nonumber\\
&\RepQuad{2} \!\!\! + \frac{\pi}{4}\!\left(\vphbigg\!\!\Sgn(n-p)\,\frac{\sqrtbig{|n-p|}}{\sqrtbig{p}}\!\FresnelCTerm[|n - p|]\!- \frac{\sqrtbig{n+p}}{\sqrtbig{p}}\!\FresnelCTerm[n+p]\!\!\!\right)\!\!\FresnelCTerm[m] \nonumber\\
& + \frac{\pi}{8}\,\frac{(n+m+p)}{\sqrtbig{m}\,\sqrt{p}}\!\LinearTerm[n+m+p]\!\!\FresnelCTerm[n+m+p] \nonumber\\
&\RepQuad{4} - \frac{\pi}{8}\,\frac{|n-m+p|}{\sqrt{m}\,\sqrtbig{p}}\!\LinearTerm[|n-m+p|]\!\!\FresnelCTerm[|n-m+p|] \nonumber\\
&\RepQuad{8} - \frac{\pi}{8}\,\frac{|n+m-p|}{\sqrt{m}\,\sqrtbig{p}}\!\LinearTerm[|n+m-p|]\!\!\FresnelCTerm[|n+m-p|] \nonumber\\
&\RepQuad{12} + \frac{\pi}{8}\,\frac{|n-m-p|}{\sqrt{m}\,\sqrtbig{p}}\!\LinearTerm[|n-m-p|]\!\!\FresnelCTerm[|n-m-p|] \nonumber\\
& - \frac{1}{8}\,\frac{(n+m+p)}{\sqrtbig{m}\,\sqrt{p}}\,\SinTerm[n+m+p]\!+ \frac{1}{8}\,\frac{|n-m+p|}{\sqrtbig{m}\,\sqrt{p}}\,\SinTerm[|n-m+p|] \nonumber\\
&\RepQuad{4} \left.\vphbigggg + \frac{1}{8}\,\frac{|n+m-p|}{\sqrt{m}\,\sqrt{p}}\,\SinTerm[|n+m-p|]\!- \frac{1}{8}\,\frac{|n-m-p|}{\sqrt{m}\,\sqrt{p}}\,\SinTerm[|n-m-p|]\!\right].
\label{eq:phase_equation_multi_mode_third_order_solution}
\end{ph}
As with the second-order result, this one reduces to the appropriate previous expressions for \(n = m = p = 1\), or for \(n = m = p = \{r, s\}\). Moreover, it is easy to recognise the generalised form of the terms proportional to \(\lambda^{n}\) and how they reduce to their single-mode counterparts. A more detailed discussion about the features and validity of these solutions for different mode counts is presented in the main text of this article.

\subsection{Phase equation with a single mode and non-zero phase}
\noindent
Having considered the multi-mode zero-phase equation, we revert back to the single mode case and introduce a non-zero phase \(\delpsi\):
\begin{align}
\varphi^{\prime\prime} (x) = 1 + \lambda \cos\big(\varphi (x) - \delpsi\big). \tag{\ref{eq:phaseSMPhEq}}
\end{align}
The first-order solution is \(\varphi_{0} = x^{2}/2\), and we substitute it into eq.~(\ref{eq:phaseSMPhEq}) to obtain the differential equation for \(\varphi_{(1)}\):
\begin{align}
\varphi_{(1)}^{\prime\prime} (x) = \lambda \cos\left(\varphi_{0} - \delpsi\right).
\end{align}
The method of solution is identical to expand the trigonometric function on the right-hand side:
\begin{align}
\varphi_{(1)}^{\prime\prime} (x) = \lambda \cos(\delpsi) \cos\!\left(\varphi_{0}\right) + \lambda \sin(\delpsi) \sin\!\left(\varphi_{0}\right).
\end{align}
We now proceed to integrate this exactly as before, only this time there are twice as many terms that need to be integrated. The final solution can be presented in a fashion similar to before, however this time we have \(\delpsi\)-dependence both in the slope and in the oscillating terms.
\begin{ph}{align}
\begin{split}
&\varphi_{1} (x) = \frac{\pi}{2} \!\LinearTerm[][\SlopeFrstOrdSMP]\np{2} \\
&\RepQuad{3} + \lambda\!\left[\vphbiggg\pi \! \LinearTerm\!\!\left(\vphbig\!\!\cos(\delpsi)\!\FresnelCTerm\!+ \sin(\delpsi)\!\FresnelSTerm\!\!\right)\!-\!\left(\vphbig\!\SinTerm[][][\!- \delpsi]\!+ \sin(\delpsi)\!\right)\!\right]
\end{split}
\end{ph}
This solution can be used again in the source equation to find the second-order correction:
\begin{ph}{align}
&\varphi_{(2)} (x) = \pm\,\pi\!\left[\vphbig \frac{1}{2}\big(\!\cos(\delpsi) + \sin(\delpsi)\big)\lambda -\!\left(\!\frac{\sqrt{2}}{4} \big(\!\cos(2\delpsi) + \sin(2\delpsi)\big) - \frac{1}{2}\big(\!\cos(\delpsi) - \sin(\delpsi)\big)\sin(\delpsi)\!\right)\!\lambda^{2}\right]\!\!\LinearTerm \nonumber\\[5pt]
&\RepQuad{38} + \frac{\pi}{8}\big(\!\cos(\delpsi) + \sin(\delpsi)\big)^{2}\lambda^{2} \nonumber\\[5pt]
&+ \lambda \!\left[\pi\!\LinearTerm[][\SlopeFrstOrdSMP]\!\times \vphbigggg\right. \nonumber\\
&\RepQuad{4} \times\!\left(\vphbigg\!\!\cos(\delpsi)\!\FresnelCTerm[][\SlopeFrstOrdSMP]\!+ \sin(\delpsi)\!\FresnelSTerm[][\SlopeFrstOrdSMP]\!\!\!\right) \nonumber\\
&\RepQuad{23} \left.\vphbigggg -\!\left(\vphbigg\!\SinTerm[][\SlopeFrstOrdSMP][\!- \delpsi]\!+ \sin(\delpsi)\!\right)\!\right] \nonumber\\
& + \lambda^{2} \!\left[\frac{\pi}{2}\!\left(\!\!\vphbig\cos(\delpsi)\!\FresnelCTerm\!+ \sin(\delpsi)\!\FresnelSTerm\!\!\right)\np{2} \vphbigggg\right. \nonumber\\
&\RepQuad{4} - \frac{\pi}{2}\!\LinearTerm[2]\!\!\left(\!\!\vphbigg\cos(2\delpsi)\!\FresnelCTerm[2]\!+ \sin(2\delpsi)\!\FresnelSTerm[2]\!\!\!\right) \nonumber\\
&\RepQuad{8} - \pi\!\LinearTerm\!\!\left(\!\!\vphbig\sin(\delpsi)\!\FresnelCTerm\!- \cos(\delpsi)\!\FresnelSTerm\!\!\right)\!\sin(\delpsi) \nonumber\\
&\RepQuad{12} \left.\vphbigggg + \frac{1}{2}\!\left(\!\vphbigg\SinTerm[2][][-2 \delpsi]\!+ \sin(2\delpsi)\!\right)\!+\!\left(\!\vphbig\CosTerm[][][-\delpsi]\!- \cos(\delpsi)\!\right)\!\sin(\delpsi)\right] \nonumber\\
\end{ph}

We can apply the same technique to find the third-order solution. The third-order correction is given below.

\begin{ph}{align}
&\varphi_{(3)} (x) = \pm\,\pi\!\left[\vphbigg\frac{1}{2}\big(\!\cos(\delpsi) + \sin(\delpsi)\big)\lambda -\!\left(\!\frac{\sqrt{2}}{4} \big(\!\cos(2\delpsi) + \sin(2\delpsi)\big) - \frac{1}{2}\big(\!\cos(\delpsi) - \sin(\delpsi)\big)\sin(\delpsi)\!\right)\!\lambda^{2}\right. \nonumber\\[5pt]
&\RepQuad{5} + \left(\!\frac{3\sqrt{3}}{16}\big(\!\cos(3\delpsi) + \sin(3\delpsi)\big) - \frac{\sqrt{2}}{2}\,\sin(\delpsi)\big(\!\cos(2\delpsi) - \sin(2\delpsi)\big) - \frac{3}{16}\big(\!\cos(\delpsi) + \sin(\delpsi)\big) \right. \nonumber\\
&\RepQuad{30} \left.\vphbigg \left.\vph + \frac{1}{8}\,\cos(2\delpsi)\big(\!\cos(\delpsi) + \sin(\delpsi)\big)\!\right)\!\lambda^{3}\right]\!\!\LinearTerm \nonumber\\
&\RepQuad{3} \;\;\; + \pi\!\left[\vphbig\frac{1}{8}\big(\!\cos(\delpsi) + \sin(\delpsi)\big)^{2}\lambda^{2} - \left(\!\frac{\sqrt{2}}{8}\big(\!\cos(2\delpsi) + \sin(2\delpsi)\big)\big(\!\cos(\delpsi) + \sin(\delpsi)\big) - \frac{1}{4} \cos(2\delpsi) \sin(\delpsi)\!\right)\!\lambda^{3}\right] \nonumber\\
&+ \lambda\!\left[\pi\!\LinearTerm[][\SlopeScndOrdSMPSymb]\!\!\left(\vphbig\!\cos(\delpsi)\!\FresnelCTerm[][\SlopeScndOrdSMPSymb]\!+ \sin(\delpsi)\!\FresnelSTerm[][\SlopeScndOrdSMPSymb]\!\!\right) \right. \nonumber\\[-3pt]
&\RepQuad{32} \left.\vphbigg -\!\left(\vphbig\!\SinTerm[][\SlopeScndOrdSMPSymb][- \delpsi]\!+ \sin(\delpsi)\!\right)\!\right] \nonumber\\[5pt]
& + \lambda^{2} \!\left[\frac{\pi}{2} \!\left(\vphbig\!\cos(\delpsi)\!\FresnelCTerm[][\SlopeFrstOrdSMPSymb]\!+ \sin(\delpsi)\!\FresnelSTerm[][\SlopeFrstOrdSMPSymb]\!\!\right)\np{2} \vphbigggg\right. \nonumber\\
&\RepQuad{4} - \frac{\pi}{2}\!\LinearTerm[2][\SlopeFrstOrdSMPSymb]\!\!\left(\!\vphbigg\cos(2\delpsi)\!\FresnelCTerm[2][\SlopeFrstOrdSMPSymb]\!+ \sin(2\delpsi)\!\FresnelSTerm[2][\SlopeFrstOrdSMPSymb]\!\!\right) \nonumber\\
&\RepQuad{8} -\,\pi\!\LinearTerm[][\SlopeFrstOrdSMPSymb]\!\!\left(\!\vphbig\sin(\delpsi)\!\FresnelCTerm[][\SlopeFrstOrdSMPSymb]\!- \cos(\delpsi)\!\FresnelSTerm[][\SlopeFrstOrdSMPSymb]\!\!\right)\!\sin(\delpsi) \nonumber\\
&\RepQuad{12} \left.\vphbigggg + \frac{1}{2}\,\SinTerm[2][\SlopeFrstOrdSMPSymb][-2 \delpsi]\!+ \CosTerm[][\SlopeFrstOrdSMPSymb][-\delpsi] \sin(\delpsi)\right] \nonumber\\[5pt]
& + \lambda^{3} \!\left[\frac{\pi}{2}\CosTerm[][][- \delpsi]\!\!\left(\vphbig\!\cos(\delpsi)\!\FresnelCTerm\!+ \sin(\delpsi)\!\FresnelSTerm\!\!\right)\np{2} \vphbigggg\right. \nonumber\\
&\RepQuad{1} - \frac{\pi}{\sqrt{2}}\!\left(\vphbigg\!\cos(2\delpsi)\!\FresnelCTerm[2]\!+ \sin(2\delpsi)\!\FresnelSTerm[2]\!\!\right)\!\times \nonumber\\[-2.5pt]
&\RepQuad{27} \times\!\left(\vphbig\!\cos(\delpsi)\!\FresnelCTerm\!+ \sin(\delpsi)\!\FresnelSTerm\!\!\right) \nonumber\\
&\RepQuad{1} - \pi\!\left(\vphbig\!\sin(\delpsi)\!\FresnelCTerm\!- \cos(\delpsi)\!\FresnelSTerm\!\!\right)\!\!\left(\vphbig\!\cos(\delpsi)\!\FresnelCTerm\!+ \sin(\delpsi)\!\FresnelSTerm\!\!\right)\!\sin(\delpsi) \nonumber\\
&\RepQuad{1} + \frac{3\pi}{8}\!\LinearTerm[3]\!\!\left(\vphbigg\!\cos(3\delpsi)\!\FresnelCTerm[3]\!+ \sin(3\delpsi)\!\FresnelSTerm[3]\!\!\right) \nonumber\\
&\RepQuad{5} + \pi\!\LinearTerm[2]\!\!\left(\vphbigg\!\sin(2\delpsi)\!\FresnelCTerm[2]\!- \cos(2\delpsi)\!\FresnelSTerm[2]\!\!\right)\!\sin(\delpsi) \nonumber\\
&\RepQuad{9} + \frac{\pi}{8}\!\LinearTerm\!\!\left(\vphbig\!\cos(\delpsi)\FresnelCTerm\!+ \sin(\delpsi)\!\FresnelSTerm\!\!\right)\!\big(2 \cos(2\delpsi)\!- 3\big) \nonumber\\
&\RepQuad{1} \left.\vphbigggg - \frac{3}{8}\,\SinTerm[3][][-3\delpsi]\!- \CosTerm[2][][-2\delpsi] \sin(\delpsi)- \frac{1}{8}\,\SinTerm[][][-\delpsi]\!\big(2 \cos(2\delpsi)\!- 3\big)\right]
\end{ph}
Details about these solutions are discussed in the main text of the article. We can now address the final problem, namely that of a phase equation with many oscillating modes, each with an independent initial phase.

\subsection{Multi-mode phase equation with non-zero phase}
\label{sec:app_phase_solution_multi_mode_nonzero_phase}
\noindent
The solutions to the phase resonance equation with an infinite number of modes and non-zero phases
\begin{align}
\bm{\varphi}^{\prime\prime} (x) = 1 + \sum_{n > 1} \lambda_{n} \cos\!\big(n\,\bm{\varphi}(x) - \delta \psi_{n}\big). \tag{\ref{eq:phaseEquation}}
\end{align}
can be found using the established algorithm. Note this is the most complete solution to the phase resonance equation. The first-order correction to the zeroth-order solutions is given by the expression below.
\begin{ph}{align}
\begin{split}
&\bm{\varphi}_{(1)} (x) = \pm\,\pi \!\left[\sum_{n}\frac{1}{2} \big(\!\cos(\delpsi_{n}) + \sin(\delpsi_{n})\big)\frac{\lambda_{n}}{\sqrt{n}}\right]\!\!\LinearTerm \\[5pt]
&\RepQuad{5} + \sum_{n} \frac{\lambda_{n}}{n}\!\left[\pi\!\LinearTerm[n]\!\!\left(\vphbigg\!\cos\!\left(\delpsi_{n}\right)\!\FresnelCTerm[n]\!+ \sin\!\left(\delpsi_{n}\right)\!\FresnelSTerm[n]\!\!\right) \vphbigggg\right.\\
&\RepQuad{27} \left.\vphbigggg -\!\left(\vphbigg\!\SinTerm[n][][-\delpsi_{n}]\!+ \sin(\delpsi_{n})\!\right)\!\right]
\end{split} \label{eq:appMMPhFrstOrdSol}
\end{ph}

Similarly, we can find the second-order correction, given here.
\begin{ph}{align}
&\bm{\varphi}_{(2)} (x) = \pm\,\pi\!\left[\sum_{n}\frac{1}{2} \big(\!\cos(\delpsi_{n}) + \sin(\delpsi_{n})\big)\frac{\lambda_{n}}{\sqrt{n}} + \sum_{n,\,m} \left(\frac{1}{8}\,\frac{|n - m|^{\sfrac{3}{2}}}{\sqrt{n} \sqrt{m}}\,\big(\cos(\delpsi_{n} - \delpsi_{m}) + \Sgn(n-m)\sin(\delpsi_{n} - \delpsi_{m})\big) \vph \right. \vphbig \right. \nonumber\\
& \RepQuad{26} -\,\frac{1}{8}\,\frac{(n + m)^{\sfrac{3}{2}}}{\sqrt{n} \sqrt{m}}\,\big(\!\cos(\delpsi_{n} + \delpsi_{m}) + \sin(\delpsi_{n} + \delpsi_{m})\big) \nonumber\\
& \RepQuad{6} \left. \vphbig \left. \vph + \frac{1}{4}\,\frac{n}{\sqrt{m}}\,\big(\!\cos(\delpsi_{n}) - \sin(\delpsi_{n})\big) \sin(\delpsi_{m}) + \frac{1}{4}\,\frac{m}{\sqrt{n}}\,\sin(\delpsi_{n}) \big(\!\cos(\delpsi_{m}) - \sin(\delpsi_{m})\big)\!\right)\!\frac{\lambda_{n}}{\sqrt{n}} \frac{\lambda_{m}}{\sqrt{m}}\right]\!\!\left(\!\frac{x}{\spi}\!\right) \nonumber\\
&\RepQuad{23} + \frac{\pi}{8} \sum_{n,\,m} \big(\!\cos(\delpsi_{n}) + \sin(\delpsi_{n})\big) \big(\!\cos(\delpsi_{m}) + \sin(\delpsi_{m})\big) \, \frac{\lambda_{n}}{\sqrt{n}} \frac{\lambda_{m}}{\sqrt{m}} \nonumber\\[5pt]
& + \sum_{n} \frac{\lambda_{n}}{n}\!\left[\pi\LinearTerm[n][\SlopeFrstOrdMMP]\!\!\left(\vphbigg\!\cos\!\left(\delpsi_{n}\right)\!\FresnelCTerm[n][\SlopeFrstOrdMMP]\!+ \sin\!\left(\delpsi_{n}\right)\!\FresnelSTerm[n][\SlopeFrstOrdMMP]\!\!\right) \vphbigggg\right. \nonumber\\
& \RepQuad{29} \left.\vphbigggg -\!\left(\vphbigg\!\SinTerm[n][\SlopeFrstOrdMMP][- \delpsi_{n}]\!+ \sin(\delpsi_{n})\!\right)\!\right] \nonumber\\
& + \sum_{n,\,m} \frac{\lambda_{n}}{\sqrt{n}}\frac{\lambda_{m}}{\sqrt{m}}\!\left[\frac{\pi}{2}\!\left(\vphbigg\!\cos\!\left(\delpsi_{n}\right)\!\FresnelCTerm[n]\!+ \sin\!\left(\delpsi_{n}\right)\!\FresnelSTerm[n]\!\!\right)\!\times \vphbigggg\right. \nonumber\\[-2.5pt]
& \RepQuad{20} \times\!\left(\vphbigg\!\cos\!\left(\delpsi_{m}\right)\!\FresnelCTerm[m]\!+ \sin\!\left(\delpsi_{m}\right)\!\FresnelSTerm[m]\!\!\right) \nonumber\\
& \RepQuad{2} + \frac{\pi}{2}\frac{\sqrt{n}}{\sqrt{m}}\!\LinearTerm[|n - m|]\!\times \nonumber\\[-2.5pt]
& \RepQuad{4} \times\!\left(\vphbigg\!\Sgn(n - m)\cos\!\left(\delpsi_{n} - \delpsi_{m}\right)\!\FresnelCTerm[|n - m|]\!+ \sin\!\left(\delpsi_{n} - \delpsi_{m}\right)\!\FresnelSTerm[|n - m|]\!\!\right) \nonumber\\
& \RepQuad{2} - \frac{\pi}{2}\frac{\sqrt{n}}{\sqrt{m}}\!\LinearTerm[n + m]\!\times \nonumber\\[-2.5pt]
& \RepQuad{10} \times\!\left(\vphbigg\!\cos\!\left(\delpsi_{n} + \delpsi_{m}\right)\!\FresnelCTerm[n + m]\!+ \sin\!\left(\delpsi_{n} + \delpsi_{m}\right)\!\FresnelSTerm[n + m]\!\!\right) \nonumber\\
& \RepQuad{2} + \pi\,\frac{\sqrt{n}}{\sqrt{m}}\!\LinearTerm[n]\!\!\left(\vphbigg\!\cos\!\left(\delpsi_{n}\right)\!\FresnelSTerm[n]\!- \sin\!\left(\delpsi_{n}\right)\!\FresnelCTerm[n]\!\!\right)\!\sin\!\left(\delpsi_{m}\right) \nonumber\\
& \RepQuad{6} - \frac{1}{2}\frac{\sqrt{n}}{\sqrt{m}}\,\SinTerm[|n - m|][][- \left(\delpsi_{n} - \delpsi_{m}\right)][\Sgn(n - m)] \nonumber\\
& \RepQuad{12} + \frac{1}{2}\frac{\sqrt{n}}{\sqrt{m}}\,\SinTerm[n + m][][- \left(\delpsi_{n} + \delpsi_{m}\right)] \nonumber\\
& \RepQuad{18} \left.\vphbigggg + \frac{\sqrt{n}}{\sqrt{m}}\,\CosTerm[n][][- \delpsi_{n}] \sin\!\left(\delpsi_{m}\right) \right] \label{eq:appMMPhScndOrdSol}
\end{ph}

Finally, we get to the point where we can derive the most complete solution to the problem we are considering in this article. The third-order solution in this case is given by eq.~(\ref{eq:phaseSolFinal}).
\begin{ph}{align}
\bm{\varphi}_{3} (x) = \frac{\pi}{2}\!\LinearTerm[][\SlopeThrdOrdMMP]\np{2}\!+ \sum_{n} \frac{\lambda_{n}}{n} \, \bm{\varrho}_{1}^{n}\!\LinearTerm[][\SlopeScndOrdMMP]\!+ \sum_{n,\,m} \frac{\lambda_{n}}{\sqrt{n}} \frac{\lambda_{m}}{\sqrt{m}} \, \bm{\varrho}_{2}^{nm}\!\LinearTerm[][\SlopeFrstOrdMMP]\!+\!\!\sum_{n,\,m,\,p}\!\!\lambda_{n} \frac{\lambda_{m}}{\sqrt{m}}  \frac{\lambda_{p}}{\sqrt{p}} \, \bm{\varrho}_{3}^{nmp}\!\LinearTerm \tag{\ref{eq:phaseSolFinal}}
\end{ph}
In this solution, \(\bm{\varsigma}_{3}\) is the third-order slope function, dependent on the resonance flux parameters \(\{\lambda_{n}\}\) and the mode numbers \(\{n\}\) (this solutions does not require the mode numbers to be consecutive). The lower-order functions \(\bm{\varsigma}_{1}\) and \(\bm{\varsigma}_{2}\) are given as part of the corrections \(\bm{\varphi}_{(1)}\) and \(\bm{\varphi}_{(2)}\), respectively.
\begin{align}
\bm{\varsigma}_{3} &= \sum_{n}\frac{1}{2} \big(\!\cos(\delpsi_{n}) + \sin(\delpsi_{n})\big)\frac{\lambda_{n}}{\sqrt{n}} \nonumber\\
& + \sum_{n,\,m} \left(\frac{1}{8}\,\frac{|n - m|^{\sfrac{3}{2}}}{\sqrt{n} \sqrt{m}}\,\big(\cos(\delpsi_{n} - \delpsi_{m}) + \Sgn(n-m)\sin(\delpsi_{n} - \delpsi_{m})\big) \vph\right. \nonumber\\
& \RepQuad{6} -\,\frac{1}{8}\,\frac{(n + m)^{\sfrac{3}{2}}}{\sqrt{n} \sqrt{m}}\,\big(\!\cos(\delpsi_{n} + \delpsi_{m}) + \sin(\delpsi_{n} + \delpsi_{m})\big) \nonumber\\
& \RepQuad{9} \left.\vph + \frac{1}{4}\,\frac{n}{\sqrt{m}}\,\big(\!\cos(\delpsi_{n}) - \sin(\delpsi_{n})\big) \sin(\delpsi_{m}) + \frac{1}{4}\,\frac{m}{\sqrt{n}}\,\sin(\delpsi_{n}) \big(\!\cos(\delpsi_{m}) - \sin(\delpsi_{m})\big)\!\right)\!\frac{\lambda_{n}}{\sqrt{n}} \frac{\lambda_{m}}{\sqrt{m}} \nonumber\\
& + \sum_{n,\,m,\,p} \left(\frac{1}{16}\,\frac{(n + m + p)^{\sfrac{3}{2}}}{\sqrt{m}\sqrt{p}}\,\big(\!\cos(\delpsi_{n} + \delpsi_{m} + \delpsi_{p}) + \sin(\delpsi_{n} + \delpsi_{m} + \delpsi_{p})\big) \vph \right. \nonumber\\
& \RepQuad{7} - \frac{1}{16}\,\frac{|n - m + p|^{\sfrac{3}{2}}}{\sqrt{m}\sqrt{p}}\,\big(\!\cos(\delpsi_{n} - \delpsi_{m} + \delpsi_{p}) + \Sgn(n - m + p) \sin(\delpsi_{n} - \delpsi_{m} + \delpsi_{p})\big) \nonumber\\
& \RepQuad{10} - \frac{1}{16}\,\frac{|n + m - p|^{\sfrac{3}{2}}}{\sqrt{m}\sqrt{p}}\,\big(\!\cos(\delpsi_{n} + \delpsi_{m} - \delpsi_{p}) + \Sgn(n + m - p) \sin(\delpsi_{n} + \delpsi_{m} - \delpsi_{p})\big) \nonumber\\
& \RepQuad{7} + \frac{1}{16}\,\frac{|n - m - p|^{\sfrac{3}{2}}}{\sqrt{m}\sqrt{p}}\,\big(\!\cos(\delpsi_{n} - \delpsi_{m} - \delpsi_{p}) + \Sgn(n - m - p) \sin(\delpsi_{n} - \delpsi_{m} - \delpsi_{p})\big) \nonumber\\
& \RepQuad{5} + \frac{1}{8}\,\frac{|n - m|^{\sfrac{3}{2}}}{\sqrt{m}\sqrt{p}}\,\big(\!\Sgn(n - m) \cos(\delpsi_{n} - \delpsi_{m}) - \sin(\delpsi_{n} - \delpsi_{m})\big) \sin(\delpsi_{p}) \nonumber\\
& \RepQuad{8} - \frac{1}{8}\,\frac{(n + m)^{\sfrac{3}{2}}}{\sqrt{m}\sqrt{p}}\,\big(\!\cos(\delpsi_{n} + \delpsi_{m}) - \sin(\delpsi_{n} + \delpsi_{m})\big) \sin(\delpsi_{p}) \nonumber\\
& \RepQuad{5} + \frac{1}{8}\,\frac{|n - p|^{\sfrac{3}{2}}}{\sqrt{m}\sqrt{p}}\,\big(\!\Sgn(n - p) \cos(\delpsi_{n} - \delpsi_{p}) - \sin(\delpsi_{n} - \delpsi_{p})\big) \sin(\delpsi_{m}) \nonumber\\
& \RepQuad{8} - \frac{1}{8}\,\frac{(n + p)^{\sfrac{3}{2}}}{\sqrt{m}\sqrt{p}}\,\big(\!\cos(\delpsi_{n} + \delpsi_{p}) - \sin(\delpsi_{n} + \delpsi_{p})\big) \sin(\delpsi_{m}) \nonumber\\
& \RepQuad{5} - \frac{1}{8}\,\frac{(n - m)}{\sqrt{m}}\,\cos(\delpsi_{n} - \delpsi_{m}) \big(\!\cos(\delpsi_{p}) + \sin(\delpsi_{p})\big) \nonumber\\
& \RepQuad{8}  + \frac{1}{8}\,\frac{(n + m)}{\sqrt{m}}\,\cos(\delpsi_{n} + \delpsi_{m}) \big(\!\cos(\delpsi_{p}) + \sin(\delpsi_{p})\big) \nonumber\\
& \RepQuad{5} - \frac{1}{8}\,\frac{(n - p)}{\sqrt{p}}\,\cos(\delpsi_{n} - \delpsi_{p}) \big(\!\cos(\delpsi_{m}) + \sin(\delpsi_{m})\big) \nonumber\\
& \RepQuad{8}  + \frac{1}{8}\,\frac{(n + p)}{\sqrt{p}}\,\cos(\delpsi_{n} + \delpsi_{p}) \big(\!\cos(\delpsi_{m}) + \sin(\delpsi_{m})\big) \nonumber\\
& \RepQuad{5} - \frac{1}{4}\,\frac{n^{\sfrac{3}{2}}}{\sqrt{m}\sqrt{p}}\,\big(\!\cos(\delpsi_{n}) + \sin(\delpsi_{n})\big) \sin(\delpsi_{m}) \sin(\delpsi_{p}) \nonumber\\
& \RepQuad{8} + \frac{1}{4}\,\frac{n}{\sqrt{m}}\,\sin(\delpsi_{n}) \sin(\delpsi_{m}) \big(\!\cos(\delpsi_{p}) + \sin(\delpsi_{p})\big) \nonumber\\
& \RepQuad{11} + \frac{1}{4}\,\frac{n}{\sqrt{p}}\,\sin(\delpsi_{n}) \big(\!\cos(\delpsi_{m}) + \sin(\delpsi_{m})\big) \sin(\delpsi_{p}) \nonumber\\
& \RepQuad{14} - \frac{1}{4}\,\sqrt{m}\,\cos(\delpsi_{n}) \cos(\delpsi_{m}) \big(\!\cos(\delpsi_{p}) + \sin(\delpsi_{p})\big) \nonumber\\
& \RepQuad{17} \left. \vph - \frac{1}{4}\,\sqrt{p}\,\cos(\delpsi_{n}) \big(\!\cos(\delpsi_{m}) + \sin(\delpsi_{m})\big) \cos(\delpsi_{p})\!\right)\!\lambda_{n} \frac{\lambda_{m}}{\sqrt{m}} \frac{\lambda_{p}}{\sqrt{p}} \label{eq:appMMPThrdOrdSol}
\end{align}
The three oscillating functions \(\{\bm{\varrho}_{i}\}\) are given below.
\begin{align}
&\bm{\varrho}_{1}^{n} (x) = \pi\!\LinearTerm[n][\SlopeScndOrdMMP]\!\!\left(\vphbigg\!\cos(\delpsi_{n})\!\FresnelCTerm[n][\SlopeScndOrdMMP]\!+ \sin(\delpsi_{n})\!\FresnelSTerm[n][\SlopeScndOrdMMP]\!\!\right) \label{eq:correctionFirstOrder} \\
&\RepQuad{28} - \left(\vphbigg\!\SinTerm[n][\SlopeScndOrdMMP][- \delpsi_{n}]\!+ \sin(\delpsi_{n})\!\right), \nonumber
\end{align}
\begin{align}
&\bm{\varrho}_{2}^{nm} (x) = \frac{\pi}{2}\!\left(\vphbigg\!\cos\!\left(\delpsi_{n}\right)\!\FresnelCTerm[n][\SlopeFrstOrdMMP]\!+ \sin\!\left(\delpsi_{n}\right)\!\FresnelSTerm[n][\SlopeFrstOrdMMP]\!\!\right)\!\times  \nonumber\\[-2.5pt]
& \RepQuad{16} \times\!\left(\vphbigg\!\cos\!\left(\delpsi_{m}\right)\!\FresnelCTerm[m][\SlopeFrstOrdMMP]\!+ \sin\!\left(\delpsi_{m}\right)\!\FresnelSTerm[m][\SlopeFrstOrdMMP]\!\!\right) \nonumber\\
& + \frac{\pi}{2}\frac{\sqrt{n}}{\sqrt{m}}\!\LinearTerm[|n - m|][\SlopeFrstOrdMMP]\!\times \nonumber\\[-2.5pt]
& \times\!\left(\vphbigg\!\Sgn(n - m)\cos\!\left(\delpsi_{n} - \delpsi_{m}\right)\!\FresnelCTerm[|n - m|][\SlopeFrstOrdMMP]\!+ \sin\!\left(\delpsi_{n} - \delpsi_{m}\right)\!\FresnelSTerm[|n - m|][\SlopeFrstOrdMMP]\!\!\right) \nonumber\\
& - \frac{\pi}{2}\frac{\sqrt{n}}{\sqrt{m}}\!\LinearTerm[n + m][\SlopeFrstOrdMMP]\!\times \nonumber\\[-2.5pt]
& \RepQuad{6} \times\!\left(\vphbigg\!\cos\!\left(\delpsi_{n} + \delpsi_{m}\right)\!\FresnelCTerm[n + m][\SlopeFrstOrdMMP]\!+ \sin\!\left(\delpsi_{n} + \delpsi_{m}\right)\!\FresnelSTerm[n + m][\SlopeFrstOrdMMP]\!\!\right) \nonumber\\
& \RepQuad{2} + \pi\,\frac{\sqrt{n}}{\sqrt{m}}\!\LinearTerm[n][\SlopeFrstOrdMMP]\!\!\left(\vphbigg\!\cos\!\left(\delpsi_{n}\right)\!\FresnelSTerm[n][\SlopeFrstOrdMMP]\!- \sin\!\left(\delpsi_{n}\right)\!\FresnelCTerm[n][\SlopeFrstOrdMMP]\!\!\right)\!\sin\!\left(\delpsi_{m}\right) \nonumber\\
& \RepQuad{6} - \frac{1}{2}\frac{\sqrt{n}}{\sqrt{m}}\,\SinTerm[|n - m|][\SlopeFrstOrdMMP][- \left(\delpsi_{n} - \delpsi_{m}\right)][\Sgn(n - m)] \nonumber\\
& \RepQuad{10} + \frac{1}{2}\frac{\sqrt{n}}{\sqrt{m}}\,\SinTerm[n + m][\SlopeFrstOrdMMP][- \left(\delpsi_{n} + \delpsi_{m}\right)] \nonumber\\
& \RepQuad{14} + \frac{\sqrt{n}}{\sqrt{m}}\,\CosTerm[n][\SlopeFrstOrdMMP][- \delpsi_{n}] \sin\!\left(\delpsi_{m}\right), \label{eq:correctionSecondOrder}
\end{align}
and
\begin{align}
&\bm{\varrho}_{3}^{nmp} (x) = \frac{\pi}{2}\,\CosTerm[n][][- \delpsi_{n}]\!\!\left(\vphbigg\!\cos\!\left(\delpsi_{m}\right)\!\FresnelCTerm[m]\!+ \sin\!\left(\delpsi_{m}\right)\!\FresnelSTerm[m]\!\!\right)\!\times  \nonumber\\[-2.5pt]
& \RepQuad{20} \times\!\left(\vphbigg\!\cos\!\left(\delpsi_{p}\right)\!\FresnelCTerm[p]\!+ \sin\!\left(\delpsi_{p}\right)\!\FresnelSTerm[p]\!\!\right) \nonumber\\
& \RepQuad{2} + \frac{\pi}{4}\!\left(\vphbigggg\!\frac{\sqrt{|n - m|}}{\sqrt{m}}\!\left(\vphbigg\!\Sgn(n - m)\cos\!\left(\delpsi_{n} - \delpsi_{m}\right)\!\FresnelCTerm[|n - m|] \right. \right. \nonumber\\[-2.5pt]
& \RepQuad{28} \left.\vphbigg + \sin\!\left(\delpsi_{n} - \delpsi_{m}\right)\!\FresnelSTerm[|n - m|]\!\!\right) \nonumber\\[-2.5pt]
& \RepQuad{3} \left. \vphbigggg - \frac{\sqrtbig{n + m}}{\sqrt{m}}\!\left(\vphbigg\!\cos\!\left(\delpsi_{n} + \delpsi_{m}\right)\!\FresnelCTerm[n + m]\!+ \sin\!\left(\delpsi_{n} + \delpsi_{m}\right)\!\FresnelSTerm[n + m]\!\!\right)\!\!\right)\!\times \nonumber\\[-2.5pt]
& \RepQuad{20} \times\!\left(\vphbigg\!\cos\!\left(\delpsi_{p}\right)\!\FresnelCTerm[p]\!+ \sin\!\left(\delpsi_{p}\right)\!\FresnelSTerm[p]\!\!\right) \nonumber\\
& \RepQuad{2} + \frac{\pi}{4}\!\left(\vphbigggg\!\frac{\sqrt{|n - p|}}{\sqrt{p}}\!\left(\vphbigg\!\Sgn(n - p)\cos\!\left(\delpsi_{n} - \delpsi_{p}\right)\!\FresnelCTerm[|n - p|] \right. \right. \nonumber\\[-2.5pt]
& \RepQuad{29} \left.\vphbigg + \sin\!\left(\delpsi_{n} - \delpsi_{p}\right)\!\FresnelSTerm[|n - p|]\!\!\right) \nonumber\\[-2.5pt]
& \RepQuad{5} \left. \vphbigggg - \frac{\sqrtbig{n + p}}{\sqrt{p}}\!\left(\vphbigg\!\cos\!\left(\delpsi_{n} + \delpsi_{p}\right)\!\FresnelCTerm[n + p]\!+ \sin\!\left(\delpsi_{n} + \delpsi_{p}\right)\!\FresnelSTerm[n + p]\!\!\right)\!\!\right)\!\times \nonumber\\[-2.5pt]
& \RepQuad{19} \times\!\left(\vphbigg\!\cos\!\left(\delpsi_{m}\right)\!\FresnelCTerm[m]\!+ \sin\!\left(\delpsi_{m}\right)\!\FresnelSTerm[m]\!\!\right) \nonumber\\
& \RepQuad{2} + \frac{\pi}{2}\,\frac{\sqrt{n}}{\sqrt{m}}\!\left(\vphbigg\!\cos\!\left(\delpsi_{n}\right)\!\FresnelSTerm[n]\!- \sin\!\left(\delpsi_{n}\right)\!\FresnelCTerm[n]\!\!\right) \sin\!\left(\delpsi_{m}\right)\times \nonumber\\[-2.5pt]
& \RepQuad{20} \times\!\left(\vphbigg\!\cos\!\left(\delpsi_{p}\right)\!\FresnelCTerm[p]\!+ \sin\!\left(\delpsi_{p}\right)\!\FresnelSTerm[p]\!\!\right) \nonumber\\
& \RepQuad{2} + \frac{\pi}{2}\,\frac{\sqrt{n}}{\sqrt{p}}\!\left(\vphbigg\!\cos\!\left(\delpsi_{n}\right)\!\FresnelSTerm[n]\!- \sin\!\left(\delpsi_{n}\right)\!\FresnelCTerm[n]\!\!\right)\!\times \nonumber\\[-2.5pt]
& \RepQuad{15} \times\!\left(\vphbigg\!\cos\!\left(\delpsi_{m}\right)\!\FresnelCTerm[m]\!+ \sin\!\left(\delpsi_{m}\right)\!\FresnelSTerm[m]\!\!\right) \sin\!\left(\delpsi_{p}\right) \nonumber\\
& \RepQuad{2} + \frac{\pi}{8}\,\frac{(n + m + p)}{\sqrt{m} \sqrt{p}}\!\LinearTerm[n + m + p]\!\times \nonumber\\
& \RepQuad{6} \times\!\left(\vphbigg\!\cos\!\left(\delpsi_{n} + \delpsi_{m} + \delpsi_{p}\right)\!\FresnelCTerm[n + m + p] \right. \nonumber\\[-2.5pt]
& \RepQuad{24} \left.\vphbigg + \sin\!\left(\delpsi_{n} + \delpsi_{m} + \delpsi_{p}\right)\!\FresnelSTerm[n + m + p]\!\!\right) \nonumber\\
& \RepQuad{2} - \frac{\pi}{8}\,\frac{|n - m + p|}{\sqrt{m} \sqrt{p}}\!\LinearTerm[|n - m + p|]\!\times \nonumber\\
& \RepQuad{6} \times\!\left(\vphbigg\!\cos\!\left(\delpsi_{n} - \delpsi_{m} + \delpsi_{p}\right)\!\FresnelCTerm[|n - m + p|] \right. \nonumber\\[-2.5pt]
& \RepQuad{17} \left.\vphbigg + \Sgn(n - m + p)\,\sin\!\left(\delpsi_{n} - \delpsi_{m} + \delpsi_{p}\right)\!\FresnelSTerm[|n - m + p|]\!\!\right) \nonumber\\
& \RepQuad{2} - \frac{\pi}{8}\,\frac{|n + m - p|}{\sqrt{m} \sqrt{p}}\!\LinearTerm[|n + m - p|]\!\times \nonumber\\
& \RepQuad{6} \times\!\left(\vphbigg\!\cos\!\left(\delpsi_{n} + \delpsi_{m} - \delpsi_{p}\right)\!\FresnelCTerm[|n + m - p|] \right. \nonumber\\[-2.5pt]
& \RepQuad{17} \left.\vphbigg + \Sgn(n + m - p)\,\sin\!\left(\delpsi_{n} + \delpsi_{m} - \delpsi_{p}\right)\!\FresnelSTerm[|n + m - p|]\!\!\right) \nonumber\\
& \RepQuad{2} + \frac{\pi}{8}\,\frac{|n - m - p|}{\sqrt{m} \sqrt{p}}\!\LinearTerm[|n - m - p|]\!\times \nonumber\\
& \RepQuad{6} \times\!\left(\vphbigg\!\cos\!\left(\delpsi_{n} - \delpsi_{m} - \delpsi_{p}\right)\!\FresnelCTerm[|n - m - p|] \right. \nonumber\\[-2.5pt]
& \RepQuad{17} \left.\vphbigg + \Sgn(n - m - p)\,\sin\!\left(\delpsi_{n} - \delpsi_{m} - \delpsi_{p}\right)\!\FresnelSTerm[|n - m - p|]\!\!\right) \nonumber\\
& \RepQuad{2} - \frac{\pi}{4} \,\frac{|n - m|}{\sqrt{m} \sqrt{p}}\!\LinearTerm[|n - m|] \sin\!\left(\delpsi_{p}\right)\times \nonumber\\[-2.5pt]
& \RepQuad{3} \times\!\left(\vphbigg\!\sin\!\left(\delpsi_{n} - \delpsi_{m}\right)\!\FresnelCTerm[|n - m|]\!- \Sgn(n - m)\,\cos\!\left(\delpsi_{n} - \delpsi_{m}\right)\!\FresnelSTerm[|n - m|]\!\!\right) \nonumber\\
& \RepQuad{2} + \frac{\pi}{4} \,\frac{(n + m)}{\sqrt{m} \sqrt{p}}\!\LinearTerm[n + m] \sin\!\left(\delpsi_{p}\right)\times \nonumber\\[-2.5pt]
& \RepQuad{9} \times\!\left(\vphbigg\!\sin\!\left(\delpsi_{n} + \delpsi_{m}\right)\!\FresnelCTerm[n + m]\!- \cos\!\left(\delpsi_{n} + \delpsi_{m}\right)\!\FresnelSTerm[n + m]\!\!\right) \nonumber\\
& \RepQuad{2} - \frac{\pi}{4} \,\frac{|n - p|}{\sqrt{m} \sqrt{p}}\!\LinearTerm[|n - p|] \sin\!\left(\delpsi_{m}\right)\times \nonumber\\[-2.5pt]
& \RepQuad{4} \times\!\left(\vphbigg\!\sin\!\left(\delpsi_{n} - \delpsi_{p}\right)\!\FresnelCTerm[|n - p|]\!- \Sgn(n - p)\,\cos\!\left(\delpsi_{n} - \delpsi_{p}\right)\!\FresnelSTerm[|n - p|]\!\!\right) \nonumber\\
& \RepQuad{2} + \frac{\pi}{4} \,\frac{(n + p)}{\sqrt{m} \sqrt{p}}\!\LinearTerm[n + p] \sin\!\left(\delpsi_{m}\right)\times \nonumber\\[-2.5pt]
& \RepQuad{10} \times\!\left(\vphbigg\!\sin\!\left(\delpsi_{n} + \delpsi_{p}\right)\!\FresnelCTerm[n + p]\!- \cos\!\left(\delpsi_{n} + \delpsi_{p}\right)\!\FresnelSTerm[n + p]\!\!\right) \nonumber\\
& \RepQuad{2} - \frac{\pi}{2} \,\frac{n}{\sqrt{m} \sqrt{p}}\!\LinearTerm[n]\!\!\left(\vphbigg\!\cos\!\left(\delpsi_{n}\right)\!\FresnelCTerm[n]\!+ \sin\!\left(\delpsi_{n}\right)\!\FresnelSTerm[n]\!\!\right)\!\sin\!\left(\delpsi_{m}\right)\sin\!\left(\delpsi_{p}\right) \nonumber\\
& \RepQuad{2} - \frac{1}{8}\,\frac{(n + m + p)}{\sqrt{m} \sqrt{p}}\,\SinTerm[n + m + p][][- \left(\delpsi_{n} + \delpsi_{m} + \delpsi_{p}\right)] \nonumber\\
& \RepQuad{5} + \frac{1}{8}\,\frac{(n - m + p)}{\sqrt{m} \sqrt{p}}\,\SinTerm[n - m + p][][- \left(\delpsi_{n} - \delpsi_{m} + \delpsi_{p}\right)][\Sgn(n - m + p)] \nonumber\\
& \RepQuad{8} + \frac{1}{8}\,\frac{(n + m - p)}{\sqrt{m} \sqrt{p}}\,\SinTerm[n + m - p][][- \left(\delpsi_{n} + \delpsi_{m} - \delpsi_{p}\right)][\Sgn(n + m - p)] \nonumber\\
& \RepQuad{11} - \frac{1}{8}\,\frac{(n - m - p)}{\sqrt{m} \sqrt{p}}\,\SinTerm[n - m - p][][- \left(\delpsi_{n} - \delpsi_{m} - \delpsi_{p}\right)][\Sgn(n - m - p)] \nonumber\\
& \RepQuad{2} + \frac{1}{4}\,\frac{(n - m)}{\sqrt{m} \sqrt{p}}\,\CosTerm[|n - m|][][- \left(\delpsi_{n} - \delpsi_{m}\right)][\Sgn(n - m)]\sin\!\left(\delpsi_{p}\right) \nonumber\\
& \RepQuad{5} - \frac{1}{4}\,\frac{(n + m)}{\sqrt{m} \sqrt{p}}\,\CosTerm[n + m][][- \left(\delpsi_{n} + \delpsi_{m}\right)]\sin\!\left(\delpsi_{p}\right) \nonumber\\
& \RepQuad{8} + \frac{1}{4}\,\frac{(n - p)}{\sqrt{m} \sqrt{p}}\,\CosTerm[|n - p|][][- \left(\delpsi_{n} - \delpsi_{p}\right)][\Sgn(n - p)]\sin\!\left(\delpsi_{m}\right) \nonumber\\
& \RepQuad{11} - \frac{1}{4}\,\frac{(n + p)}{\sqrt{m} \sqrt{p}}\,\CosTerm[n + p][][- \left(\delpsi_{n} + \delpsi_{p}\right)]\sin\!\left(\delpsi_{m}\right) \nonumber\\
& \RepQuad{2} + \frac{1}{2}\,\frac{n}{\sqrt{m} \sqrt{p}}\,\SinTerm[n][][- \left(\delpsi_{n}\right)]\sin\!\left(\delpsi_{m}\right) \sin\!\left(\delpsi_{p}\right) \label{eq:correctionThirdOrder}
\end{align}
\end{appendices}
\end{document}